\documentclass[12pt]{article}
\pdfoutput=1
\usepackage{graphicx} % Required for inserting images
\usepackage{xspace}
\usepackage{url}
\usepackage{comment}
\usepackage{amsmath}
\usepackage{amssymb}
\usepackage{fancyhdr}  
\usepackage{hepunits}
\usepackage[usenames,dvipsnames]{color}  
\usepackage{enumitem}
\usepackage{multirow}
\usepackage{lineno}
\usepackage{pdfpages}

\usepackage[a4paper, total={17cm, 22cm}, headheight=0.6cm]{geometry}
\usepackage{subcaption}  % needed for side-by-side figs
\usepackage[pdftex,bookmarks,hidelinks]{hyperref}

\usepackage[normalem]{ulem}
\def\phaseone{Phase~I\xspace} %Anne remv'd hyphen
\def\phasetwo{Phase~II\xspace} %Anne remv'd hyphen

\def\expshort{DUNE\xspace}
\def\dune{\expshort}
\def\explong{The Deep Underground Neutrino Experiment\xspace}

\def\larmass{\SI{17.5}{\kt}\xspace} % full mass in cryostat
\newcommand{\nominalmodsize}{\SI{10}{kt}\xspace} % nominal FD module size 
\newcommand{\fdfiducialmass}{\SI{40}{kt}\xspace} % total Fid mass

\def\coldbox{cold box\xspace}

\newcommand{\rms}{RMS\xspace} % Might want this small caps?
\newcommand{\threed}{3D\xspace}
\newcommand{\twod}{2D\xspace}
\newcommand{\phel}{photoelectron\xspace} 
\newcommand{\frfour}{FR-4\xspace} 
\newcommand{\efield}{\ensuremath{E} field\xspace}

\newcommand{\numu}{\ensuremath{\nu_\mu}\xspace}
\newcommand{\nue}{\ensuremath{\nu_e}\xspace}

\newcommand{\deltacp}{\ensuremath{\delta_{\rm CP}}\xspace}   % example \deltacp
\newcommand{\numubartonumubar}{
\ensuremath{\overline{\numu}\rightarrow\overline{\numu}}\xspace}

\def\argon40{${}^{40}$Ar}       
\def\Ar39{$^{39}$Ar}
\def\Cl40{$^{40}$Cl}
\def\K40{$^{40}$K}
\def\B8{$^{8}$B}
\newcommand{\lsim}{{\;\raise0.3ex\hbox{$<$\kern-0.75em\raise-1.1ex\hbox{$\sim$}}\;}}
\newcommand{\gsim}{{\;\raise0.3ex\hbox{$>$\kern-0.75em\raise-1.1ex\hbox{$\sim$}}\;}}
\newcommand{\beq}{\begin{equation}}
\newcommand{\eeq}{\end{equation}}
\newcommand{\bea}{\begin{eqnarray}}
\newcommand{\eea}{\end{eqnarray}}

\mathchardef\minus="002D

\DeclareSIUnit \c {$c$}
\DeclareSIUnit\magn{$\times$}
\DeclareSIUnit\min{min}
\DeclareSIUnit\hr{hr}
\DeclareSIUnit\hrs{hrs}
\DeclareSIUnit\week{week}
\DeclareSIUnit\month{mo}
\DeclareSIUnit\months{mos}
\DeclareSIUnit\year{yr}
\DeclareSIUnit\years{years}
\DeclareSIUnit\yr{yr}
\DeclareSIUnit\standard{std}
\DeclareSIUnit\str{sr}
\DeclareSIUnit\ppm{ppm}
\DeclareSIUnit\ppb{ppb}
\DeclareSIUnit\ppt{ppt}
\DeclareSIUnit\pe{PE}
\DeclareSIUnit\spe{SPE}
\DeclareSIUnit\pdm{PDM}
\DeclareSIUnit\ev{events}
\DeclareSIUnit\ct{counts}
\DeclareSIUnit\neutron{\mbox{$n$}}
\DeclareSIUnit\smp{samples}
\DeclareSIUnit\Sample{S}
\DeclareSIUnit\ch{ch}
\DeclareSIUnit\hit{hit}
\DeclareSIUnit\hits{hits}
\DeclareSIUnit\bin{(\mbox{5-PE}~bin)}
\DeclareSIUnit\sgm{\mbox{$\sigma$}}
\DeclareSIUnit\rms{RMS}
\DeclareSIUnit\keVee{\mbox{keV$_{e{\rm e}}$}}
\DeclareSIUnit\keVr{\mbox{keV$_{\rm nr}$}}
\DeclareSIUnit\eVee{\mbox{eV$_{\rm ee}$}}
\DeclareSIUnit\eVr{\mbox{eV$_{\rm nr}$}}
\DeclareSIUnit\ph{photon}
\DeclareSIUnit\el{\mbox{$e^-$}}
\DeclareSIUnit\pm{\mbox{PMT}}
\DeclareSIUnit\pixel{\mbox{pixel}}
\DeclareSIUnit\inch{''}
\DeclareSIUnit\foot{'}
\DeclareSIUnit\bit{bit}
\DeclareSIUnit\sample{samples}
\DeclareSIUnit\barn{barn}
\DeclareSIUnit\bara{bar}
\DeclareSIUnit\bar{bar}
\DeclareSIUnit\barg{barg}
\DeclareSIUnit\mlardepth{\mbox(meter~of~\LAr~depth)}
\DeclareSIUnit\Curie{Ci}
\DeclareSIUnit\PSI{psi}
\DeclareSIUnit\psia{psia}
\DeclareSIUnit\atm{atm}
\DeclareSIUnit\psf{psf}
\DeclareSIUnit\pcf{pcf}
\DeclareSIUnit\parsec{pc}
\DeclareSIUnit\cps{cps}
\DeclareSIUnit\slpm{\SI{}{\liter\per\minute}}
\DeclareSIUnit\rpm{rpm}
\DeclareSIUnit\mwe{\mbox{m.w.e.}}
\DeclareSIUnit\liveday{\mbox{live-days}}
\DeclareSIUnit\days{\mbox{days}}
\DeclareSIUnit\miles{\mbox{miles}}
\DeclareSIUnit\lumens{\mbox{lm}}
\DeclareSIUnit\degreeC{\mbox{$^{\circ}$C}}
\DeclareSIUnit\degreeF{\mbox{$^{\circ}$F}}
\DeclareSIUnit\electron{\mbox{$e^-$}}
\DeclareSIUnit\Euro{\mbox{\euro}}
\DeclareSIUnit\cph{cph}
\DeclareSIUnit\neq{neq}
\DeclareSIUnit\normal{\mbox{N}}
\DeclareSIUnit\USD{\mbox{\$}}
\DeclareSIUnit\Vpercm{\mbox{V/cm}}
\DeclareSIUnit\kV{\mbox{kV}}

\DeclareSIUnit \mm {\milli\meter}
\DeclareSIUnit \cm {\centi\meter}
\DeclareSIUnit \us {\micro\second}
\DeclareSIUnit \ms {\milli\second}
\DeclareSIUnit \pA {\pico\ampere}
\DeclareSIUnit \pC {\pico\coulomb}
\DeclareSIUnit \fC {\femto\coulomb}
\DeclareSIUnit \fF {\femto\farrad}
\DeclareSIUnit \pF {\pico\farrad}
\DeclareSIUnit \mV {\milli\volt}
\DeclareSIUnit \kV {\kilo\volt}
\DeclareSIUnit \V {\volt}
\DeclareSIUnit \GOhm {\giga\ohm}
\DeclareSIUnit \MOhm {\mega\ohm}
\DeclareSIUnit \ton {\tonne}
\DeclareSIUnit \kton {\kilo\tonne}
\DeclareSIUnit \kt {\kilo\tonne}
\DeclareSIUnit \Mt {\mega\tonne}
\DeclareSIUnit \eV {\electronvolt}
\DeclareSIUnit \keV {\kilo\electronvolt}
\DeclareSIUnit \MeV {\mega\electronvolt}
\DeclareSIUnit \GeV {\giga\electronvolt}
\DeclareSIUnit \km {\kilo\meter}
\DeclareSIUnit \kW {\kilo\watt}
\DeclareSIUnit \MW {\mega\watt}
\DeclareSIUnit \MHz {\mega\hertz}
\DeclareSIUnit \kHz {\kilo\hertz}
\DeclareSIUnit \mrad {\milli\radian}
\DeclareSIUnit \year {year}
\DeclareSIUnit \POT {POT}
\DeclareSIUnit \sig {$\sigma$}
\DeclareSIUnit\parsec{pc}
\DeclareSIUnit\lightyear{ly}
\DeclareSIUnit\foot{ft}
\DeclareSIUnit\ft{ft}
\usepackage[toc]{glossaries}
\makeglossaries

\newcommand{\dshort}[1]{\glsentrytext{#1}}  % doesn't provide link
\newcommand{\dshorts}[1]{\glsentryshortpl{#1}}  % doesn't provide link
\newcommand{\dfirst}[1]{\glsfirst{#1}\glsunset{#1}}

\newcommand{\dword}[1]{\gls{#1}}
\newcommand{\dwords}[1]{\glspl{#1}}

\newcommand{\newduneword}[3]{
    \newglossaryentry{#1}{
        text={#2},
        long={#2},
        name={\glsentrylong{#1}},
        first={\glsentryname{#1}},
        firstplural={\glsentrylong{#1}\glspluralsuffix},
        description={#3},
        sort={#2}
    }
}

\newcommand{\newduneabbrev}[4]{
  \newglossaryentry{#1}{
    text={#2},
    long={#3},
    shortplural={{#2}\glspluralsuffix},
    longplural={{#3}\glspluralsuffix{}},
    name={\glsentrylong{#1}{} (\glsentrytext{#1}{})},
    first={#3 (#2)},
    firstplural={#3\glspluralsuffix{} (\glsentrytext{#1}\glspluralsuffix{})},
    description={#4},
    sort={#2}
  }
}

\newcommand{\newduneabbrevs}[5]{
  \newglossaryentry{#1}{
    text={#2},
    long={#3},
    plural={#4},
    shortplural={{#2}\glspluralsuffix},
    longplural={#4},
    name={\glsentrylong{#1}{} (\glsentrytext{#1}{})},
    first={#3 (#2)},
    firstplural={#4 (\glsentrytext{#1}\glspluralsuffix{})},
    description={#5},
    sort={#2}    
  }
}

\newduneword{dword}{DUNE Word}{A term in the DUNE lexicon}

\newduneword{nasa}{NASA}{U.S. National Aereonautics and Space Administration}

\newduneabbrev{nd}{ND}{near detector}{Refers to the collection of \gls{dune} detector components 
 installed close to the neutrino source at \gls{fnal}; also a subproject of \gls{usproj} that  includes  installation, infrastructure, and the cryogenics systems for this detector} 
 
\newduneabbrev{fd}{FD}{far detector}{The \SI{70}{kt} total (\fdfiducialmass fiducial) mass \gls{lartpc} DUNE detector, composed of four \larmass total (\nominalmodsize fiducial) mass modules,  
  to be installed at the far site at \gls{surf} in
  Lead, SD, USA}

\newduneabbrev{sp}{SP}{single-phase}{Distinguishes a \gls{lartpc} technology by the fact that it operates using argon in its liquid phase only; a legacy \gls{dune} term now replaced by \gls{hd} and \gls{vd}} 

\newduneabbrev{dp}{DP}{dual-phase}{Distinguishes a \gls{lartpc} technology by the fact that it operates using argon 
 in both gas and liquid phases; sometimes called double-phase} 

\newduneabbrev{pds}{PDS}{photon detection system}{The detector 
  subsystem sensitive to light produced in the \gls{lar} }

\newduneabbrev{hvs}{HVS}{high voltage system}{The detector 
  subsystem that provides the \gls{tpc} drift field}

\newduneabbrev{tpc}{TPC}{time projection chamber}{Depending on context: (1) A type of particle detector that uses an \efield together with a sensitive volume of gas or liquid, e.g., \gls{lar}, to perform a \threed reconstruction of a particle trajectory or interaction. The activity is recorded by digitizing the waveforms of current
  induced on the anode as the distribution of ionization charge passes by
  or is collected on the electrode. (2) TPC is also used in \gls{usproj} for ``total project cost''} 

\newduneabbrev{lartpc}{LArTPC}{liquid-argon time-projection chamber}{A \gls{tpc} filled with liquid argon; 
the basis for the \gls{dune} \gls{fd} modules} 

\newduneabbrevs{apa}{APA}{anode plane assembly}{anode plane assemblies}{A unit of the \gls{sphd}
  detector module containing the elements sensitive to ionization in the \gls{lar}. 
  Each anode face has three planes of wires (two induction, one collection) to provide a 3D view, and interfaces to the cold electronics and photon detection system} 

\newduneabbrev{awg}{AWG}{American wire gauge} {U.S. standard set of non-ferrous wire conductor sizes}

\newduneabbrev{ufer}{Ufer}{concrete encased electrode} {U.S. National Electrical Code grounding method refered to as Concrete Encased Electrode}

\newduneabbrev{cro}{CRO}{charge readout}{The system for detecting
  ionization charge distributions in a 
  detector module} 
  
\newduneabbrev{lro}{LRO}{light readout}{The system for detecting
  scintillation photons in a \gls{lartpc} detector module}

\newduneabbrev{shv}{SHV}{safe high voltage}{Type of bayonet mount
connector used on coaxial cables that has additional insulation 
compared to standard BNC and MHV connectors that makes it safer
for handling \gls{hv} by preventing accidental contact with the
live wire connector in an unmated connector or plug}

\newduneabbrev{fe}{FE}{front-end}{The front-end refers to a point that is
  ``upstream'' of the data flow for a particular subsystem. 
 For example the \gls{sphd} front-end electronics is where the cold electronics
  meet the sense wires of the TPC and the front-end \gls{daq} is where the \gls{daq} meets the output of the electronics}

\newduneabbrev{ }{DAQ RU}{DAQ readout unit}{The first element in the data flow of the \gls{daq}}

\newduneabbrev{cots}{COTS}{commercial off-the-shelf}{Items, typically hardware such as 
computers, that may be purchased whole, without any custom design or fabrication and 
thus at normal consumer prices and availability}

\newduneabbrev{i2c}{I2C}{Inter-Integrated Circuit}{I$^2$C or I2C is a synchronous, 
multi-master, multi-slave, packet switched, single-ended, serial computer bus widely used 
for attaching lower-speed peripheral ICs to processors and microcontrollers in short-distance, 
intra-board communication} 

\newduneabbrev{spi}{SPI}{Serial Peripheral Interface}{The Serial Peripheral Interface is a 
synchronous serial communication interface specification used for short distance 
communication, primarily in embedded systems}

\newduneabbrev{miso}{MISO}{master in slave out}{The Master In Slave Out is a logic
signal on the \gls{spi} bus on which the data from the slave are transmitted once
a request from the master is received} 

\newduneabbrev{mosi}{MOSI}{master out slave in}{The Master Out Slave In is a logic
signal on the \gls{spi} bus on which the data from the master is transmitted} 

\newduneabbrev{uart}{UART}{Universal Asynchrous Receiver/Transmitter}{A universal 
asynchronous receiver-transmitter is a computer hardware device for asynchronous 
serial communication in which the data format and transmission speeds are configurable}

\newduneword{cr}{CR}{Capacitance-Resistance} 

\newduneword{dc}{DC}{direct coupling} 

\newduneword{ac}{AC}{Alternating Current; when used in the phrase ``AC coupling'' refers to a circuit element that filters out low-frequency components, such as constant offsets, leaving higher frequency signal components. The frequency filtering is determined both by a resistor and a capacitor}

\newduneabbrev{pll}{PLL}{Phase-Locked Loop}{A control system that generates an
output signal whose phase is related to the phase of an input signal} 

\newduneword{fifo}{FIFO}{First-In-First-Out} 

\newduneword{saci}{SACI}{\gls{slac} \gls{asic} Control Interface}

\newduneword{om3}{OM3}{Type of multi-mode fiber optic cable, typically capable of \SI{10}{Gbps} data transmission at lengths up to \SI{300}{m}}

\newduneword{om4}{OM4}{Type of multi-mode fiber optic cable, typically capable of \SI{10}{Gbps} data transmission at lengths up to \SI{550}{m}}

\newduneword{qfp}{QFP}{Quad Flat Package} 

\newduneabbrev{ams}{AMS}{analog and mixed signal}{Verilog-AMS is a derivative of the Verilog hardware description language that includes analog and mixed-signal extensions (AMS) in order to define the behavior of analog and mixed-signal systems}

\newduneabbrev{hepa}{HEPA}{High Efficiency Particulate Air}{The High Efficiency Particulate Air filters are a type of air filter that remove 99.97\% of particles that have a size greater than or equal to \SI{0.3}{\micro\meter}}  

\newduneabbrev{uvm}{UVM}{universal verification methodology}{The Universal Verification Methodology is a standardized methodology for verifying integrated circuit designs}  

\newduneword{lhc}{LHC}{Large Hadron Collider}

\newduneabbrev{lsb}{LSB}{least significant bit}{The bit with the lowest numerical value in a binary number}

\newduneabbrev{ldo}{LDO}{low-dropout regulator}{A low-dropout or LDO regulator is a \gls{dc} linear voltage regulator that can regulate the output voltage even when the supply voltage is very close to the output voltage}

\newduneabbrev{adc}{ADC}{analog-to-digital converter}{A sampling of a voltage
  resulting in a discrete integer count corresponding in some way to
  the input}

\newduneabbrev{inl}{INL}{integral non-linearity}{A commonly used measure of performance in \glspl{adc}. It is the deviation between the ideal input threshold value and the measured threshold level of a certain output code}

\newduneabbrev{dnl}{DNL}{differential non-linearity}{A commonly used measure of performance in \glspl{adc}. The DNL error is defined as the difference between an actual step width and the ideal value of one \gls{lsb}}

\newduneword{pnp}{PNP}{Type of bipolar junction transistor consistning of a
layer of N-doped semiconductor sandwiched between two layers of P-doped material}

\newduneabbrev{spice}{SPICE}{Simulation Program with Integrated Circuit Emphasis}{a general-purpose, 
open-source analog electronic circuit simulator. It is a program used in integrated 
circuit and board-level design to check the integrity of circuit designs and to 
predict circuit behavior} 

\newduneabbrev{daq}{DAQ}{data acquisition}{The data acquisition system
  accepts data from the detector \gls{fe} electronics, buffers
  the data, performs a \gls{trigdecision}, builds events from the selected
  data and delivers the result to the offline \gls{diskbuffer}}

\newduneabbrev{iov}{IOV}{interval of validity}{Interval over which something is valid}

\newduneword{calci}{CALCI}{Calibration and Cryogenic Instrumentation}

\newduneword{detmodule}{far detector module}{The entire DUNE far detector design calls for segmentation into  four modules,  each with a total/fiducial mass of approximately \SI{17}{\kton}/\SI{10}{\kton}}
  
\newduneword{module}{module}{Many aspects of the DUNE far and near detectors are modular, so ``module'' must be understood in context.  It may refer to one of the four far detector modules,  distinct portions of a subdetector as in a ``field cage module,'' a software or electronics module,  e.g., a separate framework plug-in,  and so on}

\newduneword{detunit}{detector unit}{A portion of a \gls{detmodule} may be further partitioned into a number of similar parts.   For example, the \gls{sphd} \gls{tpc} is made up of \gls{apa}  units (and other elements)}

\newduneword{diskbuffer}{secondary DAQ buffer}{A secondary
  \gls{daq} buffer holds a small subset of the full rate as
  selected by a \gls{trigcommand}. 
  This buffer also marks the interface with the DUNE Offline}

\newduneabbrev{om}{OM}{online monitoring}{Processes that run inside
  the \gls{daq} on data ``in flight,'' specifically before landing on the
  offline disk buffer, and that provide feedback on the operation of
  the \gls{daq} itself and the general health of the data it is marshalling}

\newduneabbrev{dqm}{DQM}{data quality monitoring}{Analysis of the raw
  data to monitor the integrity of the data and the performance of the
  detectors and their electronics. This type of monitoring may be
  performed in real time, within the \gls{daq} system, or in later
  stages of processing, using disk files as input}

\newduneword{dumpbuffer}{DAQ dump buffer}{This \gls{daq} buffer
  accepts a high-rate data stream, in aggregate, from an associated
  portion of a \gls{detmodule} sufficient to collect all data likely relevant to
  a potential \gls{snb}}
\newduneabbrev{etf}{ETF}{Experiment Test Framework}{\gls{wlcg} testing middleware that runs grid jobs that actively test distributed sites' services and capabilities,  and reports back to monitoring services}

\newduneabbrev{etl}{ETL}{external trigger logic}{Trigger processing
  that consumes \gls{detmodule} level \gls{trignote} information
  and other global sources of trigger input and emits
  \gls{trigcommand} information back to the \glspl{mtl}}
\newduneabbrev{daqeti}{ETI}{external trigger interface}{Interface between \glspl{mtl} and external source and sinks of relevant trigger information}

\newduneword{trignote}{trigger notification}{Information provided by
  \gls{mtl} to \gls{etl} about \gls{trigdecision} %its 
  processing}

\newduneword{trigprimitive}{trigger primitive}{Information derived by
  the \gls{daq} \gls{fe} hardware that describes a region of space (e.g.,
  one or several neighboring channels) and time (e.g., a contiguous set
  of \gls{adc} sample ticks) associated with some activity}

\newduneword{externtrigger}{external trigger candidate}{Information
  provided to the \gls{mtl} about events external to a
  \gls{detmodule} so that it may be considered in forming
  \glspl{trigcommand}}

\newduneabbrev{daqoob}{OOB dispatcher}{out-of-band trigger command
  dispatcher}{This component is responsible for dispatching a \gls{snb} dump
  command to all \glspl{daqfer} in the \gls{detmodule}}

\newduneabbrev{mtl}{MTL}{module trigger logic}{Trigger processing
  that consumes \gls{detunit} level \gls{trigcommand} information
  and emits \glspl{trigcommand}. 
  It provides the \gls{etl} with \glspl{trignote} and receives back any
  \glspl{externtrigger}}

\newduneword{octant}{octant}{Any of the eight parts into which 4$\pi$
  is divided by three mutually perpendicular axes. 
  In particular in referencing the value for the mixing angle
  $\theta_{23}$}

\newduneword{trigcandidate}{trigger candidate}{Summary information derived
  from the full data stream and representing a contribution toward
  forming a \gls{trigdecision}}

\newduneword{trigcommand}{trigger command}{Information derived from
  one or more \glspl{trigcandidate}  that directs elements of a
  \gls{detmodule} to read out a portion of the data stream}

\newduneabbrev{tcm}{TCM}{trigger command message}{A message flowing
  down the trigger hierarchy from global to local context.  Also see \gls{tpm}}

\newduneabbrev{mlt}{MLT}{module level trigger}{The \gls{daq} component responsible for producing a \gls{trigdecision} that will be used to command the readout of a detector module}

\newduneword{trigdecision}{trigger decision}{The process by which
  \glspl{trigcandidate} are converted into \glspl{trigcommand}}

\newduneabbrev{tpm}{TPM}{trigger primitive message}{A message flowing
  up the trigger hierarchy from local to global context.  Also see \gls{tcm}}

\newduneabbrev{ipc}{IPC}{inter-process communication}{A system for software elements to exchange information between threads, local processes or across a data network.  An IPC system is typically specified in terms of protocols  composed of message types and their associated data schema}

\newduneword{daqdispre}{discovery and presence}{As used in the context of the \gls{ipc}, a system that provides mechanisms for a node on a communication network to learn of the existence of peers and their identity (discovery) as well as determine if they are currently operational or have become unresponsive (presence)}

\newduneabbrev{pubsub}{PUB/SUB}{publish-subscribe communication pattern}{An \gls{ipc} communication pattern where one element, the publisher, sends data to all connected elements, the subscribers.  Each subscriber may connect to multiple publishers.  A variant is PUB/SUB with topics where a subscriber may register an identifier, the topic, to limit the information received to just an associated subset}

\newduneabbrev{eb}{EB}{event builder}{A software agent that executes \glspl{trigcommand}  for one  \gls{detmodule} by reading out the requested data}

\newduneabbrev{daqdfo}{DFO}{data flow orchestrator}{The process by which trigger commands are executed in parallel and asynchronous manner by the back-end output subsystem of the \gls{daq}}

\newduneabbrev{daqubi}{UBI}{upstream DAQ buffer interface}{The process which provides read-only access to data residing in the upstream \gls{daq} buffers to processes on the network}

\newduneabbrev{cob}{COB}{cluster on board}{An ATCA motherboard housing four RCEs}

\newduneabbrev{rce}{RCE}{reconfigurable computing element}{Data processor located outside of the cryostat on a \gls{cob} that contains \gls{fpga}, RAM and \gls{ssd} resources, responsible for buffering data, producing trigger primitives, responding to triggered requests for data and synching \gls{snb} dumps}

\newduneabbrev{bow}{BOW}{Bump On Wire}{A working name for the front-end readout computing elements used in the nominal \gls{daq} design to interface the \gls{dp}  crates to the \gls{daq} front-end computers} 

\newduneabbrev{atca}{ATCA}{Advanced Telecommunications Computing
  Architecture}{An advanced computer architecture specification developed for the telecommunications, military, and aerospace industries that incorporates the latest trends in  high-speed interconnect technologies, next-generation processors, and improved reliability, availability and serviceability} 

\newduneabbrev{utca}{$\mu$TCA}{Micro Telecommunications Computing Architecture}{The computer architecture specification followed by the crates that house charge and light readout electronics; used in the SP vertical drift and \gls{dp} technologies} 

\newduneabbrev{udp}{UDP}{user datagram protocol}{A simple,
  connectionless Internet protocol that supports data integrity
  checksums, requires no handshaking, and does not guarantee packet delivery}

\newduneabbrev{amc}{AMC}{advanced mezzanine card}{Holds digitizing
  electronics and lives in \gls{utca} crates}

\newduneabbrev{rf}{RF}{radio frequency}{Electromagnetic emissions
  that are within the (radio) frequency band of sensitivity of the detector
  electronics}

\newduneabbrev{fpga}{FPGA}{field programmable gate array}{An
integrated circuit technology that allows the hardware to be reconfigured to
execute different algorithms after its manufacture and deployment}

\newduneabbrev{fmc}{FMC}{FPGA mezzanine card}{Boards holding \glspl{fpga} and other integrated circuitry that attach to a motherboard}

\newduneabbrev{felix}{FELIX}{Front-End Link eXchange}{A
  high-throughput interface between \gls{fe} and trigger electronics
  and the standard PCIe computer bus}

\newduneword{daqpart}{DAQ partition}{A cohesive and
 coherent collection of \gls{daq} hardware and software working together to trigger and read out some portion of one detector module; it consists of an integral number of \glspl{daqfrag}. 
 Multiple \gls{daq} partitions may operate simultaneously, but each instance operates independently}

\newduneabbrev{fec}{DAQ FEC}{DAQ front-end computer}{The portion of one
  \gls{daqpart} that hosts the \gls{daqdr}, \gls{daqbuf} and
  \gls{daqds}.  It hosts the \gls{daqfer} and corresponding portion of the \gls{daqbuf}}

\newduneword{daqfrag}{DAQ front-end fragment}{The portion of one
  \gls{daqpart} relating to a single \gls{fec} and corresponding to an
  integral number of \glspl{detunit}.  See also \gls{datafrag}}

\newduneword{datafrag}{data fragment}{A block of data read out from a single \gls{daqfrag} that
span a contiguous period of time as requested by a \gls{trigcommand}}

\newduneabbrev{daqfer}{FER}{DAQ front-end readout}{The portion of a
  \gls{daqfrag} that accepts data from the detector electronics and
  provides it to the \gls{fec}}

\newduneabbrev{daqdr}{DDR}{DAQ data receiver}{The portion of the
  \gls{daqfrag} that accepts data from the \gls{daqfer}, emits
  trigger candidates produced from the input trigger primitives, and
  forwards the full data stream to the \gls{daqbuf}}

\newduneword{daqbuf}{DAQ primary buffer}{The portion
  of the \gls{daqfrag} that accepts full data stream from the
  corresponding \gls{detunit} and retains it sufficiently long for it
  to be available to produce a \gls{datafrag}}

\newduneword{daqds}{data selector}{The portion of the \gls{daqfrag}
  that accepts \glspl{trigcommand} and returns the corresponding
  \gls{datafrag}.  Not to be confused with \gls{daqdsn}}

\newduneword{daqdsn}{data selection}{The process of forming a trigger decision for selecting a subset of detector data for output by the \gls{daq} from the content of the detector data itself.  Not to be confused with \gls{daqds}}

\newduneabbrev{daqros}{DAQ RO}{DAQ readout subsystem}{The subsystem of the \gls{daq} for accepting and buffering data input from detector electronics}

\newduneabbrev{daqdss}{DAQ DS}{DAQ data selection subsystem}{The subsystem of the \gls{daq} responsible for forming a trigger decision based on a portion of the input data stream.  The majority subset of the \gls{daqtrs}}

\newduneabbrev{daqtrs}{DAQ TS}{DAQ trigger subsystem}{The subsystem of the \gls{daq} responsible for forming a trigger decision}

\newduneabbrev{daqbes}{DAQ BE}{DAQ back-end subsystem}{The portion of the \gls{daq} that is generally toward its output end.  It is responsible for accepting and executing trigger commands and marshaling the data they address to output storage buffers}

\newduneabbrev{daqtss}{DAQ TSS}{DAQ timing and synchronization subsystem}{The portion of the \gls{daq} that provides for timing and synchronization to various components}

\newduneabbrev{femb}{FEMB}{front-end mother board}{Refers to a unit of
  the \gls{sp} \gls{ce} that contains the \gls{fe} amplifier
  and \gls{adc} \glspl{asic} covering 128 channels}

\newduneword{asic}{ASIC}{application-specific integrated circuit}

\newduneword{lv}{LV}{low voltage}

\newduneabbrev{iceberg}{ICEBERG}{ICEBERG R\&D cryostat and electronics}{Integrated Cryostat and Electronics Built for Experimental Research Goals: a double-walled cryostat built and installed at \gls{fnal}  for liquid argon detector R\&D and for testing of DUNE detector components}

\newduneword{coldadc}{ColdADC}{A newly developed 16-channels \gls{asic} providing analog to digital conversion}

\newduneword{coldata}{COLDATA}{A 64-channel control and communications \gls{asic}}

\newduneword{cryo}{CRYO}{(1) Integrated ASIC including \gls{fe} circuitry providing signal amplification and pulse shaping, analog to digital conversion, and control and communication functionalities for 64 channels; (2) acryonym for cryogenic systems and cryostat work scopes in \gls{lbnf}}

\newduneword{larasic}{LArASIC}{A 16-channel \gls{fe} \gls{asic} that provides signal amplification and pulse shaping}

\newduneword{cmos}{CMOS}{Complementary metal-oxide-semiconductor}

\newduneword{nmos}{NMOS}{N-channel metal-oxide semiconductor}

\newduneabbrev{enc}{ENC}{equivalent noise charge}{The equivalent noise charge is the input charge that corresponds to a \gls{s/n}$=1$}

\newduneword{sar}{SAR}{successive approximation register}

\newduneword{protodune}{ProtoDUNE}{Either of the two initial DUNE prototype detectors constructed at \gls{cern}.  One prototype implemented \gls{sp} technology and the other \gls{dp}}
  
\newduneword{protodune2}{ProtoDUNE-II}{The second run of a ProtoDUNE detector}

\newduneword{pdsp}{ProtoDUNE-SP}{The \gls{sphd} \gls{protodune} detector constructed at \gls{cern}  in \gls{np04}}

\newduneword{pddp}{ProtoDUNE-DP}{The \gls{dp} \gls{protodune} detector constructed at \gls{cern} in \gls{np02}}

\newduneword{wa105}{WA105 DP demonstrator}{The 
3m$\times$1m$\times$1m WA105 \gls{dp} prototype detector at \gls{cern}}

\newduneword{rawevent}{DAQ event block}{The unit of data output by the
  \gls{daq}.  
  It contains trigger and detector data spanning a unique, contiguous
  time period and a subset of the detector channels}

\newduneabbrev{ssd}{SSD}{solid-state disk}{Any storage device that
  may provide sufficient write throughput to receive, both collectively and
  distributed, the sustained full rate of data from a \gls{detmodule}
  for many seconds}
\newduneabbrev{nvme}{NVMe}{Non-volatile memory express}{A specification for an interface to storage media attached via PCIe}

\newduneabbrev{hlt}{HLT}{high-level trigger}{This is actually a filter applied to data that has been triggered and aggregated in order to further reduce or characterize it}

\newduneabbrev{pid}{PID}{particle ID}{Particle identification}

\newduneword{readout window}{readout window}{A fixed, atomic and
  continuous period of time over which data from a \gls{detmodule}, in
  whole or in part, is recorded. 
  This period may differ based on the trigger that initiated the
  readout}

\newduneabbrev{zs}{ZS}{zero-suppression}{Used to delete some portion of a
  data stream that does not significantly deviate from zero or
  intrinsic noise levels. 
  It may be applied at different granularity from per-channel to per
  \gls{detunit}}

\newduneword{rc}{RC}{Depending on context, one of (1) resistive-capacitive (circuit), (2) run control, the system for configuring, starting and terminating the \gls{daq}, or (3) resource coordinator, a member of the \gls{dune} management team responsible for coordinating the financial resources of the project}

\newduneabbrev{daqccm}{CCM}{DAQ control, configuration and monitoring subsystem}{A system for controlling, configuring and monitoring other systems in particular those that make up the \gls{daq} where the CCM encompasses \gls{rc}}

\newduneword{daqrun}{DAQ run}{A period of time over which relevant data taking conditions and \gls{daq} configuration are asserted to be unchanged. 
  Multiple \gls{daq} runs may occur simultaneously when multiple \glspl{daqpart} are active. 
  This term should not be confused with DUNE experiment or beam ``runs'' that typically span many \gls{daq} runs}
\newduneword{daqrunnum}{DAQ run number}{A monotonically increasing count that uniquely and globally identifies a \gls{daqrun}}

\newduneabbrev{ccsn}{CCSN}{core-collapse supernova}{The collapse of stars more than $8\times$ as massive as the sun which produces an intense burst of neutrinos at the end of its fusion cycle in a matter of seconds which ejects the outermost stellar gas leaving behind a neutron star remnant.}

\newduneabbrev{snb}{SNB}{supernova neutrino burst}{A prompt 
  increase in the flux of low-energy neutrinos emitted in the first few seconds of a \dword{ccsn}.  It can also refer to a trigger command type that may be due to this phenomenon,
  or detector conditions that mimic its interaction signature}

\newduneabbrev{snble}{SNB/LE}{supernova neutrino burst and low
  energy}{Supernova neutrino burst and low-energy physics program}

\newduneabbrev{snews}{SNEWS}{SuperNova Early Warning System}{A global
  supernova neutrino burst trigger formed by a coincidence of \gls{snb} 
  triggers collected from participating experiments}

\newduneabbrev{pps}{1PPS signal}{one-pulse-per-second signal}{An
  electrical signal with a fast rise time and that arrives in real
  time with a precise period of one second}

\newduneabbrev{sls}{SLS}{spill location system}{A system residing at
  the DUNE far detector site that provides information, possibly
  predictive, indicating periods of time when neutrinos are being
  produced by the \gls{fnal} Main Injector beam spills}

\newduneabbrev{wib}{WIB}{warm interface board}{Digital electronics
  situated just outside a FD cryostat that receives digital data
  from the \glspl{femb} (part of \gls{ce}) over cold copper connections and sends it to the \gls{rce}
  \gls{fe} readout hardware}

\newduneabbrev{gps}{GPS}{Global Positioning System}{A satellite-based system that provides a highly accurate \gls{pps} that may be used to synchronize clocks and determine location}

\newduneabbrev{ntp}{NTP}{Network Time Protocol}{A networking protocol that allows synchronizing of clocks to within a few \si{\milli\second} of a time standard on a local network and within a few tens of \si{\milli\second} over the Internet} 

\newduneword{ptp}{PTP}{Depending on context, either p-terphenyl, a \gls{wls} material, or Precision Time Protocol, a networking protocol that allows synchronizing of clocks to within a few \si{\micro\second} of a time standard on a local network}  

\newduneabbrev{irig}{IRIG}{inter-range instrumentation group}{A standards body that defined a time-code standard for transferring timing information}

\newduneabbrev{nic}{NIC}{network interface controller}{Hardware for controlling the interface to a communication network.  Typically, one that obeys the Ethernet protocol}

\newduneabbrev{wiec}{WIEC}{warm interface electronics crate}{Crates mounted on the signal flanges that contain the \glspl{wib}}

\newduneabbrev{ptc}{PTC}{power and timing card}{Cards that provide further processing and distribution of the signals entering and exiting the \gls{sp} cryostat}

\newduneabbrev{ptb}{PTB}{power and timing backplane}{Backplane used to connect the \glspl{wib} and the \glspl{ptc} on the \gls{wiec}. Also connects the \gls{ce} flange on the cryostat penetration}

\newduneabbrev{sipm}{SiPM}{silicon photomultiplier}{A solid-state
  avalanche photodiode sensitive to single \phel signals}

\newduneabbrev{cisc}{CISC}{cryogenic instrumentation and slow controls}{Includes equipment to monitor all detector  components and  \gls{lar} quality and behavior, and provides a control system for many of the detector components}

\newduneword{fte}{FTE}{full-time equivalent. A unit of labor
  for the project. One year of work from one person}

\newduneword{art}{art}{A software framework implementing an
  event-based execution paradigm} 
  
\newduneabbrev{sam}{SAM}{sequential
  access via metadata}{A data-handling system to store and retrieve
  files and associated metadata, including a complete record of the
  processing that has used the files}

\newduneword{artdaq}{artdaq}{A data acquisition toolkit for data transfer, aggregation and processing}

\newduneword{beamline}{beamline}{A sequence of control and monitoring devices used for the formation of a directed collection of particles; also subproject within \gls{usproj}}

\newduneword{cdr}{CDR}{Depending on context, either ``conceptual design report,'' a formal project  document  that describes the experiment at a conceptual level, or ``conceptual design review,'' a formal review of the conceptual design of the experiment or of a component}  

\newduneabbrev{cf}{CF}{conventional facilities}{Pertaining to
  construction and operation of buildings and conventional infrastructure, and includes  cavern excavation}

\newduneabbrev{cp}{CP}{charge conjugation and parity}{Product of charge conjugation and parity
  transformations} 
  
\newduneabbrev{cpt}{CPT}{charge, parity, and time reversal symmetry}{product of charge, parity
  and time-reversal transformations}

\newduneabbrev{cpv}{CPV}{Charge Conjugation-Parity Symmetry Violation}{Lack of
  symmetry in a system before and after charge conjugation and parity
  transformations are applied. 
  For \gls{cp} symmetry to hold,  a particle turns into its
 corresponding antiparticle under a charge transformation,  and a parity
transformation inverts its space coordinates,  i.e. produces the mirror image}

\newduneword{doe}{DOE}{U.S. Department of Energy}

\newduneabbrev{fra}{FRA}{Fermi Research Alliance}{A joint partnership of the University of Chicago and the Universities Research Association (URA) that manages and operates \gls{fnal} on behalf of the \gls{doe}}

\newduneabbrev{dune}{DUNE}{Deep Underground Neutrino Experiment}{A leading-edge, international experiment for neutrino science and proton decay studies; refers to the entire international experiment and collaboration} 

\newduneabbrev{esh}{ES\&H}{environment, safety and health}{A discipline and specialty that studies and implements practical aspects of environmental protection and safety at work} 

\newduneabbrev{ppe}{PPE}{personnel protective equipment}{Equipment worn to minimize exposure to hazards that cause serious workplace injuries and illnesses}

\newduneabbrev{odh}{ODH}{oxygen deficiency hazard}{a hazard that occurs when inert gases such as nitrogen, helium, or argon displace room air and thus reduce the percentage of oxygen below the level required for human life}

\newduneabbrev{feshm}{FESHM}{Fermilab Environment, Safety and Health Manual}{The document that contains \gls{fnal} 's policies and procedures designed to manage environment, safety, and health in all its programs}

\newduneabbrev{fscf}{FSCF}{Far Site Conventional Facilities}{The \gls{cf} at the DUNE far detector site, \gls{surf}, including all detector caverns and support infrastructure}

\newduneabbrev{nscf}{NSCF}{Near Site Conventional Facilities}{The \gls{cf} at the DUNE near detector site, \gls{fnal}}

\newduneabbrev{surf}{SURF}{Sanford Underground Research Facility}{SURF is an underground laboratory in Lead, South Dakota, where the \gls{dune} \gls{fd} will be installed and operated. It is the deepest underground laboratory in the United States.}

\newduneabbrev{fd1c}{FD1+C}{far detector module 1 + cryogenics}{The first far detector module to be built at \gls{surf}, including integration and installation, and all cryogenics infrastructure to support FD1 and \gls{fd2}; also a subproject of \gls{usproj}} 

\newduneabbrev{fd1}{FD1}{far detector module 1}{The first DUNE far detector module to be built at \gls{surf}}

\newduneabbrev{fd2}{FD2}{far detector module 2}{The second DUNE far detector module to be built at \gls{surf}} 

\newduneabbrevs{gut}{GUT}{grand unified theory}{grand unified theories}{A class of theories that unifies the electroweak and strong forces}

\newduneabbrev{lar}{LAr}{liquid argon}{Argon in its liquid phase; it is a cryogenic liquid with a boiling point of \SI{87}{K} and density of \SI{1.4}{g/ml}}

\newduneabbrev{lbl}{LBL}{long-baseline}{Refers to the distance between the 
  neutrino source  and the \gls{fd}.  It can also refer to the distance between the near and far detectors. 
  The ``long'' designation is an approximate and relative distinction. For DUNE, this distance  (between \gls{fnal} and \gls{surf}) is approximately \SI{1300}{km}}

\newduneabbrev{lbnf}{LBNF}{Long-Baseline Neutrino Facility}{Long-Baseline Neutrino Facility; refers to the facilities that support the experiment including in-kind contributions under the line-item project. The portion of \gls{usproj} responsible for developing the neutrino beam, the far site cryostats,  and far and near site cryogenics systems, and the conventional facilities, including the excavations } 
  
\newduneabbrev{lbnf-dune}{LBNF/DUNE}{LBNF and DUNE enterprise}{Long-Baseline Neutrino Facility/Deep Underground Neutrino Experiment; refers to the overall enterprise or program including \gls{usproj}, participating international projects, and the \gls{dune} experiment and collaboration} 

\newduneword{usproj}{LBNF/DUNE-US}{Long-Baseline Neutrino Facility/Deep Underground Neutrino Experiment - United States; project to design and build the conventional and beamline facilities and the \gls{doe} contributions to the detectors. It is organized as a \gls{doe}/\gls{fnal} project and incorporates contributions to the facilities from international partners. It also acts as host for the installation and integration of the DUNE detectors} 

\newduneword{duneus}{DUNE-US}{Deep Underground Neutrino Experiment - United States; refers to the U.S. contribution to DUNE under the line-item \gls{usproj} project} 
  
\newduneabbrev{lbnc}{LBNC}{Long-Baseline Neutrino Committee}{The committee, composed of internationally prominent scientists with relevant expertise, charged by the \gls{fnal} director to review the scientific, technical, and managerial progress, plans and decisions associated with \gls{dune}}

\newduneabbrev{ncg}{NCG}{Neutrino Cost Group}{A group of internationally prominent scientists with relevant experience that is charged by the \gls{fnal} director to review the cost, schedule, and associated risks for the \gls{dune} experiment}

\newduneabbrev{mh}{MH}{mass hierarchy}{Describes the separation
  between the mass squared differences related to the solar and
  atmospheric neutrino problems (also written as \gls{mo})}

\newduneabbrev{mo}{MO}{mass ordering}{See \gls{mh}}

\newduneabbrev{mi}{MI}{Fermilab Main Injector}{An accelerator at
  \gls{fnal} that provides a beam of high-energy protons to the the \gls{usproj}  beamline}

\newduneabbrev{pot}{POT}{protons on target}{Typically used as a unit
  of normalization for the number of protons striking the neutrino
  production target}

\newduneabbrev{qa}{QA}{quality assurance}{The process of ensuring that 
the quality of each element meets requirements during design and development, and to detect and correct poor results prior to production}

\newduneabbrev{qc}{QC}{quality control}{The process (e.g., inspection, testing, measurements) 
of ensuring that each manufactured element meets its quality requirements prior to assembly or installation} 

\newduneabbrev{sm}{SM}{Standard Model}{Refers to a theory describing
  the interaction of elementary particles}

\newduneword{tdr}{TDR}{Depending on context, either ``technical design report,'' a formal project  document  that describes the experiment at a technical level, or ``technical design review,'' a formal review of the technical design of the experiment or of a component}  

\newduneabbrev{tp}{IDR}{interim design report}{An intermediate
milestone on the path to a full \gls{tdr}} 

\newduneabbrev{ckm}{CKM matrix}{Cabibbo-Kobayashi-Maskawa
  matrix}{Refers to the matrix describing the mixing between mass and
  weak eigenstates of quarks}

\newduneabbrev{cl}{CL}{confidence level}{Refers to a probability
  used to determine the value of a random variable given its
  distribution}

\newduneabbrev{pmns}{PMNS}{Pontecorvo-Maki-Nakagawa-Sakata}{A type of matrix that describes the mixing between mass and weak eigenstates of
  the neutrino}

\newduneword{hnl}{HNL}{heavy neutral lepton} 

\newduneabbrevs{cpa}{CPA}{cathode plane assembly}{cathode plane assemblies}{The component of the \gls{sphd} detector module that provides the drift HV cathode}

\newduneword{fc}{field cage}{The component of a \gls{lartpc} that contains and shapes the applied \efield}

\newduneword{cpafc}{CPA/FC}{A pair of \gls{cpa} panels and the top and bottom \gls{fc} portions that attach to the pair; an intermediate assembly for installation into the \gls{spmod} }

\newduneabbrev{topfc}{top FC}{top field cage}{The horizontal portions of the \gls{sphd} \gls{fc}   on the top of the \gls{tpc}}

\newduneabbrev{botfc}{bottom FC}{bottom field cage}{The horizontal portions of the \gls{sphd} \gls{fc} on the bottom of the \gls{tpc}}

\newduneabbrev{ewfc}{endwall FC}{endwall field cage}{The vertical portions of the \gls{fc} near the end walls}

\newduneword{gp}{ground plane}{An electrode held electrically neutral relative to Earth ground voltage; it is mounted on the \gls{fc} to protect the cryostat wall}

\newduneword{gg}{ground grid}{An electrode held electrically neutral relative to Earth ground voltage; it is installed between the cathode and the \glspl{pd} in a \gls{dpmod} to protect the \glspl{pmt}, maintaining high transparency to light}

\newduneabbrev{alara}{ALARA}{as low as reasonably
  achievable}{Typically used with regard management of radiation
  exposure but may be used more generally. It means making every
  reasonable effort to maintain e.g., exposures, to as far below the
  limits as practical, consistent with the purpose for that the
  activity is undertaken}

\newduneabbrev{ecal}{ECAL}{electromagnetic calorimeter}{A detector
  component that measures energy deposition of traversing particles (in the \gls{dune} near detector design)}

\newduneabbrev{hv}{HV}{high voltage}{Generally describes a voltage
  applied to drive the motion of free electrons through some media, e.g., LAr}

\newduneword{spmod}{SP module}{single-phase DUNE \gls{fd} module}  

\newduneword{vdmod}{vertical drift module}{vertical drift DUNE \gls{fd} module} 

\newduneword{dpmod}{DP module}{dual-phase DUNE \gls{fd} module} 

\newduneword{dsp}{DUNE-SP}{a single-phase DUNE far detector module} % 
\newduneabbrev{tcoord}{TC}{technical coordinator}{A member of the \gls{dune} management team responsible for organizing the technical aspects of the project effort; is head of \gls{tc}}

\newduneabbrev{tc}{TCN}{technical coordination}{The DUNE organization responsible for overall integration of the detector elements and successful execution of the detector construction project; areas of responsibility include general project oversight, systems engineering, \gls{qa} and safety}

\newduneabbrev{exb}{EB}{executive board}{The highest level DUNE
  decision-making body for the collaboration}

\newduneabbrev{tb}{TB}{technical board}{The DUNE organization responsible for
  evaluating technical decisions}

\newduneabbrev{rrb}{RRB}{Resources Review Board}{A part of \gls{dune}'s international project governance structure, composed of representatives of all funding agencies that sponsor the project, and of  \gls{fnal} management, established to provide coordination among funding partners and oversight of \gls{dune}}

\newduneabbrev{inc}{INC}{International Neutrino Council}{A highest-level international advisory body to the U.S. \gls{doe} and the  \gls{fnal} directorate on matters related to the  \gls{lbnf} and the  \gls{pip2} projects. This council is composed of representatives from the international funding agencies and  \gls{cern} that make major contributions the infrastructure}

\newduneabbrev{cc}{CC}{charged current}{Refers to an interaction
  between elementary particles where a charged weak force carrier
  ($W^+$ or $W^-$) is exchanged}

\newduneabbrev{dis}{DIS}{deep inelastic scattering}{Refers to interaction between
  elementary particles and a nucleus in an energy range where the
  interaction can be modeled as occurring between constituent quarks
  of one nucleon and resulting in no bulk recoil of the resulting
  nucleus}

\newduneabbrev{fsi}{FSI}{final-state interactions}{Refers to
  interactions between elementary or composite particles subsequent to
  the initial, fundamental particle interaction, such as may occur as
  the products exit a nucleus}
  
\newduneabbrev{fsint}{FSI}{far site integration}{The scope of work at the \gls{fs} for the \gls{integoff}} 

\newduneword{geant4}{Geant4}{A
  software toolkit for the simulation of the passage of particles
  through matter using \gls{mc} methods}

\newduneabbrev{genie}{GENIE}{Generates Events for Neutrino Interaction
  Experiments}{Software providing an object-oriented neutrino
  interaction simulation resulting in kinematics of the products of
  the interaction}

\newduneabbrev{mc}{MC}{Monte Carlo}{Refers to a method of numerical
  integration that entails the statistical sampling of the integrand
  function. 
  Forms the basis for some types of detector and physics simulations}

\newduneabbrev{qe}{QE}{quasi-elastic}{Refers to the 
  interaction of an elementary charged particle with a nucleus in an
  energy range where the interaction can be modeled as taking place with
  individual nucleons} 

\newduneabbrev{mou}{MoU}{memorandum of understanding}{A project management methodology that  \gls{usproj} uses to document agreement,s,  e.g., between \gls{fnal} and the Project, for how \gls{fnal}  will support the Project. More generally, a document
  summarizing an agreement between two or more parties} 

\newduneabbrev{pip2}{PIP-II}{Proton Improvement Plan II}{A \gls{fnal} project for
  improving the protons on target delivered delivered by the \gls{lbnf} neutrino production beam. 
  This is version two of this plan and it is planned to be followed by a PIP-III}
  
\newduneabbrev{sdsta}{SDSTA}{South Dakota Science and Technology
  Authority}{The legal entity that manages \gls{surf}, in Lead, S.D}
  
\newduneabbrev{sdsd}{SDSD}{Fermilab South Dakota Services Division}{A \gls{fnal}  division responsible providing host laboratory functions at SURF in South Dakota}

\newduneabbrev{firus}{FIRUS}{Facility Information Reporting Utility System}
 {Facility incident reporting systems, one at \gls{fnal} and at \gls{surf}, that monitors and reports the status of various fire, security and utility sensors} 

\newduneabbrev{bsi}{BSI}{building and site infrastructure}
 {The work package for outfitting of the \gls{lbnf} underground infrastructure}

\newduneabbrev{wbs}{WBS}{work breakdown structure}{An organizational
  project management tool by which the tasks to be performed are
  partitioned in a hierarchical manner}

\newduneabbrev{br}{BR}{branching ratio}{A fractional probability for a
  decay of a composite particle to occur into some specified set or
  sets of products}
\newduneword{bsm}{BSM}{beyond the Standard Model}

\newduneabbrev{dm}{DM}{dark matter}{The term given to the unknown
  matter or force that explains measurements of galaxy motion  that are otherwise inconsistent with the amount of mass associated
  with the observed amount of photon production}
  
\newduneabbrev{bdm}{BDM}{boosted dark matter}{A new model that describes a relativistic dark matter particle boosted by the annihilation of heavier dark matter particles in the galactic center or the sun}

\newduneabbrev{cern}{CERN}{European Laboratory for Particle Physics}{The leading particle physics laboratory in Europe and home to the \glspl{protodune} and other prototypes and demonstrators, including the \glspl{mod0}}

\newduneabbrev{dsnb}{DSNB}{diffuse supernova neutrino background}{The
  term describing the pervasive, constant flux of neutrinos due to all
  past supernova neutrino bursts}

\newduneabbrev{espp}{ESPP}{European Strategy for Particle Physics}{The
cornerstone of Europe's decision-making process for the long-term future of the field. Mandated by the \gls{cern} Council, it is formed through a broad
consultation of the grass-roots particle physics community, it
actively solicits the opinions of physicists from around the world,
and it is developed in close coordination with similar processes in
the USA and Japan in order to ensure coordination between regions and
optimal use of resources globally}

\newduneabbrev{gar}{GAr}{gaseous argon}{argon in its gas phase}
\newduneabbrev{gartpc}{GArTPC}{gaseous argon time-projection chamber}{A \gls{tpc} filled with gaseous argon}

\newduneabbrev{globes}{GLoBES}{General Long-Baseline Experiment
  Simulator}{A software package for simulating energy spectra of
  neutrino flux, interactions, and energy spectra measured after application of some
  model of a detector response)}

\newduneabbrev{snowglobes}{SNOwGLoBES}{SuperNova
Observatories with GLoBES} {From the official description: SNOwGLoBES is public software for computing interaction rates and distributions of observed quantities for \gls{snb} neutrinos in common detector materials}

\newduneword{l/e}{L/E}{length-to-energy ratio}
\newduneword{lri}{LRI}{long-range interactions}

\newduneabbrev{nc}{NC}{neutral current}{Refers to an interaction
  between elementary particles where a neutrally charged weak force carrier
  ($Z^0$) is exchanged}

\newduneabbrev{nh}{NH}{normal hierarchy}{Refers to the neutrino mass
  eigenstate ordering whereby the sign of the mass squared difference
  associated with the atmospheric neutrino problem is positive}

\newduneabbrev{ih}{IH}{inverted hierarchy}{Refers to the neutrino mass
  eigenstate ordering whereby the sign of the mass squared difference
  associated with the atmospheric neutrino problem is negative}

\newduneabbrev{no}{NO}{normal ordering}{Refers to the neutrino mass
  eigenstate ordering whereby the sign of the mass squared difference
  associated with the atmospheric neutrino problem is positive}

\newduneabbrev{io}{IO}{inverted ordering}{Refers to the neutrino mass
  eigenstate ordering whereby the sign of the mass squared difference
  associated with the atmospheric neutrino problem is negative}

\newduneabbrev{msw}{MSW}{Mikheyev-Smirnov-Wolfenstein effect}{Explains
  the oscillatory behavior of neutrinos produced inside the sun as
  they traverse the solar matter}

\newduneabbrev{nsi}{NSI}{nonstandard interaction}{A general class of
  theory of elementary particles other than the Standard Model}

\newduneabbrev{pfive}{P5}{Particle Physics Project Prioritization
Panel}{The Particle Physics Project Prioritization Panel (P5) was a
subpanel of the High Energy Physics Advisory Panel (HEPAP). It completed
its Report, a ten-year strategic plan for high energy physics in the
U.S., in 2014. This report included a recommendation that ``host a world-leading neutrino
program that will have an optimized set of short- and long-baseline neutrino oscillation experiments, and its long-term focus
is a reformulated venture referred to here as the Long Baseline
Neutrino Facility (LBNF)''}

\newduneabbrev{sme}{SME}{standard-model extension}{an effective field theory that contains the \gls{sm}, general relativity, and all possible operators that break Lorentz symmetry (Wikipedia)}

\newduneabbrev{susy}{SUSY}{supersymmetry}{Theoretical symmetry between a fermion and a boson}

\newduneabbrev{wimp}{WIMP}{weakly-interacting massive particle}{A
  hypothesized particle that may be a component of dark matter}

\newduneabbrev{ce}{CE}{cold electronics}{Analog and digital readout electronics that operate at cryogenic temperatures}

\newduneabbrev{crp}{CRP}{charge-readout plane}{An anode technology using a stack of perforated \glspl{pcb} with etched electrode strips to provide \gls{cro} in 3D; it has two induction layers and one collection layer; it is used in the SP vertical drift \gls{fd} and \gls{dp} designs} 

\newduneabbrev{dram}{DRAM}{dynamic random access memory}{A computer memory technology}

\newduneabbrev{fnal}{Fermilab}
{Fermi National Accelerator Laboratory}{U.S. national laboratory in Batavia, IL.  It is the laboratory that hosts \gls{lbnf} and \gls{dune}, and serves as the experiment's near site}

\newduneabbrev{bnl}{BNL}{Brookhaven National Laboratory}{US national laboratory in Upton, NY}

\newduneabbrev{slac}{SLAC}{SLAC National Accelerator Laboratory}{US national laboratory in Menlo Park, CA}

\newduneabbrev{lbnl}{LBNL}{Lawrence Berkeley National Laboratory}{US national laboratory in Berkeley, CA}

\newduneabbrev{anl}{ANL}{Argonne National Laboratory}{US national laboratory in Lemont, IL}

\newduneabbrev{lanl}{LANL}{Los Alamos National Laboratory}{US national laboratory in Los Alamos, NM}

\newduneword{fs}{FS}{Depending on context, one of (1) the far site, \gls{surf}, where the DUNE far detector is located; (2) ``Full Stream'' relates to a data stream that has not undergone selection, compression or other form of reduction}

\newduneabbrev{lem}{LEM}{large electron multiplier}{A micro-pattern detector suitable for use in ultra-pure argon vapor; LEMs consist of copper-clad PCB boards with sub-millimeter-size holes through which electrons undergo amplification}

\newduneabbrev{lng}{LNG}{liquefied natural gas}{Pertaining to natural gas in its liquid phase}

\newduneabbrev{mip}{MIP}{minimum ionizing particle}{Refers to a
  particle traversing some medium such that the particle's mean energy loss is  
  near the minimum}

\newduneabbrev{pd}{PD}{photon detector}{The detector
  elements involved in measurement of the number and arrival times of
  optical photons produced in a detector module} 

\newduneabbrev{pmt}{PMT}{photomultiplier tube}{A device that makes use
  of the photoelectric effect to produce an electrical signal from the
  arrival of optical photons}

\newduneabbrev{ppm}{ppm}{parts per million}{A concentration equal to one part in $10^{6}$}
\newduneabbrev{ppb}{ppb}{parts per billion}{A concentration equal to one part in $10^{9}$}
\newduneabbrev{ppt}{ppt}{parts per trillion}{A concentration equal to one part in $10^{12}$}

\newduneword{rio}{RIO}{reconfigurable input output}

\newduneabbrev{s/n}{S/N}{signal-to-noise}{signal-to-noise ratio}

\newduneword{ssp}{SSP}{\gls{sipm} signal processor}

\newduneabbrev{sbn}{SBN}{Short-Baseline Neutrino}{A \gls{fnal} program consisting of three collaborations, \gls{microboone}, \gls{sbnd}, and \gls{icarus}, to perform sensitive searches for $\nue$ appearance and $\numu$ disappearance in the Booster Neutrino Beam}

\newduneabbrev{stt}{STT}{Straw Tube Tracker}{Target/tracker system that is part of the \gls{sand} near detector.}

\newduneword{wire board}{wire board}{At the head end of the APA in the \gls{sphd} \gls{tpc}, stacks of electronics boards referred to as ``wire boards'' are arrayed to anchor the wires.  They also provide the connection between the wires and the cold electronics} 

\newduneabbrev{wls}{WLS}{wavelength-shifting}{A material or process by
  which incident photons are absorbed by a material and photons are
  emitted at a different, typically longer, wavelength}
  
\newduneabbrev{tpb}{TPB}{tetra-phenyl butadiene}{A \gls{wls} material}

\newduneword{mcnd}{MCND}{More Capable Near Detector}

\newduneabbrev{sft}{SFT}{signal feedthrough}{A cryostat penetration allowing for the passage of cables or other extended parts}

\newduneabbrev{sftchimney}{SFT chimney}{signal feedthrough chimney}{A volume above the cryostat penetration used for a signal feedthrough} 

\newduneabbrev{catiroc}{CATIROC}{charge and time integrated readout chip}{A complete read-out chip manufactured in AustriaMicroSystem designed to read arrays of 16 photomultipliers}

\newduneabbrev{wr}{WR}{White Rabbit}{A component of the timing system that forwards clock signal and time-of-day reference data to the master timing unit}

\newduneabbrev{mch}{MCH}{MicroTCA Carrier Hub}{A network switching device}

\newduneabbrev{wrmch}{WR-MCH}{White Rabbit \gls{utca} Carrier Hub}{A card mounted in \gls{utca} crate that recieves time syncronization information and trigger data packets over \gls{wr} network and disributes them to the \gls{amc} over \gls{utca} backplane} 

\newduneabbrev{wrtsn}{WR-TSN}{White Rabbit TimeStamping Node}{A unit on the \gls{wr} network that timestamps the trigger signals and sends out trigger data packets to \gls{wrmch}}

\newduneword{qap}{QAP}{quality assurance plan} 

\newduneword{ieshp}{IESHP}{integrated environmental, safety and health plan}

\newduneword{dmp}{DMP}{data management plan} 
\newduneword{qam}{QAM}{quality assurance manager} 

\newduneabbrev{dss}{DSS}{detector support system}{The system of rails suspended from the cryostat ceiling in a \gls{spmod} used to support the \gls{apa}s, \gls{cpa}s, and the \glspl{ewfc}} 

\newduneabbrev{ddss}{DDSS}{DUNE detector safety system}{The hardware system responsible for the safety of the detector, implemented either via a \gls{plc} or via custom hardware protections} 

\newduneabbrev{tco}{TCO}{temporary construction opening}{An opening in the side of a cryostat through which detector elements are brought into the cryostat; utilized during construction and installation}

\newduneabbrev{uit}{UIT}{underground installation team}{An organizational unit responsible for installation in the underground area at the \gls{surf} site}

\newduneabbrev{cmgc}{CMGC}{construction manager/general contractor}{The contracted company hired to manage overall construction, used by \gls{lbnf} at the \gls{surf} site for the \gls{fscf} construction} 

\newduneword{pdr}{PDR}{Depending on context, either ``preliminary design report,'' a formal project document  that describes the experiment at a preliminary level, or ``preliminary design review,'' a formal review of the preliminary design of the experiment or of a component} 

\newduneword{fdr}{FDR}{Depending on context, either ``final design report,'' a formal project document  that describes the experiment at a final level, or ``final design review,'' a formal review of the final design of the experiment or of a component} 

\newduneabbrev{prr}{PRR}{production readiness review}{A project management mechanism by which the production readiness is reviewed}  

\newduneabbrev{irr}{IRR}{installation readiness review}{A project management mechanism by which the plan for installation is reviewed}  

\newduneabbrev{orr}{ORR}{operational readiness review}{A project management mechanism by which the operational readiness is reviewed}  

\newduneabbrev{ppr}{PPR}{production progress review}{A project management mechanism by which the progress of production is reviewed} 

\newduneabbrev{edms}{EDMS}{engineering document management system}{A computerized document management system developed and supported at the \gls{cern} in which some \gls{dune} project and collaboration documents, drawings and engineering models are managed}  

\newduneabbrev{docdb}{DocDB}{Document DataBase}{A computerized document management system developed and supported at \gls{fnal} in which most \gls{lbnf-dune} documents are managed (docs.dunescience.org); some documents are maintained in \gls{edms}}

\newduneword{wrgm}{WR grandmaster}{White Rabbit grandmaster}

\newduneabbrev{larsoft}{LArSoft}{Liquid Argon Software}{A shared base of physics software across \gls{lartpc} experiments}

\newduneword{nova}{NOvA}{The \gls{nova} off-axis neutrino oscillation experiment at \gls{fnal}}
\newduneword{minerva}{MINERvA}{Neutrino cross sections experiment at \gls{fnal},  shut down in 2019}
\newduneword{microboone}{MicroBooNE}{A \gls{lartpc} neutrino oscillation experiment at \gls{fnal}}
\newduneword{sbnd}{SBND}{The Short-Baseline Near Detector experiment at  \gls{fnal}}
\newduneabbrev{nexo}{nEXO}{Enriched Xenon Observatory}{Experiment at Lawrence Livermore National Laboratory (U.S. national lab in Livermore, CA) searching for new physics with neutrinoless double-beta decay}
\newduneword{argoneut}{ArgoNeuT}{The ArgoNeuT test-beam experiment and \gls{lartpc} prototype at  \gls{fnal}}
\newduneword{icarus}{ICARUS}{A neutrino experiment that was located at the Laboratori Nazionali del Gran Sasso (LNGS) in Italy, then refurbished at \gls{cern} for re-use in the same neutrino beam from \gls{fnal} used by the \gls{miniboone} , \gls{microboone} and \gls{sbnd} experiments at \gls{fnal}}
\newduneword{atlas}{ATLAS}{One of two general-purpose detectors at the \gls{lhc}. It investigates a wide range of physics, from the measurements of the Higgs boson properties to searches for extra dimensions and particles that could make up \gls{dm}}

\newduneword{lbne}{LBNE}{Long Baseline Neutrino Experiment; (1) a terminated U.S. experiment that was reformulated in 2014 under the auspices of the new \gls{dune} collaboration, an internationally coordinated and internationally funded program, with \gls{fnal} as host; and (2) the former incarnation of \gls{usproj} }

\newduneabbrev{lbno}{LBNO}{Long Baseline Neutrino Observatory} {A terminated European project that, during its six-year duration, assessed the feasibility of a next-generation deep underground neutrino observatory in Europe)}

\newduneword{wirecell}{Wire-Cell}{A tomographic automated \threed neutrino event reconstruction method for \glspl{lartpc}}
\newduneabbrev{wct}{WCT}{Wire-Cell Toolkit}{A software toolkit with data flow processing components for \gls{lartpc} noise and signal simulation, noise filtering, signal processing, and tomographic \threed ionization activity imaging}
\newduneword{ftslite}{F-FTS-lite}{Light-weight version of the \gls{fnal} File Transfer system used for rapid data transfers out of the online systems}
\newduneabbrev{fts}{FTS}{File Transfer System}{A file transfer system developed at \gls{fnal} to catalog and move data to permanent storage}

\newduneword{35t}{35 ton prototype}{A prototype cryostat and \gls{sp} detector built at \gls{fnal} before the \gls{protodune} detectors}

\newduneabbrev{mcr}{MCR}{main communications room}{Space at the \gls{fd} site for cyber infrastructure}

\newduneabbrev{cuc}{CUC}{central utility cavern}{The utility cavern at the 4850L of \gls{surf} located between the two detector caverns. It contains utilities such as central cryogenics and other systems, and the underground data center and control room}

\newduneabbrev{cnc}{CNC}{Computer Numerical Control}{A precise drilling method that utilizes a rotating cutting tool to produce round holes in a stationary work piece}

\newduneabbrev{cfd}{CFD}{computational fluid dynamics}{High performance computer-assisted modeling of fluid dynamical systems}
\newduneword{vuv}{VUV}{vacuum ultra-violet}
\newduneword{tallbo}{TallBo}{A cylindrical cryostat at \gls{fnal} primarily used for developing scintillation light collection technologies for \gls{lartpc} detectors}

\newduneword{root}{ROOT}{A modular scientific software toolkit. It provides all the functionalities needed to deal with big data processing,  statistical analysis, visualization, and storage.  It is mainly written in C++ but integrated with other languages such as Python and R}

\newduneword{eos}{\textsc{Eos}}{The XRootD-based distributed file system developed by CERN}

\newduneabbrev{ehn1}{EHN1}{Experiment Hall North One}{Location at \gls{cern} of the \gls{np02} and \gls{np04} areas used for the \glspl{protodune} and for other test and prototyping activities for DUNE} 

\newduneword{led}{LED}{Light-emitting diode}
\newduneabbrev{rtd}{RTD}{resistance temperature detector}{A temperature sensor consisting of a material with an accurate and reproducible resistance/temperature relationship}
\newduneword{swc}{SWC}{Software \& Computing}
\newduneabbrev{las}{LAS}{LEM-anode Sandwich}{In the \gls{dp} technology, a \gls{lem} and its corresponding anode are mounted together in a module called a LEM-anode sandwich}

\newduneword{roi}{ROI}{region of interest}
\newduneabbrev{hpc}{HPC}{high-performance computing}{high-performance computing facilities; generally computing facilities emphasizing parallel computing with aggregate power of more than a teraflop}

\newduneword{comfund}{common fund}{The shared resources of the collaboration}
\newduneabbrev{ims}{IMS}{integrated master schedule}{A project management device consisting of linked tasks and milestones}

\newduneword{hvdb}{HVDB}{HV divider board}

\newduneword{hvft}{HVFT}{HV feedthrough}  

\newduneword{sas}{SAS}{Another term for the materials airlock; a pass-through chamber used to ensure safe transfer of materials into a clean room, avoiding contamination in both directions}

\newduneabbrev{fea}{FEA}{finite element analysis}{Simulation of a physical phenomenon using the numerical technique called Finite Element Method (FEM), a numerical method for solving problems of engineering and mathematical physics}

\newduneword{fss}{FSS}{field shaping strips}
\newduneword{lvds}{LVDS}{low-voltage differential signaling}

\newduneword{esd}{ESD}{electrostatic discharge}

\newduneabbrev{rp}{RP}{resistive panel}{Resistive panels form the constant potential surfaces for a \gls{spmod} \gls{cpa}; they are composed of a thin layer of carbon-impregnated Kapton and laminated to both sides of a \frfour sheet}

\newduneword{uhmwpe}{UHMWPE}{ultra-high molecular weight polyethylene}

\newduneword{cts}{CTS}{Cryogenic Test System}

\newduneabbrev{plc}{PLC}{programmable logic controller}{An industrial digital computer that has been ruggedized and adapted for the control of manufacturing or other processes that require high reliability, ease of programming, and process fault diagnosis} 

\newduneword{mppc}{MPPC}{\SI{6}{mm}$\times$\SI{6}{mm} Multi-Pixel Photon Counters produced by Hamamatsu\texttrademark{} Photonics K.K}

\newduneabbrev{sfp}{SFP}{small form-factor pluggable}{a particular standard for optical transceivers}

\newduneabbrev{minipod}{MiniPOD}{miniature parallel optical device}{a family of types of multi-channel optical transceivers}

\newduneword{ccc}{CCC}{configuration change command}
\newduneword{ccondc}{CCC}{code of conduct committee}

\newduneword{act}{ACT}{activation time stamp}
\newduneword{lcm}{LCM}{light calibration module}
\newduneword{lpm}{LPM}{light pulser module}
\newduneword{dac}{DAC}{digital-to-analog converter}
\newduneword{arapuca}{ARAPUCA}{A \gls{pds} design that consists of a light trap that captures wavelength-shifted photons inside boxes with highly reflective internal surfaces until they are eventually detected by \gls{sipm} detectors or are lost}
\newduneword{sarapu}{S-ARAPUCA}{Standard \gls{arapuca} design with different \gls{wls} coatings on both faces of the dichroic filter window(s) of the cell}
\newduneword{xarapu}{X-ARAPUCA}{Extended \gls{arapuca} design with \gls{wls} coating on only the external face of the dichroic filter window(s) but with a \gls{wls} doped plate inside the cell}
\newduneword{feb}{FEB}{front-end board}

\newduneabbrev{lsnd}{LSND}{Liquid Scintilator Neutrino Detector}{A scintillation detector and associated experiment located at Los Alamos National Laboratory}

\newduneabbrev{cvn}{CVN}{convolutional visual network}{An algorithm for identifying neutrino interactions based on their topology and without the need for detailed reconstruction algorithms}

\newduneword{pandora}{Pandora}{The Pandora multi-algorithm approach to pattern recognition}

\newduneabbrev{pma}{PMA}{Projection Matching Algorithm}{A reconstruction algorithm that combines \twod reconstructed objects to form a \threed representation}
\newduneabbrev{bdt}{BDT}{boosted decision tree}{A method of multivariate analysis}
\newduneabbrev{cnn}{CNN}{convolutional neural network}{A deep learning technique most commonly applied to analyzing visual imagery}
\newduneword{pdg}{PDG}{Particle Data Group}

\newduneword{pci}{PCI}{Peripheral Component Interconnect}

\newduneword{labview}{LabVIEW}{Laboratory Virtual Instrument Engineering Workbench is a system-design platform and development environment for a visual programming language from National Instruments}

\newduneword{pcb}{PCB}{printed circuit board}

\newduneword{crio}{cRIO}{Compact Reconfigurable Input Output}

\newduneword{dcs}{DCS}{Distributed Communications System}

\newduneword{opc-ua}{OPC-UA}{OPC  Unified Architecture is a machine to machine communication protocol for industrial automation developed by the OPC Foundation. OPC stands for Object Linking and Embedding for Process Control}

\newduneword{cabangle}{Cabibbo angle}{A quark mixing parameter that governs the coupling of up quarks to strange quarks}
\newduneword{valor}{VALOR}{A neutrino oscillation fitting framework that is used by \gls{t2k}; the name stands for VALencia-Oxford-Rutherford, the original three institutions that developed it}
\newduneword{cafana}{CAFAna}{Common Analysis File Analysis}
\newduneabbrev{pca}{PCA}{principal component analysis}{A statistical procedure that uses an orthogonal transformation to convert a set of observations of possibly correlated variables into a set of values of linearly uncorrelated variables called principal components (Wikipedia)}
\newduneword{numi}{NuMI}{a set of facilities at \gls{fnal}, collectively called ``Neutrinos at the Main Injector.''  The NuMI neutrino beamline target system converts an intense proton beam into a focused neutrino beam}
\newduneword{gibuu}{GiBUU}{Giessen Boltzmann-Uehling-Uhlenback Project; a unified theory and transport framework in the MeV and GeV energy regimes for elementary reactions on nuclei }
\newduneabbrev{rpa}{RPA}{random phase approximation} {an approximation method commonly used for describing the dynamic linear electronic response of electron systems (Wikipedia)}
\newduneword{t2k}{T2K}{T2K (Tokai to Kamioka) is a long-baseline neutrino experiment in Japan studying neutrino oscillations}
\newduneword{mptdet}{MPT detector}{multipurpose tracking detector}

\newduneword{lariat}{LArIAT}{The repurposed ArgoNeuT \gls{lartpc}, modified for use in a charged particle beam, dedicated to the calibration and precise characterization of the output response of these detectors}

\newduneword{captain}{CAPTAIN}{Experimental program sited at \gls{lanl} that is designed to make measurements of scientific importance to \gls{lbl} neutrino physics and physics topics that will be explored by large underground detectors}

\newduneword{dayabay}{Daya Bay}{a neutrino-oscillation experiment in Daya Bay, China, designed to measure the mixing angle $\Theta_{13}$  using antineutrinos produced by the reactors of the Daya Bay and Ling Ao nuclear power plants}

\newduneword{nuwro}{NuWro}{neutrino interaction generator}

\newduneabbrev{neut}{NEUT}{neutrino interaction generator}{A neutrino interaction simulation program library for the studies of atmospheric and accelerator neutrinos}

\newduneword{minos}{MINOS}{A long-baseline neutrino experiment, with a near detector at \gls{fnal} and a far detector in the Soudan mine in Minnesota,  designed to observe the phenomena of neutrino oscillations (ended data runs in 2012)}

\newduneabbrev{efig}{EFIG}{Experimental Facilities Interface Group}{The body that provides a means to coordinate and discuss issues related to the interfaces within and between the facility and the DUNE detectors at both the \gls{fnal} and \gls{surf} sites} 

\newduneword{ashriver}{Ash River}{The Ash River, Minnesota, USA \gls{nova} experiment far site, used as an assembly test site for \gls{dune}} 

\newduneword{ipd}{project integration director}{Responsible for integration and installation of DUNE detector deliverables. Manages the integration project} 

\newduneabbrev{jpo}{JPO}{Joint Project Office}{The framework used to facilitate close and coherent coordination across all elements with many shared management resources; 
JPO functions include systems engineering, procurement, \gls{esh}, \gls{qa}, finance, project controls, risk management, compliance, internal reviews, partner agreement management, document management, and administrative support} 

\newduneword{ifbeam}{IFbeam}{Database that stores beamline information indexed by timestamp}

\newduneabbrev{marley}{MARLEY}{Model of Argon Reaction Low Energy Yields}{Developed at UC Davis, MARLEY is the first realistic model of neutrino electron interactions on argon for enegies less than \SI{50}{MeV}. This includes the energy range important for \gls{snb} neutrinos and also solar 8--boron neutrinos}

\newduneabbrev{es}{ES}{elastic scattering}{Events in which a neutrino
elastically scatters off of another particle}

\newduneabbrev{cno}{CNO}{carbon nitrogen oxygen}{The CNO cycle (for carbon-nitrogen-oxygen) is one of the two known sets of fusion reactions by which stars convert hydrogen to helium, the other being the proton-proton chain reaction (pp-chain reaction). In the CNO cycle, four protons fuse, using carbon, nitrogen, and oxygen isotopes as catalysts, to produce one alpha particle, two positrons and two electron neutrinos}

\newduneabbrev{sdwf}{SDWF}{South Dakota Warehouse Facility}{Warehousing operations in South Dakota responsible for receiving LBNF and DUNE goods and coordinating shipments to the access shaft (Ross Shaft) at \gls{surf}}

\newduneabbrev{wms}{WMS}{warehouse management system}{Commercial software package used to track shipments and interface to freight forwarders. This includes a database for shipping}

\newduneabbrev{dcdb}{DCDB}{DUNE construction database}{Database used by DUNE to track the history and testing of all parts of each \gls{detmodule}}

\newduneabbrev{aup}{AUP}{acceptance for use and possession}{Required for beneficial occupancy of the underground areas at \gls{surf} for \gls{lbnf} and \gls{dune}}

\newduneabbrev{bms}{BMS}{building management system}{A system provided by the \gls{cf} to manage the utility (cooling, ventilation, power, etc.) and fire/life safety systems. Separate systems are provided at \gls{surf} and at \gls{fnal}} 

\newduneabbrev{fls}{FLS}{fire and life safety system}{Fire and life safety; systems designed with \gls{cf} to meet building/safety code compliance for safe facilities at \gls{surf} and at \gls{fnal}}

\newduneabbrev{sno}{SNO}{Sudbury Neutrino Observatory}{The Sudbury Neutrino Observatory was a detector built 6800 feet under ground, in INCO's Creighton mine near Sudbury, Ontario, Canada. SNO was a heavy-water Cherenkov detector designed to detect neutrinos produced by fusion reactions in the sun}

\newduneword{sk}{Super-Kamiokande}{Experiment sited in the Kamioka-mine, Hida-city, Gifu, Japan that uses a large water Cherenkov detector to study neutrino properties through the observation of solar neutrinos, atmospheric neutrinos and man-made neutrinos}

\newduneabbrev{id}{ID}{inner diameter}{Inner diameter of a tube}

\newduneabbrev{od}{OD}{outer diameter}{Outer diameter of a tube}

\newduneabbrev{rms}{RMS}{root mean square}{The square root of the arithmetic mean of the squares of a set of values, used as a measure of the typical magnitude of a set of numbers, regardless of their sign}

\newduneabbrev{orc}{ORC}{operational readiness clearance}{Final safety approval prior to the start of operation}

\newduneabbrev{gsc}{GSC group}{global safety coordination group}{DUNE group that evaluates applicable codes and standards, including international code equivalency, for the design, assembly, and installation of the \gls{fd}}

\newduneabbrev{ha}{HA}{hazard analysis}{A first step in a process to assess risk; the result of hazard analysis is the identification of the hazards present for a task or process}
\newduneword{har}{HAR}{hazard analysis report}

\newduneabbrev{tap}{TAP}{trip action plan}{A document required for any trip by a worker to the underground area at \gls{surf}, per that site's access control program; it describes the work to be accomplished during the trip} 

\newduneword{em}{EM}{emergency management}
\newduneword{ert}{ERT}{emergency response team}

\newduneabbrev{ndk}{NDK}{nucleon decay}{The hypothetical, baryon number violating decay of a proton or a bound neutron into lighter particles}

\newduneabbrev{emi}{EMI}{electromagnetic interference}{Disturbance generated by an external source that affects an electrical circuit by electromagnetic induction, electrostatic coupling, or conduction}

\newduneabbrev{pe}{PE}{photoelectron}{An electron ejected from the surface of a material by the photoelectric effect}

\newduneabbrev{spe}{SPE}{single photoelectron}{A single photoelectron}

\newduneabbrev{fwhm}{FWHM}{full width at half maximum}{Width of a distribution measured between those points at which the distribution is equal to half of its maximum amplitude}

\newduneabbrev{gdml}{GDML}{geometry description markup language}{An application-independent, geometry-description format based on XML}

\newduneabbrev{xml}{XML}{extensible markup language}{A markup language that defines a set of rules for encoding documents in a format that is both human-readable and machine-readable}

\newduneabbrev{crt}{CRT}{cosmic-ray tagger}{Detector external to the TPC designed to tag TPC-traversing cosmic ray particles}

\newduneabbrev{sn}{SN}{supernova}{Event that occurs upon the death of certain types of stars}

\newduneabbrev{wg}{WG}{working group}{A group of persons working together to achieve specified goals}

\newduneabbrev{ctsf}{CTSF}{coating, testing and storage facility}{A facility where the the \gls{dp} photon detectors will be coated, tested, and stored} 

\newduneword{rucio}{Rucio}{Data management system originally developed
by \gls{atlas} but now open-source and shared across \gls{hep}}
\newduneabbrev{doma}{DOMA}{data organization, management, and access}{data organization, management, and access efforts through the \gls{hsf}}

\newduneabbrev{hsf}{HSF}{HEP Software Foundation}{A foundation that facilitates cooperation and common efforts in high energy physics software and computing internationally} 

\newduneabbrev{wlcg}{WLCG}{Worldwide LHC Computing Grid}{Worldwide LHC
Computing Grid}
\newduneabbrev{osg}{OSG}{Open Science Grid}{Open Science Grid}
\newduneabbrev{sci}{SCI}{Scientific Computing Infrastructure}{Proposed
extension of the infrastructure component of \gls{wlcg} to other
experiments}
\newduneabbrev{csc}{CSC}{computing and software consortium}{DUNE
computing and software consortium}

\newduneword{dirac}{DIRAC}{Computing workflow management designed for LHCb and now used by many HEP experiments}

\newduneword{frp}{FRP}{fiber-reinforced plastic}
\newduneabbrev{hdpe}{HDPE}{high-density polyethylene}{A thermoplastic polymer made from petroleum commonly used to make plastic bottles}
\newduneword{hvps}{HVPS}{\gls{hv} power supply}
\newduneword{aisi}{AISI}{American Iron and Steel Institute}
\newduneword{ific}{IFIC}{Instituto de Fisica Corpuscular (in Valencia, Spain)}
\newduneabbrev{rsds}{RSDS}{radioactive source deployment system}{Proposed calibration system based on the deployment of
radioactive sources inside the \gls{dune} cryostat}
\newduneword{2p2h}{2p2h}{two particle, two hole}
\newduneabbrev{duneprism}{DUNE-PRISM}{DUNE Precision Reaction-Independent Spectrum Measurement}{a mobile near detector that can perform measurements over a range of angles off-axis from the neutrino beam direction in order to sample many different neutrino energy distributions}
\newduneword{arcube}{ArgonCube}{The name of the core part of the \gls{dune} \gls{nd}, a \gls{lartpc}}

\newduneabbrev{citf}{CITF}{cryogenic instrumentation test facility}{A facility at \gls{fnal} with small ($<$\SI{1}{ton}) to intermediate ($\sim$\SI{1}{ton}) volumes of instrumented, purified TPC-grade \gls{lar}, used for testing devices intended for use in \gls{dune}}

\newduneabbrev{3dst}{3DST}{3D scintillator tracker}{The core part of the \threed projection scintillator tracker spectrometer in the near detector conceptual design}
\newduneabbrev{3dsts}{3DST-S}{3D scintillator tracker spectrometer}{The \threed projection scintillator tracker spectrometer  in the near detector conceptual design}
\newduneabbrev{mpd}{MPD}{multi-purpose detector}{A component of the near detector conceptual design; it is a magnetized system consisting of a \gls{hpgtpc} and a surrounding \gls{ecal}}
\newduneabbrev{hpg}{HPG}{high-pressure gas}{gas at high pressure to be used in a \gls{hpgtpc}} 
\newduneabbrev{hpgtpc}{HPgTPC}{high-pressure gaseous argon TPC}{A \gls{tpc} filled with gaseous argon; a possible component of the \gls{dune} \gls{nd}}

\newduneword{src}{SRC}{short-range correlated nucleon-nucleon interactions}
\newduneword{larpix}{LArPix}{ \gls{asic} pixelated charge readout for a \gls{tpc} }
\newduneword{arclt}{ArCLight}{a light detector for the \gls{arcube} effort}
\newduneword{fhc}{FHC}{forward horn current ($\numu$ mode)}
\newduneword{rhc}{RHC}{reverse horn current (\numubartonumubar mode)}
\newduneword{mwpc}{MWPC}{multi-wire proportional chamber}
\newduneword{na61}{NA61}{CERN hadron production experiment}
\newduneword{pdnd}{ProtoDUNE-ND}{a prototype \gls{dune} \gls{nd}}

\newduneabbrev{roc}{ROC}{readout chamber}{readout chamber for gaseous argon \gls{tpc}}
\newduneabbrev{iroc}{IROC}{inner readout chamber}{inner (radial) readout chamber for gaseous argon \gls{tpc}}
\newduneabbrev{oroc}{OROC}{outer readout chamber}{outer (radial) readout chamber for gaseous argon \gls{tpc}}

\newduneword{lux}{LUX}{Large Underground Xenon (LUX) dark matter detector at \gls{surf} }

\newduneword{mjdemo}{Majorana Demonstrator}{Experiment sited at \gls{surf} that  seeks to determine whether neutrinos are their own antiparticles}

\newduneword{lz}{LZ}{Experiment sited at \gls{surf} that  seeks to detect faint interactions between galactic dark matter and regular matter}

\newduneword{mu2e}{Mu2e}{An experiment sited at \gls{fnal} that searches for charged-lepton flavor violation and seeks to discover physics beyond the \gls{sm}}

\newduneword{pdsp2}{ProtoDUNE-SP-II}{A second test run in the single-phase ProtoDUNE test stand at CERN, acting as a validation of the final
single-phase detector design}

\newduneword{osha}{OSHA}{Occupational Safety and Health Administration (USA Department of Labor) formed by the Occupational Safety and Health Act of 1970}
\newduneabbrev{pns}{PNS}{pulsed neutron source}{Calibration system based
on neutron capture gamma showers spread out in the whole detector}

\newduneabbrev{fv}{FV}{fiducial volume}{The detector volume within the \gls{tpc} that is selected for physics analysis through cuts on reconstructed event position}

\newduneword{p6}{P6}{software framework used to plan and status the resource-loaded schedule of activities associated with \gls{usproj}}

\newduneabbrev{evms}{EVMS}{earned value management system}{Earned Value Management is a systematic approach to the integration and measurement of cost, schedule, and technical (scope) accomplishments on a project or task. It provides both the government and contractors the ability to examine detailed schedule information, critical program and technical milestones, and cost data (text from the US DOE); the EVMS is a system that implements this approach}

\newduneword{core}{CORE}{CORE contributions are in either monetary units or labor hours. They can be technical components for the facility or experiment and the effort of the staff needed to produce, install, and test them;  major facilities for the experiment; or other products and services relevant for the completion of the facility or experiment} 

\newduneabbrev{ahj}{AHJ}{Authority Having Jurisdiction}{An organization, office, or individual responsible for enforcing the requirements of a code or standard, or for approving equipment, materials, an installation, or a procedure (\gls{osha})}
\newduneword{cte}{CTE}{coefficient of thermal expansion}

\newduneabbrev{opc}{OPC}{open platform communications}{Open platform communications is a series of standards and specifications for industrial telecommunication} 
\newduneword{scada}{SCADA}{supervisory control and data acquisition}
\newduneword{ln}{LN$_2$}{liquid nitrogen}
\newduneabbrev{lapd}{LAPD}{Liquid Argon Purity Demonstrator}{Cryostat at \gls{fnal}  for long-term studies requiring a large volume of argon}

\newduneabbrev{pab}{PAB}{Proton Assembly Building}{Home of several \gls{lar} facilities at \gls{fnal} }
\newduneword{hep}{HEP}{high energy physics}
\newduneword{cms}{CMS}{Compact Muon Solenoid experiment; one of two general-purpose detectors at the \gls{lhc}. }
\newduneword{alice}{ALICE}{A Large Ion Collider Experiment, at CERN}
\newduneword{gpib}{GPIB}{general purpose interface bus}

\newduneabbrev{pfparticle}{PFParticle}{particle flow particle}{Each of the individual reconstructed particles in the hierarchy (or particle flow) describing the reconstructed event interaction}

\newduneabbrev{mcparticle}{MCParticle}{Monte Carlo Particle}{Individual true simulated particle}
\newduneword{au}{AU}{astronomical unit}
\newduneword{nufit}{NuFIT 4.0}{The NuFIT 4.0 global fit to neutrino oscillation data}

\newduneword{uhv}{UHV}{ultra high vacuum}
\newduneword{lps}{LPS}{laser positioning system}

\newduneword{unicamp}{UNICAMP}{University of Campinas, Sao Paulo, Brazil}
 
\newduneabbrev{fbk}{FBK}{Fondazione Bruno Kessler}{FBK is a research non-profit entity in Trento, Italy that partners in the development of technology with applications in various fields including High Energy Physics}

\newduneabbrev{enob}{ENOB}{effective number of bits}{The effective number of bits is a measure of the dynamic range of an \gls{adc} and its associated circuitry. The resolution of an \gls{adc} is specified by the number of bits used to represent the analog value, in principle giving 2N signal levels for an N-bit signal. However, all real \gls{adc} circuits introduce noise and distortion. ENOB specifies the resolution of an ideal \gls{adc} circuit that would have the same resolution as the circuit under consideration}
\newduneabbrev{sndr}{SNDR}{signal to noise and distortion ratio}{Also known as SINAD. Ratio of the \gls{rms} signal amplitude to the mean value of the root-sum-square of all other spectral components, including harmonics, but excluding \gls{dc} levels. It is a good indication of the overall dynamic performance of an \gls{adc} because it includes all components which make up noise and distortion}
\newduneabbrev{sfdr}{SFDR}{spurious free dynamic range}{Spurious free dynamic range is the ratio of the \gls{rms} value of the signal to the \gls{rms} value of the worst spurious signal regardless of where it falls in the frequency spectrum. The worst spur may or may not be a harmonic of the original signal}
\newduneabbrev{thd}{THD}{total harmonic distortion}{Total harmonic distortion is the ratio of the \gls{rms} value of the fundamental signal to the mean value of the root-sum-square of its harmonics} 
\newduneword{tvs}{TVS}{transient voltage suppression}

\newduneword{riskprob}{risk probabilities}{The risk probability, after taking into account the planned mitigation activities, is ranked as 
 L (low $<\,$\SI{10}{\%}), 
M (medium \SIrange{10}{25}{\%}), or 
H (high $>\,$\SI{25}{\%}). 
The cost and schedule impacts are ranked as 
L (cost increase $<\,$\SI{5}{\%}, schedule delay $<\,$2 months), 
M (\SIrange{5}{25}{\%} and 2--6 months, respectively) and 
H ($>\,$\SI{20}{\%} and $>\,$2 months, respectively)}

\newduneabbrev{lbls}{LBLS}{laser beam location system}
{Auxiliary calibration system providing an independent location measurement of the ionization laser beams direction}

\newduneabbrev{lsst}{LSST}{Large Synoptic Survey Telescope}{8.4 m telescope with 3.2G-pixel camera that will start taking data in 2023}
\newduneabbrev{ska}{SKA}{Square Kilometer Array}{International radio telescope array planned to start data-taking in 2027}
\newduneabbrev{hyperk}{HyperK}{Hyper Kamiokande}{260 kt water Cerenkov neutrino detector to begin construction at Kamiokande in 2020}
\newduneword{lhcb}{LHCb}{LHC experiment dedicated to forward physics}
\newduneword{belleii}{Belle II}{B-factory experiment now running at KEK}

\newduneabbrev{ldm}{LDM}{light-mass dark matter}{Refers to dark matter particles with mass values much lower than the electroweak scale, specifically below the 1~GeV level}
 
\newduneabbrev{bnv}{BNV}{baryon-number violating}{Describing an interaction where \gls{baryonnumber} is not conserved}

\newduneword{bugey}{Bugey}{Neutrino experiment that operated at the Bugey nuclear power plant in France}

\newduneword{minosplus}{MINOS$+$}{The successor to the \gls{minos} experiment, utilizing the same detectors and beam line, but operating at higher beam energy tune than \gls{minos}, parasitic with \gls{nova}}

\newduneword{baryonnumber}{baryon number}{A quantity expressing the total number of baryons in a system minus the number of antibaryons}

\newduneword{np04}{NP04}{The CERN North Area in \gls{ehn1} intersected by the \gls{h4} hadron beamline,  the location of  \gls{pdsp} and \gls{pdsp2}; also used to refer to the 800\,t cryostat in this area}

\newduneword{np02}{NP02}{The CERN North Area in \gls{ehn1} intersected by the \gls{h2} hadron beamline, the location of  the 800\,t cryostat used for \gls{pddp} and for SP vertical drift tests and prototypes; also used to refer to the 800\,t cryostat in this area}

\newduneword{h4}{H4}{CERN North Area hadron beamline used for \gls{pdsp} and \gls{pdsp2}}  

\newduneword{h2}{H2}{CERN North Area hadron beamline used for \gls{pddp} and SP vertical drift prototypes and demonstrators}  

\newduneword{ua1}{UA1}{UA1 (Underground Area 1) was a particle detector at \gls{cern}'s  Super Proton Synchrotron (SPS). It ran from 1981 until 1990, when the SPS was used as a proton-antiproton collider, searching for traces of W and Z particles in collisions. (CERN) The UA1 dipole magnet was reused in the NOMAD experiment and currently provides the magnetic field for the \gls{t2k} ND280 detector}

\newduneword{ssc}{SSC}{The Superconducting Super Collider was to be a huge underground ring complex beneath the area near Waxahachie, Texas, USA, that would have been the world’s most energetic particle accelerator. It was begun in 1990, but canceled by the U.S. Congress in 1993 (scientificamerican.com Oct 2013)}

\newduneword{daphne}{DAPHNE}{Detector electronics for Acquiring PHotons from NEutrinos is a custom-developed warm front-end waveform digitizing electronics module derived from the readout system developed at \gls{fnal}  for the Mu2e experiment}
 
\newduneword{nersc}{NERSC}{National Energy Research Computing Facility at \gls{lbnl}}

\newduneword{integoff}{integration project}{The \gls{doe} project element that organizes the onsite teams responsible for coordinating far detector installation and detector-facility integration activities at \gls{surf} as well as near detector installation activities at \gls{fnal}.  The integration project office includes overall LBNF/DUNE systems engineering, compliance and review offices} 

\newduneabbrev{sma}{SMA}{SubMiniature version A}{Connector interface for coaxial cables
with a screw-type coupling mechanism}

\newduneword{kloe}{KLOE}{KLOE is a $e^+ e^-$ collider detector spectrometer operated at DAFNE,  the $\phi$-meson factory at Frascati, Rome.  In DUNE it will consist of a \SI{26}{cm} Pb+scintillating fiber \gls{ecal} surrounding a cylindrical open detector region that is  \SI{4.00}{m} in diameter and \SI{4.30}{m} long.  The \gls{ecal} and detector region are embedded in a \SI{0.6}{T} magnetic field created by a \SI{4.86}{m} diameter superconducting coil and a \SI{475}{tonne} iron yoke}

\newduneabbrev{ro}{RO}{review office}{An office within the \gls{integoff} that organizes reviews} 

\newduneabbrev{doecd}{CD}{critical decision}{The U.S. DOE's Order 413.3B outlines a series of staged project approvals, each of which is referred to as a critical decision (CD)}

\newduneabbrev{lbnfspac}{LBNF/DUNE-US SPAC}{LBNF / DUNE-US Strategic Project Advisory Committee}{A committee charged by the host laboratory director to provide expert, independent advice on significant issues and strategies related to \gls{usproj} project organization, management, and risks} 

\newduneabbrev{sand}{SAND}{System for on-Axis Neutrino Detection}{The beam monitor component of the near detector that remains on-axis at all times and serves as a dedicated neutrino spectrum monitor}

\newduneword{4850l}{4850L}{The depth in feet (1480 m) of the access level for the DUNE underground area at SURF; called the ``4850 level''} 

\newduneword{apb}{APB}{authorship and publications board}
\newduneword{crb}{CRB}{collaboration resources board}
\newduneword{cube}{CuBe}{beryllium copper, used to make \gls{sphd} \gls{apa} readout planes}
\newduneword{drift}{drift}{(1) refers to electron drift under the influence of an electric field in a \gls{tpc}; (2) an excavated horizontal corridors (tunnels) in the underground areas at \gls{surf}}
\newduneword{exposure}{exposure}{The integrated detector fiducial mass times beam intensity; it is proportional to the number of interactions and is used to normalize cross sections in a data sample}
\newduneword{fr4}{FR-4}{Flame-retardant fiberglass-reinforced epoxy resin laminate used in making PCBs and other detector components}
\newduneword{g10}{G-10}{Non-flame-retardant fiberglass-reinforced epoxy resin laminate used in making PCBs}
\newduneword{kapton}{Kapton}{A polyimide plastic film that is stable over a broad range of temperatures and is resistant to radiation damage}
\newduneword{shaft}{shaft}{A vertical excavation at \gls{surf} connecting with the surface}
\newduneword{winze}{winze}{A vertical excavation at \gls{surf} connecting two drifts, not connecting to the surface}
\newduneword{ib}{IB}{institutional board; all institutions participating in DUNE are represented on this board}
\newduneword{irb}{IBR}{institutional board representative}
\newduneword{sc}{SC}{depending on context, either speakers committee or scientific computing} 
\newduneword{sac}{SAC}{spokespersons advisory committee}
\newduneword{eoc}{EOC}{education and outreach committee}
\newduneword{indico}{Indico}{Web-based meeting organization tool}

\newduneabbrev{htc}{HTC}{High Throughput Computing}{Computing facilities typically consisting of large numbers of commodity servers as opposed to a single large machine. Best suited for running large numbers of independent jobs in parallel, these facilities are what is usually meant by ``grid computing''}
\newduneword{dcache}{dCache}{A distributed, highly scalable (multi-PB) storage system, usable as both a standalone system and as a high-speed frontend to a tape storage system (such as \gls{pnfs} at \gls{fnal} )}
\newduneabbrev{ifdhc}{IFDHC}{Intensity Frontier Data Handling Client}{A multi-protocol tool for data transfer and file delivery in jobs. It is able to automatically select transfer protocols based on source and destination characteristics}
\newduneabbrev{ifdh}{ifdh}{Intensity Frontier Data Handling}{The actual command invoked when using \gls{ifdhc}, on the command line, e.g. ifdh cp source\_file dest\_file}
\newduneabbrev{pnfs}{PNFS}{Pseudo Network File System}{A file system often used in large storage systems. Typically interaction is very similar to a regular NFS volume, but there can be some subtle and important differences}

\newduneword{nde}{NDE}{non-destructive evaluation} 
\newduneword{psv}{PSV}{pressure safety valve}
\newduneword{pickling}{pickling}{steel pickling and oiling is a metal surface treatment finishing process used to remove surface impurities such as rust and carbon scale from hot rolled carbon steel}

\newduneabbrev{tof}{ToF}{time of flight}{The time a particle takes to fly between two visible interactions observed in the detector. If combined with the distance traveled by the particle, for example a neutron, it can be used for energy reconstruction}

\newduneword{pep4}{PEP-4}{TPC for the Positron Electron Project 4 Collider at Stanford}

\newduneword{ndlar}{ND-LAr}{\gls{lartpc} component of the near detector based on \gls{arcube} technology}

\newduneword{ndgar}{ND-GAr}{component of the near detector with a core gaseous argon \gls{tpc} surrounded by an \gls{ecal} and a magnet}

\newduneword{sfgd}{SuperFGD}{Super Fine-Grained Detector (SuperFGD) is a 3D granular plastic scintillator detector that adopts the same technology as \dword{3dst}. It will be installed in the \dword{t2k} \dword{nd280} system . The \dword{3dst} design will inherit in large part from the SuperFGD detector}

\newduneword{nd280}{ND280}{Near Detector 280, is the \dword{t2k} magnetized near detector} 

\newduneword{fee}{FEE}{front-end electronics}

\newduneword{mpgd}{MPGD}{MicroPattern Gas Detectors}

\newduneword{rmm}{RMM}{Resistive MicroMegas}

\newduneabbrev{stv}{STV}{Single Transverse Variables}{Kinematical variables obtained by projecting the neutrino interaction onto the transverse plane}

\newduneabbrev{ccqe}{CCQE}{charged current quasielastic interaction} {An interaction where a neutrino scatters from a nucleon, producing a charged lepton and converting a neutron to a proton or vice versa}

\newduneword{sis}{SIS}{shallow inelastic scattering}

\newduneabbrev{res}{RES}{resonant scattering}{The mode of scattering where the target nucleon is excited to a resonant state and decays, typically producing one or more pions}

\newduneabbrev{hadw}{W}{invariant mass of the hadronic system}{Refers to the invariant mass of the hadronic system formed during the neutrino scatter}

\newduneabbrev{agky}{AGKY}{Andreopoulos-Gallagher-Kehayias-Yang}{A model for hadronization of non-resonant inelastic neutrino reactions used in \gls{genie}. At low invariant hadronic masses, typically less than 2.3\,GeV/c$^2$, it is a KNO-inspired empirical model anchored on several bubble chamber measurements of neutrino-induced shower characteristics. For invariant hadronic masses between 2.3 and 3.0\,GeV/c$^2$, the model transitions linearly to a \gls{genie}-tuned version of PYTHIA, which is also used for the simulation of events at higher invariant masses} 

\newduneabbrev{clas}{CLAS}{CEBAF Large Acceptance Spectrometer}{A nuclear and particle physics detector located in the experimental Hall B at Jefferson Laboratory in Newport News, Virginia, United States. It is used to study the properties of the nuclear matter by the collaboration of over 200 physicists. Of particular relevance is its study of electron interactions with nuclei, including argon} 

\newduneabbrev{e4nu}{e4nu}{Electrons for Neutrinos}{A collaboration dedicated to using JLab's electron-scattering data to deliver improved neutrino-nucleus cross sections} 

\newduneabbrev{ldmx}{LDMX}{Light Dark Matter eXperiment}{The LDMX detector concept consists of a small precision tracker, and electromagnetic and hadronic calorimeters, all with near $2\pi$ azimuthal acceptance from the forward beam axis out to $\sim40^\circ$ angle. This detector would be capable of measuring correlations among electrons, pions, protons, and neutrons in electron-nucleus scattering at exactly the energies relevant for DUNE physics} 

\newduneabbrev{mec}{MEC}{meson-exchange currents} {An nuclear effect wherein pairs or larger groups of nucleons within a nucleus are bound together through the exchange of pions or other mesons. Neutrinos and other particles can scatter from these correlated pairs}

\newduneword{imt}{IMT}{Intranuclear momentum transfer}

\newduneword{miniboone}{MiniBooNE}{The Mini Booster Neutrino Experiment,  at \gls{fnal} , was designed to fully explore the LSND result}

\newduneabbrev{tms}{TMS}{Muon Spectrometer}{A muon spectrometer for the Near Detector that will be installed for the initial running period of DUNE, before the \gls{mpd} detector component is ready} 

\newduneabbrev{cvmfs}{CVMFS}{CERN VM File System}{A distributed file system designed for scalable, high-performance distribution of software to interactive and batch computers} 

\newduneword{kerberos}{Kerberos}{A strong authentication system used by the computing resources at \gls{fnal} and \gls{cern}} 

\newduneabbrev{mrb}{MRB}{Multi Repository Build System}{A \gls{fnal}-developed build system based on \dword{cmake} that allows development and builds of code from multiple repositories} 

\newduneword{cmake}{Cmake}{CMake is an open-source, cross-platform family of tools designed to build, test and package software}

\newduneabbrev{ups}{UPS}{UNIX product support}{A software tool that sets up a consistent environment of versions of pre-installed products and their dependencies on UNIX-like platforms} 

\newduneabbrev{upd}{UPD}{UNIX product distribution}{A tool for uploading and downloading pre-built software products between local systems and centralized software distribution servers.  UPD is not frequently used on DUNE because newer tools are more convenient} 

\newduneabbrev{sso}{SSO}{single sign-on}{Used at \gls{fnal} to indicate that a group of services,  such as DocDB or the DUNE Wiki share common sign-in credentials and active sessions.  \gls{fnal}  services that say "Sign in with SSO username and password" mean to use your \gls{fnal} Services or federated username and password} 

\newduneabbrev{vo}{VO}{virtual organization}{A database containing a list of member names, certificate distinguishing information, and a list of permissions members have to access computing grid and data resources} 

\newduneabbrev{nas}{NAS}{network attached storage}{Disk storage that is available on computers but shared between them.  Relies on \gls{nfs} mounts rather than authenticated file transfer protocols.  Usually found on interactive servers to provide space for home directories, app and data storage} 

\newduneabbrev{nfs}{NFS}{network file system}{Industry-standard mechanism for mounting disks over a network.  Provides regular UNIX file and directory access} 

\newduneword{recombination}{recombination}{Electrons freed from Argon atoms will sometimes recombine with the positive argon ions, either the same ones from which they came or nearby ones.  Sometimes called ``quenching"} 

\newduneword{elife}{electron lifetime}{The attachment of electrons drifting through liquid argon to impurity molecules such as oxygen or water is parameterized by an exponential as a function of time with a time constant called the electron lifetime} 

\newduneword{gplane}{grid plane}{The uninstrumented plane of wires or electrodes on an anode plane  
facing the drift volume.   It shapes the signals and provides \gls{esd} protection} 

\newduneword{gmesh}{grounding mesh}{A metal mesh attached to the SP \gls{apa} frame between the collection-plane wires and the space inside the frame where the \gls{pd} modules are installed.  It provides electric field uniformity so the collection-plane wires all have similar fields around them} 

\newduneabbrev{bcr}{BCR}{baseline change request}{A DOE project change, part of the change management system process}

\newduneword{ncav}{North Cavern}{the location of two of the planned four DUNE far detector modules at \gls{surf}}

\newduneword{scav}{South Cavern}{the location of two of the planned four DUNE far detector modules at \gls{surf}}

\newduneword{semp}{SEMP}{systems engineering management plan}  

\newduneabbrev{tpcost}{TPC}{total project cost}{The DOE terminology for the total budget and contingency for the entire \gls{usproj} project from CD-0 to CD-4}  

\newduneabbrev{nsint}{NSI}{near site integration}{The scope of work at the near site for the \gls{integoff}} 

\newduneabbrev{opcost}{OPC}{other project costs}{The DOE project costs that support conceptual design, pre-operations commissioning, technical coordination, and power}

\newduneabbrev{moa}{MOA}{memorandum of agreement}{A project management methodology that documents an agreement between \gls{fnal} and the \gls{usproj}  Project for how \gls{fnal}  will support the project} 

\newduneabbrev{croc}{CROC}{central readout chamber}{central (radial) readout chamber for the ND \gls{gartpc}} 

\newduneabbrev{tki}{TKI}{transverse kinematic imbalance}{The imbalance among final-state particle momenta in the transverse plane to the neutrino direction; different aspects of the imbalance are sensitive to the detail of the nuclear effects in neutrino-nucleus interactions}

\newduneabbrev{iandi}{I\&I}{Integration and Installation}{One of the three project areas in the LBNF/DUNE-US \gls{doe} project, along with LBNF and DUNE-US} 

\newduneword{hmi}{HMI}{human-machine interface}

\newduneword{lhe}{LHe}{liquid helium}

\newduneword{echain}{energy chain}{mechanical machine elements used to carry and guide power to moving parts of machines or structures, as required for \gls{duneprism} to carry power, data, and utilities to and from each movable near detector component at any arbitrary position along its travel path}

\newduneabbrev{sphd}{FD1-HD}{horizontal drift detector module}{LArTPC design used in FD1 in which electrons drift horizontally to wire plane anodes (\glspl{apa}) that along with the front-end electronics are immersed in LAr}

\newduneabbrev{cru}{CRU}{charge-readout unit}{In the SP vertical drift design an assembly of the \glspl{pcbp} plus adapter boards; two to a \gls{crp}} 

\newduneword{mpv}{MPV}{most probable value}

\newduneword{pcbp}{PCB panel}{In the SP vertical drift design, one of four \glspl{pcb} of size 
1.5 $\times$ 1.7\.,m assembled into a \gls{cru}}

\newduneword{anodepln}{anode plane}{a planar array of charge readout devices covering an entire face of a detector module}

\newduneword{msps}{MSPS}{megasamples per second}

\newduneword{gpsdo}{GPSDO}{\gls{gps} disciplined oscillator}

\newduneword{tdaq}{TDAQ}{trigger and DAQ system} 

\newduneword{nios}{NIOS}{network identity operating system}

\newduneabbrev{corsika}{CORSIKA}{COsmic Ray SImulations for KAscade}{a program for detailed simulation of extensive air showers initiated by high-energy cosmic ray particles}

\newduneabbrev{scd}{SCD}{scientific computing division}{\gls{fnal}'s Scientific Computing Division}

\newduneabbrev{garsoft}{GArSoft}{Gaseous Argon Software}{A software toolkit similar to \gls{larsoft}, but targeted at the gaseous argon time projection chamber and calorimeter of \gls{ndgar}}

\newduneword{ndgarlite}{ND-GAr-Lite}{a temporary muon spectrometer consisting of the magnet and steel flux return of \gls{ndgar}, but with a simplified tracking chamber made with scintillating bars}

\newduneword{github}{GitHub}{a commercial web service providing code version management, storage,  and browsing via \gls{git}}

\newduneword{git}{git}{a distributed version-control system, commonly used to manage software}

\newduneword{xrootd}{xrootd}{a high-performance data system widely used in \gls{hep} to store and to distribute data to jobs.  It allows streaming of data}

\newduneabbrev{gpuaas}{GPUAAS}{GPU As A Service}{a technique that allows many non-GPU-enabled compute nodes to share a GPU resource by sending it work over the network and waiting for results to be returned}

\newduneword{sce}{SCE}{space charge effect}

\newduneabbrev{nest}{NEST}{noble element simulation technique}{Comprehensive simulation code modeling the excitation, ionization, and corresponding scintillation and electroluminescence processes in liquid noble elements}

\newduneword{bgr}{BGR}{A bandgap voltage reference is a circuit block in ASIC for providing stable reference voltages}

\newduneword{comsol}{COMSOL}{General-purpose simulation software based on advanced numerical methods (comsol.com)}

\newduneword{greyrm}{gray room}{ISO-8 clean room with a cleanliness level of 3.5M particles of 0.5 micron or less per cubic meter volume. ISO-8 clean rooms are referred to as grey rooms because at this level of cleanliness most standard clean room attire is not required.}

\newduneabbrev{pof}{PoF}{power-over-fiber}{a technology in which a fiber optic cable carries optical power, which is used as an energy source rather than, or as well as, carrying data; this allows a device to be remotely powered, while providing electrical isolation between the device and the power supply} 

\newduneword{ppc}{PPC}{photovoltaic power converter}
\newduneword{eelaser}{edge-emitting laser}{a laser in which  light is emitted from the edge of the substrate}
\newduneword{peek}{PEEK}{Polyether ether ketone, a colorless organic thermoplastic polymer}

\newduneword{fsm}{Finite State Machine}{a mathematical model of computation; it is an abstract machine that can be in exactly one of a finite number of states at any given time} 

\newduneword{icecube}{IceCube}{South Pole Neutrino Observatory}
\newduneword{pde}{PDE}{photon detection efficiency} 

\newduneword{pmos}{PMOS}{(see \gls{cmos}; PMOS is constructed with the p-type source and drain and an n-type substrate}

\newduneabbrev{poe}{PoE}{power-over-Ethernet}{systems that pass electric power along with data on twisted-pair Ethernet cabling} 

\newduneabbrev{daqdts}{DTS}{DUNE timing and synchronization subsystem}{The portion of the \gls{daq} that provides for timing and synchronization to various detector systems}

\newduneword{tai}{TAI}{International Atomic Time}

\newduneword{larzic}{LARZIC}{The cryogenic amplifier \dword{asic} that is the principal component of the FD2 top drift \dword{fe} analog cards}

\newduneword{dpdfd}{DPDFD}{Deputy Project Director for far detectors}

\newduneword{bsws}{BSWS}{bearing sensor wire compression seal}

\newduneword{ly}{light yield}{detected photons per unit deposited energy}

\newduneword{mtbf}{MTBF}{mean time between failures}

\newduneword{ingaas}{InGaAs}{Indium gallium arsenide is a room-temperature semiconductor commonly used as a high-speed, high-sensitivity photodetector for optical fiber telecommunications}

\newduneword{intlproj}{LBNF/DUNE Construction Project}{The international project to design and build the facilities and detectors for the \gls{lbnf-dune}; it includes the \gls{usproj} and projects at multiple international partners to manage the contributions from non-US institutions and funding agencies to design, build, and install the detector components}

\newduneword{ddmp}{DUNE Management Plan}{DUNE Management Plan}
  
\newduneword{pd2hd}{ProtoDUNE-II-HD}{The second ProtoDUNE for the HD design, using NP04 to test installation, integration, and detector component performance}

\newduneword{mod0}{Module 0}{The final pre-production instance of a detector; for the DUNE \glspl{detmodule}, the \glspl{protodune2} in the 800\,t cryostats in \gls{np02} and \gls{np04} serve this purpose}

\newduneword{vdmod0}{FD2 Module 0}{The final pre-production \gls{protodune} instance for the DUNE SP vertical drift \gls{detmodule},  it will use the 800\,t cryostat in \gls{np02} }

\newduneword{hdmod0}{FD1-HD Module 0}{The final pre-production \gls{protodune} instance for the DUNE \gls{sphd} \gls{detmodule},  it will use the 800\,t cryostat in \gls{np04} }

\newduneword{fsii}{FSII}{far site integration and installation}
\newduneword{fdc}{FDC}{Far Detector and Cryogenics Subproject}

\newduneabbrev{co}{CO}{compliance office}{a team of engineers from multiple partners that provides clear direction for designing and constructing components that will be used during integration}

\newduneword{fscfbsi}{FSCF-BSI}{\gls{usproj} subproject for far site conventional facilities, building and site infrastructure}

\newduneabbrev{poms}{POMS}{Production Operations Management System}{A workflow management system available for all DUNE users to submit and monitor grid jobs as well as view job log files}

\newduneabbrev{fifemon}{FIFEMON}{FabrIc for Frontier Experiments MONitoring}{Comprehensive suite of job and storage monitoring information available for most \gls{fnal} experiments, including DUNE}

\newduneword{larg4}{LArG4}{LArG4 is the replacement for LArsim/LArG4. LArG4 is based on \gls{artg4tk}}

\newduneword{artg4tk}{artg4tk}{artg4tk provides a general interface between \gls{geant4} and \gls{art}}

\newduneword{tde}{TDE}{top detector electronics}

\newduneword{bde}{BDE}{bottom detector electronics}

\newduneword{sigproc}{Signal Processing}{The goal of TPC signal processing is to reconstruct the distribution of ionization electrons arriving at wire planes from the digitized TPC waveform}

\newduneabbrev{compass}{COMPASS}{
Common Muon and Proton Apparatus for Structure and Spectroscopy }{a multipurpose experiment at \gls{cern}’s Super Proton Synchrotron (SPS)}

\newduneword{vd}{vertical drift}{single-phase, vertical drift \gls{lartpc} technology}

\newduneword{hd}{horizontal drift}{single-phase,  horizontal drift \gls{lartpc} technology}

\newduneabbrev{esnet}{ESnet}{Energy Sciences Network}{The \gls{doe}'s dedicated science network} 

\newduneword{project.py}{project.py}{XML-based job configuration system developed by the \gls{microboone} collaboration} 

\newduneabbrev{ppfx}{PPFX}{Package to Predict the FluX}{\gls{fnal}-supported package that implements hadron production corrections to geant4 simulations and propagates uncertainties for the NuMI  and LBNF beam lines} 

\newduneabbrev{ml}{ML}{Machine Learning}{Machine Learning}

\newduneword{datalake}{data lake}{The not-yet-realized concept of a storage service with multiple levels of quality of service in which the end user can access data without knowing the data's source location} 

\newduneabbrev{metacat}{MetaCat}{MetaCat}{Metadata Catalog, a modern replacement for the file description portion of the sam metadata catalog} 

\newduneword{g4lbnf}{g4lbnf}{LBNF neutrino beamline simulation program} 

\newduneword{Geant4Reweight}{Geant4Reweight}{Framework for evaluating and propagating hadronic interaction uncertainties in Geant4} 

\newduneword{geantnet}{G\'EANT}{G\'EANT interconnects Europe's national research and education networking (NREN) organizations with the high bandwidth, high speed and highly resilient pan-European backbone.
}

\newduneabbrev{nren}{NREN}{National Research and Education Network}{National level research computing network infrastructure} 

\newduneabbrev{vrf}{VRF}{Virtual Routing and Forwarding}{Networking overlays that provide  a logical routing infrastructure  that allows flexible traffic engineering} 

\newduneword{samvalue}{value}{A generic quantity describing a file in the \dword{sam} data catalog} 

\newduneword{samparameter}{parameter}{A user or experiment described quantity describing a file in the \dword{sam} data catalog} 

\newduneword{samproject}{SAM-project}{A server process running centrally that maintains a predefined list of files and delivers information about  their locations when asked by distributed processes. The project tracks success and failure of file processing} 

\newduneword{samconsumer}{SAM-consumer}{A client process that requests information about file locations from a \dword{samproject}, process the file and reports success or failure} 

\newduneword{samdataset}{SAM-dataset}{ A dynamic collection of files defined by queries to the \dword{sam} data catalog} 
\newduneword{metacatdataset}{MetaCat-dataset}{ A fixed but mutable collection of files defined by queries to the \dword{metacat} data catalog} 

\newduneword{samsnapshot}{SAM-snapshot}{A fixed collection of files corresponding to a \dword{sam} \dword{samdataset} at a particular point in time} 

\newduneabbrev{mql}{MQL}{\dword{metacat} Query Language} {A query language which supports queries of the \dword{metacat} data catalog, including parentage and logical functions such as union, join and subtraction} 

\newduneword{pdhd}{ProtoDUNE-HD}{\gls{protodune} with horizontal drift technology.  This refers to the \gls{sp} \gls{apa}-based prototype to run in \gls{np04} (in the  \gls{pd2} phase)} 

\newduneword{pdvd}{ProtoDUNE-VD}{ProtoDUNE with vertical drift technology.  This refers to the \gls{crp}-based prototype to run in \gls{np02}  (in the  \gls{pd2} phase)} 

\newduneword{pd2}{ProtoDUNE-II}{\gls{protodune} test runs at CERN in 2022-2023; also called \gls{mod0}}

\newduneabbrev{ucondb}{uconDB}{Unstructured Conditions Database}{Unstructured conditions database developed for \gls{fnal} fixed target experiments} 

\newduneabbrev{json}{JSON}{JavaScript Object Notation}{Open standard data interchange format that uses pair-value pairs and maps well onto python data formats such as tuples and lists} 

\newduneabbrev{dqmdb}{DQMDB}{Data Quality and Monitoring Database}{Database storing the results of data-quality monitoring} 

\newduneword{db}{DB}{database}

\newduneabbrev{vm}{VM}{virtual machine}{Emulator of a physical computer that allows multiple users to configure different operating systems  while sharing physical hardware}

\newduneword{enstore}{Enstore}{A mass storage system developed by \gls{fnal} that provides distributed access and management of data stored on tapes} 

\newduneword{stashcache}{StashCache}{A distributed caching federation that enables opportunistic users to utilize nearby opportunistic storage} 

\newduneabbrev{hdf5}{HDF5}{Hierarchical Data Format}{Data format widely used in \gls{ml}} 

\newduneword{glideinwms}{GlideinWMS}{A system of submitting pilot jobs to grid computing sites, inside of which user jobs run, presenting a uniform setup across many different sites} 

\newduneword{mars}{MARS}{The MARS code system is a set of \gls{mc} programs for detailed simulation of coupled hadronic and electromagnetic cascades, with heavy ion, muon and neutrino production and interactions} 

\newduneabbrev{fife}{FIFE}{Fabric for Intensity Frontier Experiments}{\gls{fnal} computing infrastructure for \gls{if} Experiments} % \

\newduneword{tr}{trigger record}{A data record produced by the \gls{dune} \gls{daq} system.  A Trigger Record can contain multiple interaction ``events'' or none} 

\newduneword{postgres}{Postgres}{also known as PostgreSQL, Postgres is a free and open-source relational database management system used extensively for databases in \gls{hep}}

\newduneword{fhicl}{FHICL}{Fermilab Hierarchical Configuration Language; a standard configuration language for the storage, communication, and manipulation of scientific parameter sets}

\newduneword{api}{API}{Application Programming Interface}

\newduneword{hwdb}{HWDB}{hardware database}   

\newduneword{s3}{S3}{The Amazon cloud-based commercial storage service} 

\newduneword{openstack}{OpenStack}{An open source cloud software used to deploy instances of containers}

\newduneabbrev{cric}{CRIC}{Computing Resource Information System}{a framework providing a centralized (and flexible) way to describe which resources are being used by the experiment and how}

\newduneabbrev{ccb}{CCB}{Computing Contributions Board}{a board made up of institutional representatives for larger countries and laboratories. It meets annually to negotiate collaboration contributions to computing infrastructure} 

\newduneword{garfield}{Garfield}{A simulation program developed at \gls{cern} for gaseous detectors } 

\newduneword{datatier}{data tier}{Differing data types produced in a processing sequence, for example, raw data, reconstructed, derived analysis sample, histograms} 

\newduneabbrev{gpu}{GPU}{Graphical Processing Unit}{Specialized computing hardware optimized for image processing} 

\newduneabbrev{gpvm}{dunegpvm}{DUNE General Purpose Virtual Machine}{Centrally managed virtual Linux systems at \gls{fnal} with access to network attached and \gls{pnfs} storage. Used for small-scale data analysis and algorithm development} 

\newduneword{madx}{MAD-X}{framework that provides the de facto standard scripting language to describe particle accelerators, simulate beam dynamics, and optimize beam optics at \gls{cern}}

\newduneabbrev{cpu}{CPU}{Central Processing Unit}{A computing processor, when used as a unit of processing; generally refers to a single core} 

\newduneabbrev{fft}{FFT}{Fast Fourier Transform}{An algorithm that calculates the frequency components of a time-domain waveform in a computationally efficient manner} 

\newduneword{grain}{GRAIN}{In the \gls{sand} detector, a small cryostat containing \gls{lar} installed upstream of the straw-tube tracker inside the \gls{ecal}}

\newduneword{voms}{VOMS}{Virtual Organization Membership Service (in grid computing)}

\newduneword{cry}{CRY}{A cosmic ray shower library and software tool}

\newduneword{jwt}{JWT}{\gls{json} web token}

\newduneabbrev{ferry}{FERRY}{Frontier Experiments RegistRY}{a central database that keeps track of all scientific computing users at \gls{fnal}, the experiments and groups of which they are members, and the various capabilities they are allowed}

\newduneabbrev{perfsonar}{perfSONAR}{performance Service-Oriented Network monitoring ARchitecture}{a network measurement toolkit designed to provide federated coverage of paths and to help establish end-to-end usage expectations}

\newduneword{ipv6}{IPv6}{the most recent version of the Internet communications protocol that provides an identification and location system for computers on networks and routes traffic across the Internet}

\newduneword{monit}{MONIT}{a suite of tools using open source technology used in the monitoring of the \gls{cern} IT data center and \gls{wlcg} infrastructure. }

\newduneword{garana}{GArAna}{provides a facility for making an analysis ntuple from information stored in \gls{garsoft} data products}

\newduneword{ci}{CI}{continuous integration}

\newduneword{datadis}{Data Dispatcher}{communicates between running processing jobs and the data delivery systems.  It provides file location information and basic bookkeeping on file access and transfers} 

\newduneabbrev{castor}{CASTOR}{CERN Advanced STORage manager}{a hierarchical storage system with tape and disk developed at \gls{cern}. It is being replaced by \gls{cta}} 

\newduneabbrev{cta}{CTA}{CERN Tape Archive}{a hierarchical storage system with tape and disk developed at \gls{cern}. It is replacing \gls{castor}} 

\newduneabbrev{sonic}{SONIC}{Services for Optimized
Network Inference on Coprocessors}{Framework for implementing machine learning algorithms on co-processors developed by \gls{fnal}}

\newduneword{jobsub}{JobSub}{The \gls{fnal} \gls{if} job submission client that supports user submission of complex workflows to HT-Condor}

\newduneword{hepcloud}{HEPCloud}{routes jobs to local or remote computing resources based on the policy for a particular experiment, workflow requirements, and cost and efficiency of accessing the various resources. It expands the resources available to include \gls{hpc} centers and commercial cloud resources}

\newduneword{wmsmon}{WMS monitoring}{Workflow Management System Monitoring}

\newduneword{redmine}{Redmine}{an open source code repository and issue tracking tool that was historically used by many of the \gls{fnal} general computing projects and \gls{if} experiments} 

\newduneabbrev{if}{IF}{Intensity Frontier}{refers to \gls{hep} experiments, particularly in the U.S., that rely on high luminosity instead of high energy for discovery.  Includes B-physics, neutrino, and muon experiments}

\newduneword{slack}{Slack}{a commercial business communication platform} 

\newduneword{servicenow}{ServiceNow}{a commercial enterprise workflow system used by \gls{fnal} for formal issue tracking and IT workflows such as user account preparation} 

\newduneabbrev{rse}{RSE}{Rucio Storage Element}{A storage element that is known to the DUNE \gls{rucio} instance} 

\newduneabbrev{coffea}{COFFEA}{Columnar Object Framework For Effective Analysis}{Columnar data analysis frame\-work developed at \gls{fnal}} 

\newduneword{nuisance}{Nuisance}{An open source C++ framework for studying neutrino interaction cross sections} 

\newduneabbrev{highland}{HighLAND}{High Level Analysis at a Neutrino Detector}{Analysis framework developed by the \gls{t2k} collaboration} 
 
\newduneabbrev{raii}{RAII}{Resource Acquisition Is Initialization}{a C++ programming technique that binds the lifecycle of a resource that must be acquired before use (e.g., allocated heap memory, thread of execution, open socket, open file, locked mutex, disk space, database connection—anything that exists in limited supply) to the lifetime of an object}  
 
\newduneword{fts3}{FTS3}{File Transfer Service version 3, developed by \gls{cern}, and distinct from, the \gls{fnal} File Transfer Service} 
 
\newduneword{wfms}{WFMS}{Workflow Management System}
 
\newduneword{condmeta}{conditions metadata}{defined as the information necessary to understand the context of physics data, e.g., beam data or calibrations} 
 
\newduneword{cdb}{conditions DB}{The Conditions Database stores the \gls{condmeta} needed for data processing and analysis}

\newduneword{rntuple}{RNtuple}{The next generation of the \gls{root} I/O system} 

\newduneabbrev{lan}{LAN}{Local Area Network}{Computing network confined to a relatively small geographic area} 
 
\newduneabbrev{wan}{WAN}{Wide Area Network}{Computing network that interconnects \glspl{lan} across relatively broad geographic areas} 

\newduneword{layer3}{Layer-3}{Networking protocol  {https://en.wikipedia.org/wiki/Network\_layer}}
 
\newduneabbrev{rest}{REST}{Representational State Transfer}{Standard for web interfaces}  
\newduneabbrev{faq}{FAQ}{Frequently Asked Questions}{A software system for collecting and answering the most common questions about an activity} 
 
\newduneword{spack}{Spack}{Software package manager for UNIX and mac OS systems} 

\newduneabbrev{lxplus}{LXPLUS}{Linux Public Login User Service)}{The interactive logon service to Linux for all \gls{cern} users} 

\newduneword{dunesw}{dunesw}{The base DUNE software release} 

\newduneword{posix}{POSIX}{Portable Operating System Interface - Unix standard for operating system interfaces} 

\newduneword{nutools}{NuTools}{shared code for \gls{larsoft} + NOvA + other neutrino experiments that use \gls{art}.  Includes beam and event generators and re-weighting packages}

\newduneword{singlecube}{SingleCube}{A cubical time projection chamber 30 cm on a side with a single ND-LAr pixel readout tile, used for prototyping and tests}

\newduneword{hllhc}{HL-LHC}{High-luminosity \gls{lhc}}

\newduneabbrev{sof}{SoF}{signal-over-fiber}{a technology in which a fiber optic cable carries detector output that has been converted from an electrical to an optical pulse}

\newduneword{icd}{interface document}{Interface document}

\newduneword{ipmi}{IPMI}{Intelligent Platform Management Interface}

\newduneword{grafana}{Grafana}{add def}

\newduneword{trigact}{trigger activitation}{a correlation (or cluster) in time and space within a stream of \glspl{trigprimitive}}

\newduneword{cad}{CAD}{computer aided design}

\newduneword{gkvm}{GKVM}{Gava, Kneller, Volpe, McLaughlin Supernova model}

\newduneword{daqrou}{DAQ readout unit}{The basic component of the \gls{daqros}}

\newduneword{kpp}{key performance parameter}{(KPP) Defined by \gls{doecd}-2 (U.S.) as characteristic,  function, requirement or design basis that if changed would have a major impact on the system or facility performance, schedule,  cost and/or risk}

\newduneword{ftbf}{FTBF}{Fermilab Test Beam Facility}

\newduneabbrev{toad}{TOAD}{Test stand of an Overpressure Argon Detector}{A test of a high-pressure gaseous argon TPC in the Fermilab Test Beam Facility \gls{ftbf}}

\newduneabbrev{fcss}{FCSS}{field cage support system}{The support system suspended from the cryostat ceiling in the \gls{vdmod0} used to support the \gls{fc}} 

\newduneword{trigrec}{trigger record}{a collection of data and metadata from selected detector space-time volumes corresponding to a trigger decision}

\newduneword{solarization}{solarization}{Aging caused by exposure to UV-light, which can lead to increased attenuation}

\newduneword{cam}{CAM}{control account manager}

\newduneword{msds}{MSDS}{Manufacturer Safety Data Sheet}

\newduneword{darkside}{Darkside-50}{a dual-phase argon \gls{tpc}, developed for use in a search for direct evidence of dark matter}

\newduneword{apex}{APEX}{Aluminum Profiles with Embedded X-arapuca}

\newduneabbrev{df}{DF}{dichroic filter}{Optical filter that reflects some wavelengths of light and transmits others, with almost no absorption for all wavelengths of interest}

\newduneword{qpix}{Q-Pix}{A pixel-based, 3D, readout technology based on a continuously integrating low-power charge-sensitive amplifier viewed by a Schmitt trigger}

\newduneword{lightpix}{LightPix}{Low-power, cryogenic-compatible and scalable \gls{sipm} readout electronics based on the \gls{larpix} \gls{asic}}

\newduneabbrev{mosfet}{MOSFET}{Metal–oxide–semiconductor field-effect transistor}{A type of field-effect transistor}

\newduneword{gem}{GEM}{Gaseous electron multiplier}

\newduneabbrev{thgem}{THGEM}{Thick GEM}{High-gain gaseous electron multiplier}

\newduneabbrev{solar}{SoLAr}{Solar neutrinos in Liquid Argon}{A new concept for a liquid-argon neutrino detector technology to extend the sensitivities of these devices to the MeV energy range}

\newduneabbrev{ase}{aSe}{Amorphous selenium}{A type of photoconductive material.}

\newduneabbrev{ai}{AI}{Artificial Intelligence}{A field of study in computer science which develops and studies intelligent machines.}

\newduneabbrev{slomo}{SLoMo}{Sanford Underground Low background Module}{A dedicated low background far detector module that would enhance the physics program of DUNE.}

\newduneabbrev{tmg}{TMG}{Tetra-methyl-germanium}{A photosensitive hydrocarbon capable of converting \gls{vuv} scintillation light into ionization charge in a \gls{lar} detector.}

\newduneabbrev{nldbd}{0$\nu\beta\beta$}
{neutrinoless double-$\beta$ decay}{A hypothetical nuclear transition in which a nucleus with with Z
protons decays into a nucleus with Z+2 protons and the same mass number, together with the emission of two electrons and no neutrinos.}

\newduneword{theia}{\textsc{Theia}}{Proposed hybrid detector with both Cherenkov and scintillation detection capabilities}

\newduneword{theia25}{\textsc{Theia-25}}{A 25\,kt version of the \gls{theia} detector concept that could serve as DUNE's fourth far detector module.}

\newduneabbrev{wbls}{WbLS}{Water-based Liquid Scintillator}{A scintillating material consisting of water loaded with liquid scintillator.}

\newduneabbrev{wbnd}{WbND}{Water-based Near Detector}{A possible DUNE \phasetwo near detector sub-system employing water as neutrino target.}

\newduneabbrev{lappd}{LAPPD}{Large Area Picosecond Photo-Detector}{A kind of imaging photodetector designed to provide exquisite time resolution.}

\newduneabbrev{goat}{GOAT}{Gas-argon Operation of ALICE TPC}{A test stand for \gls{ndgar} R\&D.}

\newduneabbrev{gorg}{GORG}{GEM Over-pressurized with Reference
Gases}{A test stand for \gls{ndgar} R\&D}

\newduneabbrev{gat0}{GAT0}{Gaseous Argon T0}{An Optical \gls{tpc} demonstrator in Spain. A test stand for \gls{ndgar} R\&D}

\newduneabbrev{spy}{SPY}{Solenoid with Partial return Yoke}{Magnet concept currently envisaged to magnetize \gls{ndgar}.}

\newduneabbrev{fd3}{FD3}{far detector module 3}{The third DUNE far detector module to be built at \gls{surf}}

\newduneabbrev{fd4}{FD4}{far detector module 4}{The fourth DUNE far detector module to be built at \gls{surf}} 

\newduneword{ariadne}{ARIADNE}{Charge readout technology for \gls{lartpc} dual-phase detectors based on gaseous electron multipliers and fast optical cameras.}

\newduneword{acemirt}{ACE-MIRT}{The Accelerator Complex Evolution with Main Injector Ramp and Target upgrade is a proposed set of major upgrades to the Fermilab accelerator complex aimed at an early implementation of an enhanced 2.1 MW beam for DUNE.}

\newduneabbrev{ecfa}{ECFA}{European Committee for Future Accelerators}{Committee charged with the long-range planning of European high-energy facilities: accelerators, large-scale facilities and equipment.}

\newduneabbrev{cpad}{CPAD}{Coordinating Panel for Advanced Detectors}{US panel that seeks to promote, coordinate and assist in the research and development of instrumentation and detectors for high-energy
physics experiments.}

\newduneabbrev{trl}{TRL}{Technology Readiness Level}{A method for estimating the maturity of technologies.}

\newduneabbrev{qpixlilar}{Q-Pix-LILAr}{Q-Pix Light Imaging in Liquid Argon}{A \gls{qpix} pixel coated with a type of photoconductive material, to perform integrated charge/light readout on the anode.} 

\newduneword{drd}{DRD}{ECFA Detector R\&D}

\newduneword{rdc}{RDC}{Detector R\&D collaborations}

\newduneabbrev{cevns}{CE$\nu$NS}{Coherent Elastic Neutrino-Nucleus Scattering}{A type of neutrino interaction with matter.}

\newduneabbrev{rich}{RICH}{Ring-imaging Cherenkov detector}{A detector for identifying a particle of known momentum by measurement of Cherenkov radiation.}

\newduneabbrev{alp}{ALP}{Axion-like particle}{A hypothetical pseudoscalar particle that appears in the spontaneous breaking of a global symmetry.}

\title{%
\Huge DUNE \phasetwo: \\
\Large Scientific Opportunities, Detector Concepts, Technological Solutions}
\author{The DUNE Collaboration\footnote{Editors: Sowjanya Gollapinni, Anne Heavey, Stefan S\"oldner-Rembold, Michel Sorel}}
\date{\today}

\begin{document}

\maketitle

\includegraphics[width=0.9\textwidth]{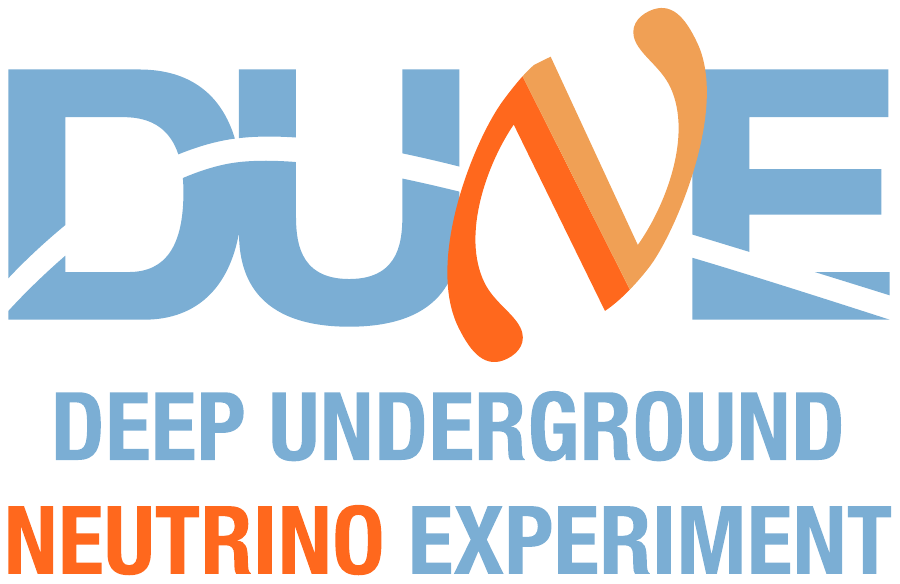}

\newpage
\includepdf[pages=-]{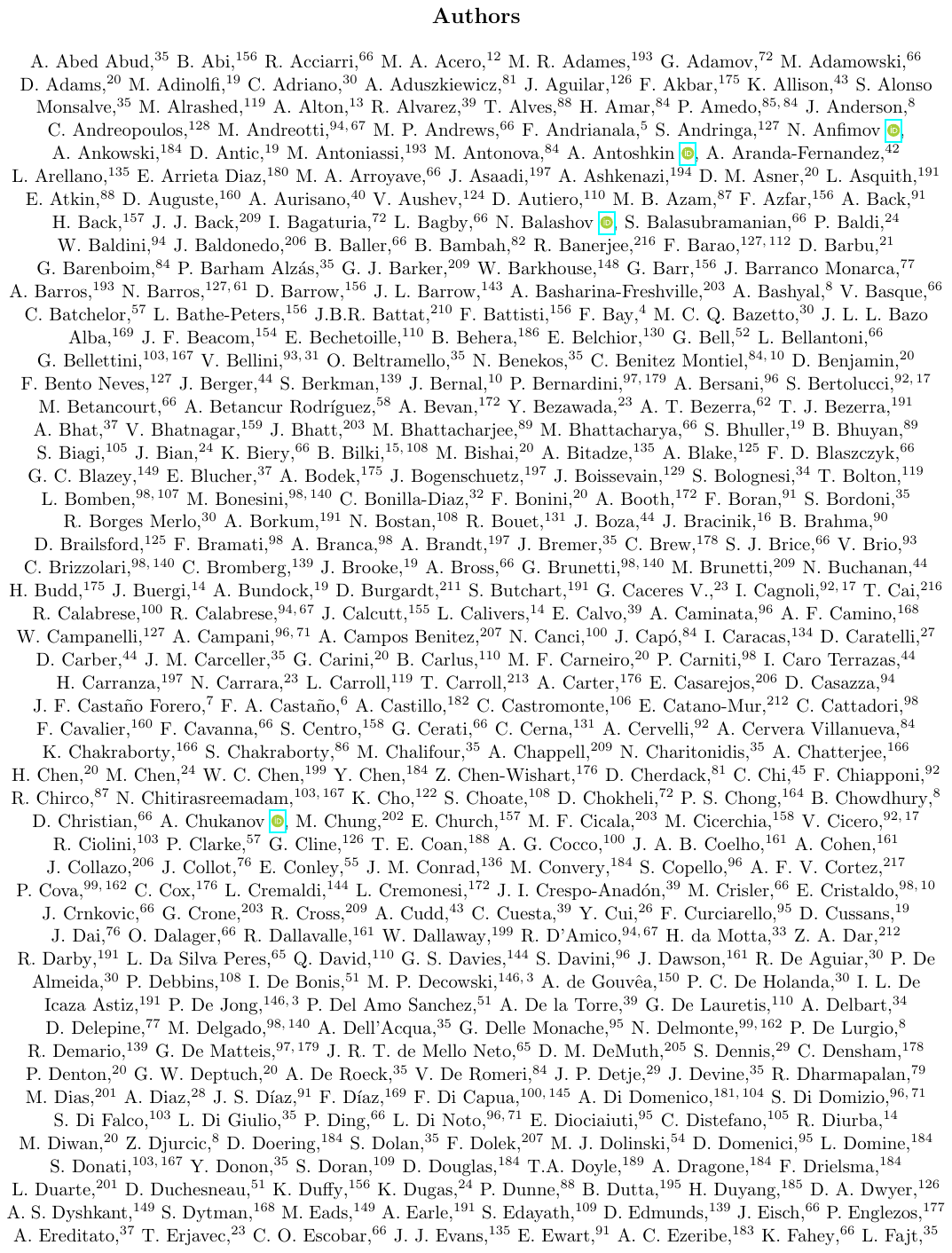}  
\newpage

\begin{abstract}
    The international collaboration designing and constructing the Deep Underground Neutrino Experiment (DUNE) at the Long-Baseline Neutrino Facility (LBNF) has developed a two-phase strategy toward the implementation of this leading-edge, large-scale science project. The 2023 report of the US Particle Physics Project Prioritization Panel (P5) reaffirmed this vision and strongly endorsed DUNE Phase~I and Phase~II, as did the European Strategy for Particle Physics. While the construction of the DUNE Phase~I is well underway, this White Paper focuses on DUNE Phase~II planning. DUNE Phase-II consists of a third and fourth far detector (FD) module, an upgraded near detector complex, and an enhanced 2.1\,MW beam. The fourth FD module is conceived as a ``Module of Opportunity'', aimed at expanding the physics opportunities, in addition to supporting the core DUNE science program, with more advanced technologies. This document highlights the increased science opportunities offered by the DUNE Phase~II near and far detectors, including long-baseline neutrino oscillation physics, neutrino astrophysics, and physics beyond the standard model. It describes the DUNE Phase~II near and far detector technologies and detector design concepts that are currently under consideration. A summary of key R\&D goals and prototyping phases needed to realize the Phase~II detector technical designs is also provided. DUNE's Phase~II detectors, along with the increased beam power, will complete the full scope of DUNE, enabling a multi-decadal program of groundbreaking science with neutrinos. 
\end{abstract}

\newpage

\tableofcontents

\newpage

\section*{Executive summary}
\label{sec:summary}
\addcontentsline{toc}{section}{Executive summary}

The preponderance of matter over antimatter in the early universe, the dynamics of the \dwords{snb} that produced the heavy elements necessary for life, the nature of dark matter, and whether protons
eventually decay -- these mysteries at the forefront of particle physics and astrophysics are key to understanding the evolution of our universe. 

The \dword{dune} will address these questions in a multidecadal science program with its world-leading \dword{lar} detector technology. The international DUNE experiment, hosted by the \dword{fnal}, is designing, developing, and constructing a \dword{nd} complex at Fermilab (the near site) and a suite of four large detector modules 1300\,km downstream at the \dword{surf} in South Dakota (the far site). These detectors will record neutrinos over a wide energy range, originating from a new high-intensity neutrino beamline at Fermilab. The modular \dword{fd} will also detect neutrinos produced by cosmic rays in the atmosphere and from astrophysical sources. The \dword{nd} and \dword{fd} will also be sensitive to a broad range of phenomena beyond the standard model (\dword{bsm}). The beamline as well as the excavations, infrastructure, and facilities for housing and supporting the DUNE detectors are provided by the \dword{lbnf}.

The DUNE Collaboration was launched in 2015, following the recommendations of the 2013 update of the European Strategy for Particle Physics~\cite{2013europeanstrategy} and of the 2014 Report of the US Particle Physics Project Prioritization Panel (P5)~\cite{2014p5report}. DUNE and \dword{lbnf} will complete this project in two phases, based on the availability of resources and the ability to reach science milestones. The latest P5 report released in December 2023 reaffirmed this vision~\cite{2023p5report}. 

The construction of the first project phase (\phaseone), funded through commitments by a coalition of international funding agencies, is well underway. Its successful completion is currently the Collaboration's main priority. Excavation at the far site is complete, and fabrication of various beamline and detector components for \phaseone is progressing well. The facilities currently being constructed by \dword{lbnf} at both the near and far sites are designed to host the full scope (\phaseone and \phasetwo) of the DUNE experiment.

The \phasetwo of DUNE that encompasses an enhanced 2.1\,MW beam, a third and fourth far detector (\dword{fd}) module, and an upgraded \dword{nd} complex, is the subject of this paper. The primary objective of \dune \phasetwo is a set of precise measurements of the parameters of the neutrino mixing matrix, $\theta_{23}$, $\theta_{13}$, $\Delta m^{2}_{32}$, and \deltacp, to establish \dword{cpv} over a broad range of possible values of \deltacp, and to search for new physics in neutrino oscillations. DUNE also seeks to detect neutrinos from low-energy astrophysical sources. The additional mass brought by the \phasetwo \dshort{fd} modules would increase the statistics of a supernova burst signal and extend DUNE's reach beyond the Milky Way. The \phasetwo \dshort{fd} module design concepts would also perform sensitive searches for new physics with solar neutrinos by lowering the detection threshold in the relevant MeV-scale energy range and by reducing the background rates in this energy regime. Finally, \phasetwo will expand DUNE's new physics discovery reach via more sensitive searches for rare processes at the \dword{nd} and \dword{fd} sites, and for non-standard neutrino oscillation phenomena.

This world-class physics program requires an increase in the statistical power of the detectors, which will be achieved by increasing the \dword{fd} target mass with the additional modules and by exceeding $2$\,MW beam power, as recommended by the 2023 P5 committee in the US. In addition, an upgraded \dword{nd} will be required to control systematic uncertainties of neutrino interactions on argon (or other selected \dword{fd} target nuclei). The design of the \phasetwo \dword{fd} modules will incorporate lessons learned from the construction of the \phaseone modules, with the goal of optimizing physics performance, reliability, and cost.

The design of the third FD module will build on that of the second, the single-phase \dword{vd} technology used for \dword{fd2}, optimizing for performance and cost. 
This design implements \dword{pcb}-based horizontal anode planes at the top and bottom of the \dword{lartpc} drift volume with a cathode plane in the middle, and a \dword{fc} that hangs from the cryostat roof on which photon detectors will be mounted. 

The fourth \dword{fd} module is conceived as a ``Module of Opportunity'', which would allow to address new physics questions, in addition to the primary science program, with more advanced technologies. R\&D for the design of the fourth module focuses primarily on optimization of readout techniques for both charge and scintillation light. It also considers the possibility of \dword{lar} doping. A possible improvement of the \dword{lar} charge-readout technology is the replacement of the \dword{pcb}-readout by pixels or by an optical-based charge readout. 
A hybrid approach to detect Cherenkov and scintillation light is also under investigation, motivated by a complementary program of low-energy physics. All technologies proposed for the Module of Opportunity are expected to provide additional and complementary \dword{cpv} sensitivity. In addition, background control is an essential ingredient of all \dshort{fd} module designs.

The \phasetwo \dword{nd} will ensure that DUNE's sensitivity to oscillation parameters is not limited by systematic uncertainties. It is optimized for highly performing \dword{pid}, low tracking thresholds for protons and pions, and acceptance over a wide range of momenta. A magnetized \dword{gartpc} at the near site, which will replace the \dword{tms}, will provide these constraints by measuring the interaction of neutrinos on argon with unprecedented precision due to its low thresholds. It offers superior discrimination between neutrinos and antineutrinos, as well as momentum determination of particles exiting the \dword{ndlar} detector. 

These plans fully align with the recommendations of the 2023 P5 report~\cite{2023p5report}, which proposes a ``second phase of DUNE (\phasetwo) with an early implementation of an enhanced $2.1$\,MW beam, a third far detector (module), and an upgraded near detector complex as the definitive long-baseline neutrino oscillation experiment of its kind,''  as well as ``research and development (R\&D) towards an advanced fourth detector.''

The \phasetwo R\&D program is a global effort with contributions from all DUNE partners. New collaborators are  also invited to participate in the development and design of the new detector technologies. Part of the R\&D described in this document is carried out within the framework of the \dword{ecfa} detector R\&D collaborations hosted by the \dword{cern} and those being formed under the umbrella of the \dword{cpad} in the US. DUNE has been designed to become the ``best-in-class" global neutrino observatory. The \phasetwo far and near detector components, and the increased beam power, will enable a new era of precision and discovery in neutrino physics.

\newpage
\section{The elements of DUNE~\phasetwo}
\label{sec:intro}

The \dword{dune} experiment at the \dword{lbnf} was conceived in 2015, following the recommendations of the 2013 update of the European Strategy for Particle Physics~\cite{2013europeanstrategy} and of the 2014 Report of the US Particle Physics Project Prioritization Panel (P5)~\cite{2014p5report}. The 2014 P5 Report recommended developing, in collaboration with international partners, a coherent long-baseline neutrino program hosted by \dword{fnal}, with a mean sensitivity to leptonic \dword{cpv} of better than three standard deviations ($\sigma$) over more than $75\%$ of the range of possible values of the CP-violating phase \deltacp. The 2014 P5 Report also recommended a broad program of neutrino astrophysics and physics Beyond the Standard Model (\dword{bsm}) as part of \dword{dune}, including demonstrated capability to search for \dwords{snb} and for proton decay. Likewise, the 2013 Update of the European Strategy for Particle Physics and its 2020 update~\cite{2020europeanstrategy} recommended that Europe and \dword{cern} (through its Neutrino Platform) continue to collaborate towards the successful completion of the \dword{intlproj}.

The DUNE Collaboration and the \dword{usproj} Project
have made substantial progress toward the realization of this enterprise, with the aim to start the scientific exploitation in 2029. Based on recent estimates, including a flux prediction using a fully-engineered neutrino beamline, the ultimate \dword{cpv} measurement goal put forward by the 2014 P5 Report can be reached with an exposure of about 1000\,kt$\cdot$MW$\cdot$yr. This can be achieved in about 15 years of physics data-taking (see Sec.~\ref{subsec:physics_lbl}), assuming that \phasetwo elements as described in this document are pursued. DUNE's neutrino astrophysics and \dword{bsm} physics programs also benefit from multidecadal operations, to realize DUNE's full physics potential and achieve its scientific goals.

For the successful implementation of DUNE and LBNF, we need to consider the full extent of the available resources and funding profiles, provide a realistic estimate of the project costs, and achieve a clear understanding of the experimental configurations and exposures that are necessary to reach various physics milestones. As a result of this exercise, the DUNE Collaboration and the \dword{usproj} Project have decided to pursue the experiment in two phases, as summarized in Table~\ref{tab:phases}. 

\begin{table}[htb]
    \centering
    %\begin{tabular}{c|c|c|c}
    \begin{tabular}{p{3.3cm}|p{4.7cm}|p{4.7cm}|p{2.3cm}} \hline
        Parameter  & \phaseone      & \phasetwo & Impact \\ \hline
        FD mass    & 2 FD modules (20\,kt fiducial) & 4 FD modules (40\,kt fiducial \dword{lar} equivalent) & FD statistics \\  \hline\hline
        Beam power & 1.2\,MW & Up to 2.3\,MW   & FD statistics \\ \hline
        ND configuration  & \dshort{ndlar}+\dshort{tms}, \dshort{sand}          & \dshort{ndlar}, \dshort{ndgar}, \dshort{sand}   & Systematics \\ \hline
    \end{tabular}
    \caption{A high-level description of the two-phased approach to DUNE. The \dword{ndlar} detector, including its capability to move sideways (\dword{duneprism}), and the \dword{sand} are present in both phases of the ND. Note that the non-argon options currently under consideration for \phasetwo near and far detectors are not shown.}
    \label{tab:phases}
\end{table}

In developing this two-phase strategy, we are guided by the 
original recommendations for the DUNE program, which remain valid and timely. The latest P5 report released in December 2023 reaffirmed this vision and strongly endorsed DUNE \phaseone and II. In its report~\cite{2023p5report}, the P5 panel reaffirmed that the highest priority in the coming decade, independent of budget scenarios, is the completion of construction of existing projects which includes \dword{lbnf} and DUNE \phaseone{}, and the \dword{pip2}. 

The panel also strongly recommended constructing a portfolio of major projects, of which the second-highest priority was a ``re-envisioned second phase of DUNE (\phasetwo) with an early implementation of an enhanced 2.1\,MW beam (\textit{aka} \dword{acemirt}), a \dword{fd3}, and an upgraded \dword{nd} complex as the definitive long-baseline neutrino oscillation experiment of its kind.'' The panel also endorsed  DUNE's fourth FD module (\dword{fd4}) concept as a ``Module of Opportunity'' and recommended exploring a range of alternative targets, including low-radioactivity argon, xenon-doped argon, and novel organic or water-based liquid scintillators, to maximize the science reach, particularly in the low-energy regime. An accelerated and expanded R\&D program in the next decade is recommended for \dword{fd4} and if budget scenarios are favorable, initiation of construction is also recommended. 

The overall project design for \phaseone is complete, and this project phase is funded through commitments by several international funding agencies and CERN. \dword{lbnf} excavation at the far site is complete, and fabrication of various beamline and detector components for \phaseone are well underway. An important component of the DUNE strategy is that the facilities constructed by LBNF at both the near and far sites are designed to support the full scope of the DUNE experiment from the beginning. During \phaseone, the facilities at the near site are thus constructed to support a $>$2~MW primary beamline and neutrino beamline, as well as a hybrid \dword{nd} for all DUNE experimental phases. Likewise, the far site design includes underground halls for four \dword{fd} modules. 

The \phaseone beamline will produce a wide-band neutrino beam with up to $1.2$\,MW beam power, designed to be upgradable to 2.4\,MW. The \phaseone \dword{nd} includes a moveable \dword{lartpc} with pixel readout called \dword{ndlar}, integrated with a downstream muon spectrometer called \dword{tms}~\cite{DUNE:2021tad}, and an on-axis magnetized neutrino detector called \dword{sand} further downstream. The \dword{ndlar}+\dword{tms} detector can be moved sideways over a range of off-axis angles and neutrino energies (\dword{duneprism} concept), for an optimal characterization of the neutrino-argon interactions. 

The \phaseone FD includes two \dword{lartpc} modules, each containing 17\,kt of liquid argon (\dword{lar}). The \dword{fd1} is a horizontal drift \dword{tpc}, as developed and operated in \dword{pdsp} at the \dword{cern} Neutrino Platform and similar in concept to the \dword{icarus}, \dword{sbnd}, and \dword{microboone} detectors~\cite{DUNE:2020txw}. The \dword{fd2} is a vertical drift TPC. Its design capitalizes on the experience with the \dword{pdvd} demonstrator at \dword{cern}. For the cryogenic infrastructure in support of the two \dword{lartpc} modules, \phaseone will include two large cryostats (one per FD module), 35\,kt of \dword{lar}, and three nitrogen refrigeration units.

While several options are under consideration for the \phasetwo components of the far and near site detectors, key elements have already been defined:
\begin{itemize}
    \item A core component of \phasetwo is a More Capable Near Detector (\dword{mcnd}). The main improvement to the \dword{nd} is the addition of a magnetized \dword{hpgtpc}, surrounded by an electromagnetic calorimeter and by a muon detector called \dword{ndgar}. \dword{ndgar} will serve both as a new muon spectrometer for \dword{ndlar}, replacing the \dword{tms} in this capacity, and as a new neutrino detector to study neutrino-argon interactions occurring in the \dword{hpgtpc}. In addition, upgrades to the \dword{ndlar} and \dword{sand} systems are considered, as well as potential ND options in the case of a non-argon technology for \dword{fd4}.
    
    \item Two additional \dword{fd} modules, \dword{fd3} and \dword{fd4}, will be added at the far site, for a total of four. The DUNE \dword{fd2} \dword{vd} technology forms the basis for the envisioned designs for \dword{fd3} and \dword{fd4}. A non-argon option such as liquid scintillator (e.g., \dword{theia}) is also under consideration as an alternative technology for \dword{fd4}. The cryogenic infrastructure at the far site will be upgraded for \phasetwo with a fourth nitrogen refrigeration unit to provide capacity for up to an additional 35\,kt of \dword{lar}. 

    \item A beam upgrade to increase the intensity to $>$2 MW. This is achieved by ACE-MIRT, which increases the frequency of beam spills by nearly a factor of two. While this is part of \phasetwo, it is possible to implement the upgrades that comprise ACE-MIRT before DUNE beam data taking begins.
\end{itemize}

The R\&D underpinning the \phasetwo concepts is performed as part of a global program. The Detector R\&D (\dword{drd}) collaborations hosted by CERN, which have been established as part of the \dfirst{ecfa} roadmap, cover many of the detector concepts under study for DUNE \phasetwo. The DRD1 collaboration focuses on gaseous detectors and the DRD2 focuses on liquid detectors. Both collaborations have been formed recently. They are expected to grow with collaborators from within and outside Europe. Both the DRD1 and DRD2 collaborations were approved in December 2023 as official CERN experiments and held their first collaboration meetings in February 2024 at CERN. The \dword{ecfa} \dword{drd} collaborations are strongly aligned with the Detector R\&D collaborations (\dwords{rdc}) being formed under the \dfirst{cpad} umbrella in the US. There is an active effort underway to coordinate activities across the CERN-hosted and US-based collaborations on synergistic areas of R\&D toward achieving common scientific and technological goals. 

This document is organized as follows. Section~\ref{sec:physics} covers the science that DUNE will pursue with \phasetwo, covering long-baseline neutrino oscillation physics, neutrino astrophysics, and \dword{bsm} physics. A progress report on DUNE's \phasetwo far and near neutrino detectors is given in Sections~\ref{sec:fd} and \ref{sec:nd}, respectively. These two sections cover our current understanding of the detector requirements to carry out the \phasetwo physics goals, and a current snapshot of the main detector design options that are under consideration. These sections also summarize the critical R\&D elements that remain to be addressed and the prototyping phases to be realized before \phasetwo detector technical designs can be finalized.

\section{DUNE \phasetwo physics}
\label{sec:physics}

This section discusses selected DUNE science drivers in the areas of long-baseline physics, neutrino astrophysics, and \dword{bsm} physics that uniquely benefit from an improved performance of \phasetwo beam and detectors. The physics sensitivities are typically shown as functions of high-level figures of merit, such as exposure time, detector mass, background levels or detector acceptance, mostly without entering into the detector technology details. Specific benefits brought by certain technologies are highlighted in Sections~\ref{sec:fd} and \ref{sec:nd}, together with the descriptions of such technologies. 

\subsection{Long-baseline neutrino oscillation physics}
\label{subsec:physics_lbl}

The DUNE experiment is designed to measure the rate of appearance of electron (anti)neutrinos ($\nu_{e}$ or $\bar{\nu}_{e}$) and the rate of disappearance of muon (anti)neutrinos 
($\nu_{\mu}$ or $\bar{\nu}_{\mu}$), as functions of neutrino energy in a wide-band beam. DUNE is sensitive to all the parameters governing $\nu_{1}-\nu_{3}$ and $\nu_{2}-\nu_{3}$ mixing in the three-flavor model: $\theta_{23}$, $\theta_{13}$, $\Delta m^{2}_{32}$ (including its sign, which is given by the neutrino mass ordering), and the phase \deltacp. 

During \phaseone, DUNE can accumulate approximately 100\,kt$\cdot$MW$\cdot$yr of data in five years. This corresponds to $\approx 400$~$\nu_{e}$ and 150~$\bar{\nu}_{e}$ candidates in the \dword{fd}, depending on the value of the oscillation parameters and assuming equal fractions of neutrino and antineutrino running. While these data sets will still be statistically limited
once \phaseone ND systematic constraints are accounted for, they are sufficient to conclusively determine the neutrino mass ordering at $>5\sigma$ significance, regardless of the true parameter values. If \dword{cpv} is nearly maximal (\deltacp $\approx \pm \pi/2$), DUNE can establish \dword{cpv} at $3\sigma$ in \phaseone. DUNE will also make measurements of the disappearance parameters $\Delta m^{2}_{32}$ and $\sin^{2}2\theta_{23}$ that improve upon current uncertainties. However, the statistics of \phaseone are too low to determine the octant of $\theta_{23}$ or to establish \dword{cpv} except in the most favorable scenarios.

\subsubsection{Goals of the oscillation physics program of \phasetwo}

The goals of the oscillation physics program of \phasetwo are high-precision measurements of all four parameters: $\theta_{23}$, $\theta_{13}$, $\Delta m^{2}_{32}$ and \deltacp, to establish \dword{cpv} at high significance over a broad range of possible values of \deltacp, and to test the three-flavor paradigm as a way to search for new physics in neutrino oscillations. Achieving these goals requires $600-1000$\,kt$\cdot$MW$\cdot$yr of data statistics, depending on the measurement. This can be achieved by operating for $6-10$ additional calendar years with a greater than $2$\,MW beam and a \dword{fd} of $40$\,kt \dword{lar} equivalent fiducial mass. Without doubling the \dword{fd} mass and the beam intensity, the additional time required would be $24-40$ years. 

\begin{figure}[htb]
    \centering
    \includegraphics[width=0.7\textwidth]{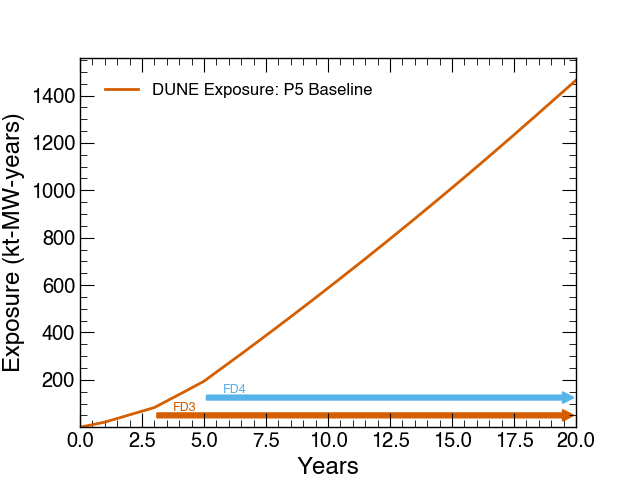}
    \caption{DUNE integrated exposure, in kt$\cdot$MW$\cdot$yr units, as a function of time assumed for the \dword{lbl} sensitivity results presented in this section. The integrated exposure is built from the beam power and \dword{fd} mass staging assumptions discussed in the text.}
    \label{fig:exposure}
\end{figure}

In particular, the long-baseline sensitivities presented below make the following assumptions concerning the time evolution of the \dword{pot} delivery, the fiducial FD mass and the \dword{nd} systematic constraints. The beam power evolution is based on the assumptions of the Fermilab Proton Intensity Upgrade Central Design Group \cite{2672528}. The average beam power during the first year of beam operations is 1.1\,MW, increasing to 1.6\,MW during year 2, thanks to \dword{acemirt} upgrades~\cite{2672528}. Further beam optimizations are assumed in subsequent years, yielding a beam power of 2.3\,MW after approximately 15 years. The assumed \dword{pot} delivery also includes a 57\% average uptime~\cite{DUNE:2020jqi}. As regards the fiducial \dword{fd} mass, the experiment is assumed to take data with two \dword{fd} modules (20\,kt fiducial mass) during the first three years. The \dword{fd3} and \dword{fd4} modules are assumed to become fully operational in years 4 and 6, respectively, to provide a nominal fiducial \dword{fd} mass of 40\,kt. Figure~\ref{fig:exposure} shows how the accumulated exposure, expressed in kt$\cdot$MW$\cdot$yr units, varies as a function of time with this assumed staging scenario. The 600 and 1000\,kt$\cdot$MW$\cdot$yr integrated exposure milestones of DUNE \phasetwo appear within reach after approximately 10 and 15 years of beam plus detector operations, respectively. Finally, it is assumed that the \phaseone \dword{nd} systematic constraints will be in effect up to year 6, with the improvements from \phasetwo \dword{nd} starting in year 7. The new constraints, as they gradually improve over the course of two years to their final values, will be applied to all past \dword{fd} data. 

\begin{figure}[htb]
    \centering
    \includegraphics[width=0.49\textwidth]{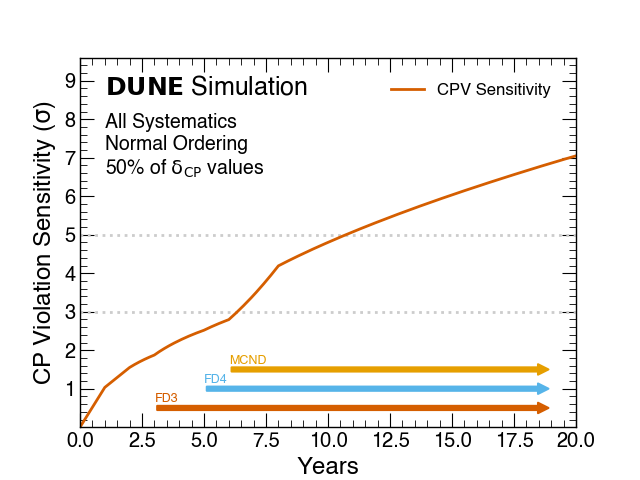} \hfill
    \includegraphics[width=0.49\textwidth]{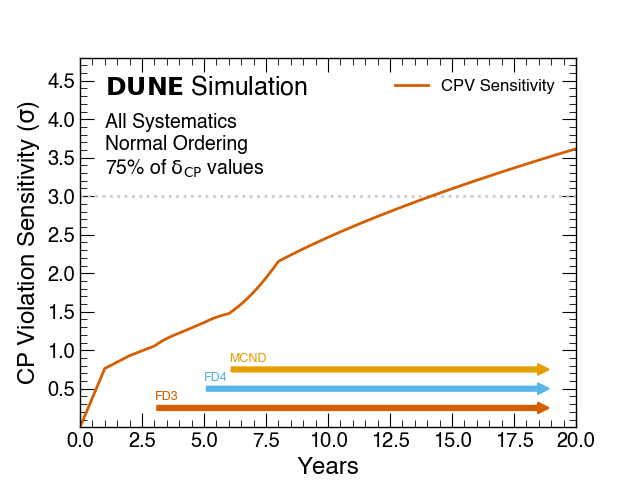}
    \caption{The significance for DUNE to establish \dshort{cpv} for 50\% (left panel) and 75\% (right) of \deltacp values as a function of running time. See text for details about the assumed staging scenario.}
    \label{fig:cpv}
\end{figure}

The precision of the high-statistics measurements is ultimately limited by the systematic uncertainties. 
To achieve DUNE's science goals will therefore require unprecedented control of systematic uncertainties. The significance for DUNE to establish \dword{cpv} is shown as a function of time in Figure~\ref{fig:cpv}. \phasetwo, with its full 1000\,kt$\cdot$MW$\cdot$yr exposure, will enable DUNE to establish \dword{cpv} at $>3\sigma$ over $75\%$ of possible \deltacp values, and measure it at a precision of $6^\circ-16^\circ$, depending on the true value.

\begin{figure}[htb]
    \centering
    \includegraphics[width=0.32\textwidth]{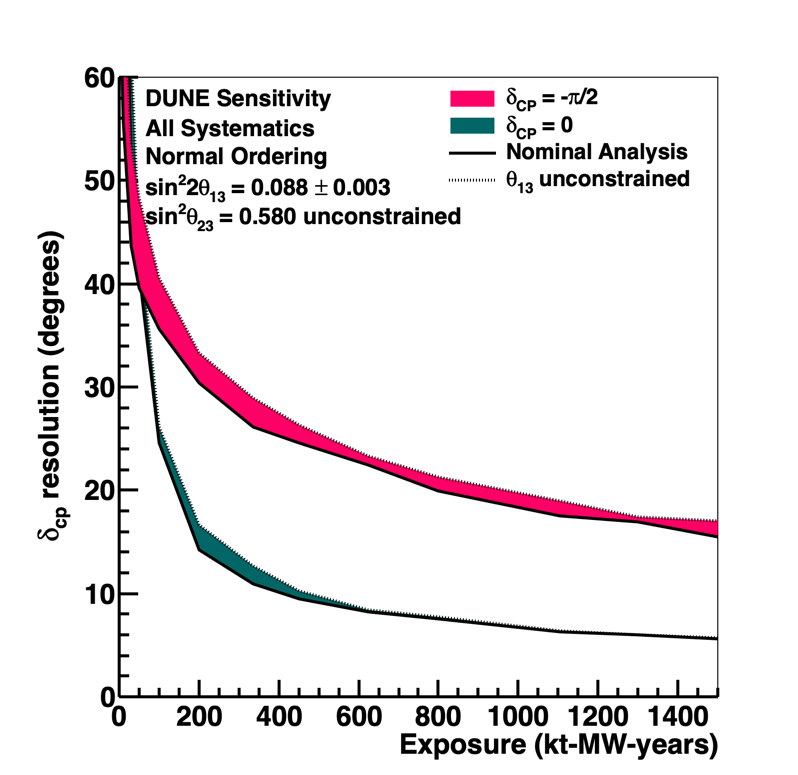}
    \includegraphics[width=0.32\textwidth]{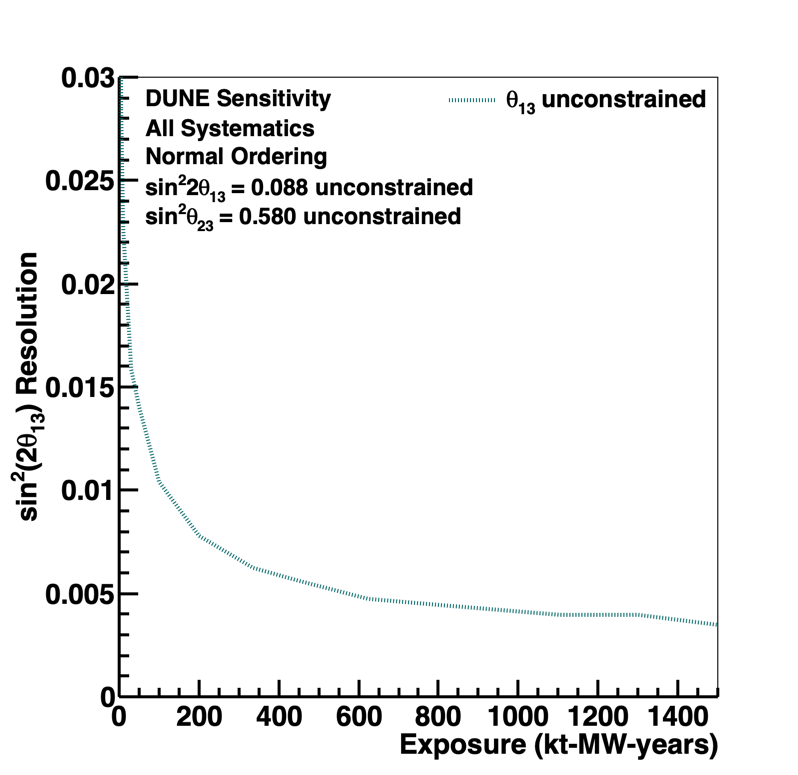}
    \includegraphics[width=0.32\textwidth]{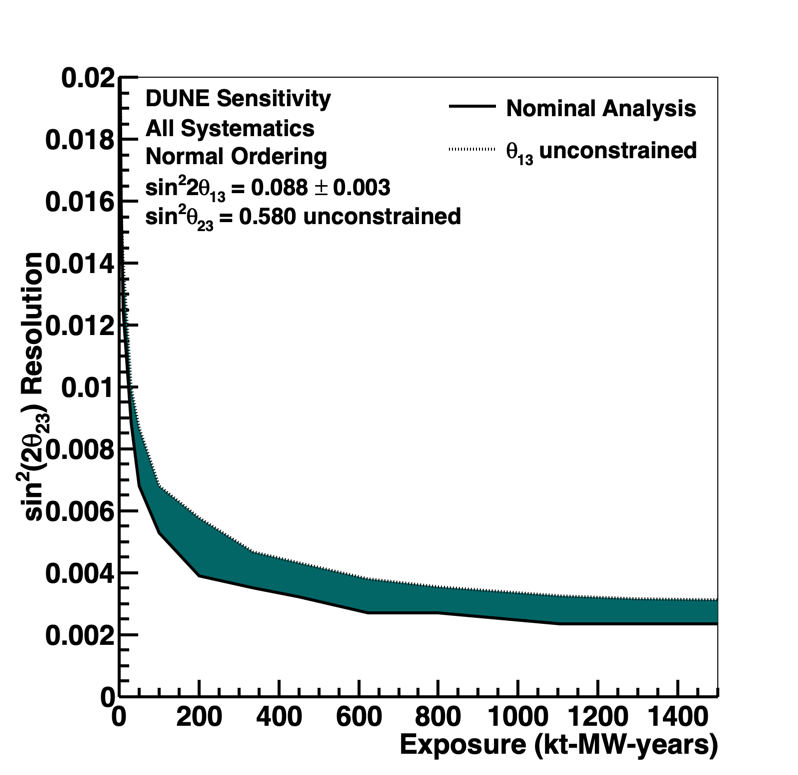}
    \caption{The resolutions to \deltacp (left), $\sin^{2}2\theta_{13}$ (center), and $\sin^{2}\theta_{23}$ (right), shown as a function of exposure in kt-MW-yrs, assuming the full constraint from the \dword{nd}, including \dword{mcnd}. The ultimate precision of DUNE requires an exposure greater than 600 kt-MW-yrs, which requires \dword{fd3} and \dword{fd4} to be built in a reasonable timescale.}
    \label{fig:resolutions}
\end{figure}

DUNE can also measure the angle $\theta_{23}$ with world-leading precision and determine the octant if it is sufficiently non-maximal. The measurements of $\theta_{13}$ and $\Delta m^{2}_{32}$ will approach the precision of the current measurement from Daya Bay~\cite{DayaBay:2022orm} and the planned measurement from JUNO~\cite{JUNO:2022mxj}, respectively, 
which are all performed with a different neutrino flavor, over a different baseline, and at a different energy. Comparing the results obtained over this wide range of conditions will provide a more complete and robust test of the three-flavor model. The resolutions to \deltacp, $\sin^{2}2\theta_{13}$, and $\sin^{2}\theta_{23}$ are shown as a function of exposure in Figure~\ref{fig:resolutions}.

DUNE is also sensitive to \dword{bsm} physics %physics beyond the \dword{sm} 
that impacts neutrino oscillations, including non-unitary mixing, non-standard interactions, violation of \dword{cpt}, and the possible existence of additional neutrino species (see Section~\ref{subsec:physics_bsm}). 

\subsubsection{The role of \phasetwo detectors}

The two additional modules, \dword{fd3} and \dword{fd4}, will provide the additional exposure and improved statistical precision that is critical to achieve the full \deltacp sensitivity (Figure~\ref{fig:cpv}). They also offer the opportunity to improve the neutrino energy reconstruction and the neutrino interaction classification with enhanced detector technology, for example through optimized charge and photon readout systems (Sec.~\ref{subsec:fd_vdoptimized}). It is crucial that the data from these modules be combined with data from \dword{fd1} and \dword{fd2}, with the systematic constraints from the \dword{nd} applied to all \dword{fd} modules. 
For this reason, the most straightforward approach is for \dword{fd3} and \dword{fd4} to be \dwords{lartpc}, so that DUNE would immediately benefit from the $\nu$-Ar measurement program of the \dword{nd}. The oscillation sensitivities presented in Figures~\ref{fig:cpv} and \ref{fig:resolutions} assume the performance of the Horizontal Drift module using an end-to-end simulation and reconstruction. The alternative \dword{lartpc} concepts being considered for \dword{fd3} and \dword{fd4} are expected to have similar, and perhaps slightly improved performance for GeV-scale beam neutrinos, so the existing simulations serve as a conservative estimate of the eventual sensitivity.

If \dword{fd4} is not a \dword{lartpc}, the impact on the long-baseline oscillation program is less straightforward. For the \dword{theia} concept discussed in Section~\ref{subsec:fd_theia}, the performance is estimated using a reconstruction based on the fiTQun package~\cite{Super-Kamiokande:2019gzr}, and a \dword{bdt}~\cite{Theia:2019non}. The analysis is less sophisticated than the DUNE TDR sensitivities~\cite{DUNE:2020jqi}, using the GLOBeS package to implement systematics as normalization shifts, but suggests that the sensitivity is comparable to what can be achieved in a single \dword{lartpc} module. It does not yet make use of the additional information potentially offered by the scintillation component, which can provide tagging of neutrons and other sub-Cherenkov threshold particles, for improved event identification and enhanced calorimetry. Combining a non-\dword{lartpc} module with \dword{lar} measurements could potentially provide a cross-check of extracted oscillation parameter values with different detector systematics. A non-\dword{lar} \dword{fd4} would require a dedicated \dword{nd} to constrain neutrino cross section uncertainties and detector response on the \dword{fd4} nuclear target to a similar precision as the \dword{lartpc} constraints. Near detector options for non-\dword{lar} \dword{fd} modules are discussed in Sec.~\ref{subsec:theiand}.

The role of \dword{mcnd} (Section~\ref{subsec:ndgar}) is to ensure that DUNE can achieve the required level of systematic uncertainties for \phasetwo and to ensure that the results are not systematically limited. It would replace the \dword{tms} with a detector that has its own standalone physics capabilities, including constraining neutrino-argon cross section uncertainties and expanding the \dword{bsm} reach of the \dword{nd}, while also measuring muons exiting the \dword{ndlar} detector. Further study of the ultimate performance of the \phaseone \dword{nd} is important for scoping the \phasetwo \dword{nd}.

%%%%%%%%%%%%%%

\subsection{Neutrino astrophysics and other low-energy physics opportunities}
\label{subsec:physics_astro}

DUNE's broad physics program includes the detection of neutrinos from astrophysical sources in the MeV energy range~\cite{DUNE:2020ypp}, primarily neutrinos from the sun and a \dword{snb}. With argon as active material, DUNE will be primarily sensitive to the astroparticle $\nu_e$ flux for energies below 100\,MeV and above 5\,MeV due to the relatively large $\nu_e$ \dword{cc} cross section for the process: $\nu_e+^{40}\mathrm{Ar}\rightarrow e^-+^{40}\mathrm{K}^\ast$. DUNE will be unique in this regard, making the experiment highly complementary
to existing and proposed experiments for the next decades aiming for similar astrophysical neutrino measurements in the 10s of MeV regime: JUNO~\cite{PhysRevD.94.023006}, Hyper-Kamiokande~\cite{Hyper-Kamiokande:2021frf}, and dark matter detectors~\cite{Lang:2016zhv,DarkSide20k:2020ymr}. 
This section highlights the improvements a \dword{lartpc} \dword{fd} module from \phasetwo would bring. It also includes a summary of the advantages of a water-based liquid scintillator, that would be sensitive primarily to the $\bar{\nu}_e$ flux, within the \phasetwo program. 

\begin{figure}[htb]
    \centering
    \includegraphics[width=0.8\textwidth]{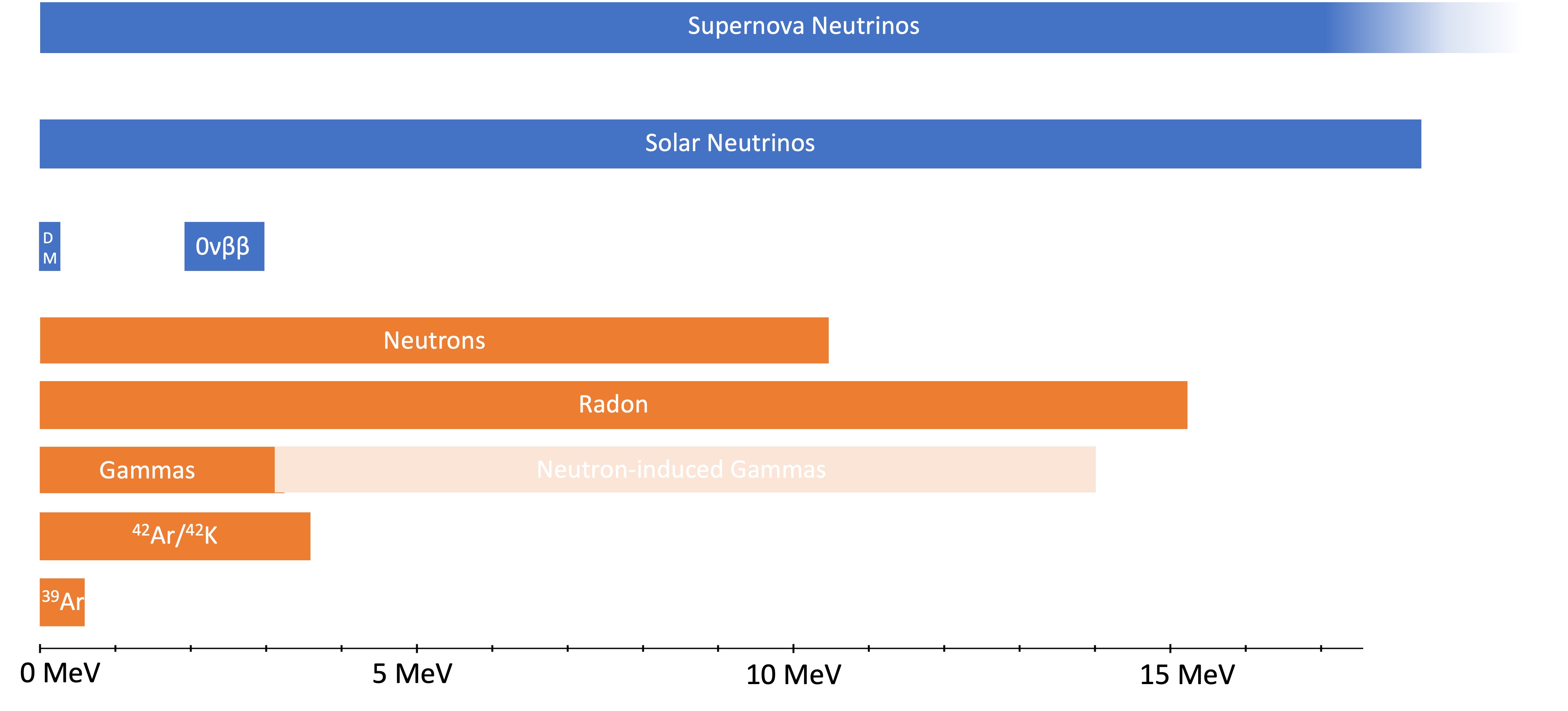}
    \caption{Detectable energy ranges in DUNE \dword{lartpc} \dword{fd} modules for potential $<20$~MeV signatures of astrophysical or other non-beam-related origin (blue), and for relevant, overlapping backgrounds (orange). The DM signature refers to \dword{wimp} direct dark matter searches. Adapted from~\cite{Bezerra:2023gvl}.} 
    \label{fig:astro_summary}
\end{figure}

DUNE \phaseone will be sensitive to solar and \dword{snb} neutrinos, achieving an energy resolution of $(10-20)\%$. The visible energy threshold will be $>5$~MeV for \dword{snb} neutrinos and higher for solar neutrinos, which do not arrive in a short pulse. Large \dword{lartpc} detectors have demonstrated much lower charge thresholds, $\sim$\,100~keV~\cite{PhysRevD.109.052007}, but \phaseone sensitivity will be limited to $>5$~MeV due to light collection performance and radiological backgrounds. Thus, innovation on these fronts can lower visible energy thresholds for astrophysical neutrinos by as much as two orders of magnitude while also improving energy resolution. Lower thresholds would also fundamentally expand the low-energy physics opportunities with DUNE \phasetwo. Figure~\ref{fig:astro_summary} shows a selection of potential signatures of astrophysical or other beam-unrelated origin with DUNE \phasetwo, together with relevant backgrounds. 

\subsubsection{SNB neutrinos} 
\begin{figure}[htb]
    \centering
    \includegraphics[width=0.8\textwidth]{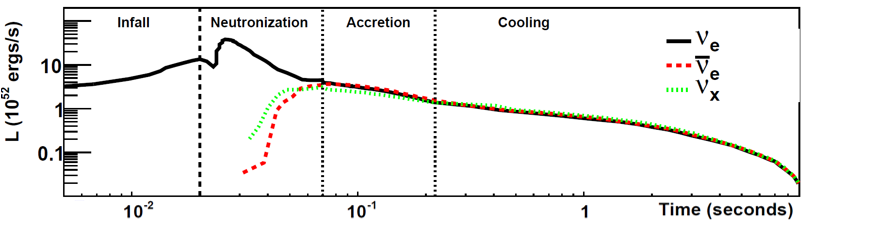}
    \includegraphics[width=0.43\textwidth]{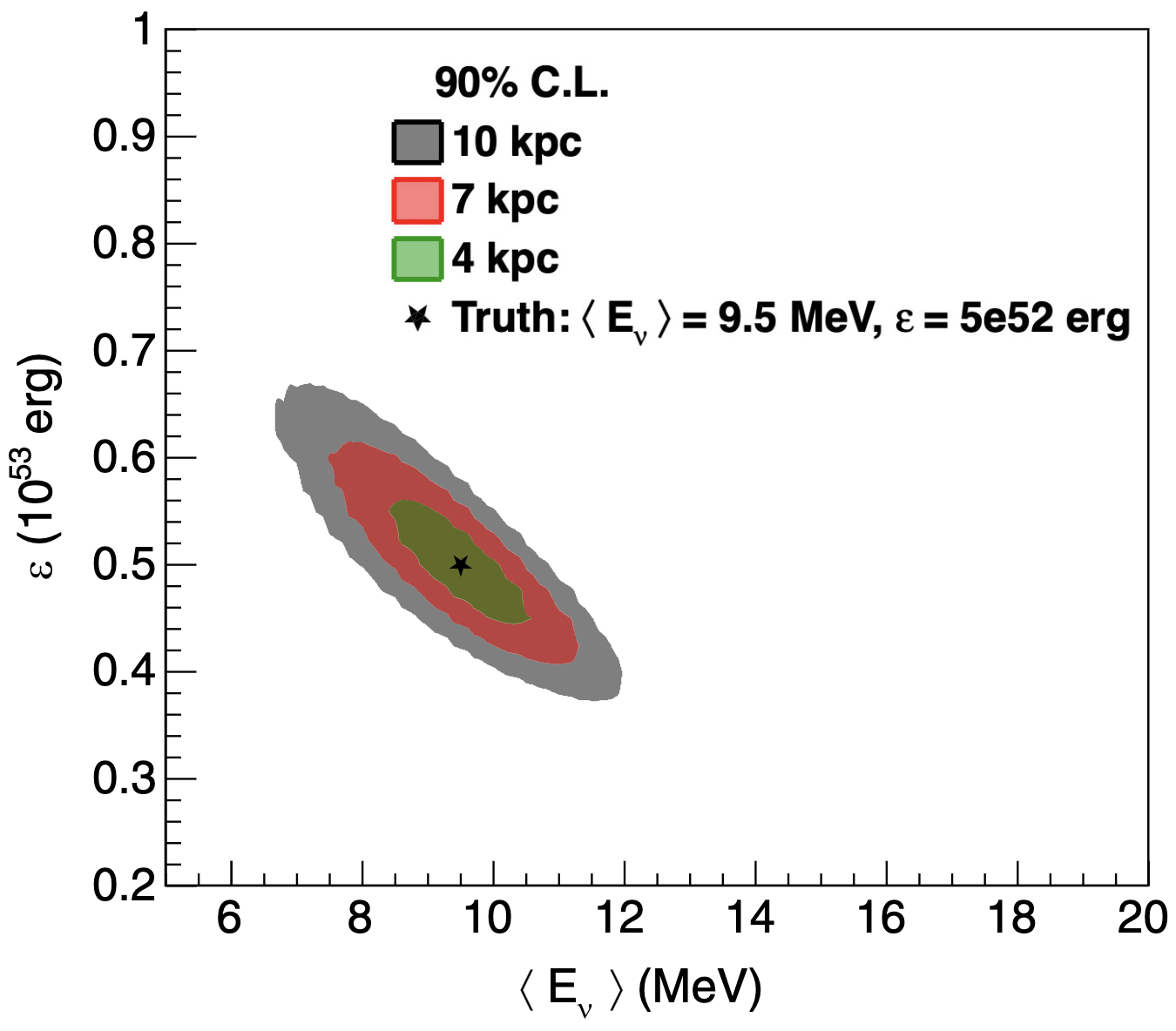} \hfill
    \includegraphics[width=0.56\textwidth]{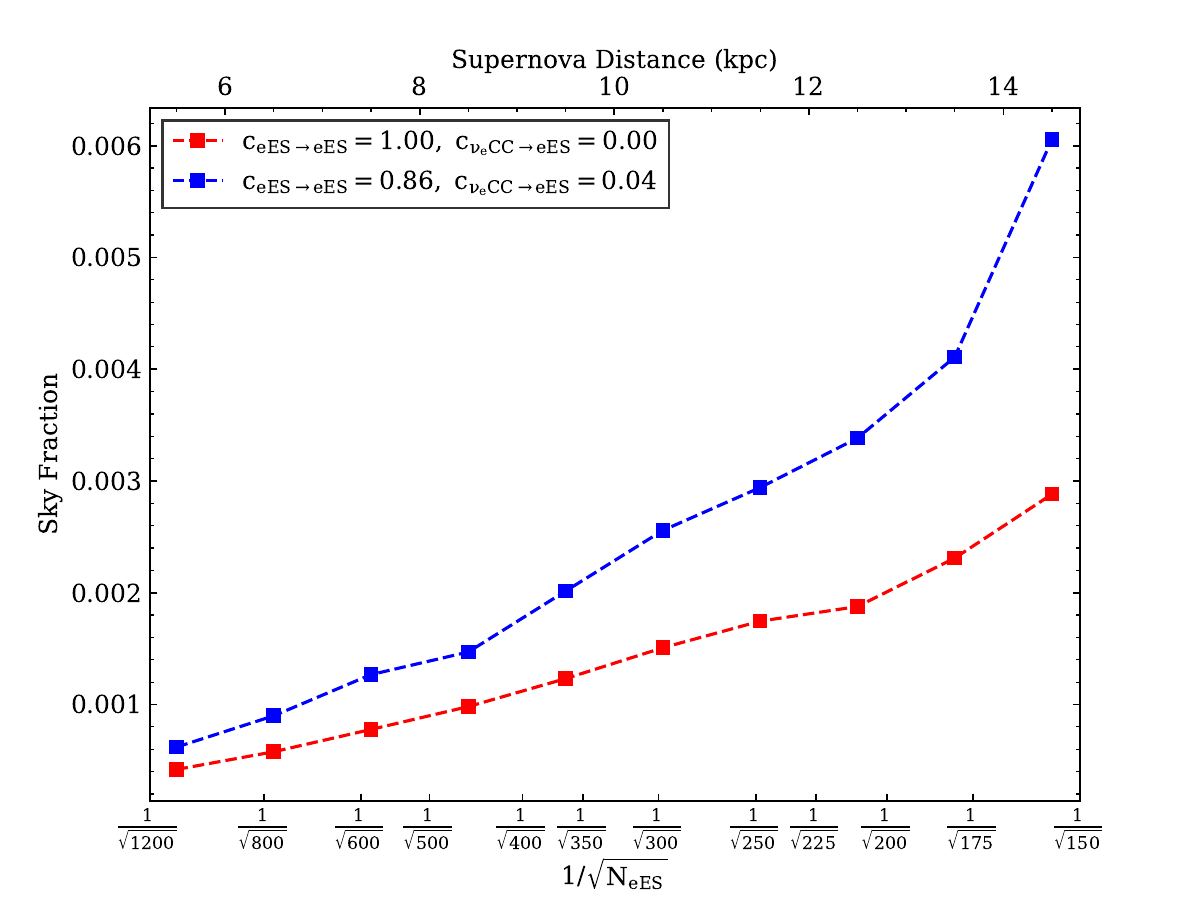}
    \caption{The total luminosity of neutrinos released during \dword{ccsn} from~\cite{PhysRevLett.104.251101} (top). Sensitivity regions in mean neutrino energy (related to the temperature of supernovae) and neutrino luminosity space for three different supernova distances from~\cite{DUNE:2023rtr} (bottom left). DUNE's reconstruction of the direction of a \dword{snb} in terms of the fraction of sky allowed at 68$\%$ confidence, as a function of recorded number of neutrino-electron scattering events as well as the corresponding supernova distance from~\cite{DUNE:2024ptd} (bottom right).}
    \label{fig:SNsensitivityregions}
\end{figure}

DUNE will be part of a collaborative, multi-messenger network of neutrino and optical telescopes studying the next galactic \dword{ccsn}. With different flavor sensitivity from other large experiments, DUNE will provide complementary information about the collapse. DUNE's $\nu_e$ sensitivity is most striking at the earliest times of the \dword{snb}, which is dominated by $\nu_e$ emission from neutronization in the stellar core, as shown in the top panel of Figure~\ref{fig:SNsensitivityregions}. During \phaseone, DUNE will already be able to detect neutrinos from a \dword{ccsn} with energies $>5$~MeV \cite{DUNE:2020zfm}, albeit with lower statistics. The larger mass of \phasetwo represents a significant step in extending the detector's sensitivity to a \dword{snb} signal, since the burst trigger efficiency, reconstruction of the supernova direction (bottom right panel of Figure~\ref{fig:SNsensitivityregions}), and precise measurement of the supernova spectral parameters (bottom left panel of Figure~\ref{fig:SNsensitivityregions}))are dominated by the number of neutrino interactions in the detectors. While the expected event rate varies significantly among supernova models, the 40\,kt (fiducial) DUNE detector would be expected to observe $\approx\,3000$ neutrinos from an \dword{snb} at a distance of 10\,kpc, just beyond the center of the Milky Way~\cite{DUNE:2020zfm}. 

The modular nature of DUNE's \dword{fd} makes the experiment ideal for \dword{snb} triggering and contributing to \dword{snews}~\cite{SNEWS:2020tbu}. Each \dword{fd} module will independently forward a \dword{ccsn} alert to \dword{snews} which will be made available to optical astronomers. Deployment of additional detectors in \phasetwo will effectively ensure that at least one \dword{fd} module is operational whenever a \dword{snb} arrives at the detector. With all modules taken together, the increased mass from \phasetwo will allow DUNE to trigger on further supernovae, increasing coverage in the neighborhood beyond the Milky Way (e.g., in the Large and Small Magellanic Clouds).

As the neutrino signal escapes a core-collapse supernova hours before the first optical signal, pin-pointing the source of the \dword{snb} is fundamentally important to facilitate optical observation of the initial stages of the supernova. The electron tracks from neutrino-electron elastic scattering, a sub-dominant interaction channel for \dword{snb} detection, are nearly parallel to the incoming neutrino flux and can thus be used to reconstruct the neutrino direction. Supernova pointing leverages the excellent tracking capabilities of a \dword{lartpc} to identify neutrino-electron scattering events yielding a pure sample of this low-rate channel. The increased mass from \phasetwo is critical for realizing DUNE's full potential. Using a typical flux model, the full 40\,kt detector would expect 326 neutrino-electron scattering events compared to 163 in \phaseone~\cite{DUNE:2024ptd}. This reduces the field-of-view for optical follow-up searches following a \dword{snb} signal from 0.26$\%$ to 0.14$\%$ of the total sky at 1\,$\sigma$ as shown in the bottom right panel of Figure~\ref{fig:SNsensitivityregions}. A \phasetwo module with pixelated charge readout would also improve \dword{snb} pointing resolution by improving the 3D reconstruction of low-energy electron tracks. 

\begin{figure}[htb]
    \centering
    \includegraphics[width=0.8\textwidth]{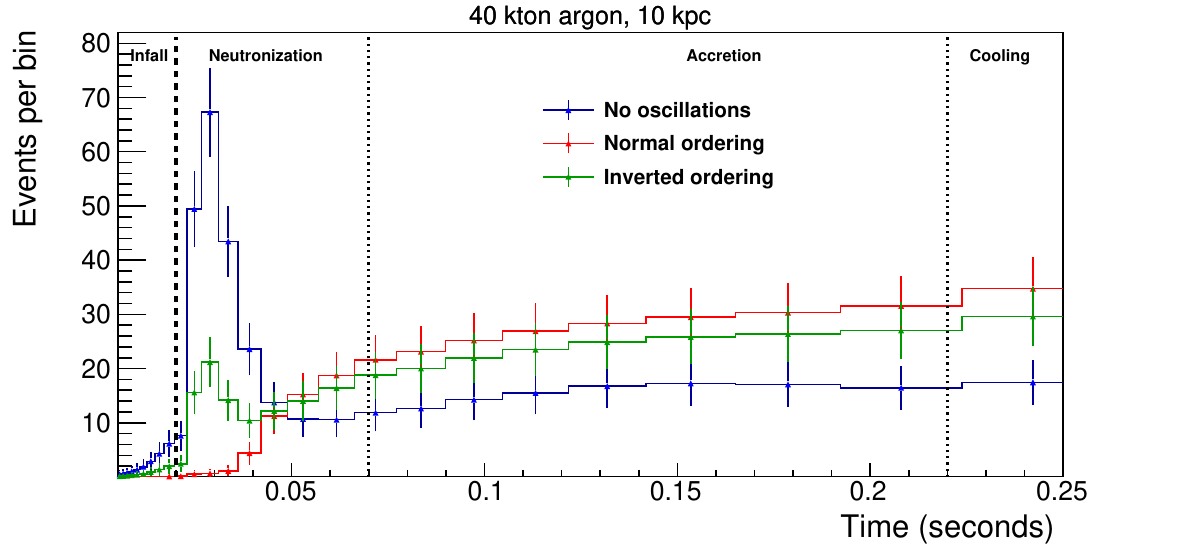}
    \caption{Expected event rates from~\cite{DUNE:2020zfm} as a function of time for the electron-capture supernova model in \cite{Hudepohl:2009tyy} and for 40\,kt of argon during early stages of the burst. Shown are  the event rate for the unrealistic case of no flavor transitions (blue) and the event rates including the effect of matter transitions for the normal (red) and inverted (green) mass orderings. Error bars are the expected statistical uncertainty in each (varying) time bin.}
    \label{fig:early_time}
\end{figure}

The neutrino mass ordering has a strong impact on the expected signal at early times (neutronization burst) when electron-type neutrinos dominate the neutrino flux at production (see Figure~\ref{fig:early_time} for expected event rates in DUNE). Neutrino flavor transformations can be induced by neutrino-neutrino scattering and collective modes of oscillation~\cite{Dasgupta:2007ws}. These effects will leave imprints on the neutrino signal and can be used to study these phenomena experimentally. These effects will test fundamental neutrino properties by measuring the neutrino self-interaction strength~\cite{Chang:2022aas}. DUNE will also provide competitive constraints on the absolute neutrino mass via measurements of the time of flight from the supernova to Earth~\cite{Pompa:2022cxc}. 

With argon's $\nu_e$ flavor sensitivity, DUNE will uniquely probe the neutrino component of the \dword{dsnb}~\cite{PhysRevD.109.023024} -- \phasetwo will be critical for such a low-rate search by increasing argon mass.

With significant increase of the photodetector area, e.g., $10\%$ coverage with \dwords{sipm}~\cite{Bezerra:2023gvl}, and use of $^{39}$Ar-depleted argon, a DUNE \phasetwo \dword{lartpc} module would be sensitive to faint light flashes from \dword{cevns} interactions on argon during a \dword{snb}. The ``CEvNS glow" from a supernova would be observed as an increase of low-PE flashes observed through the duration of the burst. As a \dword{nc} process, this channel is sensitive to all neutrino flavors, thus giving orthogonal information to the MeV-scale $\nu_e$~\dword{cc} interactions.  Importantly, CEvNS glow allows DUNE to determine the supernova neutrino fluence independent of neutrino oscillation uncertainties. 

A water-based liquid scintillator module (e.g., \dword{theia}) would sacrifice part of the \dword{snb} $\nu_e$ statistics for a significant increase in $\bar{\nu}_e$ events through inverse $\beta$ decay, with a threshold of $\sim$2~MeV. Such a module would detect approximately 5000 $\bar{\nu}_e$ interactions from a \dword{snb} at 10\,kpc distance. The scintillation light would provide a tag for neutrons to allow separation between inverse $\beta$ decay events and directionally-sensitive elastic scattering reactions. Using the high light output from the scintillation, such a module would also be sensitive to pre-supernova neutrinos~\cite{Mukhopadhyay_2020}, alerting neutrino experiments to an upcoming \dword{snb}. 

\subsubsection{Solar neutrinos} 
After more than a half century of study, there remain important open questions in particle and astrophysics that solar neutrino measurements can potentially resolve. This is due in large part to the precision tracking capabilities of the DUNE \dword{lartpc} detectors. In addition, because of argon's dominant $\nu_e$ \dword{cc} interaction channel, the detected energy will correlate strongly with the incoming neutrino energy. With these advantages, DUNE promises excellent potential for improved measurements~\cite{Capozzi:2018dat}. Initial studies suggest DUNE \phaseone can select a sample of $^8$B solar neutrinos that would improve upon current solar measurements of $\Delta m^2_{21}$~\cite{Super-Kamiokande:2023jbt} via the precise measurement of the day-night flux asymmetry induced by Earth matter effects. DUNE \phaseone can also make the first observation at $>$5$\sigma$ of the ``hep" flux produced via the $^3$He + p $\to$ $^4$He + e$^+$ + $\nu_e$ nuclear fusion. 

The energy resolution of DUNE \phaseone, $(10-20)\%$, could be improved down to $\approx2\%$ through improvements to the \phasetwo \dword{pds}. This radically improves determination of solar neutrino parameters. The measurement of the solar mass splitting $\Delta m^2_{21}$ requires precisely measuring the energy dependence of the neutrino oscillation pattern. A single module with $2\%$ resolution would make this measurement better than four modules with \phaseone energy reconstruction performance.

For beam-unrelated \dword{fd} events, reconstructed photon flashes are used to select events within the detector fiducial volume and to correct ionization charge loss along drift, greatly suppressing backgrounds and improving energy reconstruction, respectively. In \phaseone, DUNE's solar neutrino reach will be limited by non-perfect light flash reconstruction and by radiological backgrounds, primarily neutron capture on argon, which can also confuse light-charge matching. As described in Section~\ref{sec:fd}, upgrades to the photon detection can increase DUNE light yield in \phasetwo by a factor of five or more. This will make light flashes from solar neutrino signals more apparent and improve vertex reconstruction from the flash to simplify the light-charge matching algorithm. Neutron capture on $^{40}$Ar could be vetoed by rejecting optical flashes that reconstruct near the 6.1\,MeV $Q$-value and further mitigated by installation of passive shielding. Together, these would reduce the solar neutrino detection threshold for a \phasetwo \dword{fd} module. DUNE's $\nu_e$~CC signal makes it ideal for measuring the energy dependence of the solar electron-neutrino survival probability, $P_{ee}$. With a visible energy threshold at or below 5\,MeV for solar neutrinos, possible with some technology choices outlined in Section~\ref{sec:fd}, DUNE will probe the upturn in $P_{ee}$. This is the transition region between the low-energy regime where vacuum solar oscillations dominate and the high-energy regime where the oscillation probability is determined by \dword{msw} matter effects~\cite{Smirnov:2004zv} inside the sun. The same MSW matter effects, but in the earth, are central to DUNE's measurements of neutrino mixing parameters with long-baseline oscillation measurements. A significant increase in photodetector coverage would allow measurements of \dword{cno} solar neutrinos that could distinguish between solar metallicity models~\cite{Bezerra:2023gvl}. 

A \dword{fd} module based on \dword{theia} would also be sensitive to solar neutrinos, through the neutrino-electron scattering channel. The \dword{theia} technology is sensitive to both scintillation and Cherenkov light from low-energy neutrinos -- thus simultaneously providing low thresholds and event directionality. Such a module could possibly probe the solar neutrino transition region to even lower energies, near 2\,MeV.  

\subsubsection{Other low-energy physics opportunities}

Through heavy fiducialization and improved energy resolution from increased photodetector coverage, DUNE's low-energy physics can reach beyond astrophysical neutrinos in \phasetwo. 

Planning is underway for future \dword{wimp} dark matter experiments. These liquid noble detectors will scale up the current technologies active target masses with the goal to reach the so-called “neutrino fog”, that is the cross-section below which the potential discovery of a dark matter signal is slowed due to the uncertainty in the irreducible background from the coherent elastic scattering of astrophysical neutrinos with nuclei~\cite{OHare:2021utq}. Leading upcoming projects include the xenon-based XLZD experiment~\cite{Baudis:2024jnk}, and the argon-based DarkSide-20k~\cite{Agnes:2023izm} and ARGO experiments. A DUNE \dword{fd} module could perform a competitive \dword{wimp} dark matter search complementary to future argon dark matter experiments~\cite{Church:2020env}. Particularly interesting would be the sensitivity to the annual modulation of the \dword{wimp} signal, which due to DUNE’s large target mass would allow a rapid confirmation of any observed signal in the current so-called generation-2 experiments~\cite{Bezerra:2023gvl}. A 10\% photodetector coverage with \dwords{sipm} could give the necessary threshold of $\sim$100~keV. Such low thresholds would also require operating the detector module with underground argon depleted in the $^{39}$Ar isotope, for background mitigation.

By introducing a large mass fraction of $^{136}$Xe or $^{130}$Te to a \phasetwo \dword{fd} module based on \dword{lartpc} or \dword{theia} technology, respectively, DUNE could also perform \dword{nldbd} searches. The neutrinoless double $\beta$ decay community has laid out a strategic plan~\cite{Adams:2022jwx} that calls for a diverse R\&D program with sensitivity beyond next-generation ton-scale experiments. DUNE \phasetwo can contribute toward this long-term \dword{nldbd} effort, in connection with, and as a possible evolution of, the existing \dword{nldbd} program. A DUNE \dword{theia} module loaded with $^{130}$Te could be a natural evolution of the loaded liquid scintillator technique currently pursued by SNO+~\cite{SNO:2015wyx} and KamLAND-Zen~\cite{KamLAND-Zen:2024eml}. Similarly, a $^{136}$Xe-doped DUNE \dword{lartpc} module with sufficiently good energy resolution $(\sigma)$ of order 
$2\%$~\cite{Mastbaum:2022rhw} could expand on liquid xenon TPC strategies currently employed by nEXO~\cite{nEXO:2021ujk}. Such a \dword{lartpc} \dword{nldbd} module would also require sourcing very large amounts of underground argon, in order to mitigate $^{42}$Ar-induced backgrounds.

Finally, thanks to its $\bar{\nu}_e$ sensitivity, a \dword{theia} module would also observe geo-neutrinos and reactor neutrinos~\cite{Theia:2019non}. 

%%%%%%%%%%%%%
\subsection{Physics beyond the Standard Model}
\label{subsec:physics_bsm}

DUNE has discovery sensitivity to a diverse range of physics Beyond the Standard Model (\dword{bsm}), which is complementary to those at collider experiments and other precision experiments. \dword{bsm} physics accessible at DUNE may be divided into three major areas of research: rare processes in the beam observed at the \dword{nd} (for example heavy neutrinos, light dark matter, or new physics that could enhance neutrino trident production), rare event BSM particle searches at the \dword{fd} (for example inelastic boosted dark matter, nucleon decays), and non-standard neutrino oscillation phenomena (for example sterile neutrino mixing, non-standard interactions). In the following, we give examples where DUNE \phasetwo will bring additional unique sensitivity with respect to \phaseone and to the performance of current experiments~\cite{DUNE:2020fgq}.

\subsubsection{Rare event searches at the near detector}

\begin{figure}[tb]
    \centering
    \includegraphics[width=1.0\textwidth]{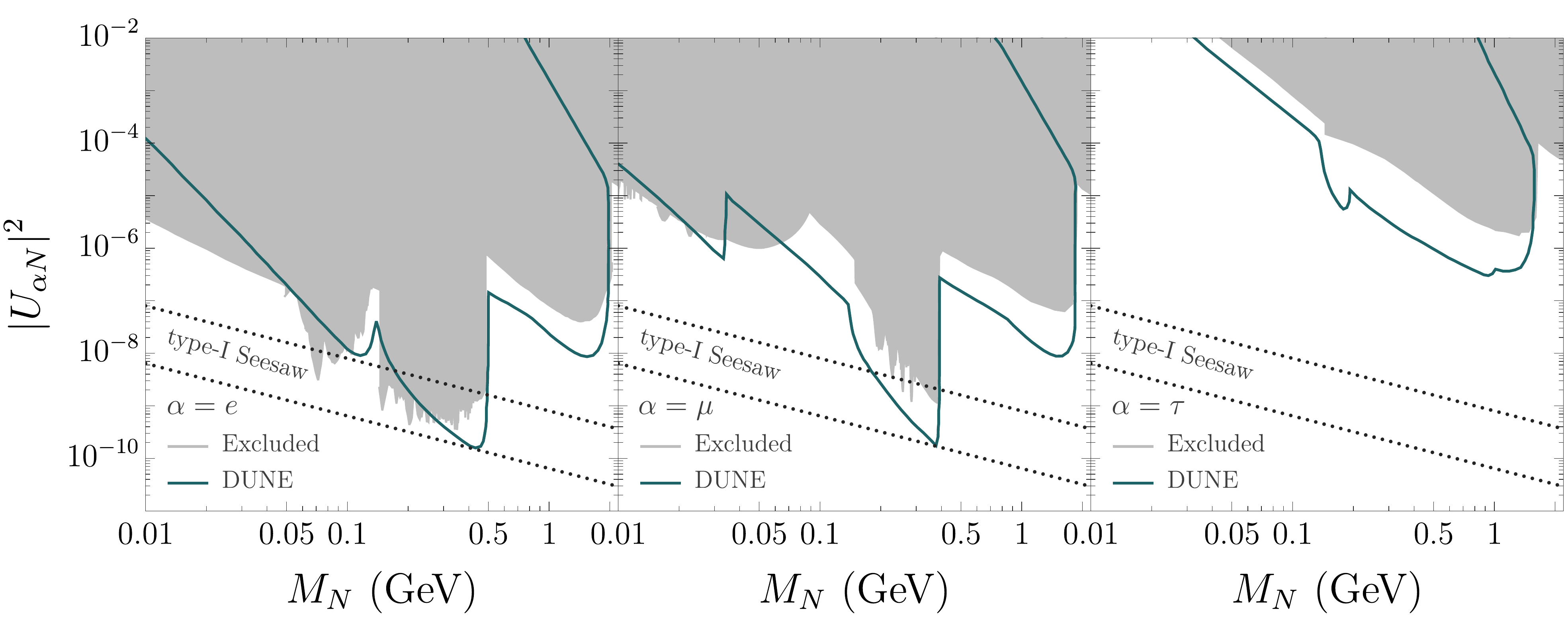}
    \caption{Expected DUNE \dword{nd} sensitivity at 90\% CL to the mixing $\vert U_{\alpha N}\vert^2$ as a function of the \dword{hnl} mass $M_N$, for a total of $7.7\cdot 10^{21}$ \dword{pot}, and combining all the possible \dword{hnl} decay channels leading to visible states in the detector. Backgrounds are assumed to be negligible. Results are shown for a \dword{hnl} coupled exclusively to: $e$ (left panel), $\mu$ (middle panel), and $\tau$ (right panel). The dotted gray lines enclose the region of parameter space where a Type I seesaw model could generate light neutrino masses in agreement with oscillation experiments and upper bounds coming from the latest KATRIN results on $\beta$-decay searches. A negligible background level after cuts and a signal selection efficiency of 20\% was assumed for this analysis. Figure adapted from~\cite{Coloma:2020lgy} to include the latest excluded areas from existing results; obtained with the \texttt{HNLimits}~\cite{Fernandez-Martinez:2023phj} package. All sensitivity curves and currently excluded areas assume Dirac neutrinos, and that the \dword{hnl} only couples to one of the charged leptons as indicated by the flavor index of the panel, while the other two mixings are set to zero.}
    \label{fig:hnl}
\end{figure}

The high intensity and high energy of the \dword{lbnf} proton beam enables DUNE to search for a wide variety of long-lived, exotic particles that are produced in the target and decay in the \dword{nd}. Heavy neutral leptons (\dwords{hnl}) and \dwords{alp} are examples of well-motivated searches that can be carried out in DUNE. Low density detectors are best suited for such a search, because the signal scales with the detector volume while the background (predominantly due to Standard Model neutrino interactions) scales with detector mass. In \phaseone, SAND can perform such searches with a density of $\sim 0.2$ g/cm$^2$. In \phasetwo, the \dword{ndgar} substantially improves the reach for these searches with even lower density and a larger volume. Signal efficiencies and background rates after selection cuts are found to improve significantly in \phasetwo, thanks to \dword{ndgar}~\cite{Coloma:2023oxx}. Background rates generally depend on the decay final state. In some cases background-free searches appear possible, for example for channels involving pairs of muons in the final state.

For \dwords{hnl}, the decay rates are proportional to $\lvert U_{\alpha N}\rvert^2$ in the case of single dominant mixing, where $\alpha = e,\mu,\tau$ and the matrix $U_{\alpha N}$ specifies the mixing between the active SM neutrinos $\alpha$ and the new heavy states $N$. 
Figure~\ref{fig:hnl} shows the combined sensitivity for \dword{hnl} decay channels probing each individual mixing matrix element $\lvert U_{\alpha N}\rvert^2$, as a function of the \dword{hnl} mass and under the assumption of no background (result adapted from~\cite{Coloma:2020lgy}).  As can be seen from the figure, DUNE is world-leading at masses below $m_\tau$, complementary to the LHC heavier mass searches. In addition, DUNE may have the potential to explore further the portion of parameter space predicted by Type I seesaw models.  Other phenomenological studies~\cite{Krasnov:2019kdc,Ballett:2019bgd,Berryman:2019dme,Breitbach:2021gvv} confirm the potential of the DUNE \phasetwo ND for \dword{hnl} searches. 

\dword{ndgar} is also sensitive to \dwords{alp} with masses between 20~MeV and 2~GeV \cite{Kelly:2020dda,Coloma:2023oxx}. DUNE is expected to improve over present constraints on \dword{alp} particles particularly for \dword{alp} masses below the kaon mass. For a wide range of \dword{alp} lifetimes, the sensitivity improvement would span many orders of magnitude. 

Another rare event search at the ND is neutrino trident production, that is, the production of a pair of oppositely-charged leptons through the scattering of a neutrino on a heavy nucleus. Neutrino trident production is a powerful probe of \dword{bsm} physics in the leptonic sector~\cite{Altmannshofer:2014pba,Ballett:2019xoj}. The \dword{sm} expectation is that the \dword{nd} will collect approximately a dozen of these rare events per ton of argon per year~\cite{Ballett:2018uuc,Altmannshofer:2019zhy}. To date, only the dimuon final-state has been observed, although with considerable uncertainties. The main challenge in obtaining a precise measurement of the dimuon trident cross-sections ($\nu_{\mu}\to\nu_{\mu}\mu^+\mu^-$ and $\bar{\nu}_{\mu}\to\bar{\nu}_{\mu}\mu^+\mu^-$) at DUNE will be the copious backgrounds, mainly consisting of \dword{cc} single-pion production events ($\nu_{\mu}N\to\mu\pi N'$), as muon and pion tracks can be easily confused. \dword{ndgar} will tackle this search by improving muon-pion separation through $dE/dx$ measurements in the \dword{hpgtpc} and the calorimeter system, and \dword{ndgar}'s magnetic field will significantly improve signal-background separation by tagging the opposite charges of the two muons in the final state. 

\subsubsection{Rare event searches at the far detector}

\phasetwo will also enhance \dword{bsm} searches at the \dword{fd}, and in particular searches that are expected to be nearly background-free at the scale of the experiment's full exposure. In such cases, the decay or scattering rate sensitivity will be inversely proportional to the FD exposure (in kt$\cdot$yr), and added exposure in \phasetwo FD modules will be significant. Background-free (or quasi-background-free) searches at the FD may include baryon-number-violating processes. For example, current estimates \cite{DUNE:2020fgq} for the $p\to K^+\bar{\nu}$ search yield a mean background rate expectation of 0.4 events for a 400\,kt$\cdot$yr exposure at the FD.

\subsubsection{Non-standard neutrino oscillation phenomena}

DUNE is sensitive to neutrino oscillation scenarios beyond the standard three-flavor picture, including sterile neutrinos, non-standard interactions, and PMNS non-unitarity. These searches rely on both the ND and FD, and require high precision and very large exposures, such that both the \phasetwo ND and FD are important.

In addition to searching for BSM modifications to the muon and electron neutrino signals, DUNE also has a unique capability to search for tau neutrino appearance because the broadband LBNF beam has significant flux above the $\sim 3.5$~GeV tau charged-current threshold. Searches for tau appearance would enable DUNE to directly constrain the tau elements of the PMNS matrix, and also to search for anomalous $\nu_{\tau}$ appearance, which may point to mixing with \dwords{hnl} or non-standard interactions~\cite{DeGouvea:2019kea,Ghoshal:2019pab}. This would become particularly interesting if hints of non-unitarity are observed in the muon and electron channels. The $\tau$ lepton from beam $\nu_{\tau}$ \dword{cc} interactions is not directly observable in the DUNE detectors due to its short $2.9\times 10^{-13}$\,s lifetime. However, the final states of $\tau$ decays ($\sim65\%$ into hadrons, $\sim18\%$ into $\nu_\tau+e^-+\bar{\nu}_e$, and $\sim17\%$ into $\nu_\tau+\mu^-+\bar{\nu}_\mu$) can be detected.

The \dword{lbnf} beamline is designed such that the target and the horn focusing system can be replaced. The \dword{intlproj} will provide targets and horns designed for 1.2~MW operation, in the standard low-energy beam tune configuration optimized for \dword{cpv} measurements. New \dword{acemirt} upgrades designed for $>$1.2~MW operation will be needed, and could provide the capability to run with a higher-energy beam tune optimized for detection above the $\tau$ production threshold. Studies indicate that the $\nu_{\tau}$ charged-current interaction rate will more than double at the FD in this case, compared to the standard LBNF beam optimized for \dword{cpv}~\cite{Rout:2020cxi}. Running with the high-energy tune is not currently planned, but could provide further physics reach for DUNE in \phasetwo.

At the ND, the baseline is far too short for $\nu_\mu\rightarrow\nu_\tau$ oscillations to occur within a three-flavor scenario. However, $\nu_\tau$ originating in rapid oscillations driven by sterile neutrinos could be detected. A challenge is that for the tau decay to muon channel, a large fraction of the signal is at very high energy. In \phaseone, the \dword{tms} cannot reconstruct momentum by curvature, and is limited to measuring $T_{\mu} < \sim6$~GeV by range. It may be possible to search for anomalous tau appearance in SAND, which is sensitive to higher energy muons by curvature but has much smaller target mass. In \phasetwo, \dword{ndgar} provides a magnetized spectrometer for \dword{ndlar} which can reconstruct very high-energy muons by curvature. With one year of data with the \phasetwo ND, preliminary studies show that DUNE's reach in this channel extends beyond the present strongest limits from NOMAD~\cite{NOMAD:2001xxt}.

\section{The DUNE phase II far detector}
\label{sec:fd}

\subsection{Introduction}

The primary objective of the \dune \phasetwo \dword{fd} is to increase the fiducial mass of DUNE to at least the originally planned 40\,kt \dword{lar}-equivalent mass. For long-baseline neutrino oscillations, it is critical that all four FD modules be compatible with the systematic constraints of the \dword{nd}. Non-\dword{lar} options for the \phasetwo FD would require corresponding additions or changes to the ND complex in order to achieve a comparable level of systematic uncertainty. These options are described in Section~\ref{subsec:theiand}.
For MeV-scale physics, the additional mass would double the number of expected neutrino interactions in a \dword{snb} and extend the reach to supernovae beyond the Milky Way. An exposure of hundreds of kt-yrs is required to improve upon oscillation parameter measurements with solar neutrinos. Most \dword{bsm} searches also require very long exposures to be competitive.

Enhancements to the detector design have the potential to improve the DUNE program by lowering the threshold for MeV-scale neutrinos or by reducing the background rates in this energy regime. \phasetwo also presents opportunities to expand the DUNE science program to new areas while preserving the essential core measurement capabilities. The design of the \phasetwo FD modules will also incorporate lessons learned from the construction of \phaseone, in particular the \dword{vd} \dword{fd2}, optimizing performance and cost. 
%%%%%%%%%%%%%%

\subsection{The vertical drift detector design}
\label{subsec:fd_fd2}

\begin{figure}[htbp]
\centering
\includegraphics[width=0.6\linewidth]{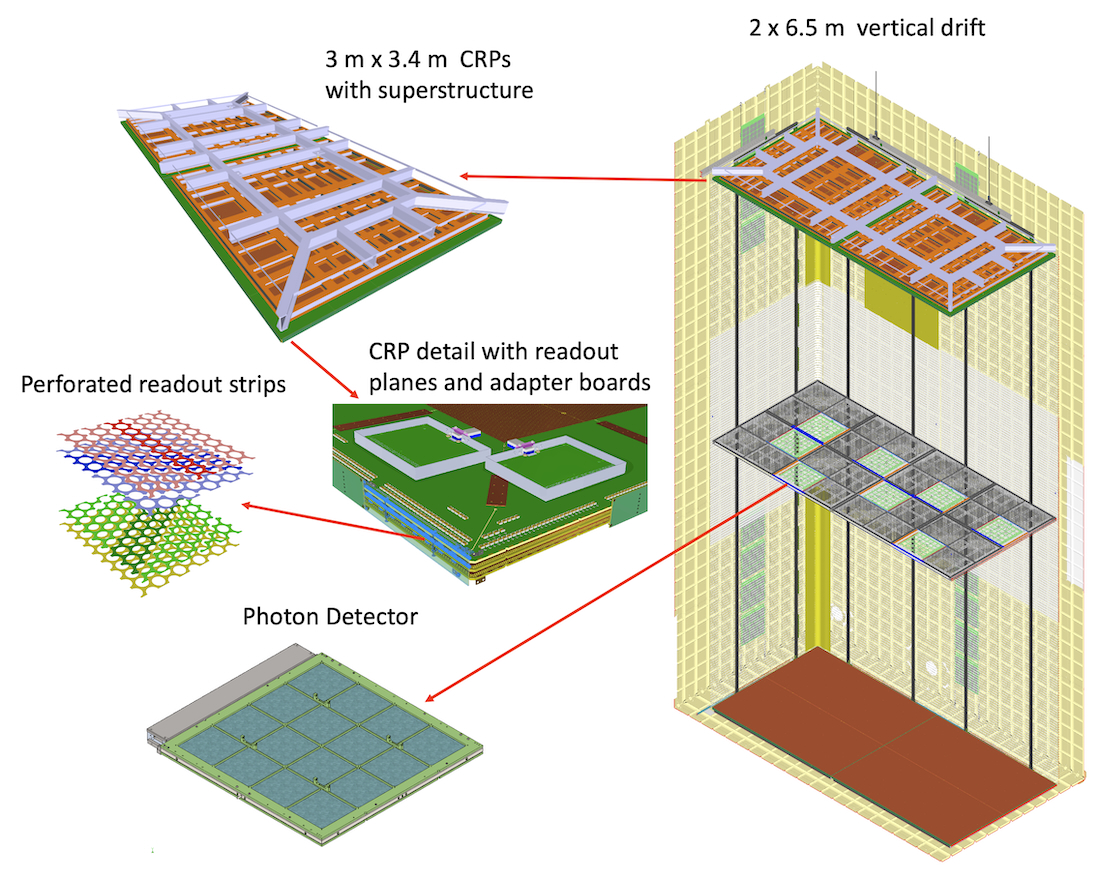}
\caption[Conceptual vertical drift design with \dword{pcb}-based charge readout] 
{Schematic of the vertical drift \dword{fd2} concept with \dword{pcb}-based charge readout. Corrugations on cryostat wall shown in yellow; \dword{pcb}-based CRPs (brown, at top and bottom with superstructure in gray for top CRPs); cathode (violet, at mid-height with openings for photon detectors); \dword{fc} modules (white) hung vertically around the perimeter (the portions near the anode planes are 70\% optically transparent); photon detectors (light green at right), placed in the openings on the cathode and on the cryostat walls, around the perimeter in the vertical regions near the anode planes. Updated from \cite{DUNE:2023nqi}.}
\label{fig:vertical_drift}
\end{figure}

\begin{figure}[htbp]
\centering
\includegraphics[width=0.7\linewidth]{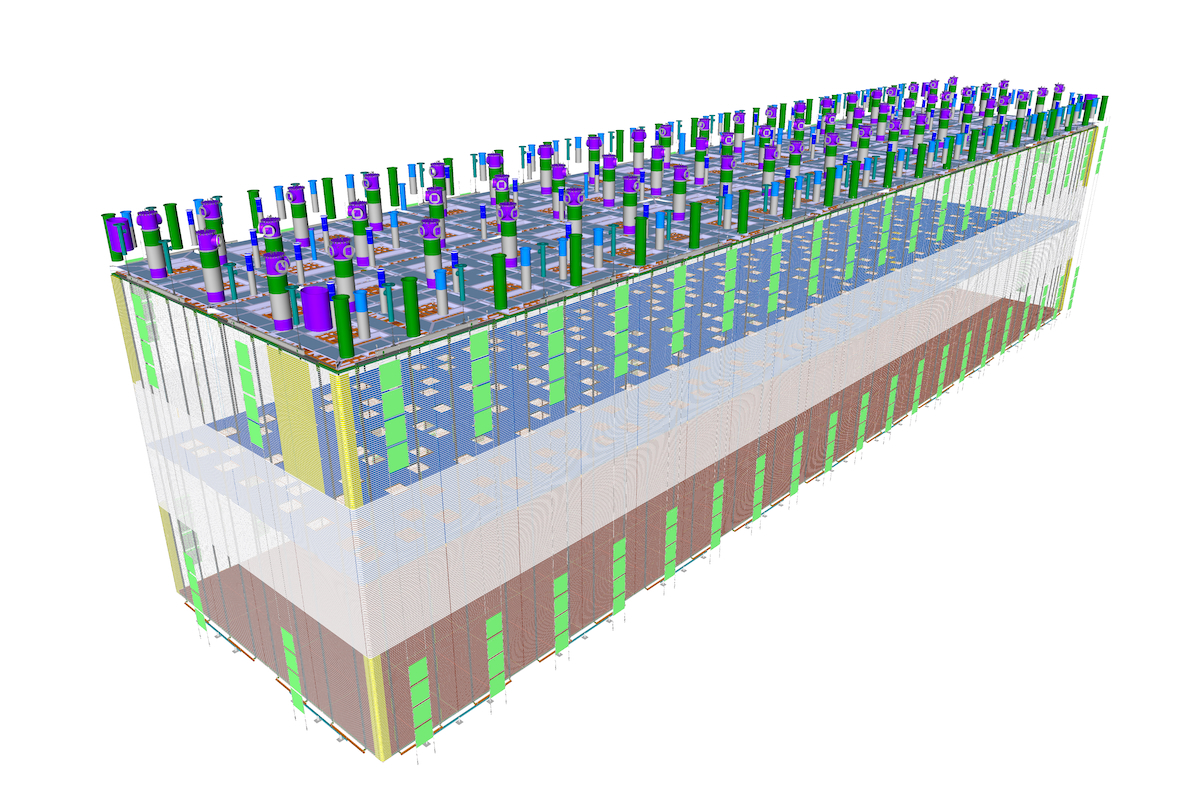}
\caption[Vertical drift detector]{Perspective view of the \dword{vd} \dword{fd2} detector. From \cite{DUNE:2023nqi}.} 
\label{fig:FD2-Detector}
\end{figure}

The single-phase \dword{vd} technology as implemented in \dword{pdvd} and planned for \dword{fd2}~\cite{DUNE:2023nqi} (Figures~\ref{fig:vertical_drift} and \ref{fig:FD2-Detector}) draws from the strengths of the \dshort{dune} prototypes \dword{pddp}~\cite{DUNE:2018mlo} and \dword{pdsp}~\cite{DUNE:2021hwx} as well as the previous \dword{hd} detectors \dword{icarus} and \dword{microboone}. 
Relative to the well-established \dword{sp} \dword{hd} design that is based on large wire plane assemblies, the vertical drift design simplifies detector construction and installation, reducing overall detector costs. The \dword{vd} module uses most of the same structural elements as the 
\dword{pddp} design (e.g., \dwords{crp} to form the anode planes, and the \dword{fc} that hangs from the cryostat roof), and is constructed of modular elements that are much easier to produce, transport, and install.

The cathode at the vertical mid-plane of the detector is suspended from the top \dshort{crp} support structure and subdivides the detector into two vertically stacked, 6.5\,m high drift volumes, with \dshort{crp} readout for both the top and bottom drift volumes. The top \dshorts{crp} are suspended from ports on the cryostat roof 
whereas the bottom \dshorts{crp} are supported by feet on the cryostat floor. 

The important features of the vertical drift design, particularly in comparison with the \dword{fd1} horizontal drift design, are:
\begin{itemize}
\item maximizing the active volume\footnote{For reference, the \dword{fd2} detector model yields an active volume of 10,586~m$^3$, a 5.6\% increase over the estimated \dword{fd1} active volume of 10,021~m$^3$.};
\item high modularity of detector components;
\item simplified anode structure based on standard industrial techniques;
\item simplified cold testing of instrumented anode modules in modest size cryogenic vessels;
\item \dword{fc} structure independent of the other detector components;
\item extended drift distance;
\item reduction of dead material in the active volume;
\item allowance for improved light detection coverage; 
\item simplified and faster installation and \dword{qa}/\dword{qc} procedures; and
\item cost-effectiveness.
\end{itemize}

\subsubsection{Charge readout planes (anodes)}
\label{subsubsec:fd_fd2_crp}

\begin{figure}[htbp]
\centering
\includegraphics[width=0.7\textwidth]{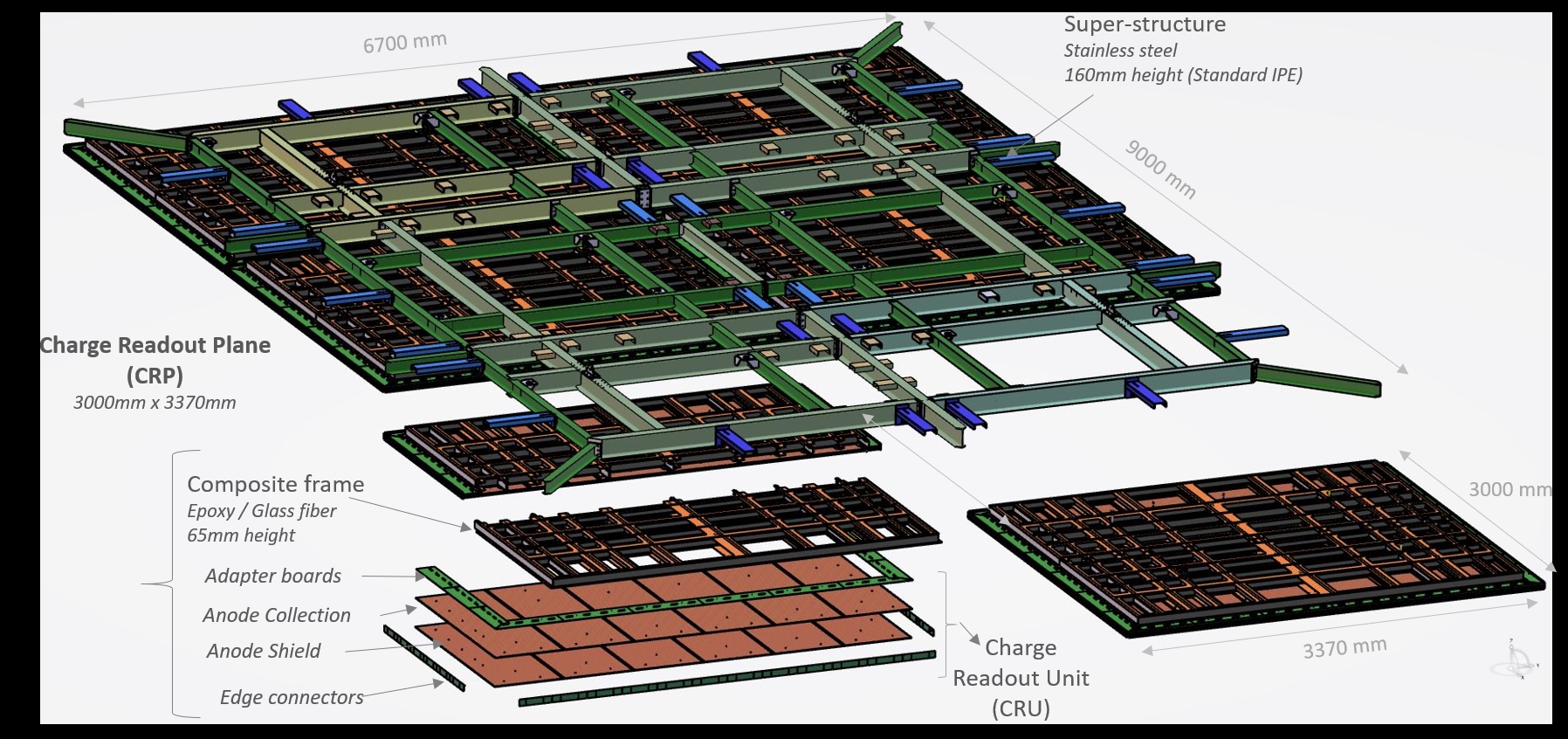}
\caption[Exploded view of a top superstructure and CRPs]{A top superstructure (green structure on top) that holds a set of six \dshorts{crp}, and below it an exploded view of a \dshort{crp} showing its components: the \dwords{pcb} (brown), adapter boards (green) and edge connectors that together form a \dshort{cru}, and composite frame (black and orange). From \cite{DUNE:2023nqi}.}
\label{fig:supercrp2ch1}
\end{figure}

The baseline design \dshort{fd2} anodes, illustrated in Figure~\ref{fig:supercrp2ch1}, provide three-view charge readout via two induction planes and one collection plane. The anodes are fabricated from two double-sided, perforated, 3.2\,mm thick printed circuit boards (\dwords{pcb}), that are connected mechanically, with their perforations aligned, to form \dwords{cru}. A pair of \dshort{cru}s is attached to a composite frame to form a \dword{crp}; the frame provides mechanical support and planarity. The holes allow the electrons to pass through to reach the collection strips. Each anode plane consists of 80 \dshorts{crp} in the same layout. The \dshorts{crp} in the top drift volume, operating completely immersed in the \dshort{lar}, are suspended from the cryostat roof using a set of superstructures, and the bottom \dshorts{crp} are supported by posts positioned on the cryostat floor. The superstructures hold either two or six \dshorts{crp}, and allow adjustment, via an externally accessible suspension system, to compensate for possible deformations in the cryostat roof geometry.

The \dshort{fd2} top and bottom drift volumes implement different \dword{cro} electronics.
The top anode is read out via the top drift electronics (\dword{tde}), based on the design used in \dshort{pddp}, that comprises both cold and warm components housed in \dwords{sftchimney}. These chimneys penetrate the cryostat roof, allowing the components to be fully accessible for repair or upgrade. 
The bottom detector electronics (\dword{bde}), on the other hand, implements the same \dword{ce} used in the \dword{hd} \dshort{fd1}, which features local amplification and digitization on the \dshort{crp} in the \dshort{lar}, thereby maximizing the signal-to-noise.

\subsubsection{High-voltage system}
\label{subsubsec:fd_fd2_hv}

\begin{figure}[htbp]
\centering
\includegraphics[width=0.7\textwidth]{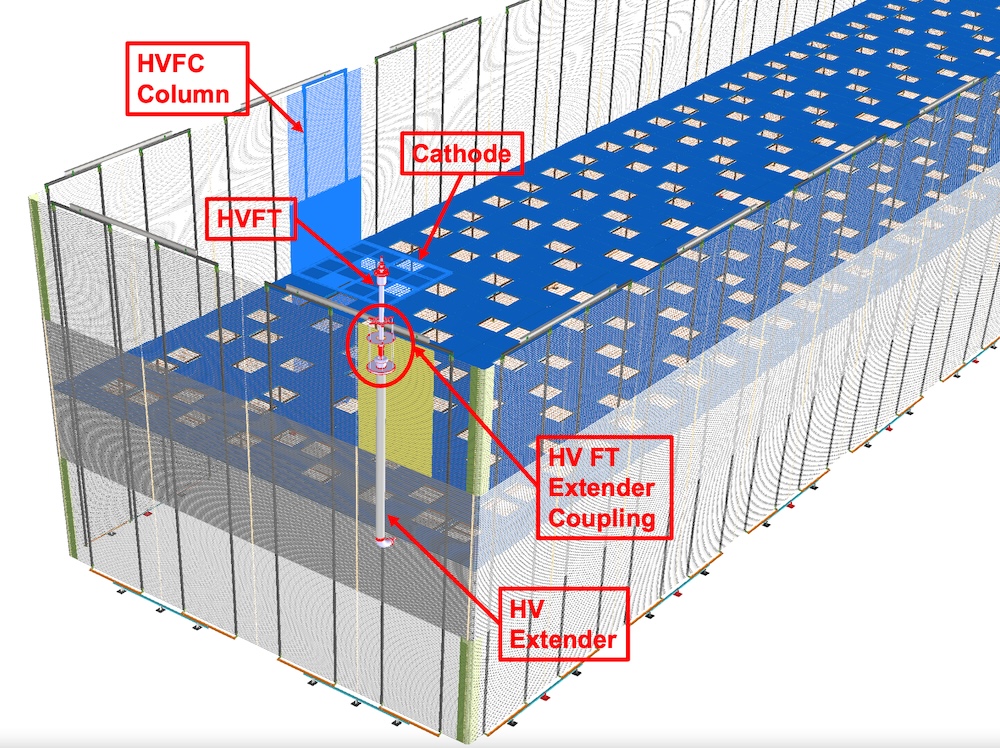}
\caption[HV system components inside the cryostat] 
{A bird's-eye view of the \dshort{fc}, with one full-height \dshort{fc} column (highlighted in cyan) that extends the entire height, the \dshort{hv} feedthrough and extender (in the foreground), and the cathode (with one cathode module highlighted in cyan and \dword{xarapu} modules installed on cathode). From \cite{DUNE:2023nqi}.}
\label{fig:vd_hvsch1}
\end{figure}

The \dword{fd2} has a horizontal cathode plane placed at detector mid-height, held at a negative voltage, and horizontal \dwords{anodepln} (biased at near-ground potentials) at the top and bottom of the detector, which together provide a nominal uniform \efield of 450\,V/cm. The main \dword{hvs} components are illustrated in Figure~\ref{fig:vd_hvsch1}.

The \dshort{hvs} is divided into two systems: (1) supply and delivery, and (2) distribution. The supply and delivery system consists of a negative high voltage power supply (\dword{hvps}), \dword{hv} cables with integrated resistors to form a low-pass filter network, a \dshort{hv} feedthrough (\dword{hvft}), and a 6\,m long extender inside the cryostat to deliver $-$294\,kV to the cathode. The distribution system consists of the cathode plane, the \dword{fc}, and the \dshort{fc} termination supplies. The cathode plane is an array of 80 cathode modules, each with the same footprint as a \dword{crp}, formed by highly resistive top and bottom panels mounted on fiber-reinforced plastic (\dword{frp}) frames. The modular \dshort{fc} consists of horizontal extruded aluminum electrode profiles stacked vertically at a 6\,cm pitch. A resistive chain for voltage division between the profiles provides the voltage gradient between the cathode and the top-most and bottom-most field-shaping profiles.

In addition to the primary function of providing uniform \efield{}s in the two drift volumes, both the cathode and the \dshort{fc} designs are tailored to accommodate \dword{pds} modules (Section~\ref{subsubsec:fd_fd2_pds}) since it is not possible to place them behind the \dshort{anodepln}, as in the \dshort{sphd} design. Each cathode module is designed to hold four double-sided \dword{xarapu} \dshort{pds} modules that are exposed to the top and bottom drift volumes through highly transparent wire mesh windows. Along the walls, the \dshort{fc} is designed with narrow (15~mm width) profiles in the region within 4\,m of the \dshort{anodepln} to provide 70\% optical transparency to single-sided \dshort{pds} modules mounted on the cryostat membrane walls behind them, and conventional (46~mm width) profiles within 2.5\,m of the cathode plane.  

\subsubsection{Photon detection system}
\label{subsubsec:fd_fd2_pds}

The \dshort{fd2} module will implement \dword{xarapu}~\cite{machado2018x, Brizzolari:2021akq} \dword{pds} modules. Functionally, an \dshort{xarapu} module is a light trap that captures wavelength-shifted photons inside boxes with highly reflective internal surfaces until they are eventually detected by \dwords{sipm}. An \dshort{xarapu} module has a light collecting area of approximately 600~$\times$~600\,{\rm mm$^2$} and a light collection window on either one face (for wall-mount modules) or on two faces (for cathode-mount modules). The wavelength-shifted photons are converted to electrical signals by 160 \dshorts{sipm} distributed evenly around the perimeter of the \dword{pd} module. Groups of \dshorts{sipm} are electrically connected to form just two output signals, each corresponding to the sum of the response of 80 \dshorts{sipm}.

Since their primary components are almost identical to those of \dshort{sphd}, only modest R\&D was required for the \dshort{fd2} \dshort{pds} modules. The primary differences were to optimize the module geometry and the proximity of the \dshorts{sipm} to the \dword{wls} plates.  Both of these are more favorable in \dshort{fd2}, leading to more efficient light collection onto the \dshorts{sipm}. 
As discussed in Section~\ref{subsubsec:fd_fd2_hv}, the design has the \dshorts{pd} mounted on the four cryostat membrane  walls and on the cathode structure, facing both top and bottom drift volumes. This configuration produces approximately uniform light measurement across the entire \dshort{tpc} active volume. 

Cathode-mount \dshorts{pd} are electrically referenced to the cathode voltage, avoiding any direct path to ground. While membrane-mount \dshorts{pd} adopt the same copper-based sensor biasing and readout techniques as in \dshort{sphd}, cathode-mount \dshorts{pd} required new solutions to meet the challenging constraint imposed by \dword{hvs} operation. The cathode-mount \dshort{pds} are powered using non-conductive \dword{pof} technology~\cite{Arroyave:2024vgj}, and the output signals are transmitted through non-conductive optical fibers, \dword{sof}, thus providing voltage isolation in both signal reception and transmission. 

%%%%%%%%%%%%%%
\subsection{Optimized charge and photon readouts for \phasetwo\ vertical drift FD modules}
\label{subsec:fd_vdoptimized}

Several variations on the vertical drift design are under consideration to improve the performance and/or reduce the cost. Potential improvements can be broadly grouped into two classes, the charge readout and the photon readout systems.

Optimizations of the charge readout system include options to improve the production processes and reduce the cost of the \dwords{crp}, possible optimizations of strip pitch and length, and channel count (Section~\ref{subsubsec:fd_vdoptimized_crp}). Other options are to replace the strip-based \dword{crp} with a pixel based \dword{crp} (Section~\ref{subsubsec:fd_vdoptimized_pixels}) or with an optical readout based on electroluminescence (Section~\ref{subsubsec:fd_vdoptimized_ariadne}).

The leading criteria for selecting an optimized technology for a \phasetwo photon readout system are performance enhancement at low incremental costs, and the ability to leverage minimum-risk development of solutions already demonstrated and adopted for \phaseone. One of the most attractive options is the proposed \dword{apex} concept (Section~\ref{subsubsec:fd_vdoptimized_apex}), in which \dwords{pd} are integrated into the \dword{fc}. The \dword{apex} concept makes use of the \dword{pof} and \dword{sof} technologies developed for \dword{fd2} and opens up the opportunity to greatly extend the optical coverage. Another optimization under consideration is the addition of photon detection to a pixel-based \dword{crp} (Section~\ref{subsubsec:fd_vdoptimized_lightpixels}).

%%%%%%%%%%%%%%%%%%%%
\subsubsection{Optimized photon readout with \dshort{apex}}
\label{subsubsec:fd_vdoptimized_apex}

The \dword{apex} concept (Aluminum profiles with embedded \dshort{xarapu}) integrates a large-area photon detection system into the detector module's \dword{fc}. \dword{apex}
is a simplified, lightweight, and low(er)-cost photodetector solution for optimizing photon readout that increases the active optical coverage of the \dword{lar} target volume. This solution is derived from the well-established \dword{xarapu} technology with \dword{sipm} photosensors developed for \dword{fd2}. Concrete examples of physics topics enabled by an improved light detection system are given in Sec.~\ref{subsec:physics_astro}, in the context of the neutrino astrophysics program of DUNE.

\begin{figure}[htbp]
\begin{center}
\includegraphics[width=0.7\textwidth]{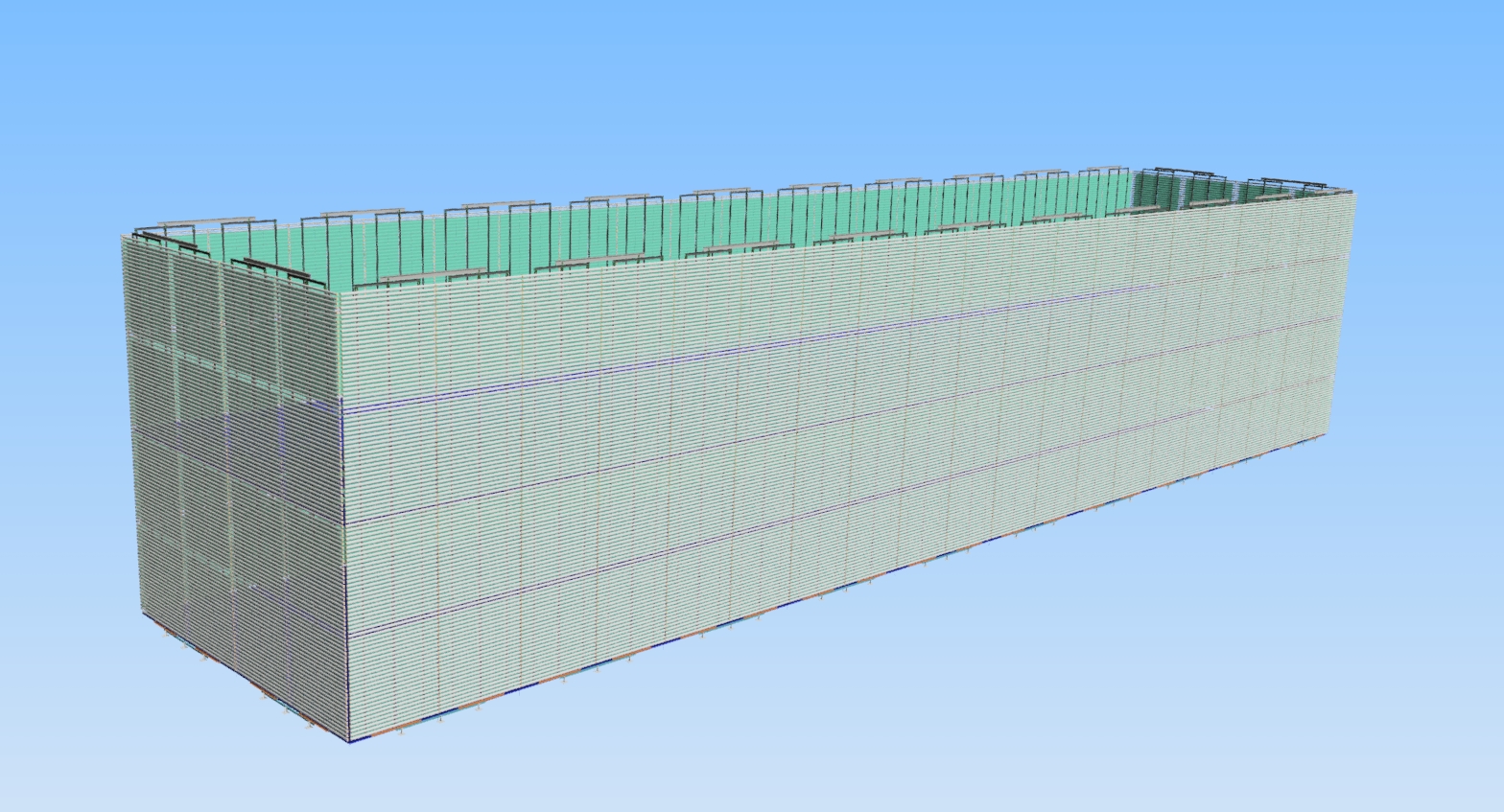}
\caption{A bird's-eye view of the \dword{fc} with integrated large-area photon detection system, \dword{apex}. The \dword{fc} structure is constructed of modules vertically stacked in groups of four hanging around the \dword{lar} drift volume perimeter from the top.}
\label{FC-APEX-FD3}
\end{center}
\end{figure}

The \dword{fc} covers the four vertical sides of the VD \dword{lartpc} active volume between the top anode and bottom anode planes, and thus offers the largest available surface for extended optical coverage, i.e., \dword{apex} can provide up to $\sim\,60\%$ coverage of the surface enclosing the \dword{lartpc} active volume if the \dword{fc} walls are fully instrumented, as shown in Fig.~\ref{FC-APEX-FD3}. In the \dword{apex} concept, no \dword{pd} modules are installed on the cathode. The \dword{pd} readout electronics would need to be referenced to the (high) voltage level of the \dword{fc} electrode profile on which the \dword{pd} module is installed, and therefore would require electrical isolation. Power and signal transmission can be established via non-conductive optical fibers by using the \dword{pof} and \dword{sof} technologies developed for the \dword{fd2} photon detectors that are integrated into that FD module's \dword{hv} cathode plane. These technologies are described in Section~\ref{subsubsec:fd_fd2_pds}. They have been demonstrated to work reliably for electrical isolation with noise immunity and long-term stability in \dword{lar}.

\begin{figure}[htbp]
\begin{center}
\includegraphics[width=0.9\textwidth]{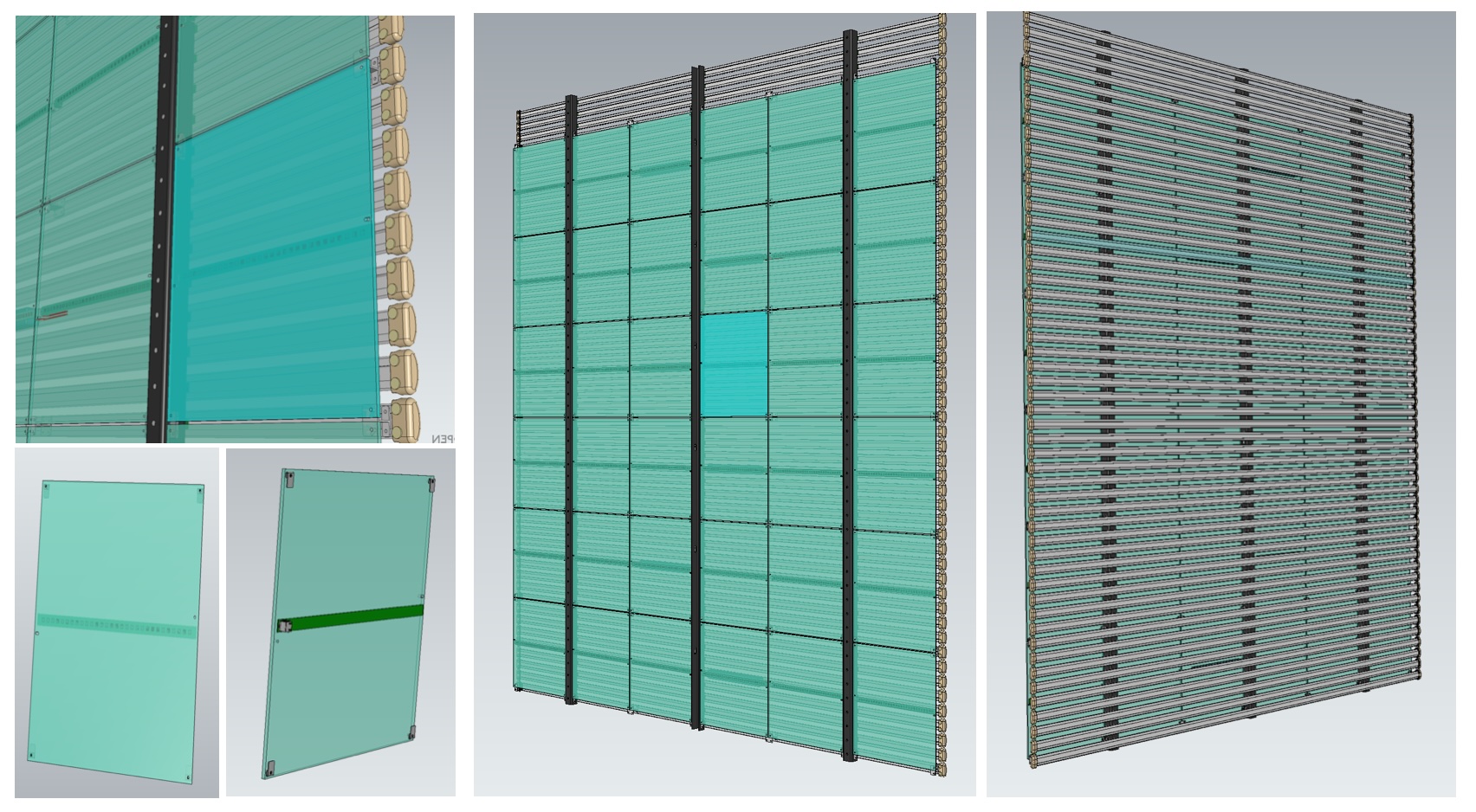}
\caption{An \dword{apex} panel. Top left: one PD module installed on \dshort{fc} profiles. Bottom left: a \dword{pd} module equipped with a \dword{sipm} strip at the center. Center: front view of an APEX panel showing the $6\times6$ array of \dword{pd} modules mounted on an \dshort{fc} module. Right: a back view of the \dshort{fc} module showing its aluminum profile structure.}
\label{APEX-panel}
\end{center}
\end{figure}

\dword{apex} keeps the same \dword{fc} structure as designed for \dword{fd2} (see Figure~\ref{fig:vd_hvsch1}), which includes 24 \dshort{fc} supermodules, each made up of eight $3.0\times 3.2\,{\rm m}^2$ \dword{fc} modules, for a total of 192 modules. A \dword{fc} module consists of horizontal extruded aluminum C-shaped electrode profiles (3\,m long and 6\,cm wide) stacked vertically at a 6\,cm pitch and mounted on vertical FR4 I-beams. An \dword{apex} panel, illustrated in Figure~\ref{APEX-panel}, is a standard \dword{vd} \dword{fc} module instrumented with a $6\times6$ array of thin, large-area ($\sim\,50\times50{\rm \,cm}^2$) \dword{xarapu}-type \dword{pd} modules installed onto the \dword{fc} structure and fully covering it, as shown in Figure~\ref{APEX-panel}. Hydrodynamic simulations are under development to understand the potential impact of \dword{apex} panels on the \dword{lar} recirculation. 

\begin{figure}[htbp]
\begin{center}
\includegraphics[width=0.7\textwidth]{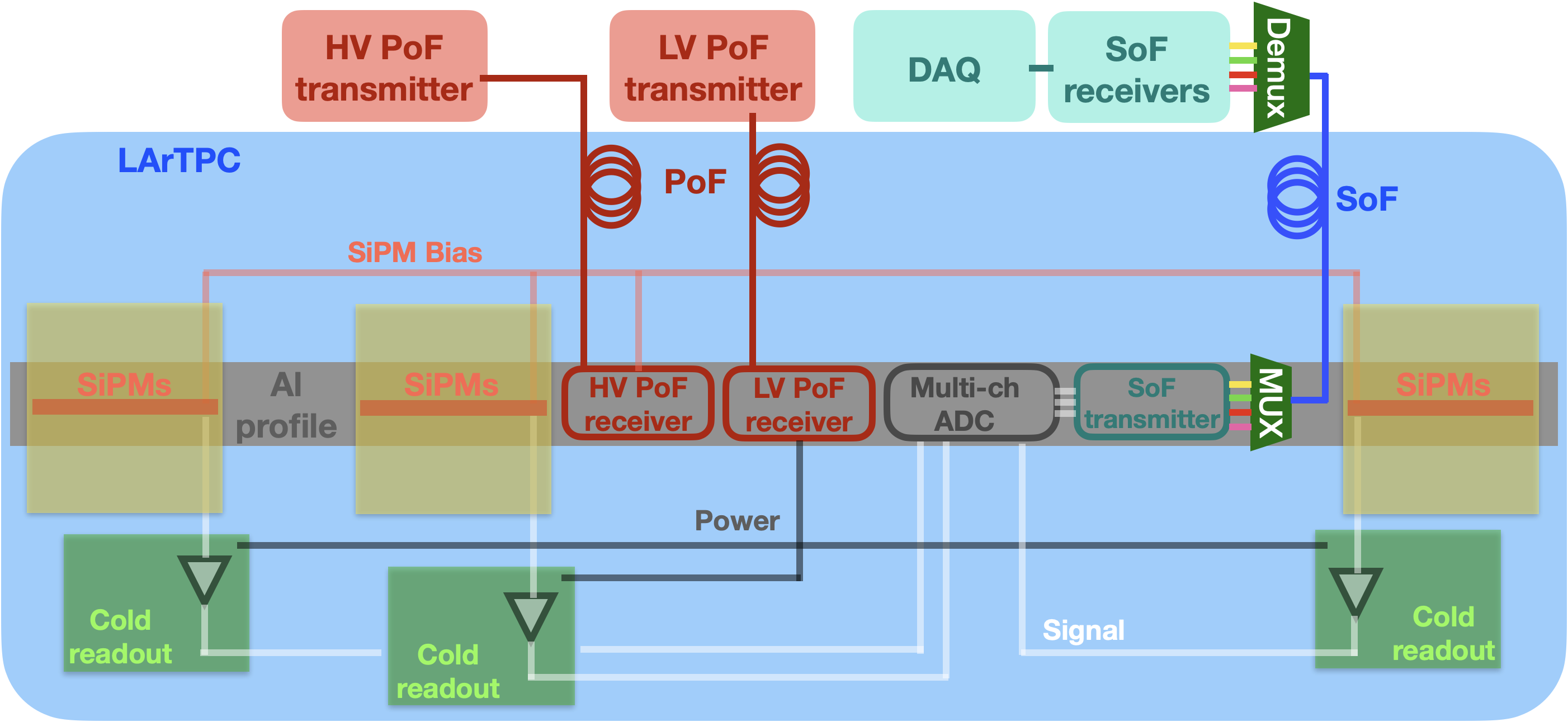}
\caption{APEX cold readout concept: a row of six \dword{pd} modules (three of the six are not shown in order to display the readout elements) in an \dword{apex} panel forming an electrically isolated single readout system.}
\label{APEX-CE}
\end{center}
\end{figure}

Six \dword{pd} modules constitute a horizontal row of the \dword{apex} panel. Each \dword{pd} module vertically spans about nine \dshort{fc} profiles and is mechanically fastened and electrically referenced to the profile at its mid-height (the fifth of nine). The cavity of this profile houses and provides Faraday shielding for the cold electronics readout boards for all six of the \dword{pd} modules in that row of the array, providing signal conditioning and digitization in cold. Several \dword{pof} receivers and an \dword{sof} transmitter (driver and laser diode) at the center of the $\sim$3\,m long profile receive power and transmit signal, respectively, for the PD modules in the row via optical fibers. The signals from the six PD modules are multiplexed and transmitted over a single optical fiber to the (warm) receivers and \dword{daq}, as schematically represented in the block diagram of Figure~\ref{APEX-CE}. The fibers are routed through the penetration at the top of the cryostat using the central vertical I-beam of the \dword{fc} structure as conduit. Each row of six \dword{pd} modules in an \dword{apex} panel thus forms an electrically isolated system.

The \dword{pd} module, the basic unit of the \dword{apex} array, is a simplified version of the light trap \dword{xarapu} concept used in the \dword{fd1} and \dword{fd2} \dword{pds}, designed to be a lightweight ($\sim$1.8\,kg), compact object suitable for efficient mass production. Two \dword{wls} stages, \#1 and \#2 -- with a \dword{df} between them -- convert, transmit, and trap incident \dword{lar} scintillation light under the dichroic filter layer. These components are contained in solid PMMA (transparent acrylic) slabs that are 6\,mm thick, with a surface area of $\sim\,50\times50\,{\rm cm}^2$, illustrated in Figure~\ref{APEX-panel}, bottom left. The \dword{df} layer is deposited directly on the front plane of the acrylic substrate, and the \dword{wls} \#1 coating on top of the \dword{df}. Chromophore molecules embedded in the substrate PMMA (\dword{wls} \#2) matrix shift transmitted light to a wavelength above the \dword{df} cutoff. Light trapping is optimized by ultra-high reflectivity non-metallic thin film lamination (e.g., Vikuiti ESR) of the acrylic slab edges and backplane. An array of \dwords{sipm} mounted on a flex \dword{pcb} are optically bonded to the acrylic surface, as shown in Figure~\ref{APEX-panel}, bottom left. Photons trapped by reflection in the slab are eventually absorbed by the photosensors, producing electronic signals. We estimate that 80 large-area \dwords{sipm} per module, with high photon detection efficiency (\dword{pde}), ganged together into one readout channel, will be sufficient to reach an overall detector efficiency of $\epsilon_D\simeq2\%$. 

Assembly of \dword{apex} panels is expected to be simple. To assemble one, aluminum \dword{fc} profiles are first assembled to form a \dword{fc} module, electronics boards are positioned in the profiles, and \dword{pd} modules are connected to the boards then fastened to the profiles to complete the \dword{apex} panel. An \dword{apex} assembly can be built out of bulk materials (aluminum profiles and acrylic plates) with low-radioactive content. For this reason, even an extended \dword{pds} coverage compared to \phaseone modules is not expected to be a dominant contributor to the internal background budget discussed in Sec.~\ref{subsec:fd_backgrounds}.

\begin{figure}[htbp]
\begin{center}
\includegraphics[width=0.60\textwidth]{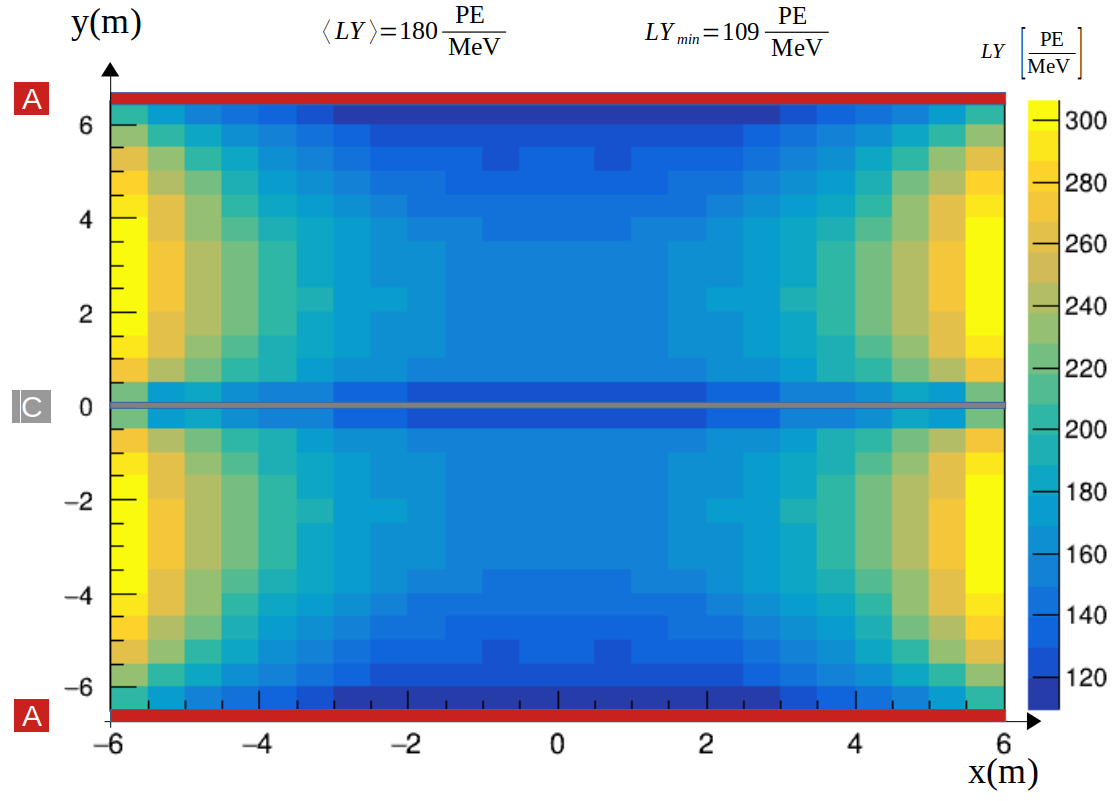}
\caption{Map showing the expected \dword{ly} in the central $(x, y)$ transverse plane at $z = 0$ for the \dword{fc}-extended coverage \dword{apex} photon detection system. Dimmer regions are present near the anode planes (with no \dwords{pd} on them) and at mid-height (near the non-instrumented cathode plane).}
\label{APEX-LY}
\end{center}
\end{figure}

A \dword{geant4} simulation was performed for a \phasetwo FD module with \dword{apex}, assuming 55\% optical coverage of the \dword{lar} volume. The light yield $LY(x,y,z)$ of the system was evaluated, i.e., the number of \dwords{pe} collected per unit of deposited energy anywhere in the \dword{lar} volume, assuming 2\% detection efficiency of the \dword{xarapu} module and standard \dword{lar} scintillation light emission and propagation parameters. The simulation resulted in an average value $\langle LY\rangle=180$~PE/MeV across the detector volume, with a minimum of $LY_{\mathrm{min}}=109$~PE/MeV near the anode planes, thanks to the extended optical coverage of the \dword{apex} system. Such a light yield would be about a factor of 5 higher than the \dword{fd2} one, where simulations indicate an average light yield of $39$~PE/MeV and a minimum light yield of $16$~PE/MeV~\cite{DUNE:2023nqi}. Figure~\ref{APEX-LY} shows the \dword{ly} map in the transverse plane at the center of the FD module long axis. \dword{apex}-specific studies on light-only and charge+light calorimetric performance, impact of non-uniform light collection, and achievable \dword{pds} thresholds in the presence of background flashes, are in progress.

A series of prototypes are planned to fully develop the \dword{apex} concept. A first round of prototyping, carried out at CERN, has studied the impact on the drift field uniformity of placing insulating material between \dword{fc} electrodes. The (expected) observation of a slow buildup of static charge on the surface of the insulating material may, counter-intuitively, allow reduction of the number of \dword{fc} electrodes, with a larger pitch. The current (2024) focus is on a second (ton-scale) TPC prototype at \dword{cern} that will be instrumented with up to eight full-size \dword{pd} modules, primarily for mechanical and cryogenic tests. Additionally, a \dword{pd} module prototype with a full electronic chain, including \dword{pof} and \dword{sof} systems, will be constructed and tested in parallel before being integrated into this prototype. A larger-sized \dword{apex} demonstrator in a several cubic meter \dword{lar} cryostat, with $\mathcal{O}$(100) \dword{sof} and \dword{pof} in/out fibers, will be a third-stage prototyping goal in 2024-2025, likely at Fermilab. Finally, a full-sized \dword{apex} \dword{pd}-instrumented \dword{fc} will be deployed in the VD \dword{protodune} cryostat at CERN, along with the proposed \dword{ariadne} optical readout (Section~\ref{subsubsec:fd_vdoptimized_ariadne}).

%%%%%%%%%%%%%%

\subsubsection{Strip-based charge readout}
\label{subsubsec:fd_vdoptimized_crp}

The \dword{pcb}-based \dword{vd} anodes, called \dwords{crp}, are made up of two stacked \dwords{pcb}, providing three projective views. The \dword{pcb} face directly opposite the cathode has a copper guard plane to absorb any unexpected discharges. The reverse side of this \dword{pcb} is etched with strips that form the first induction plane. The other \dword{pcb} has strips on the side facing the inner \dword{pcb} forming the second induction plane, and has the collection plane strips on its reverse side~\cite{DUNE:2023nqi}. The \dwords{pcb} are supported by composite frames and mechanically connected using \dword{peek} spacers. \dwords{crp} have been successfully demonstrated in the 50\,L test stand and at full scale in the vertical drift \coldbox. They have been installed in \dword{pdvd} in the NP02 cryostat at \dshort{cern} and will be deployed in the \dword{fd2} cryostat at \dword{surf} (see Figure~\ref{fig:supercrp2ch1}).

This system has already been optimized for deployment in \dword{fd2}, and as such forms a reference solution also for \dword{fd3}. Additional optimizations should be explored for \dword{fd3} to reduce cost or to improve performance further. There are various ways in which the strip-based charge readout might be re-optimized that would impact the strip pitch, length, and orientation. A more concrete optimization plan will be developed after assessing \dword{pdvd} performance in 2025.

Additional considerations to be explored include the \dword{crp} fabrication techniques, including faster production of the \dword{pcb} itself and simpler quality control. Some of these ideas have been tested in the 50\,L test stand at \dword{cern}, particularly techniques to reduce \dword{pcb} hole misalignments.

In addition to the \dword{crp}, the readout electronics need some re-optimization. Depending on the timescale for construction of FD3, some of the existing electronics production lines may no longer be available. This may require re-design of components associated with readout of the top and/or bottom \dwords{crp}. For example, a potential optimization of the \dshort{fd2} \dword{bde} includes porting the \dword{larasic}, a custom pre-amplifier and shaping \dword{asic}, from the 180\,nm to the 65\,nm production process\footnote{The 180\,nm and 65\,nm processes are advanced lithographic techniques used in semiconductor fabrication; the dimension refers to feature size.} to mitigate the risk of losing access to the 180\,nm process. The two other \dword{bde} custom \dwords{asic}, \dword{coldadc} and \dword{coldata}, are already using the 65\,nm process. Other potential optimizations for the \dshort{fd2} \dword{bde} design include reducing the cable length for signal and power (it is 27\,m for the current \dshort{fd2} design), simplifying detector installation.

%%%%%%%%%%%%%%

\subsubsection{Pixel-based charge readout}
\label{subsubsec:fd_vdoptimized_pixels}

A pixel-based readout would replace the multi-layer strip-based readout with a single-layer grid of charge-sensitive pixels at mm-scale granularity. Instrumenting each pixel with a dedicated electronics channel would achieve a \dword{lartpc} with true and unambiguous \threed readout, where information on the third spatial dimension is provided by the \dword{lartpc} drift time.
This is advantageous compared to conventional wire-based two-dimensional readout, especially for higher-multiplicity interactions of $O$(GeV) neutrinos. In interactions with several charged particles in the final state, it is possible for tracks to overlap in one 2D projection, making it more difficult to reconstruct. Similarly, straight-line tracks in strip-based readout have pathological angles where the track is recorded entirely along a single strip.
Given channel densities of $\mathcal{O}$(10$^5$) pixels per m$^2$ of anode, pixel readout would require operation at $\mathcal{O}$(100)~$\mu$W power consumption per channel, including amplification, digitization and multiplexing. This is necessary in order to avoid excessive heating of the \dword{lartpc} detector, operating near \dword{lar} boiling point.  
Significant progress has been made in recent years in the development of pixel readout for \dwords{lartpc}, overcoming issues with excessive waste heat, as well as demonstrating cryo-compatibility, $\mathcal{O}$(10$^4$) digital multiplexing, and cost-effective, scalable production.
Two options for readout are discussed below, \dword{larpix} and \dword{qpix}. 

%%%%%%%%%%%%%%
\bigskip
\noindent{\bf\dshort{larpix} Readout}
\bigskip

\noindent\dword{larpix} \cite{Dwyer:2018phu} is a complete pixel readout system for \dwords{lartpc}, consisting of 6400-channel pixel anode tiles, cryogenic-compatible data and power cabling, and a multi-tile digital controller with an integrated operating system. It has been developed as the baseline technology of \phaseone \dword{ndlar}. The system relies on the \dword{larpix} \dword{asic}, a 64-channel detector system-on-a-chip that includes analog amplification, self-triggering, digitization, digital multiplexing, and a configuration controller.

The \dword{larpix}-v1 \dword{asic} demonstrated that waste heat could be controlled through a custom low-power amplifier and channel self-triggering, where the digitization and digital readout are dormant until a signal is detected on the pixel. The \dword{larpix}-v2 \dword{asic} incorporated a variety of improvements to facilitate large-scale production of pixel anodes, including Hydra-IO, a novel programmable chip-to-chip data routing technique to improve system reliability in the inaccessible cryogenic detector environment.

\begin{figure}[htbp]
\centering
\includegraphics[width=0.4\textwidth]{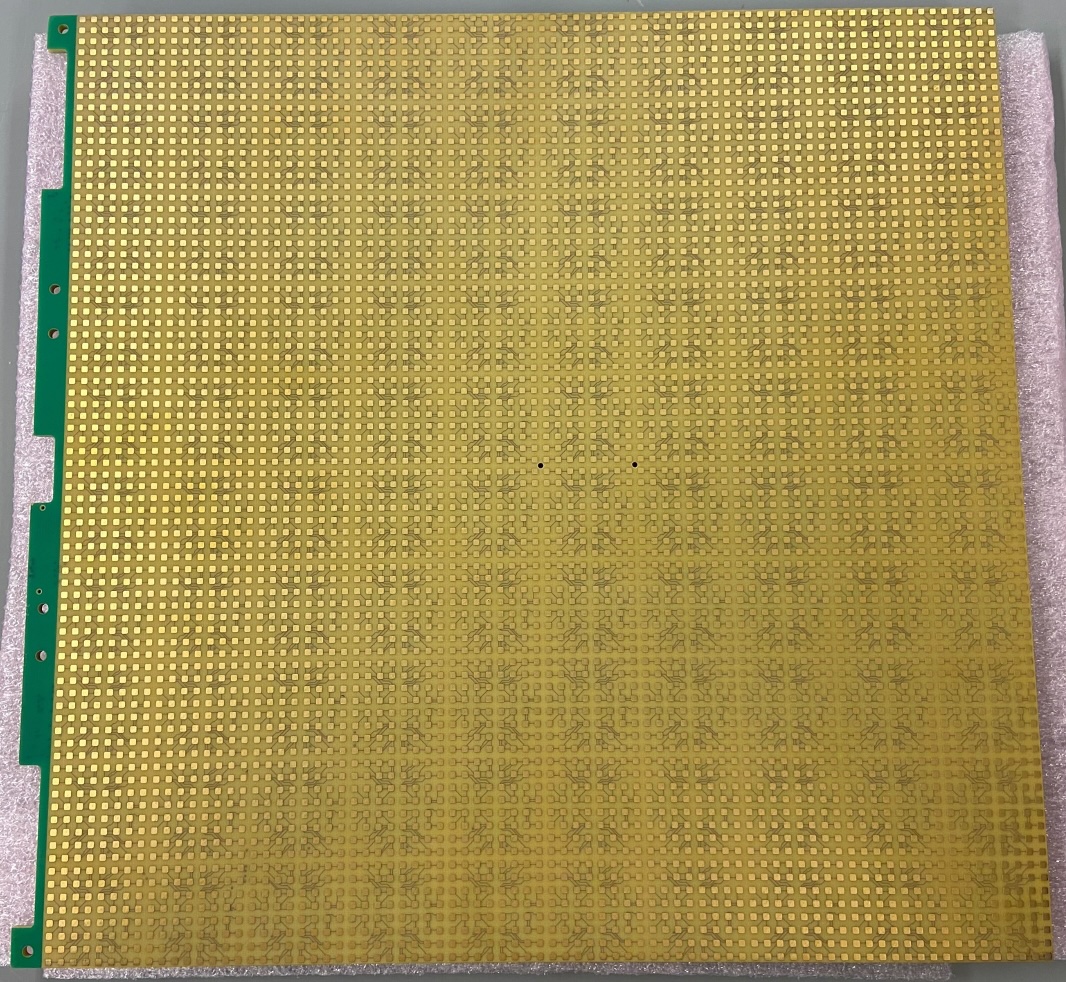}
\includegraphics[width=0.4\textwidth]{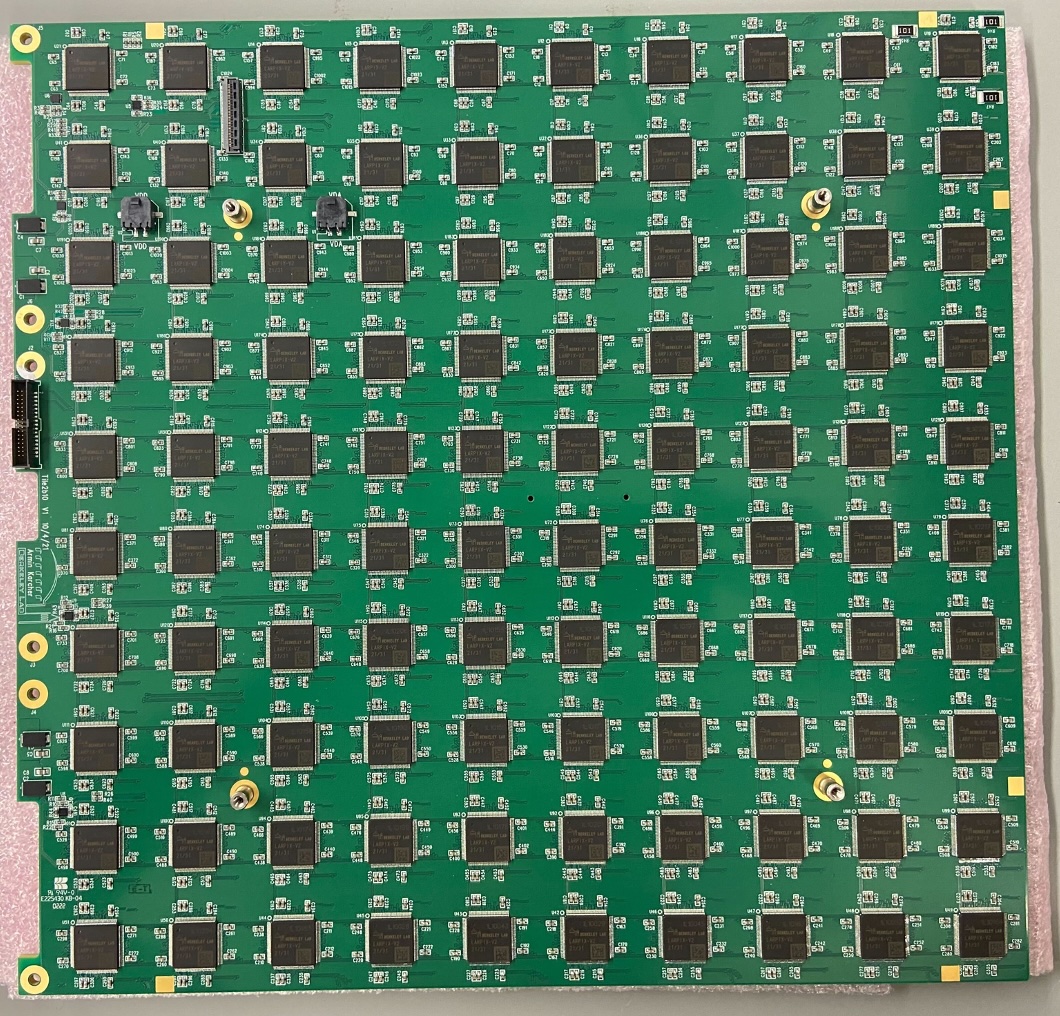}
\caption[\dword{larpix}-v2 pixel anode tile] 
{Left: Prototype \dword{larpix}-v2 anode tile 32~cm in length by 32~cm in height, with 6400 gold-plated charge sensitive pixel pads at 3.8~mm pitch driven by (right) 100 \dword{larpix}-v2 \dwords{asic}.}
\label{fig:larpix_tile}
\end{figure}

The current 32$\times$32~cm$^2$ \dword{larpix} pixel tile (Figure~\ref{fig:larpix_tile}) has 6400 charge-sensitive pixels at 3.8~mm pitch and can be configured and read out via a single set of differential digital input and output wires. The design leverages standard commercial techniques for \dword{pcb} production to realize a \dword{lartpc} anode, achieving 800~e$^-$~equivalent noise charge per channel on the sensitive TPC-facing side of the tile, while powering and communicating with 100 \dword{larpix} \dwords{asic} on the back side. The power consumption achieved by the current \dword{larpix} tile is 14~W/m$^2$, ensuring that the heat flux from the anode is lower than the one from the cryostat walls.

Data acquisition is controlled by the Pixel Array Controller and Network (PACMAN) card, responsible for delivering power and communication to the tiles. A single compact controller is currently capable of driving $\mathcal{O}$(10) pixel tiles (e.g., $\mathcal{O}$(10$^5$) pixels). It includes a CPU with integrated operating system and programmable logic similar to a \dword{fpga}. The controller is designed to mount on the room-temperature side of a \dword{lartpc} cryostat feedthrough, and incorporates power filtering and ground isolation to ensure the integrity of the low-noise environment within the detector.

\begin{figure}[htb]
\centering
\includegraphics[width=0.35\textwidth]{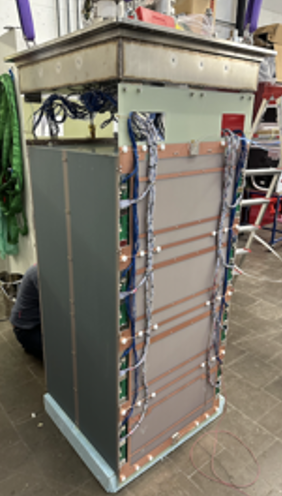}
\includegraphics[width=0.55\textwidth]{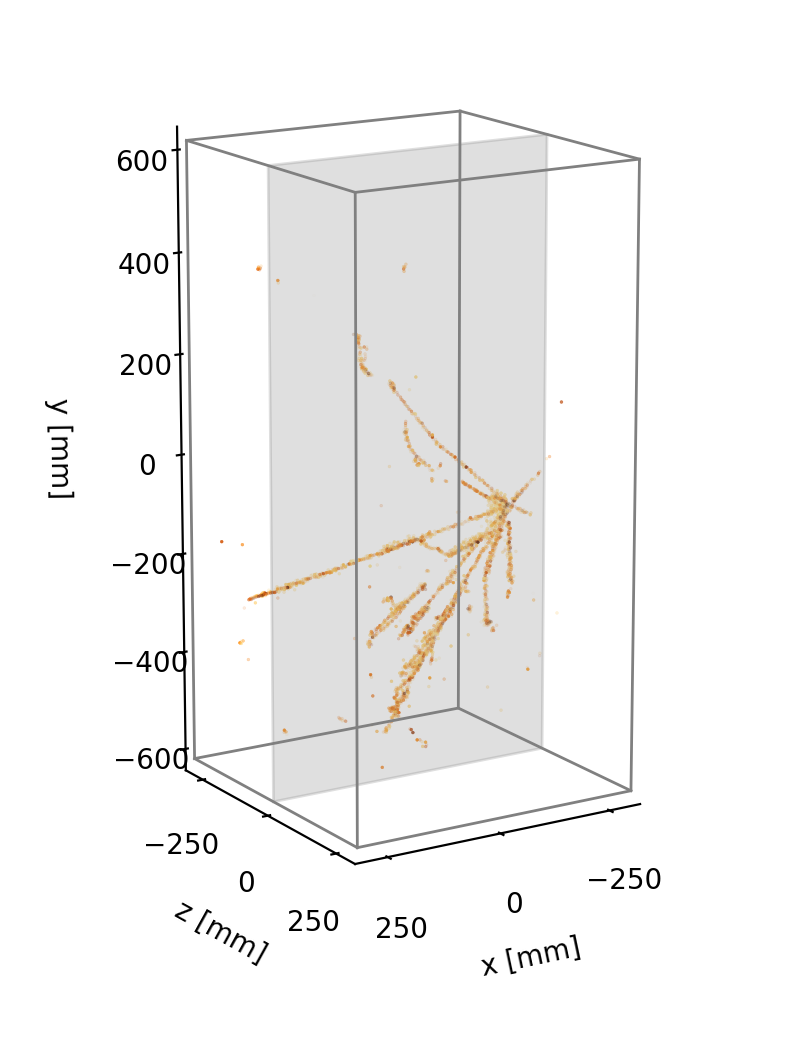}
\caption[\dword{larpix} in the DUNE ND prototype] 
{Left: A photograph of one of the four ton-scale \dword{lartpc} modules for the 2x2 Demonstrator, a prototype of the DUNE \dword{nd}. Right: an example cosmic ray imaged in true \threed using a 102,400-channel \dword{larpix}-v2 system in this module (right).}
\label{fig:larpix_ndproto}
\end{figure}

Approximately 80 \dword{larpix}-v2 pixel tiles have been produced as part of the current prototyping program for the DUNE \dword{nd}. Sets of 16 pixel tiles have been used to instrument each of the four ton-scale \dword{lartpc} modules of the 2$\times$2 Demonstrator (Figure~\ref{fig:larpix_ndproto}), a prototype of the modular \dword{lartpc} design planned for the \dword{nd}. Each module has been operated at the University of Bern, and has been used to image over 100~million cosmic ray events.

The \dword{larpix}-v2 development program has achieved its goal of a scalable design. All components are produced via commercial vendors using traditional electronics production techniques, and are ready for integrated testing; no additional assembly is required. \dword{larpix}-v2 system production costs, including all cabling, controllers, and power supplies, are approximately \$10k per square meter.

Operation of prototype \dword{larpix} anodes, in either vertical or horizontal orientations, show that natural convection provides sufficient heat dissipation to mitigate argon phase transition (bubble formation or boiling). In particular, the 2x2 Demonstrator~\cite{DUNE:2024fjn}, a prototype of the DUNE \dword{nd}, is constructed of multiple fiberglass boxes with a very low perforation, approximately 1\% of the surface, and rather limited spaces for convective heat dissipation, yet it shows no issues with thermal management and has achieved purity in excess of 2~ms. Future work includes a demonstration of \dword{larpix} anode heat dissipation in a configuration similar to the \dword{fd2} design.

Assuming completion of the development program of \dword{larpix} for the DUNE \dword{nd}, \dword{larpix} would already meet most of the requirements for deployment in a future \dword{fd} module. The development and integration of a high-speed, $\mathcal{O}$(1)\,GHz, 16-to-1 digital multiplexer would significantly reduce the number of cables and feedthroughs, making deployment in a \dword{fd} much more feasible. Tests of a large-scale \dword{larpix} prototype in the ProtoDUNE-VD system at the \dword{cern} Neutrino Platform are important to validate the integration and interfaces with the other aspects of the \dword{vd} design.

%%%%%%%%%%%%%%
\bigskip
\noindent{\bf\dshort{qpix} Readout}
\bigskip

\begin{figure}[htbp]
\begin{center}
\includegraphics[width=1.0\textwidth]{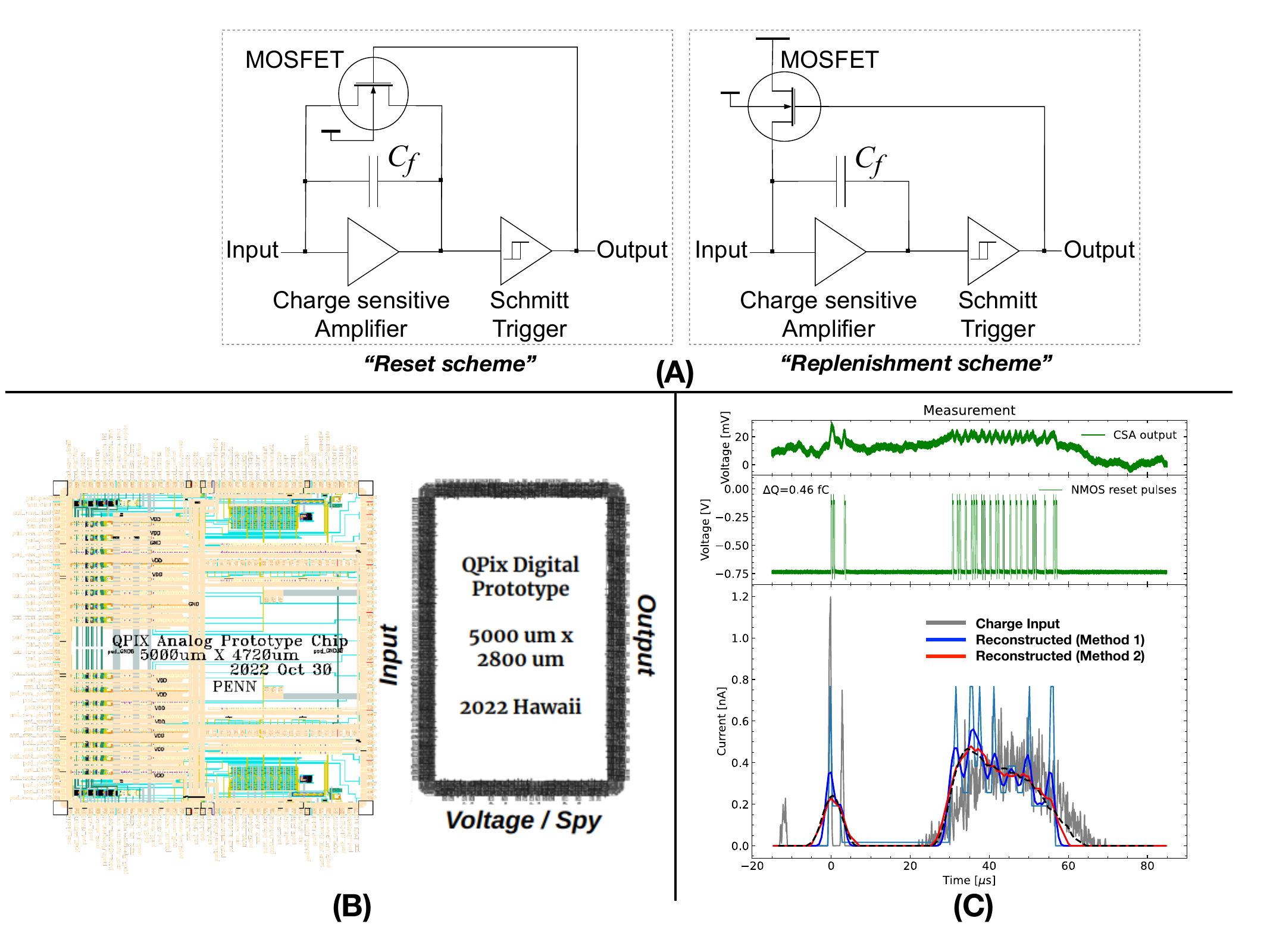}
\caption{A) Schematic of the basic concepts of the \dword{qpix} circuits for the reset and replenishment schemes. B) Left: Schematic of the 16-channel analog front-end and Right: schematic of the 16-channel digital design. C) Current waveform reconstructed using a discrete-component implementation of the Q-Pix replenishment scheme at a charge threshold of 0.46~fC ($\sim 2875 e^-$) and reconstructed using different digital filtering based on analysis. Images A and C are adapted from \cite{Miao:2023ivo}.}
\label{QPixCharge}
\end{center}
\end{figure}

\noindent 
\dword{qpix} is a novel pixel-based technology for low-threshold, high-granularity readout that is expected to improve reconstruction relative to projective-based readout, at a much reduced data throughput. It is ideally suited to the low data rate readout environment of the DUNE \dword{fd} modules. The basic concepts of the \dword{qpix} circuit \cite{Nygren:2018rbl} are shown in Figure \ref{QPixCharge} (A). The input pixel is envisioned to be a simple circular trace connected to the \dword{qpix} circuit via a \dword{pcb}. The circuit begins with the ``Charge-Integrate/Reset'' (CIR) circuit. This charge-sensitive amplifier continuously integrates incoming signals on a feedback capacitor until a threshold on a Schmitt trigger (regenerative comparator) is met. When this threshold is met, the Schmitt trigger starts a rapid ``reset'' enabled by a \dword{mosfet} switch, which drains the feedback capacitor and returns the circuit to a stable baseline, at which point the cycle is free to begin again. To mitigate any potential charge loss, an alternative design known as the ``replenishment'' scheme has also been evaluated. In contrast to the reset architecture, the \dword{mosfet} now functions as a controlled current source such that when the Schmitt trigger undergoes a transition, the \dword{mosfet} replenishes a charge of $\Delta Q =I\cdot\Delta t $, where $\Delta t$ is the reset pulse width or discharge time. 

Both the ``reset'' and ``replenishment'' schemes capture and store the present time of a local clock within one \dword{asic}. This changes the basic quantum of information for each pixel from the traditional ``charge per unit of time'' to the difference between one clock capture and the next sequential capture, the Reset Time Difference (RTD). This new unit of information measures the time to integrate a pre-defined charge. Physics signals will produce a sequence of short, $\mathcal{O}(\mu s)$, RTDs. On the other hand, in the absence of a signal, the quiescent input current from $^{39}$Ar and other radiogenic or cosmogenic backgrounds would be small, producing long, $\mathcal{O}(s)$, RTDs. Signal waveforms can be reconstructed from RTDs by exploiting the fact that the average input current and the RTD are inversely correlated.

\dword{qpix} has shown that this architecture can enhance the physics capabilities of a large-scale \dword{lartpc} through its ability to provide full 3D information of the events, as opposed to the three 2D projections provided by \dword{fd1}/\dword{fd2} readout. The first of these demonstrations shows the improved reconstruction enabled by a pixel based detector when compared to a projective-based readout for DUNE multi-GeV neutrino interactions \cite{Adams:2019uqx}. This analysis showed enhanced efficiency and purity across all neutrino interaction types analyzed and the ability to reconstruct the topology and content of the hadronic system (including number of final state protons, charged, and neutral pions). Moreover, through an analysis of supernova neutrino interactions and a simulation of the \dword{qpix} architecture, it was shown that \dword{qpix} can significantly enhance the low-energy neutrino capabilities for kiloton-scale \dwords{lartpc}. Specifically, \dword{qpix}: i) enhances the efficiency of reconstructing low-energy supernova neutrino events over the nominal wire based readout, ii) allows for a high-purity and high-efficiency identification of supernova neutrino candidates, and iii) affords these enhancements at data rates $10^6$ times less for the same energy threshold~\cite{Q-Pix:2022zjm}.

A number of prototypes are currently under construction and evaluation to demonstrate the \dword{qpix} readout architecture. These include designs in both \dword{cmos} 180\,nm and 130\,nm, evaluation of the architecture using discrete \dword{cots} components, as well as extensive digital prototyping using \dwords{fpga}. The 180\,nm design is shown schematically in Figure~\ref{QPixCharge} (B). It consists of a 16 channel analog chip implemented with the replenishment architecture, and a 16 channel digital chip. On the other hand, Figure~\ref{QPixCharge} (C) shows the implementation of the replenishment architecture for the \dword{qpix} readout using \dword{cots} discrete components. This prototype was able to demonstrate the fidelity of reconstructing input from an arbitrary waveform generator with a replenishment threshold of 0.46\,fC with replenishment pulse widths $\sim 300 - 600$\,ns and linear responses to replenishment up to 2\,MHz rates. This demonstration provides confidence that the architecture proposed will be capable of meeting the performance needs of future large scale \dwords{lartpc}.  The consortium of universities and labs working on this project expect both small and large scale demonstrator ($\mathcal{O}(1000 - 100,000)$ pixel) \dwords{lartpc} in the coming next few years.

%%%%%%%%%%%%%%

\subsubsection{Optical-based charge readout}
\label{subsubsec:fd_vdoptimized_ariadne}

The optical-based readout shares the same physics benefit as the pixel-based charge readout solutions (Sec.~\ref{subsubsec:fd_vdoptimized_pixels}) in providing a native 3D readout. This technology has also demonstrated the best spatial resolution of any \dword{lartpc} readout option so far, with $\simeq$ 1.1~mm per pixel~\cite{Hollywood:2019loi}. The data-driven readout with native zero suppression yields a very efficient raw data storage, of relevance for \dword{snb} physics. The overall optical gain and the low-noise readout environment enable low-threshold ($\simeq$500 e$^-$ per pixel) operation, supporting DUNE's MeV-scale neutrino astrophysics program. From the technical point of view, the (off-cryostat) optical readout also benefits from simplicity of access, greatly simplifying maintenance and upgrade operations. Finally, depending on the granularity versus cost trade-off chosen, significant cost savings compared to other readout technologies may also be present.

The optical charge readout with fast cameras was developed within the \dword{ariadne} program and represents a cost-effective and powerful alternative approach to the existing charge readout methodology. As first demonstrated in the one-ton dual-phase \dword{ariadne} detector, the secondary scintillation (S2) light produced in \dword{thgem} holes can be captured by fast Timepix3 (TPX3) cameras to reconstruct the primary ionization track in \threed{}.

\begin{figure}[htbp]
\centering
\includegraphics[width=0.78\textwidth]{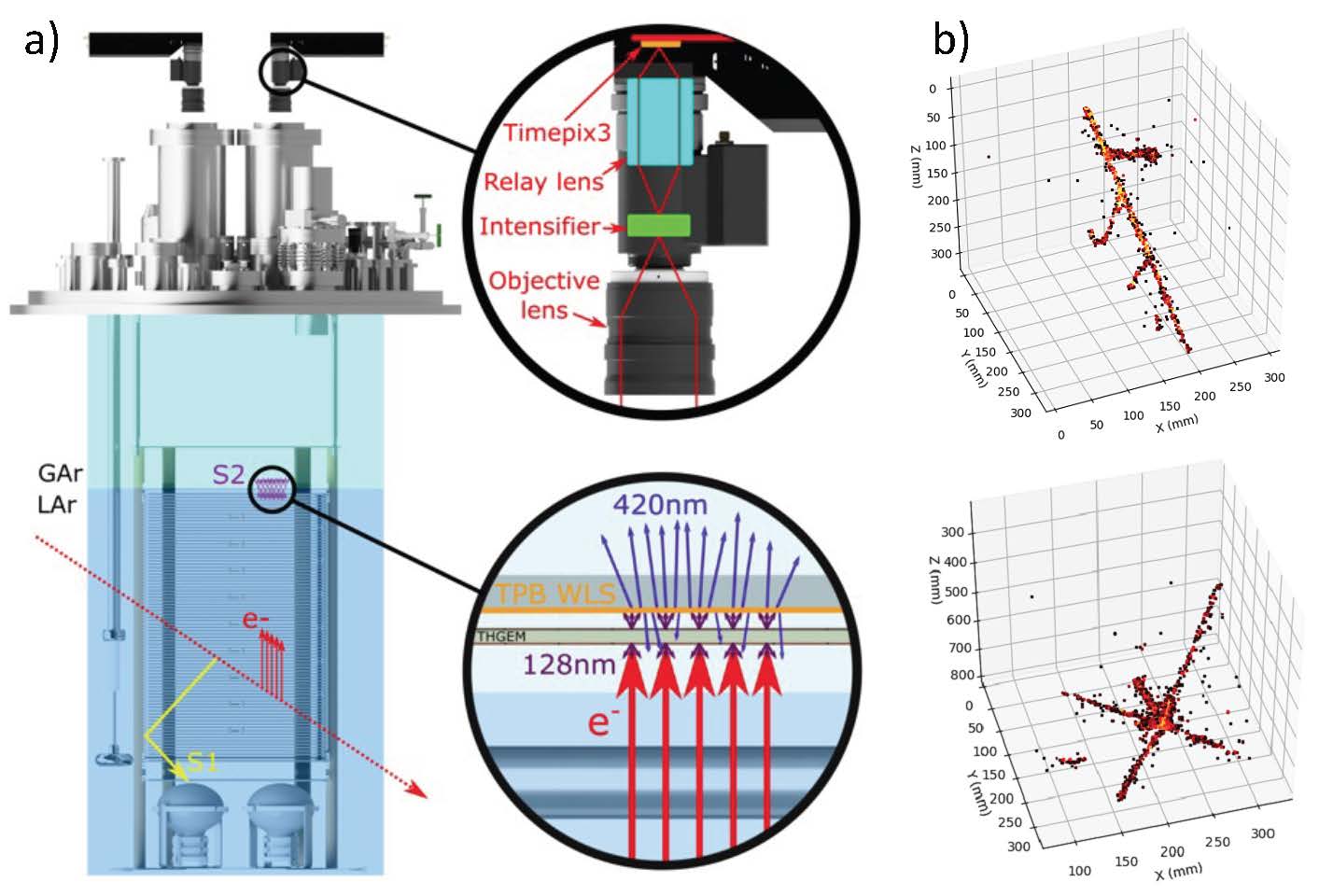}
\caption[\dword{ariadne} detection principle]{(a) Detection principle of dual-phase optical TPC readout with TPX3 camera, first demonstrated in the one-ton \dword{ariadne} detector. (b) \dword{lar} interactions from cosmic-ray muons. Figures taken from~\cite{Lowe:2020wiq}.}
\label{fig:ariadne_principle}
\end{figure}

The operation principle of a dual-phase optical TPC readout with a TPX3 system is shown in Figure~\ref{fig:ariadne_principle}a. When a charged particle enters the \dword{lar} volume, it causes prompt scintillation light (S1) and ionization. The free ionization electrons are drifted in a uniform electric field to the surface of the liquid. A higher field induced between an extraction grid and the bottom electrode of the \dword{thgem} extracts the electrons to the gas phase. Once in gas, the electrons are accelerated within the 500\,$\mu$m holes of the \dword{thgem} at a field set between 22 and 31\,kV/cm. As well as charge amplification, secondary scintillation (S2) \dword{vuv} light is produced. The light is shifted with a \dword{tpb} coated sheet to 430\,nm and then detected by cameras mounted on optical viewports above the \dword{thgem} plane. 

Originally, the optical readout was tested with EMCCD cameras within the one-ton \dword{ariadne} detector at the T9 charged-particle beamline at CERN~\cite{Hollywood:2019loi}, and later was upgraded with fast TPX3. The TPX3 camera assembly boosts the S2 light signal and simultaneously measures Time over Threshold (ToT) and Time of Arrival (ToA) information with 10-bit resolution. ToT allows accurate calorimetry and ToA gives accurate timing (1.6\,ns resolution). The TPX3 chip then sends a packet containing information that allows for full \threed reconstruction using a single device. The high readout rate (up to 80\,Mhits/s), natively \threed raw data, and low storage due to zero suppression make TPX3 ideal for optical TPC readout.

The TPX3 camera system was first tested in low-pressure CF4 gas within the \dword{ariadne} 40\,l TPC prototype~\cite{Roberts:2018sww}; following this demonstration, a TPX3 camera was mounted on \dword{ariadne} and particle tracks from cosmic-ray showers were successfully imaged in \threed for the first time (Figure~\ref{fig:ariadne_principle}b)~\cite{Lowe:2020wiq}. The cameras are shown to be sensitive even to pure electroluminescence light generated at the lower end of the \dword{thgem} field; this mitigates difficulties often faced when trying to operate \dwords{thgem} at a higher field, where there can be issues with stability. Use of cameras has additional benefits, such as ease of upgrade as they are externally mounted. Thus, they are decoupled from TPC and acoustic noise, and large areas can be covered with one camera, bringing both cost and operational benefits.

\begin{figure}[htbp]
\centering
\includegraphics[width=\textwidth]{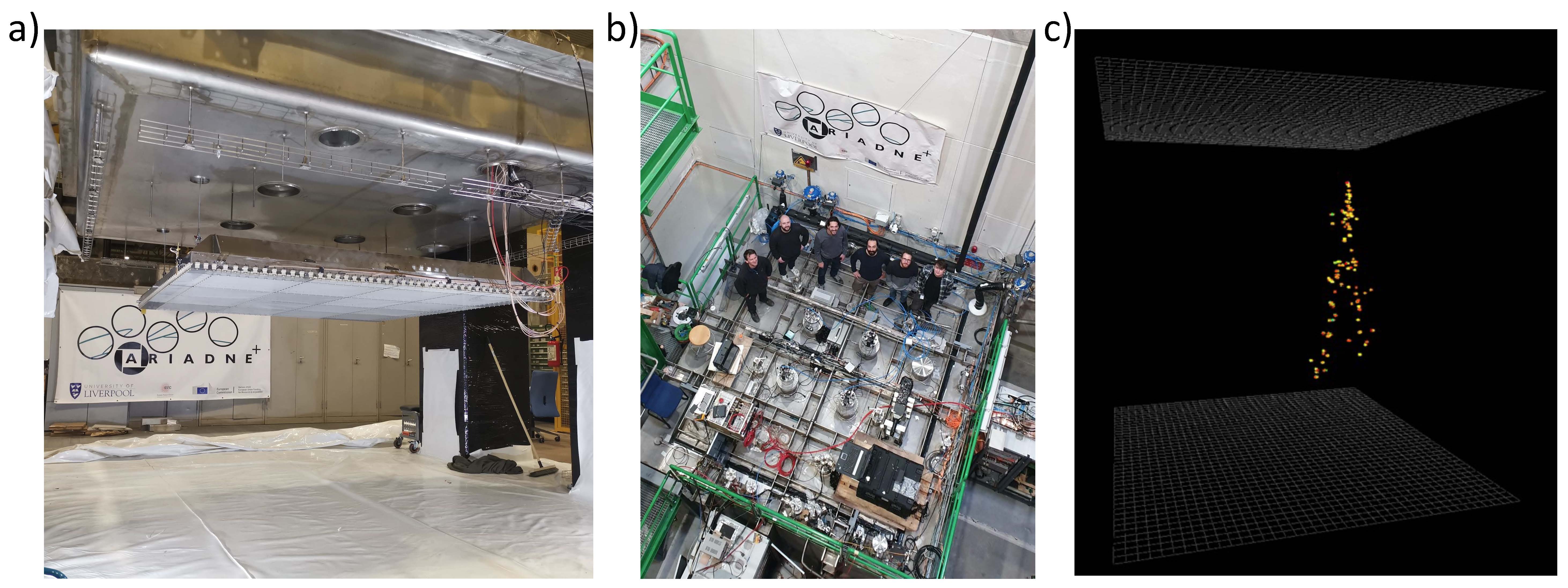}
\caption[\dword{ariadne} plus]{(a) The light readout plane under the cryostat lid; (b) the \dword{ariadne}$^{+}$ team on top of the cryostat; (c) a recorded image of an interaction in \dword{lar}.}
\label{fig:ariadne_plus}
\end{figure}

To demonstrate this technology further and at a scale relevant to the 10\,kt (fiducial) \dword{fd} modules, a larger-scale test (\dword{ariadne}$^{+}$) was recently performed~\cite{Amedo:2739360} at CERN. Four cameras, each imaging a 1$\times$1\,m$^2$ field of view, were employed. One camera utilized a novel \dword{vuv} image intensifier, eliminating the need for a wavelength shifter. The test also showcased a light readout plane (LRP) comprising sixteen, 50$\times$50\,cm$^2$ surface area, glass \dwords{thgem}. The 
novel manufacturing process for the glass \dwords{thgem} allows for mass production at large scale~\cite{Lowe:2021nkz}. Stable operation was achieved, and cosmic-ray muon data from both the visible and \dword{vuv} intensifiers were collected. An image of the detector setup is shown in Figure~\ref{fig:ariadne_plus} and results are published in~\cite{Lowe:2023pfk}.  

The TPX3Cam camera and image intensifiers are commercially available, and a proposal to instrument the ProtoDUNE \dword{np02} cryostat with optical readout is underway, with testing anticipated to take place in 2025/2026. Further R\&D into custom optics and characterization of the next generation Timepix4 (TPX4) cameras, which are anticipated to become commercially available by the end of 2025, can offer further benefits. Another promising ongoing R\&D effort is a TPX4 camera with an integrated image intensifier~\cite{Fiorini:2018gwc}. One of these devices will be tested in the near future within the \dword{ariadne} one-ton detector. Given the current progress of the TPX4 camera system, partial TPX4 instrumentation in \dword{np02} is anticipated.

%%%%%%%%%%%%%% 
\subsubsection{Integrated charge and light readout on anode}
\label{subsubsec:fd_vdoptimized_lightpixels}

DUNE is also pursuing the integration of both light and charge detection modes on the anode into a single detector element. If such a device could be made sensitive both to \dword{vuv} photons at reasonable quantum efficiency and to ionization electrons, this would transform the way noble element detectors collect and process both the charge and light signals. 
A detection element of this kind would offer: i) intrinsic fine-grained information for both charge and light, providing accurate matching between charge and light information; ii) a significant enhancement in the amount of light collected near the anode and much improved uniformity of response, through increased surface area coverage; and iii) simplification in the design and operation of noble element detectors. The technologies under investigation are described in this section, \dword{solar}, \dword{lightpix}, and \dword{qpixlilar}.

\paragraph{SoLAr}
The \dword{solar} technology \cite{Parsa:2022mnj} is based on the concept of a monolithic, light-charge, pixel-based readout to achieve a low energy threshold with excellent energy resolution ($\approx 7\%$ at few-MeV neutrino energies \cite{Capozzi:2018dat}) and background rejection through pulse-shape discrimination. 

The \dword{solar} readout unit (SRU) under development is a pixel tile based on \dword{pcb} technology that embeds charge readout pads located at the focal point of the \dword{lartpc} field-shaping system to collect drifting charges, and highly efficient \dword{vuv} \dwords{sipm} to collect photons in thousands of microcells operated in Geiger mode. In order to maintain a uniform electric field, novel monolithic \dword{vuv} \dword{sipm} sensors need to be developed with these features that have charge readout pads and highly efficient UV-light sensitive microcells.

In 2020, a joint research program between \dword{lar} detector scientists and an industrial partner (Hamamatsu Photonics) delivered a \dword{sipm} that reached a record efficiency ($15\%$ \dword{pde}) for 128\,nm light at the argon boiling point (87\,K). Nearly at the same time, the first integrated system for multiplexing the \dword{sipm} signal was commissioned and operated inside strong electric fields. In 2021 a further development with Hamamatsu Photonics produced a new \dword{sipm} with through-silicon vias that will enable the combination of light detection with the charge readout required for \dword{solar}.

\begin{figure}[htbp]
\centering
\includegraphics[width=0.8\textwidth]{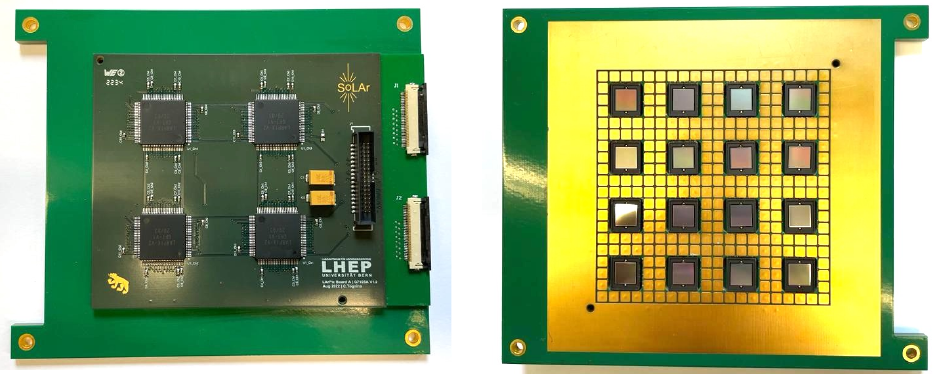}
\caption{The small-scale \dword{solar} prototype PCB tested at the University of Bern, front and back sides. The anode consists of a 7$\times$7\,cm$^2$ readout area with 16 \dword{vuv} \dwords{sipm} (the LAr-facing side, at right) and four LArPix-v2a chips on the backplane, at left. The charge pixel pads are 3\,mm in size and are placed at a 3.5\,mm pitch. The \dwords{sipm} have a 6$\times$6 mm$^2$ sensitive area and are placed at a 17.5\,mm pitch. }
\label{fig:solar_prototype}
\end{figure}

In the \dword{solar} preparatory phase (2021-2022), combined light-charge collection was demonstrated using small-size prototypes~\cite{Anfimov:2024pwg}. The prototypes were operated successfully and have demonstrated that the principle of combining charge and light readout is possible (Figure~\ref{fig:solar_prototype}). Simulations accounting for light propagation effects have shown that a $7\%$ energy resolution can be achieved at typical solar neutrino energies (5--20~MeV) and using the scintillation signal only by replacing anode planes with a pixelated readout integrating a light-sensitive area covering $\sim 10\%$ of the surface. This system would enhance the amount of collected light by a factor of five compared with \dword{fd1}, reducing the frequency at which low-energy background gets incorrectly reconstructed to the (higher) energy region of interest of the signal. The authors of~\cite{Capozzi:2018dat} have studied this remarkable 
impact of energy resolution in the background budget of a \dword{lartpc} using conventional readout in a membrane cryostat. 

The combination of shielding and a $7\%$ energy resolution gives access to the $5$--$10$~MeV region, where most of the $^8$B neutrinos ($^8$B $\to$ $^8$Be$^{\ast}$ + e$^+$ + $\nu_e$) reside, by greatly reducing the dominant background from neutrons and $^{42}$K above 5\,MeV visible energies. The energy resolution is instrumental for sharpening the 17\,MeV cutoff of the $^8$B neutrino spectrum, which lies just below the ``hep'' cutoff of 18.8\,MeV, and opens a 1.8\,MeV window that allows observation of a pure sample of hep  neutrinos ($^3$He + p $\to$ $^4$He + e$^+$ + $\nu_e$)~\cite{Kubodera:2004zm}. Light collection outside the anode is ensured by \dword{xarapu} tiles, for a total coverage of $(8-10)\%$. 

The latter provide the appropriate light yield without resorting to xenon doping, thus  preserving the pulse-shape discrimination power of liquid argon. Pulse-shape discrimination is further enhanced with respect to any existing \dword{lartpc} by the unique performance of the SRU and the increase of collected light.

Finally, \dword{solar} will implement neutron shielding embedded directly in the cryostat walls, delivering a novel membrane-based cryogenic system that also suppresses environmental background to the limit where the only residual background is generated inside the \dword{lartpc}. This will provide a radiopure environment and reduce external neutron background in the 1-4\,MeV region by three orders of magnitude.

\paragraph{\dshort{lightpix}}
A variant of the \dword{larpix} \dword{asic} has been designed for scalable readout of very large arrays of \dwords{sipm}. Called \dword{lightpix}, this \dword{asic} reuses much of the \dword{larpix} system design to provide a system that can read out $>$10$^5$ individual \dwords{sipm} in a cryogenic environment at costs far below \$1 per channel. \dword{lightpix} may be useful for instrumenting a future far detector \dword{pds} with higher quantities of \dwords{sipm} than the \dword{fd1}/\dword{fd2} \dword{pds} design. This could be used as a readout unit in conjunction with an anode-based light pixel solution, e.g., \dword{solar}. 

\dword{lightpix} prototypying in combination with \dword{vuv}-sensitive \dwords{sipm} is underway. The first-generation 64-channel \dword{lightpix}-v1 \dword{asic} includes a custom low-power time-to-digital converter (TDC) with sub-ns resolution to enable precise measurement of photon arrival times.  It also implements programmable digital coincidence logic for the suppression of dark counts, particularly useful for room-temperature detector applications. The \dword{lightpix}-v1 \dword{asic} was used to demonstrate particle detection in two small-scale prototype \dword{vuv}-scintillation detectors: a 16-channel system integrated into a \dword{larpix} pixel tile \dword{lartpc} detector at LBNL, and a 300-channel system for readout of a high-pressure gaseous helium detector at UC-Berkeley/LBNL. A second-generation \dword{asic}, \dword{lightpix}-v2, is in fabrication.  Changes include a new front-end amplifier optimized for use with larger (higher-capacitance) \dwords{sipm}, as well as a charge-integrator for use in higher-occupancy environments.

\begin{figure}[htbp]
\includegraphics[width=0.95\textwidth]{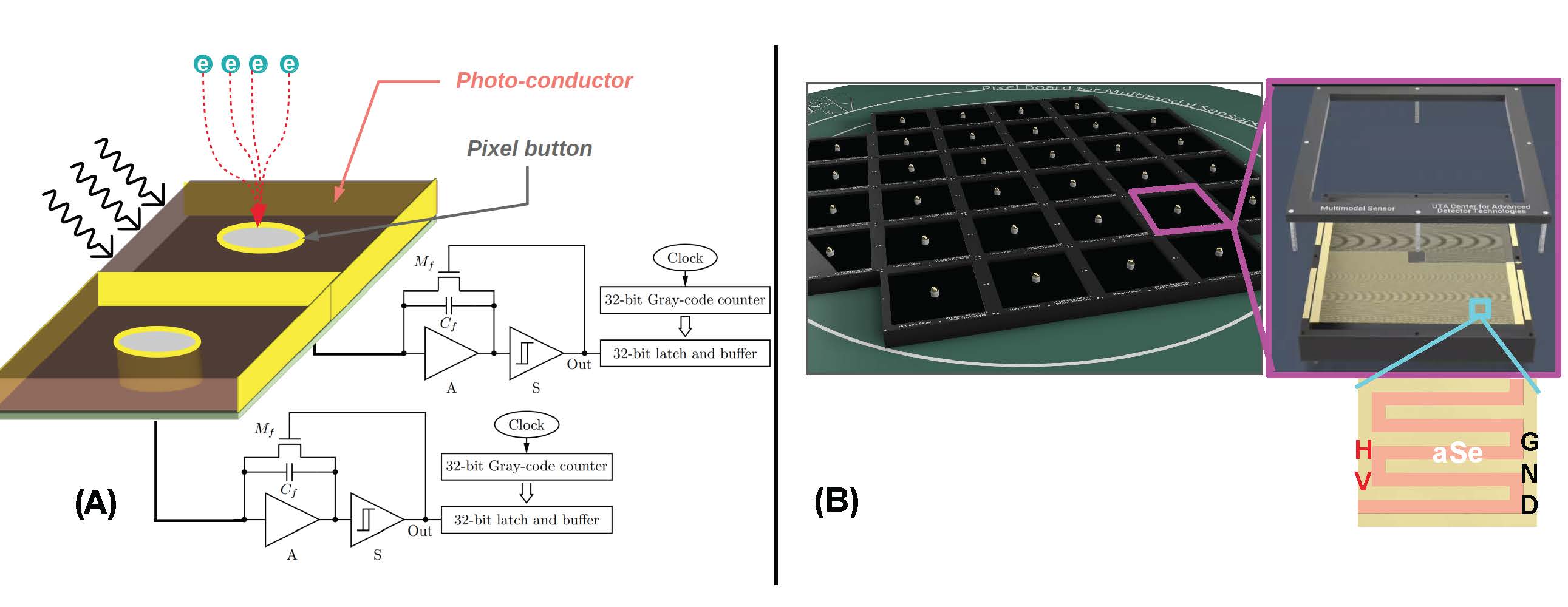}
\caption{A) Schematic design for the \dword{qpixlilar} integrated charge and light readout, from \cite{Rooks:2022vrl}. B) Example design for a multimodal (charge and light) pixel where interdigitated electrodes (IDE, not shown) are deposited around the central ionization collection pixel. Once the photoconductor creates single electrons from photon conversion, IDEs define a region of high electric field where avalanche multiplication of those single electrons occurs, producing detectable signals. }
\label{QPixLight}
\end{figure}

\paragraph{\dshort{qpixlilar}}
The \dword{qpix} consortium is pursuing a different integrated charge and light readout system on the anode, by coating a charge readout pixel with a type of photo-conductive material that, when struck by a \dword{vuv} photon, would generate a signal (charge) that could be detected by the same charge readout scheme considered for the ionization charge. The \dword{qpixlilar} concept is shown schematically in Figure~\ref{QPixLight} (A). Moreover, with the proper choice of photoconductor, such a device could have a broad photon wavelength response, thus offering detection of the full spectrum of light produced in noble element \dwords{tpc}. 

Three such photo-conductive materials have been explored in recent R\&D: \dword{ase},  zinc oxide (ZnO), and organic photodiodes (OPDs). Their application in a liquid argon environment is currently under investigation. Initial studies of constructing a multimodal pixel detector have recently focused on utilizing \dword{ase}, as the ability to prototype and test it was the simplest. The first study on \dword{ase} in cold used commercially manufactured \dwords{pcb} to demonstrate that these \dword{ase}-based interdigitated electrodes (IDE) are sensitive to VUV light at cryogenic temperatures, are cryo-resistant, and are able to maintain argon purity~\cite{Rooks:2022vrl}. Designs for a multimodal pixel are shown in Figure~\ref{QPixLight} (B) where an IDE is deposited around the central ionization collection pixel. This design provides a straightforward way to apply a local electric field to the \dword{ase}, to enable charge gain, and to instrument the area between the charge collection pixels.  
 
More recent studies have pushed this capability further to characterize the performance of such a design to a low photon flux ($\mathcal{O}(100)$~photons), at high electric fields ($>\,70$\,V/$\mu$m) and at cryogenic temperatures. These results continue to show promise. Further R\&D into \dword{ase}-based devices, as well as other photoconductors, is anticipated to be an area of active research in the near future.

%%%%%%%%%%%%%%
\subsection{Liquid-argon doping}
\label{subsec:fd:doping}

A promising direction for expanding DUNE capabilities in a \phasetwo \dword{fd} module is to introduce dopants to the \dword{lar}, creating a %target 
detector medium other than pure argon. Generically, these \dword{lar} ``doping'' techniques may be used to modify the detector response in a desirable way or to introduce new target materials of interest into the bulk detector volume. These approaches are analogous to widely-used strategies in e.g., scintillator or solid state detectors, where secondary fluors are sometimes added to scintillators to shift photon wavelengths, or elements like gadolinium or lithium are added to increase neutron or neutrino cross sections, respectively. A key constraint for \dword{lar} dopants is that they must not interfere with the operation of the \dword{lartpc} by introducing electronegative impurities that non-negligibly degrade the \dword{lar} transparency to electrons. Furthermore, as for any large-scale detector with complex physics of signal generation, the relevant microphysics (including radiative and non-radiative molecular energy transfer, electron-ion recombination, and scintillation-light production) must be well understood through a robust R\&D program to adequately assess the scalability to the DUNE \dword{fd} scale. Several such avenues are being explored that would enhance the charge or light detection capabilities, or introduce new signals of interest. This section discusses two promising potential additives that have been previously demonstrated in large-scale \dwords{lartpc}. The first category is liquid xenon, which is of interest at low concentrations for impact on the scintillation light signal, and at higher concentrations as a signal source. The second includes photosensitive dopants that convert scintillation light to ionization charge. 

These \dword{lar} dopants can be particularly impactful for DUNE \phasetwo prospects to broaden the low-energy physics program, targeting signals in the MeV to keV energy range. Detection of signals in this regime can enhance the GeV-scale neutrino oscillation physics program through enhanced neutrino energy reconstruction \cite{Friedland:2018vry}. In combination with techniques that lower radioactive and external backgrounds, the detection of keV--MeV signals also provide sensitivity to a broad array of previously inaccessible signals spanning \dword{bsm} physics and low-energy neutrino astrophysics. Examples include low-energy solar and \dword{snb} neutrinos, searches for rare decays such as \dword{nldbd}, exotic physics such as fractionally-charged particles, and dark matter scattering. A more complete list can be found in~\cite{Andringa:2023aax}. An expanded program in these areas would complement DUNE's program while leveraging the large mass and deep-underground location.

\subsubsection{Liquid xenon}

Liquid xenon is a potential additive to the \dword{lar} in DUNE \phasetwo, either at a low (\dword{ppm}) or high (up to the percent level) concentration. The loading techniques~\cite{Gallice:2021ykz} and stability conditions~\cite{Bernard:2022zyf} of \dword{lar}+LXe mixtures have been explored across this broad range of concentrations.

At low concentrations, the presence of xenon impacts the production of scintillation light in \dword{lar}, acting as a highly efficient wavelength shifter that converts the 128\,nm primary scintillation wavelength in argon to a longer 178\,nm wavelength. This has several advantages, including reduced Rayleigh scattering, improved light detection uniformity, a narrowing of the scintillation timing distribution, and a reduction in energy losses to impurities such as nitrogen. Such losses would result from non-radiative energy transfers involving the long-lived triplet state of Ar, transfers that are suppressed with the introduction of xenon. This leads to a much improved robustness of the scintillation light yield against \dword{lar} impurities, without appreciable impact on the charge signal. In \dword{pdsp} xenon doping up to $\sim20$\,ppm verified the enhancements to optical response and the recovery of light yield in the presence of impurities~\cite{Gallice:2021ykz}. Xenon doping at a 10\,\dword{ppm} level is already assumed in the \dword{fd2} \phaseone module \cite{DUNE:2023nqi}.

At higher concentrations, up to the percent level, xenon may also be of interest as a signal source. $^{136}$Xe is a candidate isotope for \dword{nldbd} which, if observed, would establish the Majorana nature of the neutrino and demonstrate a violation of lepton number conservation~\cite{Dolinski:2019nrj}. The introduction of xenon, either in its natural form (8.9\% $^{136}$Xe) or enriched to $^{136}$Xe, into a large-scale, deep-underground \dword{lartpc} detector could provide an opportunity to search for this important decay mode~\cite{Mastbaum:2022rhw}. Mitigation of important backgrounds ($^{39}$Ar, $^{42}$Ar, neutrons) near the 2.458\,MeV $Q$-value for this decay are consistent with the requirements of other potential low-energy physics goals considered for DUNE \phasetwo, as described in Section~\ref{subsec:fd_backgrounds}. A key challenge for a competitive search is the massive procurement of xenon, at a level exceeding the world's current production by more than one order of magnitude, and possibly xenon enrichment at the same scale~\cite{Avasthi:2021lgy}. Another crucial challenge is achieving an energy resolution at the percent level for MeV-scale electrons; photosensitive dopants, discussed in the following, provide one avenue toward achieving this.

\subsubsection{Photosensitive dopants}

In a typical pure-\dword{lar} \dword{tpc}, the energy deposited by charged particles is ultimately divided between ionization electrons, drifted in the electric field and detected at the anode plane, and scintillation light, detected by a photon detection system. The photon signal, which is produced promptly with ns-scale timing, is used for \threed event position reconstruction as well as triggering and absolute timing of neutrino interactions. In a \dword{lartpc} doped with a photosensitive dopant, the scintillation signal would be converted to ionization charge, effectively transferring the full deposited energy into that channel. Potential photosensitive dopants under consideration are a class of hydrocarbons with work functions on the order of the \dword{lar} (or Xe-doped \dword{lar}) \dword{vuv} primary scintillation photon energy (7--9\,eV). A dopant of this kind would convert scintillation light into ionization electrons very efficiently  with minimal loss of spatial resolution.

The use of such dopants in \dwords{lartpc} can offer benefits to both the GeV- and MeV-scale physics programs of DUNE \phasetwo. In general, the transfer of deposited energy into the ionization channel leads to an enhancement of the ionization charge, which is measured with excellent efficiency in a \dword{lartpc}. This enhancement is particularly pronounced in regions of high energy deposition, improving prospects for particle identification using charge calorimetry. Furthermore, \dword{lar} with photosensitive dopants exhibits a significantly more linear relationship between deposited and visible charge, reducing the scale and uncertainties of corrections related to electron-ion recombination effects. 

The general impact of such dopants in large-scale \dwords{lartpc} in practice was studied by the \dword{icarus} Collaboration. ICARUS performed a long-term test  of the \dword{tmg} dopant in a three-ton \dword{lartpc} exposed to cosmic rays and $\gamma$ sources, observing a clear enhancement in the ionization charge signal, and a significantly more linear response in reconstructed to deposited charge \cite{Cennini:1995ve}. Importantly, this test also demonstrated long-term stability in realistic \dword{lartpc} operating conditions. A complete and detailed model of the microphysics of energy transfer between \dword{lar} and candidate photosensitive dopants will require a comprehensive assessment of potential dopants and their ionization response across a broad range of signal energies for the \phasetwo program.

The enhancements provided by photosensitive dopants are particularly notable for improving energy resolution at low energies, e.g., to capture point-like signals at or below the MeV scale. A significant challenge with measuring such signals is the efficient collection of small amounts of scintillation light. In the \phaseone design, a limited photon detection efficiency of order $\mathcal{O}$(0.1\%) may limit the capabilities of DUNE to extract spectral information regarding MeV-scale signals, and thus to perform energy-based background mitigations. Ideally, a detector would measure both the ionization and scintillation anti-correlated signals to measure a precise total energy, as in the case of the EXO-200 experiment \cite{EXO-200:2019bbx} and as also investigated in \dword{lariat} \cite{LArIAT:2019gdz}. 

In principle, a large \dword{lartpc} can achieve percent-level energy resolution for MeV-scale signals of interest, but this would require detection of tens of percent of the scintillation photons, a level of efficiency impractical with current and near-future designs. Meanwhile, by converting the isotropic scintillation light into directional ionization charge, photosensitive dopants at the \dword{ppm} level could allow the full energy to be measured with high efficiency by the TPC charge detection system. In this sense, charge alone would provide a precise energy measurement, analogous to a correlated charge/light measurement. Previous work in the context of \dword{lar} calorimeters has also considered the impact of several candidate dopants on MeV-scale $\alpha$ particles, demonstrating substantial charge enhancements for low-energy events with relatively large scintillation signals~\cite{Anderson:1985ub}. Straightforward future R\&D using $\beta$ or $\gamma$ sources to study the low-energy electromagnetic response can further clarify the impact and achievable energy resolution for MeV-scale signals of interest for beam neutrino energy measurements, low-energy neutrino astrophysics, and keV- to MeV-scale \dword{bsm} signatures. To fully realize the potential of photosensitive dopants in \dwords{lartpc}, studies to explore the microphysics involved, and R\&D to determine the optimal dopant types and concentrations, will also be needed.

%%%%%%%%%%%%%%
\subsection{Hybrid Cherenkov plus scintillation detection} 
\label{subsec:fd_theia}
 
The \dword{theia} hybrid Cherenkov+scintillation detection concept is motivated by a science program of low-energy astroparticle, rare event, and precision physics (Section~\ref{s:theiaphysics}). It also contributes to the overall \dword{cpv} sensitivity (Sec.~\ref{subsec:physics_lbl}). The envisioned 25\,kt \dword{theia} detector offers good particle and event identification at both low and high energies, coupled with a target of high radio-purity, no inherent radio isotopes, and excellent neutron shielding. This allows the detector to probe physics that requires low threshold and low background. 

\subsubsection{Hybrid detection concept}
\label{subsubsec:hybrid}

A detector of the envisioned hybrid design would separate Cherenkov and scintillation light by the use of a novel liquid scintillator~\cite{Yeh:2011zz}, fast timing, and spectral sorting. Cherenkov light offers electron/muon discrimination at high energy via ring imaging and sensitivity to particle direction at low energy.  The scintillation signature offers improved energy and vertex resolution, \dword{pid} capability via species-dependent quenching effects on the time profile, and low-threshold (sub-Cherenkov-threshold) particle detection.
The combination boasts an additional handle on \dword{pid} from the relative intensity of the two signals.

This detector design, being developed as \dword{theia}, would offer excellent energy resolution for high-energy neutrino interactions (better than 10\% neutrino energy resolution has been achieved with preliminary algorithms), along with access to a rich program of low-energy, rare-event, and precision physics.

\begin{figure*}[ht!]
\centering
\includegraphics[width=0.7\textwidth]{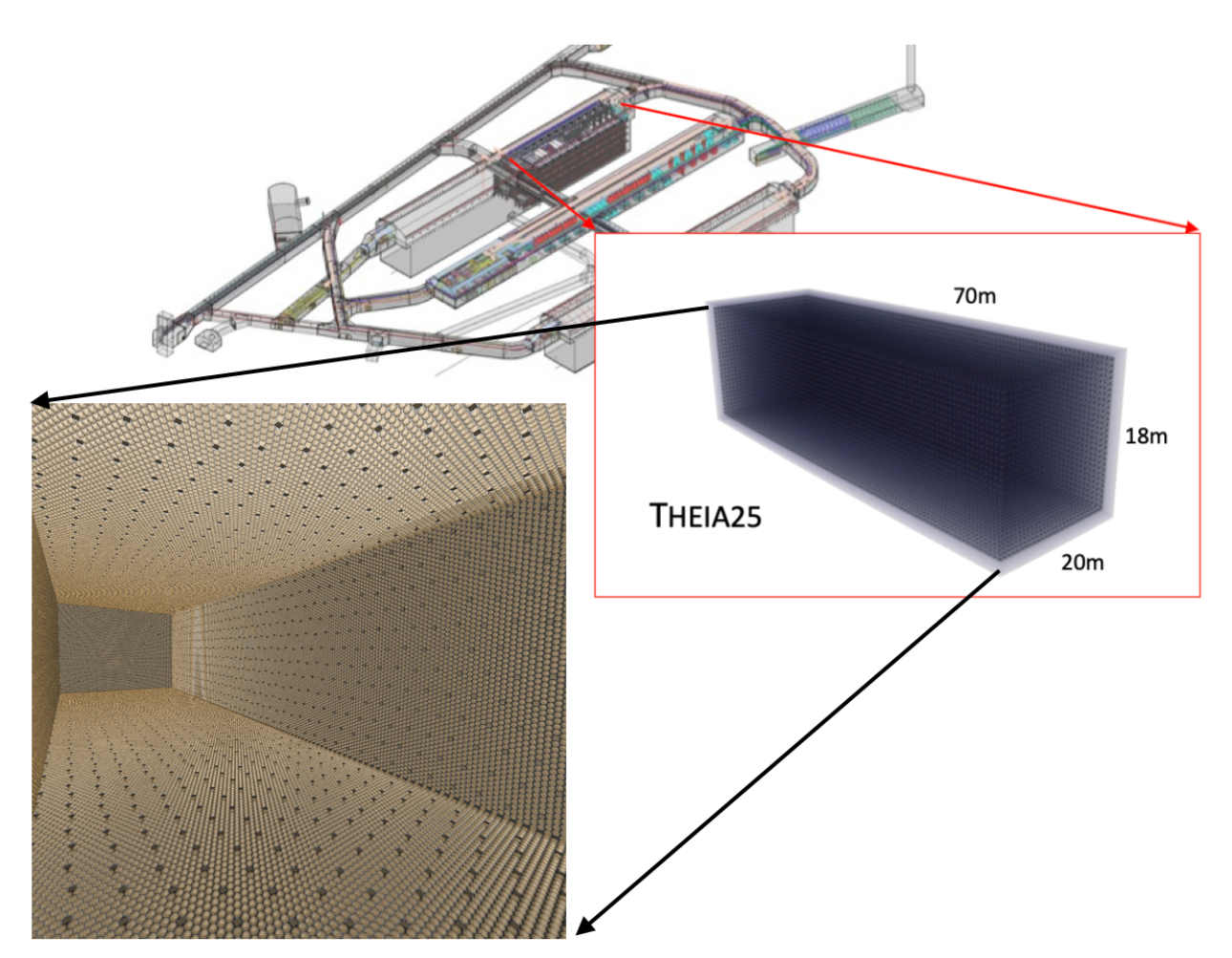}
\caption{Illustration of \dword{theia25} sited in a DUNE FD cavern, with an interior view of the \dword{theia25} concept modeled using the Chroma optical simulation package~\cite{chroma}. Taken from~\cite{Theia:2019non}.}
\label{f:detector25}
\end{figure*}

This is likely a cost-effective option, particularly among those designed to broaden the physics program, thanks to the relatively simple and well-understood detector design that omits both cryostat and \dword{fc}. 
The \dword{theia} detector concept is shown in Figure~\ref{f:detector25}.

\subsubsection{\textsc{Theia} physics program}
\label{s:theiaphysics}

\dword{theia} will seek to make leading measurements over as broad a range of neutrino physics and astrophysics as possible. The scientific program includes: 
\begin{itemize}
    \item observations of solar neutrinos -- both a precision measurement of the \dword{cno} flux, and a probe of the \dword{msw} transition region;
    \item determination of neutrino mass ordering and measurement of the neutrino \dword{cp}-violating phase \deltacp; 
    \item observations of diffuse supernova neutrinos, and sensitivity to neutrinos from an \dword{snb} with directional sensitivity;
    \item sensitive searches for nucleon decay in modes complementary to \dword{lar}; and, ultimately,
    \item a search for \dword{nldbd}, with sensitivity reaching the normal ordering regime of neutrino mass phase space ($m_{\beta\beta}\simeq 6$\,meV).
\end{itemize}

Table~\ref{tab:theia_physics} summarizes the physics reach of \dword{theia25}. The full description of the analysis in each case can be found in~\cite{Theia:2019non}. 

\begin{table*}[htbp]
\centering
\caption{ Projected \dword{theia} physics reach, from Ref.~\cite{Theia:2019non}.  Exposure is listed in terms of the fiducial volume assumed for each analysis. The total detector volume assumed is $70 \times20 \times 18$~m$^3$. For \dword{nldbd}, the target mass assumed is the mass of the candidate isotope within the fiducial volume (assumed to be housed within an inner containment vessel). Limits are given at the $90\%$~CL. 
\label{tab:theia_physics}}
\begin{tabular}{l l l} 
\hline\noalign{\smallskip}
 {\bf Physics Goal} & {\bf  Reach} & {\bf  Exposure (Assumptions) } \\
\noalign{\smallskip}\hline\noalign{\smallskip}
    Long-baseline oscillations 	& 	Equivalent to 10-kt 	&	127 kt-MW-yr  \\
     	& 	\,  \dword{lar} module	&	  \\
   Supernova burst			&	$<2^\circ$ pointing accuracy	& 25-kt detector, 10~kpc distance 	\\
 								& 5,000 events  &  \\
   DSNB 	&	$ 5 \sigma$ discovery	&	125 kt-yr (5 yr)  \\
    CNO neutrino flux	&	$< $ 10\%	& 62.5 kt-yr  (5 yr, 50\% fid. vol.)	 \\
     Reactor neutrino detection	&	2000 events	&	100 kt-yr (5 yr, 80\% fid. vol.)  \\
     Geo neutrino detection 	&	2650 events	& 100 kt-yr (5 yrs, 80\% fid.~vol.)	 \\
     \dword{nldbd}               	&    	$T_{1/2} > 1.1\times10^{28}$~yr&     211 ton-yr $^{130}$Te	 \\
     Nucleon decay $p\rightarrow \overline{\nu}K^{+}$  	&	$\tau/B>1.11\times10^{34}$~yr	& 170 kt-yr (10 yr, 17-kt fid.~vol.)	\\
     Nucleon decay $p\rightarrow 3{\nu}$  	&	$\tau/B>1.21\times10^{32}$~yr& 170 kt-yr (10 yr, 17-kt fid.~vol.)	\\
\noalign{\smallskip}\hline
\end{tabular}
\end{table*}

\subsubsection{Technology readiness levels}\label{s:theiatrl}

The \dword{theia} reference design makes use of a number of novel technologies to achieve successful hybrid event detection. This design would be used to enhance the Cherenkov signal by reducing and potentially delaying the scintillation component.  The use of angular, timing, and spectral information offers discrimination between Cherenkov and scintillation light for both low- and high-energy events.
Fast photon detectors -- such as the 8'' PMTs now manufactured by Hamamatsu, which have better than 500\,ps transit time spread --  will be coupled with spectral sorting achieved via use of dichroic filters~\cite{Kaptanoglu:2018sus}.

Successful separation of Cherenkov and scintillation light has been demonstrated even in a standard scintillator like LAB-PPO~\cite{Caravaca:2016fjg} with the use of sufficiently fast photon detectors, and will be even more powerful when coupled with the spectral sorting capabilities envisioned for \dword{theia}.

Radiopurity levels exceeding the requirements for the \dword{theia} low-energy program have been successfully demonstrated by water Cherenkov experiments (SNO) and scintillator experiments (Borexino).

Further optimization of the design could be achieved by considering deployment of \dwords{lappd}~\cite{Minot:2018zmv,Lyashenko:2019tdj}, for improved vertex resolution, or slow scintillators~\cite{Guo:2017nnr,Biller:2020uoi} to provide further separation of the prompt Cherenkov component from the slower scintillation. A more complete discussion of the relevant technology is provided in~\cite{Klein:2022tqr}.

The R\&D for \dword{theia} will be completed with the successful operation of a number of technology demonstrators currently under construction: (i) a one-ton test tank and a 30-ton \dword{wbls} deployment demonstrator at \dword{bnl} will demonstrate the required properties and handling of the scintillator; (ii) a low-energy performance demonstrator, \dword{eos}, at \dword{lbnl}~\cite{Anderson:2022lbb} will demonstrate the performance capabilities of the scintillator, fast photon detectors, and spectral sorting; and (iii) a high-energy demonstration at ANNIE, at \dword{fnal}, will validate GeV-scale neutrino detection using hybrid technology~\cite{ANNIE:2017nng}.  These detectors are all currently operational or under commissioning.

%%%%%%%%%%%%%%
\subsection{Background control}
\label{subsec:fd_backgrounds}

The potential to enhance the physics scope of DUNE \phasetwo with lower energy thresholds has been attracting significant attention within the wider community \cite{Andringa:2023aax}. The proposed ideas tend to rely on two enhancements over the \phaseone program: greater control of radioactive backgrounds and improved energy resolution at lower energies. DUNE is well placed to improve the lower-energy physics scope, first because of the depth of the \dword{fd}, which is well shielded from cosmic-induced backgrounds. Secondly, the sheer size of the detector volumes allows for significant fiducialization to reduce external backgrounds originating within the SURF cavern rock and shotcrete. \phasetwo designs that minimize material in the active regions, such as the \dword{fd2} or \dword{dp} designs, are most favorable for low-background physics due to reduced risk of radioactive backgrounds. 

We can define two natural physics target energy regions. While these targets are motivated by the intrinsic backgrounds in argon-based detector modules (Sections~\ref{subsec:fd_fd2} through~\ref{subsec:fd:doping}), most of this background control discussion also applies to water-based detectors (Section~\ref{subsec:fd_theia}), as detailed in the following.

The first background target extends the energy threshold down to about 5\,MeV. With careful control of neutron, $\gamma$, and radon related backgrounds, combined with improvements in the low-energy readout, an extended \dword{snb} neutrino program can be envisaged, with improved reach in terms of supernova distance sensitivity (to the Magellanic Clouds), for elastic scatters with improved directionality, and to the (softer energy) early or late parts of the supernova neutrino flux. A low-energy threshold could also allow a precision solar neutrino program to explore solar-reactor oscillation tensions and non-standard interactions. In an argon-based detector, this 5\,MeV threshold is set by the intrinsic $^{42}$Ar-$^{42}$K decay chain, as shown in Figure~\ref{fig:astro_summary}.

The second background target would extend the energy threshold to even lower values of $\sim$1 MeV or less. With such a low threshold, ambitious but high-reward physics measurements would include: solar \dword{cno} measurements; searches for \dword{nldbd} with xenon loading; and even high-mass weakly interacting massive particle dark matter detection could be possible \cite{Church:2020env}. In an argon-based detector, this could be accomplished only by using underground sources of argon, thus largely suppressing the intrinsic $^{42}$Ar activity.  

This section outlines some of the most significant radioactive backgrounds and identifies paths to reduce them in \phasetwo detector modules.
 
\subsubsection{External neutrons and photons}

In DUNE \phaseone, the dominant background to low-energy \dword{snb} neutrinos will be from external neutrons, that is neutrons originating from outside the detector (\dword{surf} cavern rock and shotcrete). These neutrons are primarily of radiogenic origin. When captured in the \dword{lar}, they can produce 6.1\,MeV or 8.8\,MeV $\gamma$ cascades which Compton scatter or pair produce electrons that directly mimic the \dword{cc} neutrino signals. On the other hand, in a water-based detector, neutron captures on free hydrogen would result in lower-energy gammas of 2.2\,MeV. 

To remove external neutrons in argon-based detectors, passive shielding can be deployed, as first suggested in \cite{Capozzi:2018dat}. A layer of 40\,cm of water, or 30\,cm of polyethylene or borated polyethylene, is sufficient to attenuate the neutron flux from spontaneous fission or ($\alpha$, n) reactions in the rock by 3 orders of magnitude, making it subdominant. A shield of this size fits within the warm support structure of the cryostat. Alternative approaches would involve modifications to the cryostat design, for example, layering the insulating foam with neutron-capturing materials such as gadolinium-doped acrylic. These same measures will ameliorate the cavern $\gamma$ background originating from $^{238}$U and $^{232}$Th natural decay chains. 

In a water-based detector, no passive neutron shield around the detector would be necessary. Excellent neutron shielding via detector fiducialization would be reached in this case thanks to the plentiful free hydrogen available as part of the detector target.

Spallation-induced neutron and cosmogenic background events are also possible, though they are primarily short-lived and expected to be orders of magnitude less than the radiogenic backgrounds. A full study of these backgrounds in~\cite{Zhu:2018rwc} show that these can be further reduced by tagged-muon-proximity event selections.

\subsubsection{Internal backgrounds from detector materials}

After the external cavern neutrons, the most significant source of neutron background comes from contaminants in materials within the detector, from ($\alpha$, n)-induced reactions within the cryostat and other components. Photons produced in these events can also distort reconstructed quantities due to light flash or charge blip backgrounds, particularly when close to the readout, such as the cryostat-mounted light sensors. Dark matter experiments have successfully managed such backgrounds with careful material selection programs, using radioactive assay techniques to select favorable materials for detector construction, and to ensure quality assurance during production and installation processes. The world-leading argon-based dark matter detectors have lowered backgrounds by five orders of magnitude below the DUNE \phaseone target. To maintain the sub-dominance of these internal backgrounds relative to externals removed by shielding, a DUNE \phasetwo argon-based detector module will require a less stringent reduction target of three orders of magnitude on detector components such as cryostat stainless steel \cite{Bezerra:2023gvl}. 

For a \dword{theia}-type module, internal backgrounds within the fiducial volume are driven by the cleanliness of the target itself. The chemical purity of the water-based liquid scintillator target is $10^{-17}$\,g/g in both uranium and thorium contaminants, which is considered achievable by improving target material purification techniques \cite{Theia:2019non}.

\subsubsection{Intrinsic backgrounds from unstable isotopes in the target}

Argon extracted from the atmosphere contains two background isotopes that can limit sensitivity at the lowest energy for any argon-based detector: $^{39}$Ar, with a decay Q-value of 565\,keV; and $^{42}$Ar, with a decay Q-value of 599\,keV and its daughter isotope $^{42}$K, which decays with a Q-value of 3525\,keV. The $^{42}$Ar-$^{42}$K chain sets a lower limit of about 5\,MeV, dependent on the ultimate low-energy resolution, for low-threshold physics with atmospheric argon. Dark matter experiments have successfully extracted argon from underground sources, which are depleted in both $^{42}$Ar and $^{39}$Ar \cite{DarkSide:2018kuk}. These experiments show reduction factors of order 1400 for $^{39}$Ar and have seen no $^{42}$Ar. The currently only known source of underground argon is too small to fill a detector the scale of DUNE, but work is ongoing to identify new, larger sources which can be used cost-effectively. Recent estimates of the potential reduction of $^{42}$Ar in underground-sourced argon is expected to be eight orders of magnitude \cite{Poudel:2023mtx}. 

Intrinsic argon background contributions would be absent in a \dword{theia}-type module. The most abundant radioactive isotope in this target material would be $^{14}$C. With a decay Q-value of 156\,keV, $^{14}$C would not be a relevant background for any of the low-energy physics measurements and searches discussed in Section~\ref{subsec:physics_astro} in connection with a \dword{theia} module.

\subsubsection{Radon background}

Radon gas has high mobility, emanates from all detector materials, and can diffuse easily throughout the entire detector volume. This background can mimic directly low-energy neutrinos, through ($\alpha$, $\gamma$) reactions and misidentified $\alpha$ events in the detector. It also produces daughter products which can plateout on internal components such as the photon detector system, distorting the low-energy reconstruction. Several approaches should be adopted to control this background, including: direct removal of radon in the purification system using an inline radon trap; selection of detector materials for low radon emanation; surface treatments to contain or remove radon sources; removal of a significant emanation source from dust by controlling and cleaning to higher cleanliness standards than in \phaseone; removal of radon from air during installation to lower the risk of plateout backgrounds when the detector is open; and analysis techniques such as $\alpha$ tagging by pulse shape discrimination. 

\subsubsection{The SLoMo concept}

\begin{figure}[htbp]
\includegraphics[width=0.95\textwidth]{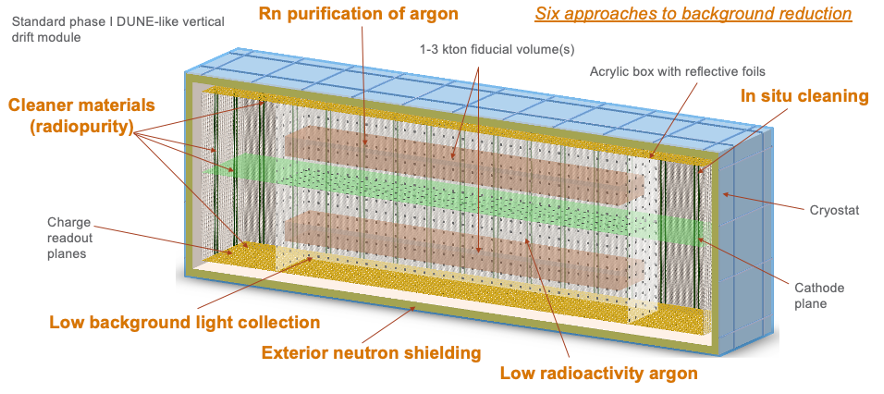}
\caption{Design for \dword{slomo}, highlighting background control methods required to achieve goals.}
\label{fig:SLoMo}
\end{figure}

One proposed design for a low-background, argon-based, \phasetwo far detector is the \dword{slomo}. This design, shown in Figure~\ref{fig:SLoMo} provides a path to lower background levels using the techniques outlined above, reducing most background sources by three orders of magnitude below the expected \phaseone levels. This is combined with a significant increase in light coverage within the detector using high quantum efficiency, DarkSide-style, \dword{sipm} tiles \cite{Consiglio:2020fgk} to increase the energy resolution and pulse shape discrimination power at lower energies. 

To get this light coverage, \dword{slomo} aims to densely instrument an interior 1--2\,kt of highly fiducialized underground argon (UAr) in the center of a \dword{vd}-like detector. The structure on which to mount the DarkSide \dword{sipm} modules is not fixed in this design, though we propose light-tight acrylic walls covered in wavelength-shifting foils. Reference~\cite{Bezerra:2023gvl} shows that a $20\%$~\dword{sipm} coverage, combined with charge detection by existing VD \dwords{crp} and viewing an inner volume, should easily lead to a sub-2\% energy resolution (sigma) at 2\,MeV. This feature, along with the negligibly low amount of $^{42}$Ar in UAr, makes possible \dword{nldbd} studies with xenon loading, should this be a program DUNE wishes to pursue. Reducing $^{42}$Ar to very low levels also allows detectable energy spectra from a supernova to reach well into the region where $\nu_e-e$ elastic scattering dominates and thus pointing, in principle, is improved. A 20\% \dword{sipm} module coverage would come at an affordable cost and would detect enough photons to allow pulse shape discrimination. The combination of high \dword{sipm} coverage, low neutron background, fiducialization and radon control would allow competitive \dword{wimp} dark matter searches in \dword{slomo}. Solar \dword{cno} investigations would also become possible, as would observation of further phenomena, such as the ``supernova glow'' \cite{scholberg_2019_3464639}. This design is outlined in \cite{Bezerra:2023gvl}, where the significant physics gains are further explained.

\subsubsection{Research and development requirements}

All the \phasetwo options to lower the energy threshold require a radioactive background budget to be specified and low-background techniques to be deployed to ensure it is achieved. The R\&D required to achieve these goals includes:
\begin{itemize} 
    \item Large-scale materials and assay campaigns, scaling up material selection techniques used by low-background fundamental physics experiments to the kt scale.
    \item Cleanliness requirements and approaches for the kt scale.
    \item Radon control in detector liquids, including emanation assay and control at the FD scale.
    \item Low-background photon detection systems, developing new designs that can increase the light detection efficiency without overwhelming a background budget.
    \item Background model and simulation campaign for physics sensitivity analyses.
    \item Novel analysis techniques, such as pulse shape discrimination, to remove background events.
     \item Compact shield designs for argon-based modules, that can fit in the limited DUNE cavern space or within the cryostat structure.
     \item New sources of underground argon, capable of filling an argon-based DUNE FD module cost-effectively.
\end{itemize}

%%%%%%%%%%%%%%

\subsection{Toward detector concepts for \phasetwo FD modules}
\label{subsec:fd_fd3_fd4}

\begin{table*}[htbp]
\centering
\begin{tabular}{|p{2.5cm}|p{7.5cm}|p{5.0cm}|} \hline
\textbf{Technology} & \textbf{Prototyping Plans} & \textbf{Key R\&D Goals} \\ \hline 

\textbf{\dword{crp}} (Sec.~\ref{subsubsec:fd_vdoptimized_crp}) & 
\textbf{2024:} Cold Box tests at \dword{cern}. \newline \textbf{2025-2026:} \dword{pdvd} at \dword{cern}. & Port LArASIC to 65~nm process \\ \hline

\textbf{\dword{apex}} (Sec.~\ref{subsubsec:fd_vdoptimized_apex}) & 
\multirow{2}{7.5cm}{\textbf{2024:} 50\,L \& 1-ton prototypes at \dword{cern}. \newline \textbf{2024-2025:} $\mathcal{O}$(100)-channel demonstrator at \dword{fnal}. \newline \textbf{2025-2028:} \dword{pdvd} at \dword{cern}.} &
Mechanical integration of \dword{apex} \dword{pd} in \dword{fc} \\ \cline{3-3}
 & & Signal conditioning, digitization and multiplexing in cold \\ \hline

\textbf{\dword{larpix}}, \textbf{\dword{lightpix}} (Secs.~\ref{subsubsec:fd_vdoptimized_pixels} and \ref{subsubsec:fd_vdoptimized_lightpixels}) &
\multirow{3}{7.5cm}{\textbf{2024:} 2x2 \dshort{nd} demonstrator at \dword{fnal}. \newline \textbf{2024-2025:} Cold Box tests at \dword{cern}. \newline \textbf{2026-2028:} ProtoDUNE at \dword{cern}.} & Micropower, cryo-compatible, detector-on-a-chip \dword{asic} \\ \cline{3-3}
 & & Scalable integrated 3D pixel anode tile \\ \cline{3-3}
 & & Digital aggregator \dword{asic} and \dword{pcb} \\ \hline

\textbf{\dword{qpix}}, \textbf{\dword{qpixlilar}} (Secs.~\ref{subsubsec:fd_vdoptimized_pixels} and \ref{subsubsec:fd_vdoptimized_lightpixels}) &
\multirow{3}{7.5cm}{\textbf{2024:} Prototype chips in small-scale demonstrator. \newline \textbf{2025-2026:} 16 channels/chip prototypes in ton-scale demonstrator at ORNL. \newline \textbf{2026-2027:} Full 32-64 channel ``physics chip''.} & Charge replenishment and measurement of reset time \\ \cline{3-3}
 & & Power consumption \\ \cline{3-3}
 & & R\&D on \dword{ase}-based devices and other photoconductors \\ \hline

\textbf{\dword{ariadne}} (Sec.~\ref{subsubsec:fd_vdoptimized_ariadne}) &
\multirow{3}{7.5cm}{\textbf{2024:} Glass \dword{thgem} production at Liverpool. \newline \textbf{2025-2026:} \dword{pdvd} at \dword{cern}.} & Custom optics for TPX3 camera \\ \cline{3-3}
 & & Light Readout Plane design with glass-THGEMs \\ \cline{3-3}
 & & Characterization of next-generation TPX4 camera \\ \hline

\textbf{\dword{solar}} (Sec.~\ref{subsubsec:fd_vdoptimized_lightpixels}) &
\multirow{2}{7.5cm}{\textbf{2024:} Small-size prototypes at Bern. \newline \textbf{2025-2028:} Mid-scale demonstrator at Boulby.} & Development of \dword{vuv}-sensitive \dwords{sipm} \\ \cline{3-3}
 & &  \dword{asic}-based readout electronics \\ \hline

\textbf{Hybrid Cherenkov+ scintillation} (Section~\ref{subsubsec:hybrid}) &  
\multirow{3}{7.5cm}{\textbf{2024-2025:} Prototypes at BNL (1- \& 30-ton), LBNL (\dword{eos}), \dword{fnal} (ANNIE). \newline \textbf{2025-26:} BUTTON at Boulby.} & \dword{theia} organic component manufacturing \\ \cline{3-3}
 & & \dword{theia} {\it in situ} purification \\ \cline{3-3}
 & & Spectral photon sorting (dichoicons) \\ \hline

\end{tabular}

\caption{Prototyping plans and key R\&D goals for the main \phasetwo FD technologies under consideration.} 
\label{tab:FDRandDSummary}
\end{table*}

The DUNE \dword{fd2} vertical drift technology forms the basis for the reference design for \dword{fd3} and \dword{fd4}. As such, the R\&D for FD3 and FD4 is primarily focused on upgraded photon detector and charge readout systems for the vertical drift layout. Most of these candidate systems are either further developments of the current systems or replacements based on technologies that are already under active R\&D or in early prototyping phases. A non-\dword{lar} option such as \dword{theia} is also under consideration as an alternative technology for FD4. A summary of technologies under consideration, both \dword{lar} and non-\dword{lar} options, for these FD modules, along with R\&D status and plans, is provided in Table~\ref{tab:FDRandDSummary}. 

The detector technologies described in the previous sections (Section~\ref{subsec:fd_fd2} through~\ref{subsec:fd_backgrounds}) will form the building blocks to define full detector designs for both FD3 and FD4. These technologies are not standalone, and most of them can be combined or integrated together, as shown in Table~\ref{tab:fd3_fd4_options}. It is important to note that the check marks in the table for \dword{fd3} are not solely driven by \dword{trl} since several other technologies listed (e.g., \dword{ariadne}, \dword{larpix}) are also technically mature. 

The choices for \dword{fd3} are primarily motivated by how straightforward the proposed upgrades are to implement without requiring major modifications to the baseline \dword{fd2} design on which \dword{fd3} will be based. This is an important consideration since in the case of \dword{fd3}, the DUNE collaboration is aiming to meet the technically limited schedule, which calls for \dword{fd3} installation to start no later than 2029. However, as the other technologies listed in Table~\ref{tab:FDRandDSummary} evolve, they may demonstrate that they meet our \dword{fd3} requirements. If so, and if timelines can be met, they will remain under consideration for \dword{fd3} (with the exception of \dword{theia}). Therefore, their continued R\&D in view of \dword{fd3} is encouraged.

While full detector solutions will be defined in the forthcoming years through dedicated design reports, the following outlines the high-level detector concepts currently under consideration by the DUNE collaboration. 

\begin{table*}[htbp]
\centering
\begin{tabular}{|p{6.5cm}ccp{5.5cm}|}\hline
\textbf{Technology} & \multicolumn{2}{c}{\textbf{Option for}}& \textbf{Can integrate with
} \\
& \textbf{\dshort{fd3}} & \textbf{\dshort{fd4}} & \\ \hline

\textbf{\dword{crp}} (strip-based charge readout)  & \checkmark & \checkmark & \dword{apex} \\ \hline

\textbf{\dword{apex}} (\dword{xarapu} light readout on \dword{fc} with \dwords{sipm}) & \checkmark & \checkmark & \dword{crp}, \dword{larpix}, \dword{qpix}, \dword{ariadne}, \dword{solar} \\ \hline

\textbf{\dword{larpix}}, \textbf{\dword{lightpix}} (pixel charge and light readout) &  & \checkmark & \dword{apex}, \dword{solar} \\ \hline

\textbf{\dword{qpix}}, \textbf{\dword{qpixlilar}} (pixel charge and light readout) &  & \checkmark & \dword{apex}, \dword{solar} \\ \hline

\textbf{\dword{ariadne}} (dual-phase with optical readout of ionization signal) & & \checkmark & \dword{apex}  \\ \hline

\textbf{\dword{solar}} (integrated charge and light pixel readout) &  & \checkmark & \dword{apex}, \dword{larpix}, \dword{qpix}  \\ \hline

\textbf{Hybrid Cherenkov + scintillation} & & \checkmark &  N.A. \\\hline

\end{tabular}
\caption{\dword{lartpc} integration of the detector technologies currently being considered for the \phasetwo \dword{fd} modules. 
Here, ``FD3" refers to the FD3 reference design requiring only minimal modification to the \dword{fd2} \dword{vd} design. The ``FD4" options could also become options for \dword{fd3} over time.} 
\label{tab:fd3_fd4_options}
\end{table*}

For \dword{fd3}, we envisage a vertical drift \dword{lartpc} that is similar in concept to \dword{fd2} (Section~\ref{subsec:fd_fd2}). We do not anticipate any changes to the \dword{fd2} high voltage system, with two drift volumes of 6.5\,m drift length each. The \dwords{crp} at both anodes would feature three \twod projective views of the events, obtained from two double-sided perforated \dwords{pcb} stacked together, similar to \dword{fd2}. Continued R\&D beyond the current \dword{crp} design for \dword{fd2} will focus on optimizations of strip pitch, length and orientation, as well as on streamlining \dword{crp} construction techniques (Section~\ref{subsubsec:fd_vdoptimized_crp}). Upgrades to \dword{fd2} charge readout electronics are possible, such as the adoption of the 65\,nm process for the fabrication of all \dword{fd3} \dwords{asic}. 

The \dword{fd3} \dword{pds} would be composed of \dword{xarapu}-based \dword{pd} modules read by \dwords{sipm} using \dword{pof} and \dword{sof}, similar to \dword{fd1} and \dword{fd2}. The installation location (\dword{fc}, cathode and/or membrane wall), optical coverage, and \dword{pd} module design of the \dword{fd3} \dword{pds} will be determined through \dword{apex} technology R\&D (Section~\ref{subsubsec:fd_vdoptimized_apex}). This R\&D will also determine the reference solution for \dword{pds} readout electronics, particularly whether analog optical signals will be transmitted outside the cryostat as in \dword{fd2}, or a digital optical transmission solution will be adopted. Background control could include incremental improvements over \dword{fd2} protocols, but no dedicated passive shields (beyond the cryostat itself) nor underground argon (Section~\ref{subsec:fd_backgrounds}) would be deployed. We envisage \dword{lar} doping as for \dword{fd2}, via the addition of trace (\dword{ppm}-level) amounts of liquid xenon (Section~\ref{subsec:fd:doping}).

The concepts for \dword{fd4} introduce further improvements. The concept for the reference design is a vertical drift \dword{lartpc} with a central cathode and two anodes with pixel-based readouts. The projective readout of \dwords{crp} would be replaced by a native \threed charge readout system, either employing charge pixels (see \dword{larpix} and \dword{qpix} technologies in Section~\ref{subsubsec:fd_vdoptimized_pixels}) or through an optical-based charge readout (see \dword{ariadne} technology in Section~\ref{subsubsec:fd_vdoptimized_ariadne}). The anode pixels may also serve as scintillation light detection units (see the \dword{solar}, \dword{lightpix} and \dword{qpixlilar} options in Section~\ref{subsubsec:fd_vdoptimized_lightpixels}). The symmetric \dword{tpc} configuration may in principle allow for implementation of different pixel-based solutions at the top and bottom anodes, depending on the R\&D outcome and available resources, as is the case for the different top and bottom electronics adopted in \dword{fd2}. 

A single-drift \dword{lartpc} solution for FD4 with a unique \dword{ariadne}-based anode plane on top and the cathode placed at the bottom of the detector is also possible. This solution would require upgrades to the \dword{hv} system to accommodate a longer (13\,m) drift. Commercial 600~kV power supplies with fluctuations in the output voltage that are sufficiently small for this application already exist. On the other hand, R\&D would  be needed to scale up the high-voltage feedthrough design currently being used in the \dword{pdvd} demonstrator, in order to adapt to the larger diameter high-voltage cable and to the higher voltage values. Scintillation light detection away from the anodes would be performed with \dword{xarapu} modules further improved from \dword{fd3} (see Section~\ref{subsubsec:fd_vdoptimized_apex}). Compact shield designs and greater control of radioactive backgrounds
would be explored for \dword{fd4}, given that an important goal for this module would be to extend the physics scope to lower energy thresholds. 

A hybrid detector module capable of separately measuring scintillation and Cherenkov light (see \dword{theia} technology in Section~\ref{subsec:fd_theia}) would provide a fully complementary detection technology for \dword{fd4} compared to FD1-3, and currently forms the basis of the alternative \dword{fd4} concept being explored by the DUNE Collaboration. This module would be designed for both high-precision Cherenkov ring imaging and long-baseline neutrino oscillation sensitivity, and a rich program across a broad spectrum of physics topics in the MeV-scale energy regime, see Table~\ref{tab:theia_physics}. This can be achieved via either a phased approach, with both the light yield of a water-based liquid scintillator target and the coverage of fast-timing photosensors increasing over time in order to broaden the physics program, or with a high light-yield liquid scintillator and sufficient Cherenkov separation to preserve the Cherenkov purity from the start. Near detector options for non-\dword{lar} \dword{fd} modules are discussed in Section~\ref{subsec:theiand}.

The \dword{daq} system for \dword{fd3} and \dword{fd4} will be based on the same architecture designed for the first two \dword{fd} modules. The DUNE timing system will be extended to these modules, facilitating inter-module synchronization and triggering. Most likely, the \dword{daq} will pursue the use of Ethernet and standard protocols for the readout interface to the detector electronics. Raw data will be stored using the same file format, easing the integration with offline computing. The configuration, control, and monitoring system will be re-used, with customizations as needed. Use of the existing \dword{daq} software will allow us to focus efforts on the aspects that may be implemented differently, and to take advantage of advances in computing technologies. For example, the trigger and data filter may evolve to rely on more sophisticated data processing techniques and technologies, such as \dword{ai}, particularly at low energies. 

 Decisions on the technology choices for FD3 and FD4 are expected to come no later than 2027 and 2028, respectively. As noted earlier in this section, the reference designs for FD3 and FD4 are upgraded versions of the vertical drift \dword{lartpc} technology, with \dword{theia} serving as an alternative technology choice only in the case of FD4. 
 
The final design milestones for \phasetwo \dword{fd} modules are driven by the number and extent of the upgrades planned. For example, in the case of an \dword{fd2}-like module where the only upgrades are optimization of \dwords{crp} and the \dword{apex} light system, one can envision being ready for \dword{fdr} by 2028 in a technically limited schedule. In this scenario, the earliest start for installation of FD3 can be anticipated in 2029 with completion of installation and filling in 2034. Alternatively, if one were to implement pixel-based upgrades such as \dword{larpix}, \dword{qpix}, \dword{solar}, or \dword{ariadne} (top anode plane only), the \dword{fdr} milestone would likely be delayed until at least 2030. 

An asymmetric \dword{dp} \dword{vd} \dword{lartpc} for FD4, with a single drift volume instrumented via a single \dword{ariadne} readout plane, would require significant changes to the \dword{hv} system. It is possible to reach the \dword{fdr} milestone for this option by 2031-32. In the case of the \dword{theia} option, a \dword{fdr} milestone no earlier than 2033 is anticipated. The ProtoDUNEs at CERN will continue to serve as important platforms to demonstrate several of these technologies and their potential for integration.

\section{The DUNE \phasetwo near detector}
\label{sec:nd}

In \phasetwo, DUNE will have accumulated \dword{fd} statistics of several thousand oscillated electron neutrinos, resulting in statistical uncertainties at the few-percent level on the number of electron appearance events. To reach the physics goals of DUNE, a similar level of systematic uncertainty must be achieved, which requires precise constraints from the \dword{nd}. To understand the needs of the \phasetwo \dword{nd}, we must first understand the expected performance of the \phaseone \dword{nd}, which consists of two measurement systems, \dword{ndlar}+\dword{tms}, and \dword{sand}. In Section~\ref{subsec:nd_motivations}, we describe why the \phaseone \dword{nd} is critical for DUNE physics, discuss limitations inherent to its design, and outline the \phasetwo requirements that are needed to provide improved constraints on the argon-based \dword{fd} data sample. This is a difficult challenge as the ultimate performance of the \phaseone \dword{nd} is not yet understood, and will depend on analysis techniques developed over the coming decade. Section~\ref{subsec:ndgar} describes a detector concept that meets the design motivations of Section~\ref{subsec:nd_motivations}. Further improvements may come from upgrades to the \phaseone \dword{nd} components, see Section~\ref{subsec:nd_phaseonend_improvements}. Near-detector options to constrain possible non-argon \dword{fd} data samples are discussed in Section~\ref{subsec:theiand}.

\subsection{Design motivations} 
\label{subsec:nd_motivations}

The \phaseone \dword{ndlar}+\dword{tms} detector is designed to measure neutrino interactions on the same nuclear target as the \dword{fd}, and with a detector response similar to the FD. Neutrino energy in the \dword{fd} is estimated by summing the lepton energy with the hadronic energy. The FD measures the muon energy by range, and the energies of all other particles calorimetrically, in both cases exploiting energy deposits occurring in \dword{lar}. The \dword{ndlar}+\dword{tms} detector also measures muons by range, and other particles calorimetrically. It is able to reconstruct the same observables as the FD, and measures them with essentially the same resolutions, in an unoscillated beam. This capability is the core requirement of the DUNE \dword{nd}, and will be a critically important constraint for all DUNE long-baseline measurements. The \dword{ndlar}+\dword{tms} system moves off-axis (via \dword{duneprism}) to collect data at different fluxes, and directly constrains the energy dependence of neutrino cross sections. \dword{sand} is permanently on-axis, and measures neutrino cross sections on various nuclear targets while also monitoring the beam. \dword{sand} has a \dword{lar} target, so that it can also measure cross section ratios, including on argon.

The dimensions of \phaseone \dword{ndlar} are driven by containment of electrons and hadrons, rather than by event rate, so that they can be measured calorimetrically in the same way as in the FD. To minimize cost, the dimensions have been chosen to be as small as possible while maintaining full coverage of the neutrino-argon phase space. However, this means that the acceptance is non-uniform and depends on the event kinematics, complicating the calorimetric energy measurement. Beam-induced muons will be reconstructed by the \dword{ndlar}+\dword{tms} combined system. \dword{tms} is able to provide sign selection of muons, which is especially important to reject wrong-sign backgrounds in antineutrino mode. However, \dword{ndlar} itself is not magnetized, so the sign selection is only possible for the muons that enter \dword{tms} ($\gtrsim$800\,MeV kinetic energy), and there is no sign selection for other particles. While \dword{tms} will provide muon momentum and sign reconstruction for the energy region relevant for long-baseline oscillation physics, the design is such that muons above 6~GeV kinetic energy will not be ranged out nor sign-selected.

The main purpose of the \dword{sand} detector will be to monitor the neutrino beam, but it will also be capable of making independent measurements of the neutrino flux and flavor content. This additional capability adds robustness to the \dword{nd} complex, enabling better control over systematics and background.
\phaseone \dword{sand} will be able to measure the sign of all charged particles in its low-density CH$_{2}$ tracker, but not generally for hadrons produced in the argon target. The \dword{sand} tracker will also measure neutrino cross sections on carbon and hydrogen targets.

To constrain neutrino-argon interaction modeling, it is useful to identify specific exclusive processes. Of those, about two thirds of neutrino interactions in DUNE will have pions in the final state. \dword{ndlar} is an excellent detector for identifying pions and protons when they are above threshold, and do not undergo strong interactions inside the detector. However, many events in the DUNE energy range have pions with hundreds of MeV kinetic energy, which travel several interaction lengths and frequently scatter, transferring some energy to the atomic nuclei that is then not seen in a calorimetric energy reconstruction. On the other hand, below threshold pions decay to final state particles, including neutrinos, leading to large fractions of their rest mass not being visible calorimetrically. Also, protons below 300\,MeV/c are impossible to detect in \dword{ndlar} because they deposit all of their energy over a range of only a few mm, producing highly saturated ionization charges recorded on a single pixel. Predictions on the multiplicity of such low-momentum protons from neutrino-argon interaction models are particularly uncertain.

These \phaseone limitations motivate the design of the \phasetwo \dword{nd}. Specifically, the \phasetwo \dword{nd} should have, when compared to the \phaseone \dword{nd}:

\begin{itemize}
    \item argon as the primary target nucleus,
    \item improved \dword{pid} across a broad range of energies and angles,
    \item lower tracking thresholds for protons and pions,
    \item minimal secondary interactions in the tracker volume,
    \item $4\pi$ acceptance over a wide range of momenta, and
    \item magnetization to achieve sign selection over a broader muon momentum range.
\end{itemize}

Employing an argon target will ensure that constraints from the \phasetwo \dword{nd} can be applied directly to the argon-based \dword{fd} without any extrapolation in atomic number. 

A broad acceptance and high \dword{pid} efficiency will enable exclusive final states to be identified, which will improve the constraints on neutrino interaction modeling. Low thresholds will make the \phasetwo \dword{nd} highly sensitive to nuclear effects. Magnetization will ensure sign selection at all energies and angles, for both charged leptons and charged pions. 

In the event that one of the \phasetwo \dword{fd} modules consists of a neutrino target material that is not argon-based and of a detector technology other than \dword{lartpc}, such as the \dword{theia} detector concept described in Section~\ref{subsec:fd_theia}, the requirements for the \phasetwo \dword{nd} complex will need to be expanded to account for the additional target material(s) and the different neutrino detection method.

%%%%%%%%%%%%%

\subsection{\phasetwo improved tracker concept}
\label{subsec:ndgar}

%========
\begin{figure}[ht]
\centering
\includegraphics[width=0.49\textwidth]{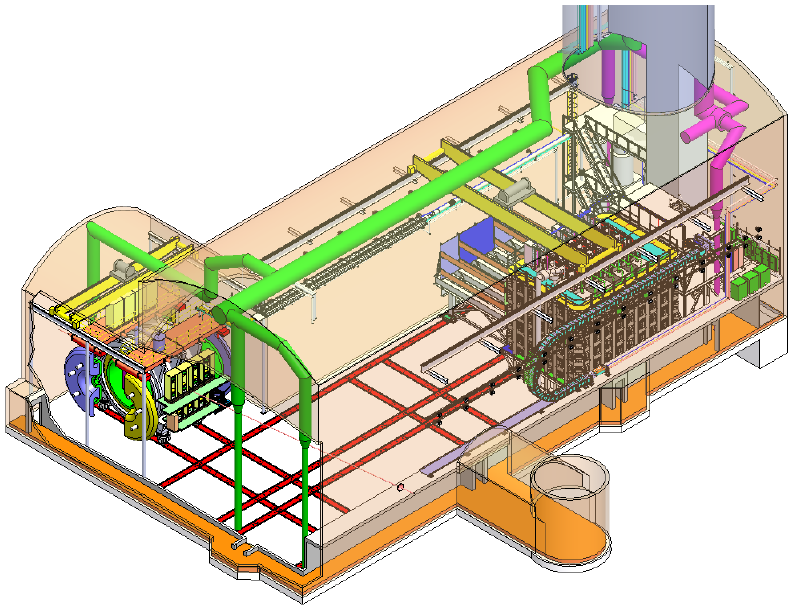} \hfill
\includegraphics[width=0.49\textwidth]{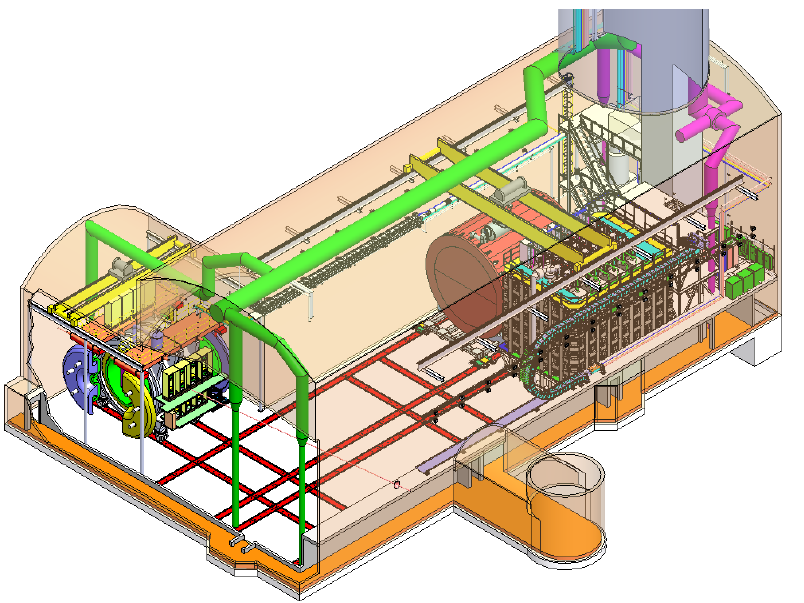} 
\caption{Layout of the envisaged \phaseone (left) and \phasetwo (right) \dword{nd} suite. The neutrino beam enters from the bottom-right corner, and exits at the top-left corner, of the drawings. The \dword{sand} detector is shown at its permanent on-axis location, while all other detectors upstream are shown at their maximum off-axis location.}
\label{fig:ndlayout}
\end{figure}
%============

\begin{figure}[ht]
     \centering
     \begin{subfigure}{0.6\textwidth}
         \centering
         \raisebox{0.7cm}{\includegraphics[width=\textwidth]{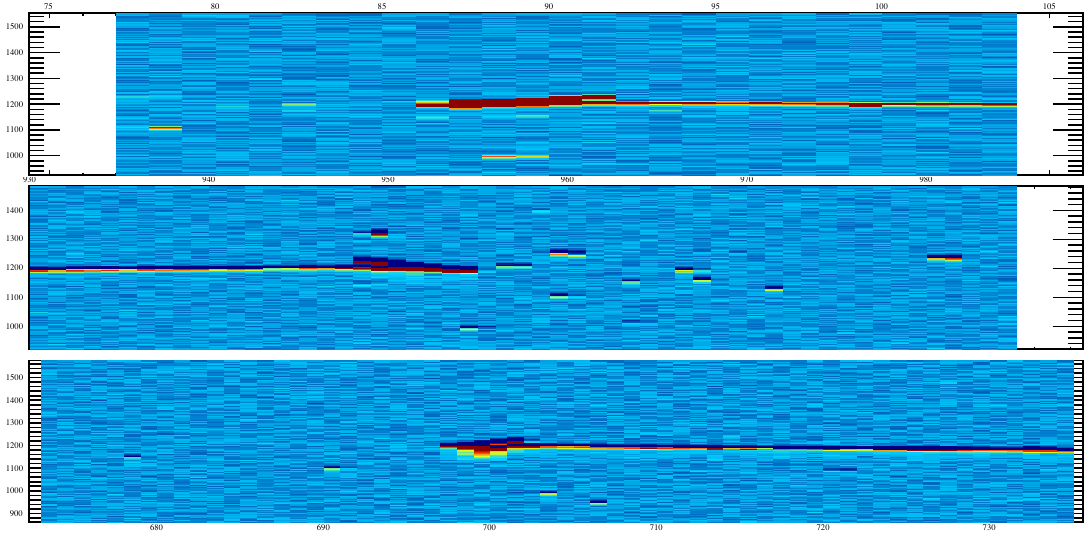}}
     \end{subfigure} \hfill
     \begin{subfigure}{0.35\textwidth}
         \centering
         \includegraphics[width=\textwidth]{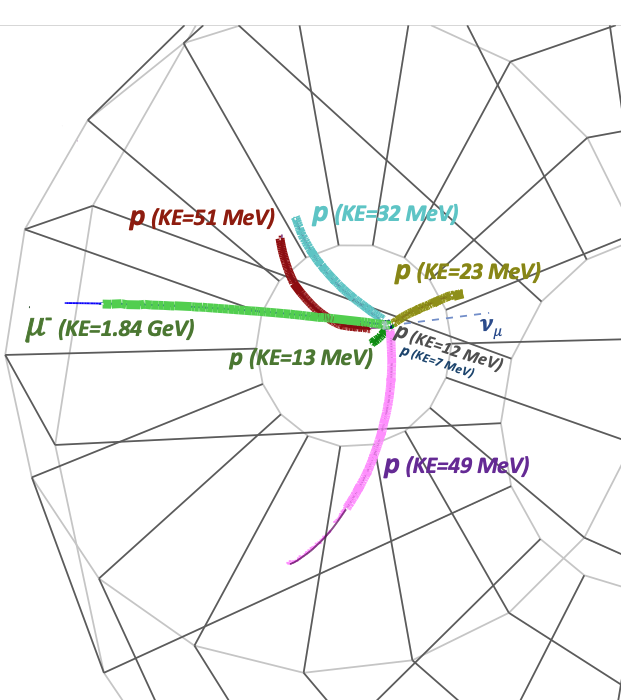}
    \end{subfigure}
        \caption{The same \dword{cc} $\nu_{\mu}$ event with seven low energy protons (kinetic energies ranging from 7 to 51 MeV) simulated in a \dword{lartpc} (left) and a \dword{gartpc} (right). The \dword{lartpc} event display shows time ticks versus channel number for the three projective views of the event. The \dword{gartpc} reconstruction algorithm finds all eight tracks in the event (seven proton tracks and one muon track), although only six are visible by eye in this view. All proton tracks travel 2.4~cm or less in \dword{lartpc}. From \cite{DUNE:2022yni}.}
        \label{fig:lar_vs_gar_event}
\end{figure}

For \phasetwo, an improved tracker concept based on a \dword{gartpc} would replace \dword{tms} downstream of \dword{ndlar}. Drawings for the envisaged layouts of the \phaseone and \phasetwo detector suites are shown in Figure~\ref{fig:ndlayout}. A \dword{gartpc} can reconstruct pions, protons and nuclear fragments with lower detection thresholds than a \dword{lartpc} can. Figure~\ref{fig:lar_vs_gar_event}, which compares the same simulated event in each, shows this. The protons travel a much longer distance and can be more clearly separated in gaseous argon.  The \dword{gartpc} is also less susceptible to confusion of primary and secondary interactions, since secondary interactions occur infrequently in the lower-density gas detector. If the TPC is inside a magnetic field, it can better distinguish neutrinos and antineutrinos and can determine the momenta of particles whose trajectories are not contained in the detector. It can also measure neutrino interactions over all directions, unlike the \dword{ndlar}, which loses acceptance at high angles with respect to the beam direction. Therefore, a \dword{gartpc} detector system at the near site, called \dword{ndgar} in the following, provides a valuable and complementary data sample to better understand neutrino-argon interactions. 

%========
%============

However, a drawback of a \dword{gartpc} is the lower neutrino event rate in a given volume due to the lower density. One way to improve this is to use high-pressure argon gas. A cylindrical volume with a diameter and length both of roughly 5\,m, and gas at 10\,bar, would have a fiducial mass of nearly one ton of argon, yielding approximately one million neutrino interactions per year. The trade-off between sufficient target mass and low detector density has not been optimized, but would nonetheless be adjustable during operations by setting the detector pressure.

\begin{figure}[ht]
\centering
\includegraphics[width=0.6\textwidth]{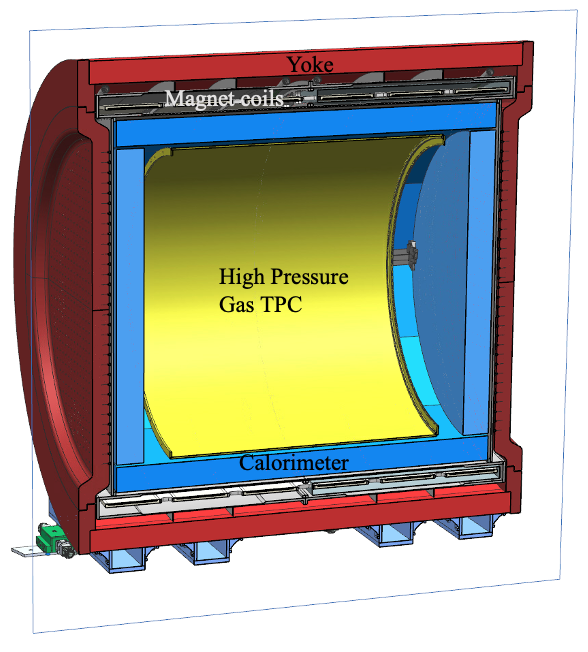}
\caption[\dword{ndgar} schematic]{Cutaway view of the full \dword{ndgar} detector system, showing the \dword{hpgtpc}, the calorimeter, the magnet, and the iron yoke. The detectors for the muon-tagging system are not shown.}
\label{fig:ndgarlayout}
\end{figure}

As illustrated in Figure~\ref{fig:ndgarlayout}, the reference design concept for the \dword{ndgar} detector comprises:
\begin{enumerate}
\item a pressurized \dword{gartpc}, 
\item a surrounding calorimeter,
\item a magnet, and 
\item a muon-tagging system.
\end{enumerate}

A \dword{pds} may also prove necessary to reduce pileup and to provide the event $t_{0}$ for the drift time determination in events that do not reach the calorimeter. It would also help improve the track matching between the TPC and the external calorimeter and muon systems~\cite{Saa-Hernandez:2024tvl}. All these subsystems are described in the following. The entire \dword{ndgar} system will move perpendicularly to the beam direction together with \dword{ndlar}, as part of the \dword{duneprism} concept.

%========
%============

This detector concept is motivated by the considerations in Section~\ref{subsec:nd_motivations}. It also affords significant opportunities to study \dword{bsm} physics, as discussed in Section~\ref{subsec:physics_bsm}. 

\subsubsection{Charge readout of TPC}
\label{tpc_charge}

 A variety of techniques can be used to amplify and collect the ionization electrons after their drift to the TPC anode, and the options currently under consideration for \dword{ndgar} are briefly presented here. In all cases, a high-pressure gas mixture with high argon content ($>$90\% molar fraction) is envisaged. A 96:4 Ar:CH$_4$ mixture was tested successfully during \dword{ndgar} R\&D~\cite{Ritchie-Yates:2023zrq,2681627}. Therefore, non-Ar components in the \dword{ndgar} target will contribute at the few percent level at most to the overall event rate in the TPC. In order to extract pure $\nu$-Ar interactions, such percent-level corrections can be made with high accuracy using Transverse Kinematic Imbalance techniques~\cite{Lu:2015hea}, joint \dword{ndgar}-\dword{sand} fits, and Monte Carlo-based  estimates. High-pressure argon gas mixtures have been used in the past, such as in the PEP-4 detector at SLAC~\cite{Grupen:1999by}, which used a (flammable) gas mixture of 80:20 Ar:CH$_4$, operated at 8.5\,atm. A known challenge for high-pressure gas detectors is that the gas amplification gain decreases as the gas pressure increases.   For the DUNE \dword{ndgar}, R\&D to ensure adequate stability and gain in a non-flammable gas is underway. 

\paragraph{Multi-Wire Proportional Chambers}
To collect sufficient event statistics, the \dword{hpgtpc}, at the core of \dword{ndgar}, must be both large and capable of functioning under high pressures. A TPC of the size used in the ALICE experiment at CERN~\cite{Alme:2010ke} may be adequate in terms of size, but only if the gas inside is pressurized to approximately 10\,atm. As a result of the recent upgrade of ALICE’s readout system to gaseous electron multipliers (\dwords{gem}), the previously operated ALICE
multi-wire proportional chambers have become available. They were previously operated in ALICE at 1\,atm, hence their operation needed to be assessed within a high-pressure argon gas environment. 

Two test stands, one each in the UK and the US, called the \dword{goat} and the \dword{toad}, respectively, are being used to test the ALICE chambers under high pressure. \dword{goat} used a pressure vessel rated to 10\,atm. It tested an ALICE inner chamber for its achievable gas gain at various pressure set points, amplification voltages, and gas mixtures~\cite{Ritchie-Yates:2023zrq}. \dword{toad} had previously tested an ALICE outer chamber for its achievable gas gain up to 5\,atm. Currently, it is being commissioned in the \dword{fnal} \dword{ftbf} for data-taking in a test beam and for performing a full detector slice test of the electronics and \dword{daq}. In both test stands, wire-based readout chambers have been tested in high-pressure environments, demonstrating that they can provide reasonable gas gains when they operate at or above the high voltage values they were subject to in ALICE. Despite this, the long-term operation and stability of these chambers at such high voltages remain to be investigated. There are also plans to test charge readout systems based on micro-pattern gas detectors, such as \dwords{gem}, as described below. 

\paragraph{TPC Readout with GEMs} 
In a TPC using \dwords{gem} or ``thick GEMs'' (\dwords{thgem}), the ionization drift electrons enter the \dword{thgem} holes and are accelerated in a high electric field. At sufficiently high fields, this acceleration causes the electrons to further ionize the gas medium, resulting in a Townsend avalanche. This exponentially increases the number of electrons and therefore the signal size. 

Typically, GEMs and \dwords{thgem} are produced starting from double copper-clad substrates, either by photolithography of kapton in the case of the former, or \dword{cnc} drilling of epoxy laminates/\frfour in the latter. We propose to use a new type of \dword{thgem} made out of glass, as also proposed in the context of the optical-based charge readout option for the \phasetwo \dword{fd} (Section~\ref{subsubsec:fd_vdoptimized_ariadne}). These glass \dwords{thgem} developed at Liverpool (UK) are fabricated using a new masked abrasive machining process. The innovation allows for customization of glass \dwords{thgem}, where both substrate and electrode materials can be tailored to our requirements, which include high stiffness, low outgassing, and resilience to damage from discharges. 

The amplified electrons from the \dwords{thgem} would be read out on a segmented anode to allow for tracking reconstruction. Borosilicate glass and fused silica are isotropic and homogeneous substrate materials that can be machined to typical \dword{thgem} thicknesses while remaining sturdy and potentially providing better surface finishes than \frfour{}-based \dwords{thgem}. Their transparency, made possible by indium tin oxide (ITO) electrodes, makes them suitable for optical imaging of primary ionization, as demonstrated up to 1.5\,bar with cosmic ray imaging at estimated optical gains up to 10$^6$~\cite{Amedo:2023pmh}. 

Future optimization of glass \dwords{gem} may include enhancements in light collection with wavelength-shifting substrates~\cite{Kuzniak:2021brl}, wavelength-shifting coatings~\cite{Leardini:2024ovj}, or diamond-like carbon (DLC) coatings for stability~\cite{Leardini:2022rfj,Tesi:2023ale}. R\&D toward a \dword{thgem}-based readout is ongoing in Spain, where a 10-bar full \threed Optical TPC (\dword{gat0}) is under commissioning. In addition, R\&D toward a GEM-based readout for \dword{ndgar} is currently underway in the US with the \dword{gorg} test stand, currently testing a triple-GEM stack.

\paragraph {TPC Readout electronics} 

Due to the high-pressure nature of this detector, readout electronics must be developed that can operate inside the pressure vessel to minimize the analog signal path. The electronics must also be zero-suppressed and compatible with the existing DUNE \dword{daq} infrastructure for the \phaseone \dword{nd}. Readout electronics has traditionally been one of the cost drivers of TPCs. While the pixel size to be used in the final detector module has not been determined, detectors like ALICE had 700k channels. With this number of channels, work is needed to ensure that the electronics system is cost-effective. 

R\&D work is underway in the UK and US to deliver such electronics. A prototype system using the SAMPA \dword{asic}, developed for the ALICE TPC upgrade and the sPHENIX detector, plus \dword{fpga}-based control and aggregation, is already in hand. This solution, scaled up to the full \dword{ndgar} detector, is expected to be much cheaper than the ones adopted for ALICE and sPHENIX, thanks to the much lower data rates. If full \threed optical tracking is ultimately adopted, the readout electronics would align with the technical proposal described in Section~\ref{subsubsec:fd_vdoptimized_ariadne} for \dword{fd3} and \dword{fd4}.

\subsubsection{Calorimeter concept}   

The \dword{gartpc} will excel in measuring charged particle tracks, but to first order is blind to neutral particles. As such, it is important to have a system that can detect them. At neutrino energies of a few GeV, these are mostly photons (for example from $\pi^0$-decays) and neutrons (from nuclear break-up) in the kinetic energy range from $\lesssim 100$\,MeV to $\sim$GeV. The photon energy will be determined calorimetrically, while the neutron energy can be determined by measuring the time of flight between the production vertex and a nuclear re-scatter in the calorimeter~\cite{Emberger:2022}. Both photons and neutrons will be key to measuring nuclear effects that will influence the relationship between true and reconstructed neutrino energy, and the dynamics of the neutrino interactions. 

It is also the case that the \dword{hpgtpc} should occupy the largest possible volume, and the calorimeter has to surround the TPC. As such, it has a rather large surface area even for modern particle physics detectors. Optimizing this detector to achieve the physics goals while still being affordable is a key task of the gaseous argon detector group.

\begin{figure}
    \centering
    \includegraphics[width=0.2\textwidth]{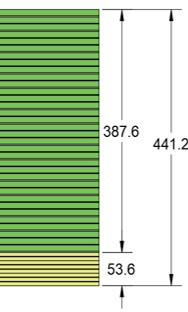}
    \caption{Possible structure of a combined tile and strip layer sampling calorimeter. The (yellow) tile/absorber layers are oriented towards the TPC; they are followed by (green) strip/absorber layers. Dimensions are in mm.}
    \label{fig:ECALStack}
\end{figure}

A possible affordable technology with the required performance is based on a plastic scintillator sampling calorimeter that is constructed from active tile layers using a combination of the technology developed by the CALICE R\&D Collaboration~\cite{Sefkow:2018rhp} and the more traditional scintillator strip, \dword{wls} fiber, and \dword{sipm} readout combination, used in neutrino experiments such as the \dword{t2k} near detector. A preliminary structure of the calorimeter is illustrated in Figure~\ref{fig:ndgarlayout}. Further details of the potential layout of the barrel detector are shown in Figure~\ref{fig:ECALStack}. 

A potential barrel geometry consists of 60 layers with the following layout:
\begin{itemize}
    \item eight inner layers of 2\,mm copper $+$ 5\,mm of 2.5 $\times$ 2.5~cm$^2$ tiles + 1\,mm \frfour, and
    \item 52 layers of 2\,mm copper $+$ 5\,mm of cross-strips 4\,cm wide
\end{itemize}

\begin{figure}
    \centering
    \includegraphics[width=0.51\textwidth]{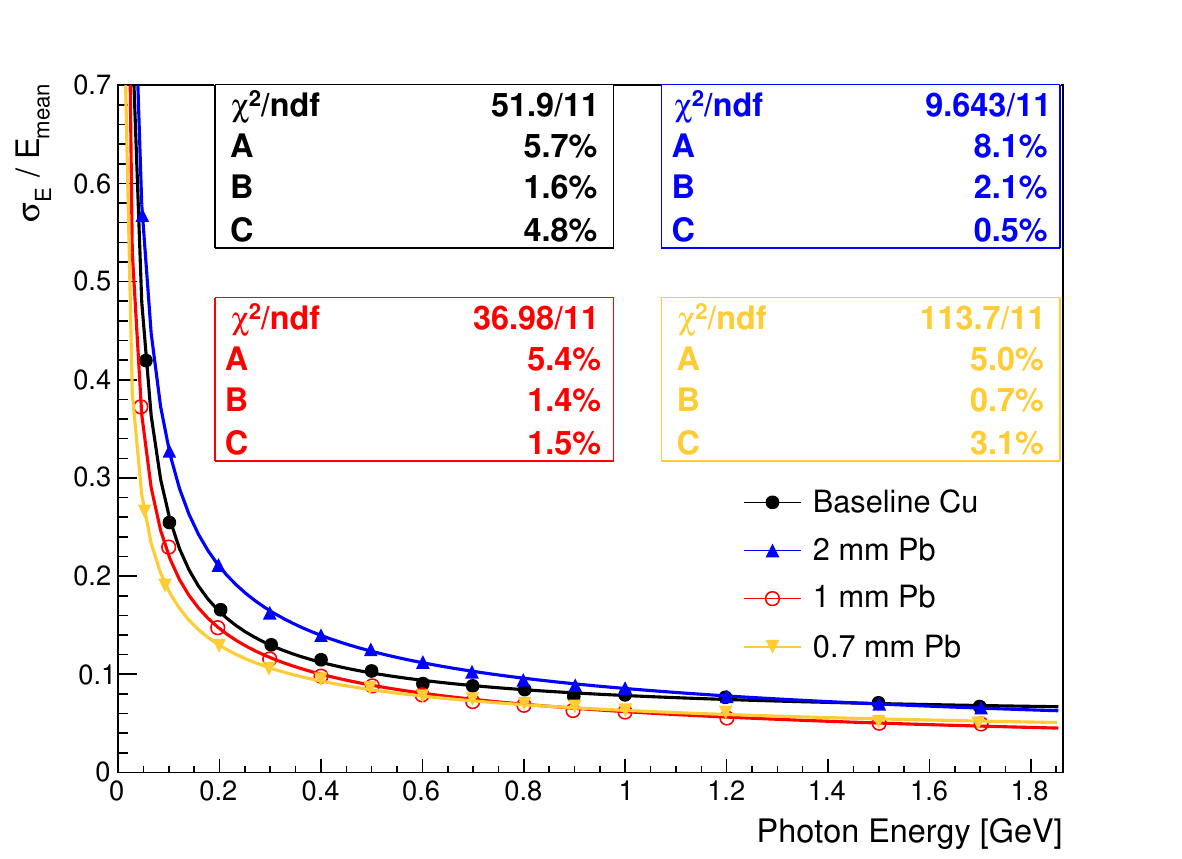} \hfill
    \includegraphics[width=0.47\textwidth]{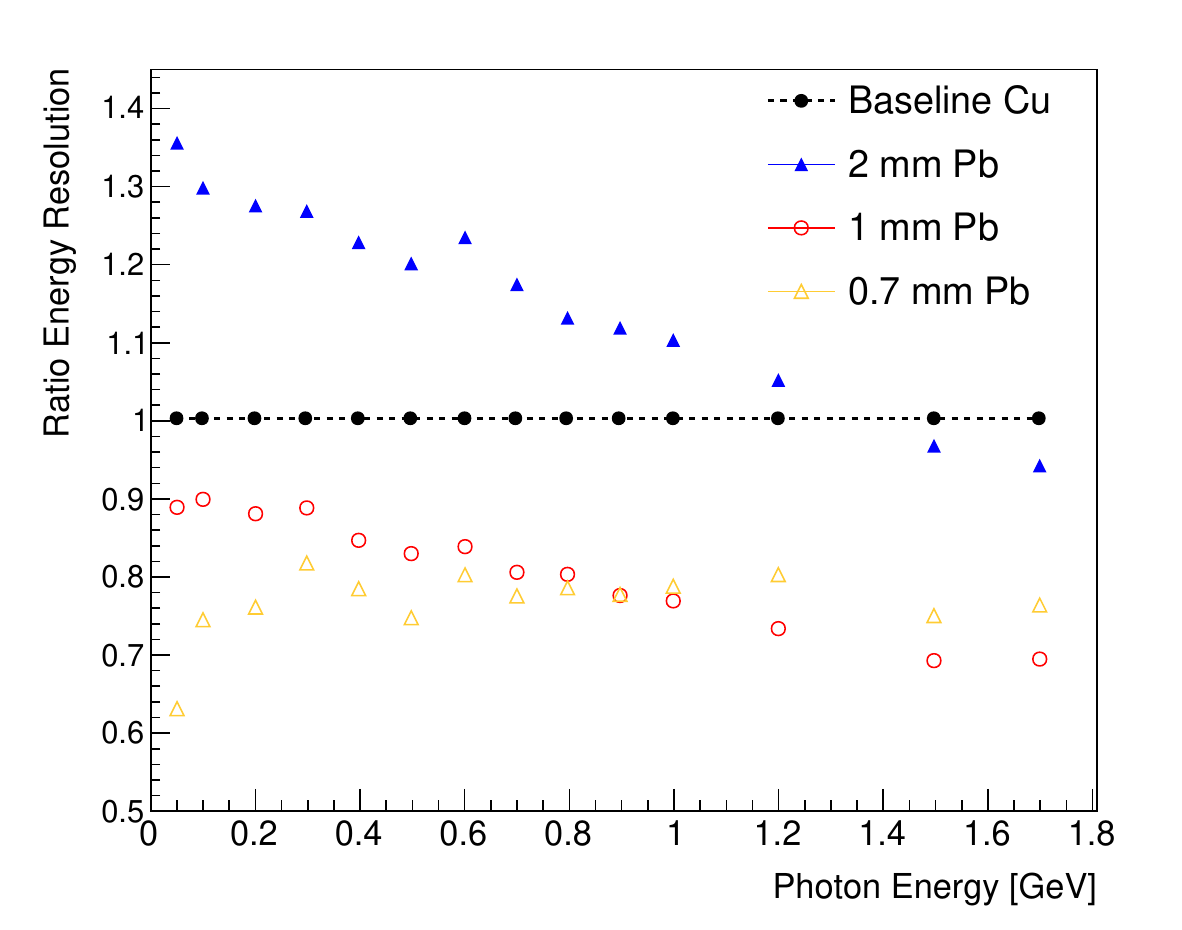}
    \caption{Left: energy resolution as a function of photon energy for different absorber configurations. The function $\frac{\sigma_E}{E}=\frac{A}{\sqrt{E}}\oplus\frac{B}{E}\oplus C$ has been fitted to the simulated data, where the $\oplus$ symbols refer to a quadrature sum of the three terms. Right: Ratio of the energy resolution for the different absorber configurations. The best resolution is achieved using thin lead absorber. The overall depth of the \dword{ecal} has been kept constant. More details can be found in \cite{DUNE:2021tad}.}
    \label{fig:ECALEres}
\end{figure}

A possible barrel calorimeter depth is about 44\,cm. The initial performance evaluation for photons, based on a preliminary design that was investigated, is summarized in Figure~\ref{fig:ECALEres}. 

For the present study, copper has been chosen as the absorber material, as initial studies have shown that this material provides a good compromise between calorimeter compactness, energy, and angular resolution. It also allows for the removal of heat generated by the electronics in the tile layer. There are several possible readout \dwords{asic} on the market to determine the time and charge of the \dword{sipm} signals, one possibility being the KLauS ASIC~\cite{Yuan:2019lub}.

\subsubsection{Magnet concept}

To achieve the physics goals, the TPC volume of the \dword{nd} must be magnetized in order to measure the momenta of muons and other particles, and to determine the sign of their charge. The magnetized system will analyze both the tracks originating from \dword{ndlar} and penetrating from upstream, and the tracks produced within the magnetic volume by neutrino interactions. 

The need to make the magnet as compact as possible, thus minimizing the material at the downstream end of the TPC and in front of the calorimeter, suggests an integrated design in which the magnet structure serves also as a pressure vessel for the TPC gas volume. The magnetic design described in~\cite{Bersani:2022gfr}, and summarized here, fulfills these requirements and is cost-effective. 

The magnet system consists of a superconducting solenoid surrounded by an iron return yoke. The superconducting solenoid cryostat serves not only as a pressure vessel body for the \dword{hpgtpc}, but also as support for it and the calorimeter elements located in its bore.   Additionally, the design of the iron magnet yoke uses the mechanical strength of the yoke’s pole faces to eliminate the large domed heads that would normally be required for a large-diameter pressure vessel. 

\begin{figure}[htbp]
    \centering
    \includegraphics[width=0.5\textwidth,angle=-90]{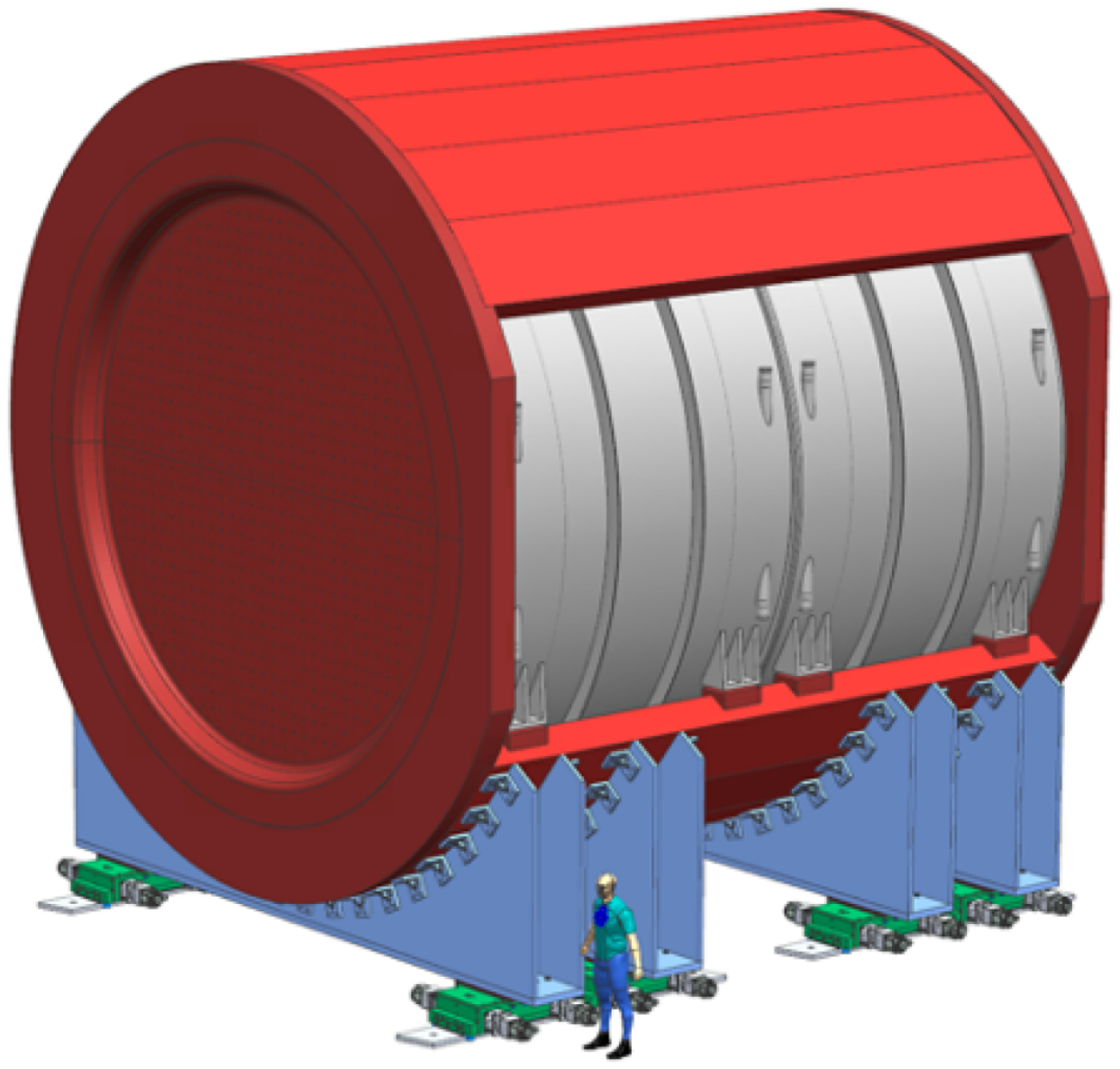}
    \caption{The \dword{spy} magnet system. The hole in the yokes is on the upstream side, to minimize material traversed by tracks originating from neutrino interactions in \dword{ndlar}.}
    \label{fig:SPY-1}
\end{figure}

Another important design requirement for \dword{ndgar} is the ability to accurately measure the momentum of muons that originate in \dword{ndlar}. This requirement limits the amount of material allowed on the upstream side of \dword{ndgar} and motivates an unconventional and asymmetrical iron yoke design. An iron yoke that eliminates a portion of the iron along the upstream face has been developed and designed, and is called \dword{spy}. Figure~\ref{fig:SPY-1} illustrates the \dword{spy} magnet system. 

The following lists the main requirements and technological features of this design in more detail:

\begin{itemize}
    \item The momentum analyzing power of the \dword{ndgar} assembly (magnet + TPC) must provide at least 3\% momentum resolution for the muons originating within \dword{ndlar}. Additionally, for particles produced as a result of neutrino interactions in the \dword{hpgtpc}, the resolution in neutrino energy reconstruction must be at least as good as that of the DUNE \dword{fd}. 
    
    \item The magnetic field uniformity, thanks to recent and relevant improvements in the capability of event reconstruction, is not required to be very high. A $\pm$10\% tolerance within the magnetic volume is expected to be sufficient, provided that a very accurate map for the magnet ``as built'' is measured. However, it is worth emphasizing that the magnet design fully described in~\cite{Bersani:2022gfr} greatly exceeds the $\pm$10\% requirement, and should offer a $\pm$2\% variation over the whole volume.  Careful studies were also done to evaluate and minimize the magnetic forces between \dword{ndgar} and \dword{sand}. 
    
    \item The superconductor will be co-extruded in high-purity aluminum to provide quench protection. The current reference solution for cable material is one based on niobium titanium. Higher temperature superconductors such as MgB$_2$ will also be considered as part of an R\&D program currently in progress. 
    
    \item The \dword{ndgar} assembly must provide good acceptance for muons exiting \dword{ndlar} and must fit within the space constraints imposed by the \dword{nd} hall design and by \dword{duneprism}. 
    
    \item The \dword{ndgar}'s magnet system must present as little material as possible in the path of the muons exiting from \dword{ndlar}. A similar requirement holds in the downstream face of the yoke, to assist in the discrimination of muons from pions.

    \item The vacuum cryostat must be capable of providing mechanical support and a cryogenic environment for the superconducting coils. The inner wall of the vacuum cryostat must be sufficiently strong to serve as the outer wall of the pressure vessel for the \dword{hpgtpc}, and to support the weight of the calorimeter.

    \item The carbon steel of the return yoke must provide a uniform 0.5\,T magnetic field over the full length of the solenoid, and limit fringe fields to the $\leq$0.01\,T levels required by the experiment and by the co-existence with \dword{sand}. It must also provide flat carbon steel pole tips for the magnet return yoke that match the magnetic field boundary conditions at the ends of the solenoid, and provide the mechanical support for the pressure vessel end flanges.
    
\end{itemize}

 The main parameters achieved in the current design~\cite{Bersani:2022gfr} are summarized in Table~\ref{table:spymagnet}.

\begin{table}[htbp]
\begin{center}
    \begin{tabular}{|p{4.cm}|p{2.1cm}|p{8.9cm}|}
    \hline
    Parameter & Requirement & Notes \\
    \hline
    Central field  & 0.5\,T  &  \\ \hline
    Field uniformity  & $\pm$10\%  &  Current design achieves $\pm$2\% \\ \hline
    Ramp time to full field  & 30\,min  &  \\\hline
    Stray field  & $\leq$0.01\,T  & Stray field in \dword{sand} negligible, in \dword{lar} \dword{fv} $\simeq$ 10\,G \\ \hline
    Bore diameter & 6.73\,m  & Reduction possible with TPC and \dword{ecal} optimizations \\\hline
    Coils diameter  & 7.85\,m  & Cryostat diameter at stiffening rings \\ \hline
    Solenoid length  & $\simeq$ 7.8\,m  &  \\\hline
    Solenoid weight  & $\simeq$ 150\,t  &  \\\hline
    Yoke total weight  & $\simeq$ 757\,t  &  \\\hline      \end{tabular}
\end{center}
\caption{List of \dword{spy} parameters according to the current reference design.}
\label{table:spymagnet}
\end{table}

\subsubsection{Muon system}

A \dword{gartpc} \dword{nd} system will need to be outside the calorimeter to improve pion/muon separation. The muon tagger would likely implement a well established technology, such as a coarsely instrumented scintillator detector. It is not likely to require substantial R\&D, but will need engineering effort.

\subsubsection{Light detection options}
\label{subsec:ndgar-light}

Enabling time-tagging by the TPC would provide an absolute determination of the vertex position and interaction time, thus simplifying the matching with external detectors and enabling reconstruction of interactions whose by-products range out before reaching them. The only demonstrated technique to accomplish this in TPCs relates to primary scintillation, which due to the complexity of the system would require significant R\&D, primarily to study the level of localization needed in order to associate time information with a given interaction. The choice of gas mixture for the detector will also be key, as different mixtures have very different scintillation properties \cite{Saito:2002ch,Santorelli:2020fxn}.

Although pure argon gas emits scintillation light copiously at a level of 20000 photons/MeV, gases employed historically in TPCs for accurate tracking in magnetic fields do not \cite{Gonzalez-Diaz:2017gxo}. The recent demonstration of strong (and fast) wavelength shifting in the Ar/CF$_4$ system, with yields in the range 700--1400 photons/MeV \cite{IGFAE_yields, Amedo:2023far}, opens up the possibility of ns-level time tagging for energy deposits down to at least 5\,MeV~\cite{Saa-Hernandez:2024tvl}. A mere 1\% CF$_4$ addition (per volume) seems sufficient to achieve this performance while keeping the electron diffusion at 3.6\,mm for a 5\,m drift (compared to 20\,mm for pure argon) for a 200\,kV cathode bias. These values are even below those expected for a conventional Ar/CH$_4$ (90/10) mixture.

In view of the requirements for single-photon detection and magnetic field compatibility, and given the spectral range of the scintillation, two technologies are of particular interest, as described below.  A third, light readout of secondary scintillation at the amplification stage, is also an option that could be explored, based on an Ar/CH$_4$ (99/1) mixture (Section~\ref{tpc_charge}). 

\paragraph{\dwords{sipm}}

\dwords{sipm} are well suited for detection in the visible range, and several ganging schemes are currently available for large area coverage (e.g.,~\cite{DIncecco:2017bau}). Silicon suffers from high dark rate at room temperature and, in fact, simulations point to the need for cryogenic operation (-25\,C) to reach MeV-thresholds in \dword{ndgar}. Methods to do this are under study. A comprehensive R\&D program has been laid out and is led by Spain. A conceptual description of the ganging scheme and active-cryostat concept proposed, along with proof-of-principle demonstrations for both, can be found in~\cite{Saa-Hernandez:2024tvl}.

\paragraph{\dwords{lappd}}

Large-Area Picosecond Photo-Detectors (\dwords{lappd}) are novel photosensors based on microchannel plate technology. With sensitive regions of order 20$\times$20\,cm and sub-cm position resolution, this type of detector is a good candidate for covering large areas. \dwords{lappd} are tolerant of magnetic fields and handle sub-ns timing with an excellent signal-to-noise ratio, and without any cooling requirements. They were recently demonstrated to work for neutrino detection by the ANNIE experiment. Ideal coverage could be achieved with about 100 \dwords{lappd}, which could be delivered within a few years at current production rates. Depending on the choice of quencher/wavelength shifter, modifying the photocathode composition or a wavelength shifter coating might be required. 

Studies to optimize this system are required, and include determining optimal coverage, the best photocathode design, and enclosures to allow the \dwords{lappd} to operate at high pressure.

\subsubsection{R\&D and engineering road map}

R\&D will be necessary for this \phasetwo improved tracker concept, starting in 2024 and lasting for several years. It will be important to fully define the detector requirements, then aim for a technical design report in the late 2020s, and be ready to begin construction in the early 2030s. The essential R\&D and design work needed for \dword{ndgar} includes, but is not limited to, the following items:

\paragraph{\dword{ndgar} magnet} INFN Genova is pursuing R\&D on using magnesium diboride (MgB$_{2}$) superconducting cables, which have a higher critical temperature and do not face some of the challenges of co-extruding NbTi superconducting cables with high-purity %pure 
aluminum.

\paragraph{\dword{ndgar} TPC charge readout and electronics test stands} Several R\&D efforts are already underway, as described in Section~\ref{tpc_charge}. Examples include the \dword{goat}, \dword{toad}, \dword{gorg}, and \dword{gat0} test stands. Charge readout TPCs 
are a mature technology, with gas mixtures identified that give sufficient gain. The current R\&D priority is to test the full readout chain, from amplification technology to readout electronics, in a high-pressure test stand, using a non-flammable gas with a high argon fraction, to ensure adequate stability and gain. Electron diffusion measurements will also be performed for the same gas mixtures. 

Concerning the amplification stage, current testing has been done with wire chambers. However, modern TPCs such as those for ALICE and sPHENIX use \dwords{gem} to achieve better stability of operation and higher gains. For this reason, R\&D for a \dword{gem}/\dword{thgem}-based amplification stage has started in the context of the DUNE \phasetwo \dword{nd}, as well. The stability of proposed wavelength-shifting gases such as Ar/CF$_4$ (99/1) must be studied, as they are low-quenched. Conventional GEM, \dword{thgem}, glass-GEMs, glass-Micromegas and wire chamber amplification stages are all currently under evaluation in Spain, the UK, and the US. 

Once detailed requirements on tracking performance are established, further R\&D on readout electronics and on the segmentation of the charge readout pads/strips will be pursued accordingly. The TPC charge readout R\&D work is currently ongoing in the UK (GEM work and readout electronics), Spain (glass-GEM, SiPMs $+$ TPX3 cameras), and the US (readout electronics and test stands) using sources and test beams. 

\paragraph{Light detection in \dword{ndgar} TPC} 

The realization of light readout at the scale of the \dword{ndgar} TPC poses important engineering challenges in relation to photosensor technology, \dword{hv} integration, and good light collection. Dedicated physics studies are needed to establish the best design path towards the optimization of the detection thresholds, time-tagging performance, and photosensor coverage. Both light readout options discussed above have R\&D needs, for example cooling control for \dwords{sipm} and operation at high pressure for \dwords{lappd}. Groups in Spain are performing R\&D on the optimization of the \dword{sipm}-based optical readout concept: required coverage, use of reflectors and light collectors, \dword{sipm} channel ganging and cooling schemes. R\&D on \dwords{lappd} is also underway in the US.

Also, the outstanding tracking performance of \dword{ndgar} needs to be guaranteed while ensuring that the photosensor plane not be blinded during the avalanche multiplication process in the anode region. This will require an additional R\&D step targeting the minimization of photon-feedback, as it is customary for instance in ring-imaging Cherenkov detector applications \cite{Blatnik:2015bka}.

\paragraph{\dword{ndgar} calorimeter} R\&D was underway in Germany on coupling fibers to \dwords{sipm} to maximize light collection and uniformly illuminate the \dword{sipm} face.  Studies must also be done to optimize the calorimeter design, including the number of strip and tile layers, and their granularity. A cost-effective readout electronics system must also be developed.  

\paragraph{\dword{ndgar} calibration systems, field cage, and gas systems} 
Engineering work is also required to design the infrastructure and support services for the \dword{ndgar} detector. The ALICE detector featured a high-performance TPC of similar size~\cite{Alme:2010ke}, therefore that design can potentially be used as a starting point. For example, a laser calibration system that can uniformly illuminate the drift volume could provide the required accurate monitoring of drift velocity variations and inhomogeneities within the volume. 
Such a system must be designed in close connection with the \dword{hv} \dword{fc}. The movable \dword{ndgar} will require design of a mechanically robust \dword{fc} with mechanical end-cap structures. A buffer region in between the \dword{fc} and pressure vessel will be needed to degrade the high voltage, and this may require the use of an additional insulating gas. 

The detector performance depends crucially on the stability and quality of the gas in the drift region, therefore it will be necessary to develop a system to control and monitor the gas mixture in the drift volume. Control operations include pressurization, recirculation, purification, and evacuation of the gas. The current design of the magnet system does not incorporate a method for evacuation~\cite{spy}. However, modifying it to function under both pressure and vacuum conditions is well understood. Generally, achieving vacuum is desirable to facilitate the reduction and monitoring of O$_{2}$ and H$_{2}$O impurities (as well as other unforeseen contaminants). It is also worth considering purifying argon gas in the gas handling system -- which might yield similar results -- although evacuation could potentially be faster.

%%%%%%%%%%%%%

\subsection{Improvements to \phaseone near detector components}
\label{subsec:nd_phaseonend_improvements}

As part of the \phasetwo program, possible enhancements and improvements to the 
exisiting \phaseone components of the \dword{nd} are being considered. 
This section discusses such possible improvements to the \phaseone detector components \dword{ndlar} and \dword{sand}.

\subsubsection{\phasetwo \dshort{ndlar} detector}

\dword{ndlar} is the \dword{lar} component of the DUNE \dword{nd} complex. With the intense neutrino flux and high event rate at the \dword{nd}, traditional, monolithic, projective wire readout \dwords{lartpc} would be stretched beyond their performance limits. To overcome this hurdle, \dword{ndlar} will be fabricated out of a matrix of smaller, optically isolated TPCs, read out individually via a pixelated readout. The subdivision of the volume into many smaller TPCs allows for shorter drift distances and times. This and the optical isolation lead to fewer problems with overlapping interactions.

The \dword{ndlar} design consists of 35 optically separated \dword{lartpc} modules, which allows for independent identification of $\nu$-Ar interactions in an intense beam environment using optical timing. Each TPC consists of a \dword{hv} cathode, a low-profile \dword{fc} that minimizes the amount of inactive material between modules, a light collection system, and a pixel-based charge readout.

One key aspect of \dword{ndlar} operation is the ability to cope with many neutrino interactions in each spill. The \dword{lbnf} neutrino beam consists of a 10\,$\mu$s wide spill, which leads to $\mathcal{O}$(50) $\nu$ interactions per spill in \phaseone and $\mathcal{O}$(100) in \phasetwo. Given the relatively low expected cosmic ray rate while the beam is on (estimated to be  $\sim$0.3/spill at 60\,m depth), this beam-related pile-up is the primary challenge confronting the reconstruction of the \dword{ndlar} events. The \threed pixel charge signal will be read out continuously. The slow drifting electrons (with charge from the cathode taking $\sim$300\,$\mu$s to travel the 50\,cm drift distance) will be read out with an arrival time accuracy of 200\,ns and a corresponding charge amplitude within a $\sim$2\,$\mu$s-wide bin. This coupled with the beam spill width gives a position accuracy of 16\,mm. While this is already good spatial positioning, the \dword{ndlar} light system will provide an even more accurate time tag of the charge as well as the ability to tag subclusters and spatially disassociated charge depositions resulting from neutral particles, such as neutrons, that come from the neutrino interaction. Thus, the \dword{ndlar} light system has a different role from that in the \dword{fd}, as it must time-tag charge signal subclusters to enable accurate association of all charge to the proper neutrino event, and to reject pile-up of charge from other neutrino signals.

The current \dword{ndlar} design being implemented for \phaseone satisfies the general requirements of DUNE for \phasetwo in terms of increased beam power and lifetime of detector components. Nonetheless, additional potential modifications to \dword{ndlar} that might enhance its capabilities are under consideration. Given that the \dword{ndlar} uptime during DUNE operations is an important factor to take into consideration, those modifications can be divided into two categories: \dword{ndlar} upgrades that imply modifications to the inner detector hardware and thus require emptying the \dword{lar}, and those that do not. In the former, more disruptive, category, current ideas under exploration include: improvements to neutron detection methods by upgrading optical detectors with $^6$Li-glass scintillator, replacement of charge tiles of a module with smaller pixels and lower threshold, use of photosensitive dopants, and use of radiopure underground argon. In the latter (less disruptive) category, possible upgrade options span the following: doping of argon with xenon, upgrade of the off-detector electronics, addition of a rock muon tracker in front of \dword{ndlar}, and use of an additional calibration system based on $^{222}$Rn injection. A decision on these possible \dword{ndlar} upgrade paths will come after the \phaseone \dword{ndlar} detector is commissioned. 

%%%%%%%%%%%%% 

\subsubsection{\phasetwo SAND detector} 

\dword{sand} is a multipurpose detector composed of a superconducting solenoid, a high-performance \dword{ecal}, a light tracker, and an active \dword{lar} target called \dword{grain}. The magnet and the \dword{ecal} were part of the \dword{kloe} detector at INFN Frascati and will be refurbished for \phaseone, without the need for upgrades during \phasetwo. The tracker, based on straw tubes, will be a completely new detector capable of reconstructing charged particle tracks in the magnetic field. Major upgrades for the tracker are not foreseen for \phasetwo.  

\dword{grain} is an innovative \dword{lar} detector that will employ a completely new readout technique, using only scintillation light for track reconstruction. This task is accomplished by cameras with light sensors made of a matrix of \dwords{sipm} and optical elements, such as special lenses or Coded Aperture Masks. The \dword{grain} project is very challenging because, due to the low efficiency of light sensors to \dword{vuv} scintillation light, the number of photons detected and used by the reconstruction algorithms is low.

For \phasetwo \dword{grain}, the goal is to enhance light collection by improving the \dword{sipm} \dword{pde} in the \dword{vuv} range. For this purpose, we are developing with ``Fondazione Bruno Kessler" (FBK-Trento) Backside Illuminated \dwords{sipm} (BSI \dwords{sipm}). In this architecture, the light entrance window is on the back of the silicon, while all the metallic contacts are on the front side. This will allow us to improve the fill factor and optimize the anti-reflective coating on the entrance window. It is planned to substitute all the \dword{grain} matrices of traditional Front Side \dwords{sipm} with the BSI ones for \phasetwo, if they will be available and mature in time.

%%%%%%%%%%%%%

\subsection{Near-detector options for non-argon far detector modules}
\label{subsec:theiand}

In the event that one of the \phasetwo \dword{fd} modules consists of a neutrino target material that is not argon-based, such as the \dword{theia} detector concept described in Section~\ref{subsec:fd_theia}, the \phasetwo \dword{nd} complex will need to provide measurements of neutrino interactions on those same target nuclei. 
Several options are under consideration for modifying the \phaseone suite of ND sub-detectors to make such measurements, including modifying the \phaseone \dword{sand} to incorporate oxygen and water targets, embedding %\dword{wbls} 
liquid scintillator targets within the \dword{ecal} of the \dword{gartpc}, and constructing a new, dedicated, water-based near detector. While they introduce identical or similar nuclear targets, these particular options do not establish a functionally similar detector at the near site that would also mitigate detector-related uncertainties at the far detector, analogous to \dword{ndlar} for the argon-based \dword{fd} modules.

\subsubsection{Oxygen and water targets in SAND}

The \dword{sand} detector is equipped with a modular \dword{stt} with target layers that are designed to be individually replaceable with different materials. A total of 78 thin planes, each about $1.6\%$ of a radiation length $X_0$, of various passive materials are alternated and dispersed throughout active layers, which are made of four straw planes, to guarantee the same acceptance to final state particles produced in (anti)neutrino interactions. The \dword{stt} allows minimizing the thickness of individual active layers and to approximate the ideal case of a pure target detector -- the targets constitute about 97\% of the mass -- while keeping the total thickness of the stack comparable to one radiation length and an average density of about $0.17$\,g/cm$^3$. The lightness of the tracking straws and the chemical purity of the targets, together with the physical spacing among the individual target planes, make the vertex resolution ($\ll 1$\,mm) less critical in associating the interactions to the correct target material. The average momentum resolution expected for muons is $\delta p /p \sim 3.5\%$ and the average angular resolution better than 2\,mrad. The momentum scale can be calibrated to about 0.2\% using reconstructed $K_ 0 \to \pi^+\pi^-$ decays. 

\begin{figure}[htbp]
    \centering
    \includegraphics[width=0.95\textwidth]{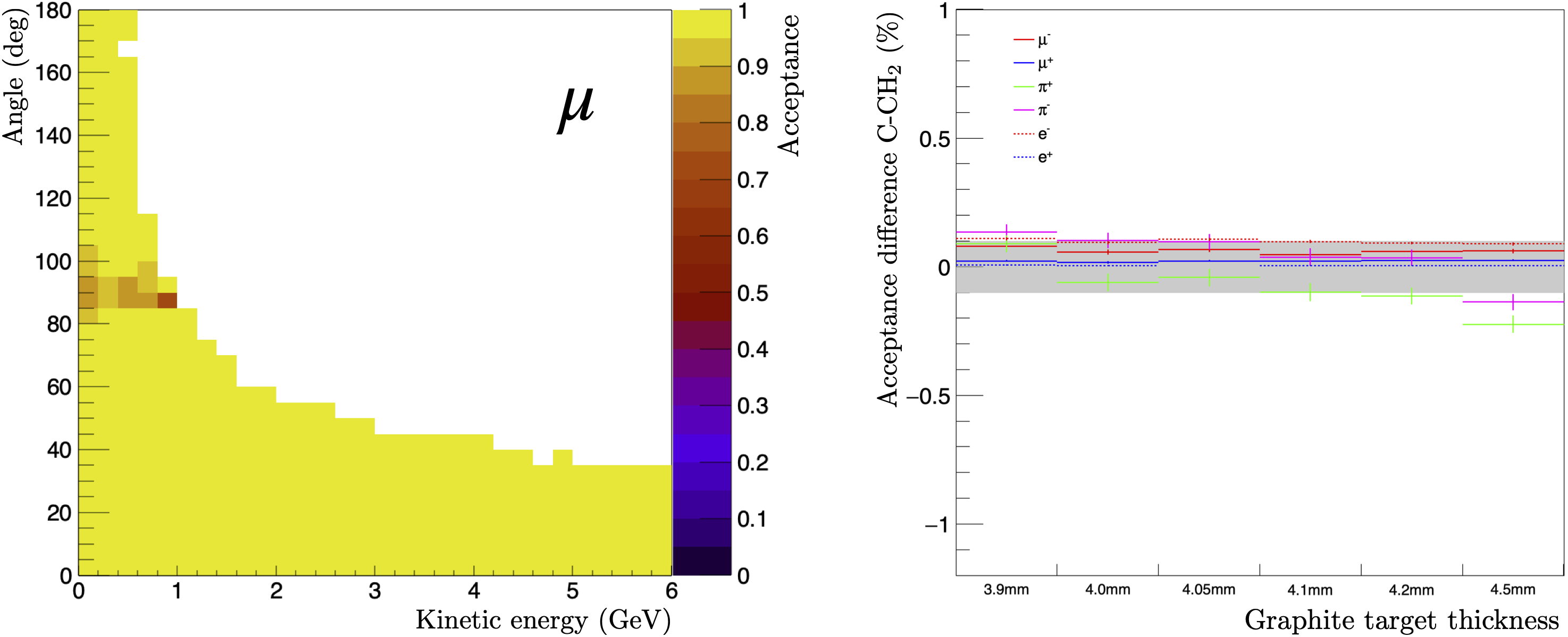}
    \caption{Left: Muon acceptance for $\nu_\mu$ CC interactions in a forward horn current (\dword{fhc}) beam in \dword{sand}. Right: Discrepancy in acceptance between the CH$_2$ and C targets in \dword{sand}.}
    \label{fig:SAND-TargetAcceptance}
\end{figure}

The \dword{stt} is optimized for the “solid” hydrogen technique, in which $\nu(\bar \nu)$ interactions on free protons are obtained by subtracting measurements on 
dedicated graphite (C) targets from those on polypropylene (CH$_2$) targets~\cite{Petti:2022bzt,Petti:2019asx,Duyang:2019prb}. The default target configuration in \phaseone includes 70 CH$_2$ targets and eight C targets. The use of a distributed target mass within a low-density tracker results in an approximately uniform acceptance over the full $4\pi$ angle, as shown in Figure~\ref{fig:SAND-TargetAcceptance}. 
The acceptance disparity between different targets can be kept within $10^{-3}$ for all particles (Figure~\ref{fig:SAND-TargetAcceptance}) due to their thinness and their alternation throughout the detector volume. The subtraction procedure between different materials can then be considered model-independent within these uncertainties. Furthermore, the detector acceptance effectively cancels in comparisons between the selected interactions on different target nuclei.

The ND can operate with both oxygen and water targets concurrently by replacing some of the initial CH$_2$ targets with polyoxymethylene (CH$_2$O, acetal) planes with equivalent thickness, i.e., in terms of radiation length and nuclear interaction length $\lambda_I$. Interactions on oxygen are obtained from a subtraction between CH$_2$O and CH$_2$ targets, while interactions on water are obtained from a subtraction between CH$_2$O and C targets~\cite{Petti:2023osk}. 

To this end, 4.5\,mm thick acetal slabs can be used, corresponding to about 0.016\,$X_0$ and 0.008\,$\lambda_I$. The oxygen content by mass within acetal dominates at $53.3\%$. By replacing only 20 polypropylene targets (out of 70) with the equivalent CH$_2$O targets, we obtain an oxygen target mass of about $760$\,kg and a water target mass of about 850\,kg. Assuming an exposure of $3\times 10^{21}$\,\dword{pot}, corresponding to about two years with the \phaseone beam intensity and to about one year with the \phasetwo beam, we expect to collect $3\times 10^6$ $\nu_\mu$ \dword{cc} events with the \dword{fhc} beam and $1 \times 10^6$ $\bar \nu_\mu$ CC events with the \dword{rhc} beam on oxygen. The subtraction procedure introduces an increase of about $40\%$ in the statistical uncertainties with respect to the use of ideal targets. For $3\times 10^{21}$\,\dword{pot}, the resulting statistical bin-to-bin uncertainties in the $\nu$-nucleus cross-sections as a function of neutrino energy are comparable to the expected systematic uncertainties introduced by the \dword{stt} momentum scale uncertainty of $0.2\%$~\cite{Petti:2023osk}.

\subsubsection{Liquid scintillator targets in the ND-GAr calorimeter}

\begin{figure}[htbp]
    \centering
    \includegraphics[width=0.5\textwidth]{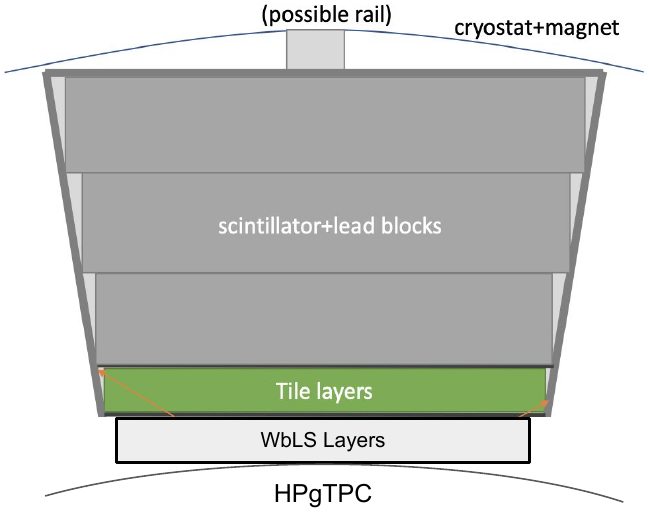}
    \caption{An upstream segment of the \dword{gartpc} \dword{ecal}, with the beam is pointing downward. The \dword{theia} layers constitute the most downstream portion of the \dword{ecal}, and particles produced in these layers via neutrino interactions are tracked in the downstream \dword{hpgtpc}.} 
    \label{fig:ndgarwbls}
\end{figure}

The \dword{gartpc} described in Section~\ref{subsec:ndgar} is capable of supporting active \dword{theia}-type targets within the downstream portion of the upstream \dword{ecal}, as shown in Figure~\ref{fig:ndgarwbls}. The \dword{theia} layers consist of X and Y bars (similar to the NOvA configuration discussed in the next section), and interactions in %the \dword{wbls} 
these layers produce particles that enter the high-pressure gas TPC where they are precisely tracked. Neutral particles are measured by the surrounding \dword{ecal}, and the active \dword{theia} layers provide an additional measure of low-energy particles near the interaction vertex (``vertex activity''). For neutrino interactions within the \dword{gartpc}, the \dword{theia} layers will form an initial low-density section of the \dword{ecal} that can provide fast timing for particles exiting the TPC.

There is sufficient space within the \dword{ecal} to include \dword{theia} layers with a thickness of at least 10\,cm, which would provide more than a ton of target mass. This would produce $\mathcal{O}$(1M) charged current $\nu_\mu$ interactions in a 14 week run on-axis, and $\mathcal{O}$(10k) charged current $\nu_\mu$ interactions in a two week run at the furthest off-axis position, both of which would be expected to occur within a nominal DUNE yearly run.

\subsubsection{Water-based near detector}

It is possible to install a detector specifically designed to make measurements for a water-based \dword{fd} module in the DUNE \dword{nd} hall. If a new \dword{gartpc} is built, it will serve as the downstream spectrometer for \dword{ndlar}, allowing \dword{tms} to be used as a downstream spectrometer for a dedicated \dword{wbnd}. Any configuration of the \dword{nd} suite will be subject to the space limitations imposed by the near site infrastructure, as completed before the beginning of beam operations. Two possible options for a \dword{wbnd} are a NOvA-style \dword{nd}, or a LiquidO \dword{nd}, both discussed below.

\begin{figure}[htbp]
    \centering
    \includegraphics[width=0.9\textwidth]{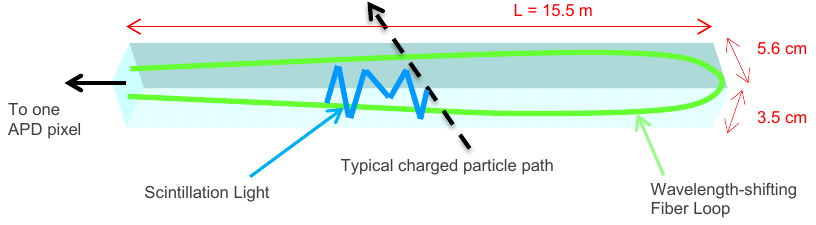}
    \caption{The fundamental unit cell of the detector of the NOvA near detector; the cells are arranged in horizontal and vertical layers.}
    \label{fig:novand}
\end{figure}

\paragraph{NOvA-style near detector}
The NOvA \dword{nd} consists of individual cells, as shown in Figure~\ref{fig:novand}, arranged in horizontal and vertical layers. The cells consist of PVC extrusions filled with liquid scintillator, and a wavelength-shifting fiber collects the light and guides it to the avalanche photodiode (APD) for readout~\cite{Talaga:2016rlq}.

The NOvA near detector design could be used to construct a \dword{wbnd} by replacing the NOvA scintillator with \dword{theia} \dword{wbls}. Its cell size and scintillator fraction %of the \dword{wbls} 
would have to be tuned to ensure a high muon reconstruction efficiency. This type of detector would also be capable of a calorimetric measurement of the hadronic energy in the neutrino final state, including better sensitivity to neutrons than a \dword{lar} detector, due to the presence of free hydrogen in the target material. This detector technology is well established and would require minimal additional R\&D.

\begin{figure}[htbp]
    \centering
    \includegraphics[width=0.8\textwidth]{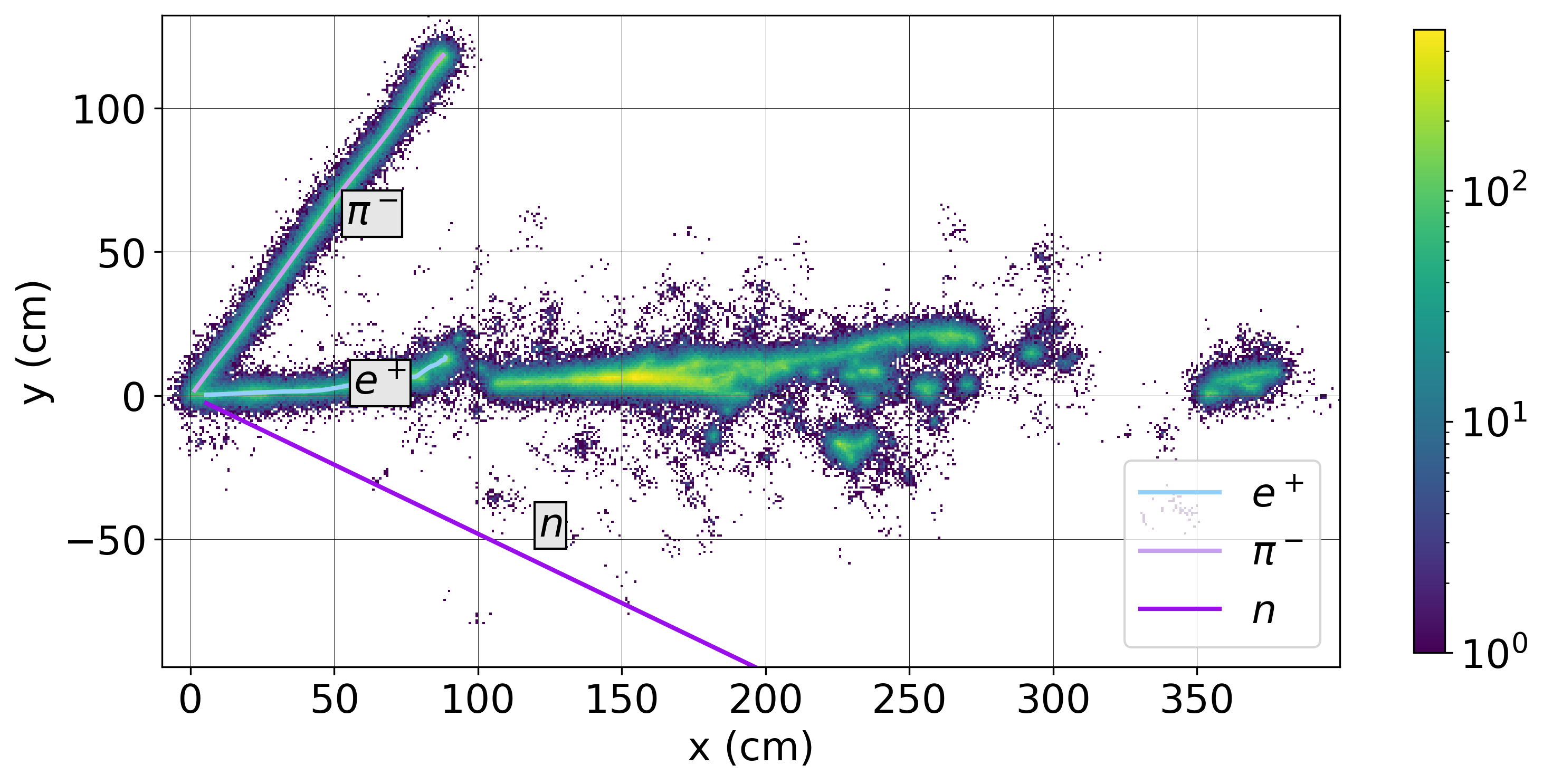}
   \caption{Illustration of a simulated 2\,GeV electron neutrino interaction in a LiquidO-style \dword{nd} with a 1~cm fiber pitch. The image shows sub-cm spatial resolution and excellent particle ID can be achieved.}
    \label{fig:LiquidOEvent}
\end{figure}

\paragraph{LiquidO near detector}
A promising new detector concept for the DUNE \dword{nd} is based on using opaque scintillators with millimeter-scale scattering length to produce high-resolution images of neutrino interactions~\cite{LiquidO:2019mxd,LiqO-appetizer}. 
The scintillation photons are stochastically confined close to the point of production via scattering, and a lattice of wavelength-shifting fibers at $\simeq$1~cm pitch is used to extract the light. This technology, called LiquidO, removes the need for manual segmentation: the lattice of fibers is constructed first, and then the opaque scintillator poured in around the fibers. Substantially better spatial resolution per readout channel is achieved by using the profile of the light detected across multiple fibers. Figure~\ref{fig:LiquidOEvent} shows a \dword{cc} muon neutrino event as imaged with a LiquidO technology \dword{nd}. Furthermore, and importantly for a potential \dword{theia} DUNE \phasetwo \dword{fd} module, the scintillator isotopic composition can be varied by exchanging the scintillator material, e.g., oil-based scintillators can be swapped with water-based ones. A design analogous to the \dword{t2k} Super-FGD detector is envisaged for DUNE with the fibers running in all three perpendicular directions, allowing fine-grained precision tracking and excellent calorimetry. The hydrogen-rich nature of organic or water-based scintillators, together with their fast timing, is advantageous for neutron time-of-flight measurements and for particle detection in high-rate environments. 

%%%%%%%%%%%%%

\section*{Acknowledgements}
% Standard piece for all FNAL-based experiments
%
This document was prepared by the DUNE collaboration using the
resources of the Fermi National Accelerator Laboratory 
(Fermilab), a U.S. Department of Energy, Office of Science, 
HEP User Facility. Fermilab is managed by Fermi Research Alliance, 
LLC (FRA), acting under Contract No. DE-AC02-07CH11359.
%
% Funding agencies, alphabetical by country, then alphabetical by agency name
%
This work was supported by
CNPq,
FAPERJ,
FAPEG and 
FAPESP,                         Brazil;
CFI, 
IPP and 
NSERC,                          Canada;
CERN;
M\v{S}MT,                       Czech Republic;
ERDF, 
H2020-EU and 
MSCA,                           European Union;
CNRS/IN2P3 and
CEA,                            France;
INFN,                           Italy;
FCT,                            Portugal;
NRF,                            South Korea;
CAM, 
Fundaci\'{o}n ``La Caixa'',
Junta de Andaluc\'ia-FEDER,
MICINN, and
Xunta de Galicia,               Spain;
SERI and 
SNSF,                           Switzerland;
T\"UB\.ITAK,                    Turkey;
The Royal Society and 
UKRI/STFC,                      United Kingdom;
DOE and 
NSF,                            United States of America.
%
% Acknowledgement of NERSC if those resources were used
%
%This research used resources of the 
%National Energy Research Scientific Computing Center (NERSC), 
%a U.S. Department of Energy Office of Science User Facility 
%operated under Contract No. DE-AC02-05CH11231.
%
\noindent Fermilab Report Number: FERMILAB-TM-2833-LBNF

% this is added just after last chapter

\cleardoublepage
\printglossaries

\cleardoublepage
\cleardoublepage
\bibliographystyle{utphys} 
% To understand the style chosen, see:
% https://arxiv.org/hypertex/bibstyles/ (very bottom -- additions) and https://www.sharelatex.com/learn/Bibtex_bibliography_styles
\bibliography{common/references} % references.bib

\end{document}